\documentclass[onecolumn,amsmath,amssymb,12pt,superscriptaddress,nofootinbib]{revtex4}
\pdfoutput=1
\usepackage[latin1]{inputenc}
\usepackage[english]{babel}
\usepackage{amssymb}
\usepackage{amsmath,bm}
\usepackage{amsthm}
\usepackage[]{graphicx}
\usepackage{subcaption}
\usepackage{tensor}
\usepackage{color}
\usepackage{cancel}
\usepackage{setspace}
\usepackage{fancyhdr}
\usepackage[bookmarks,linktocpage, colorlinks=true, plainpages = false, citecolor = blue,  linkcolor=blue, urlcolor = blue, filecolor = blue]{hyperref} 
\usepackage{framed}
\usepackage{graphicx}
\usepackage{tikz}

\usepackage{newfloat,algcompatible}
\usepackage[size=small]{caption}
\usepackage{etoolbox}

\AtBeginEnvironment{algorithm}{\noindent\hrulefill\par\nobreak\vskip-5pt}
\usepackage{newfloat}

\DeclareFloatingEnvironment[
    fileext=loa,
    listname=List of Algorithms,
    name=ALGORITHM,
    placement=tbhp,
]{algorithm}
\DeclareCaptionFormat{algorithms}{\vskip-15pt\hrulefill\par#1#2#3\vskip-6pt\hrulefill}
\begin{document}

\allowdisplaybreaks
\begin{titlepage}

\title{Oscillatory path integrals for radio astronomy}

\author{Job Feldbrugge}
\email{jfeldbrugge@perimeterinstitute.ca}
\affiliation{Perimeter Institute, 31 Caroline St N, Ontario, N2L 2Y5, Canada}
\author{Ue-Li Pen}
\email{pen@cita.utoronto.ca}
\affiliation{Canadian Institute for Theoretical Astrophysics, University of Toronto, M5S 3H8, ON, Canada}
\author{Neil Turok$^1$}
\email{nturok@perimeterinstitute.ca}

\begin{abstract}
 \noindent
We introduce a new method for evaluating the oscillatory integrals which describe natural interference patterns. As an illustrative example of contemporary interest, we consider astrophysical plasma lensing of coherent sources like pulsars and fast radio bursts in radioastronomy. Plasma lenses are known to occur near the source, in the interstellar medium, as well as in the solar wind and the earth's ionosphere. Such lensing is strongest at long wavelengths hence it is generally important to go beyond geometric optics and into the full wave optics regime. Our computational method is a spinoff of new techniques two of us, and our collaborators, have developed for defining and performing Lorentzian path integrals. Cauchy's theorem allows one to transform a computationally fragile and expensive, highly oscillatory integral into an exactly equivalent sum of absolutely and rapidly convergent integrals which can be evaluated in polynomial time. We require only that it is possible to analytically continue the lensing phase, expressed in the integrated coordinates, into the complex domain. We give a first-principles derivation of the Fresnel-Kirchhoff integral, starting from Feynman's path integral for a massless particle in a refractive medium. We then demonstrate the effectiveness of our method by computing the interference patterns of  Thom's caustic catastrophes, both in their ``normal forms" and within a variety of more realistic, local lens models, over all wavelengths. Our numerical method, implemented in a freely downloadable code, provides a fast, accurate tool for modeling interference patterns in radioastronomy and other fields of physics. 
\end{abstract}
\maketitle
\end{titlepage}

\tableofcontents


\section{Introduction}
Interference is one of the most universal phenomena in nature. In classical physics, the linear superposition of sound waves, surface waves, radio waves, light or gravitational waves all exhibit the same characteristic patterns of constructive and destructive interference. Interference is also fundamental to quantum physics. The basic quantum amplitudes describing particles or fields are most elegantly formulated as path integrals --  sums over trajectories weighted by the phase factor $e^{i {\cal S}/\hbar}$, with ${\cal S}$ the action and $\hbar$ Planck's constant. As ubiquitous as interference and interference patterns are, they are generally hard to compute. The oscillatory integrals involved are only conditionally and not absolutely convergent, meaning they converge slowly and artefacts such as dependence on unphysical cutoffs may be hard to avoid. Likewise, if the integrals are performed iteratively, as is often the only practicable method, conditional convergence is in general insufficient to guarantee uniqueness, since the order in which partial integrals are taken can affect the result. 

In quantum mechanics, these difficulties run deep. In fact, so far they have thwarted all efforts to rigorously define nontrivial real-time Feynman path integrals, even in non-relativistic quantum mechanics~\cite{Feynman:1965}. The only available existence proofs involve a Wick rotation from real, Lorentzian time to imaginary, Euclidean time, which maps the phase factor to a real Boltzmann weight (for a recent review see, {\it e.g.}, \cite{Klauder:2010}). Unfortunately, securing mathematical rigour this way comes at a high price: the system's dynamics can only be described in imaginary time instead of real time where experiments and observations actually take place. Analytic continuation back to real time is often only possible for certain quantities, such as perturbative S-matrix elements and, even then, is often hard. Furthermore, for some theories, including general relativity and quantum condensed matter models with a ``sign problem," {\it e.g.} the Hubbard model, the Wick rotation trick does not work. 

This paper represents a step towards a new, broadly applicable method for defining and computing Lorentzian path integrals. Here, we study the  interference of relativistic waves, emitted from coherent sources and propagating through a region in which the refractive index varies in space, {\it i.e.}, a lens. As we shall show, the quantum mechanical path integral amplitude reduces, in this case, to an ordinary, finite dimensional integral. 

The study of optical interference patterns dates back over two centuries, long predating Maxwell's equations, but remains of enduring interest. Starting in the 1970's, Berry, Nye and collaborators studied examples of ``diffraction catastrophes'' -- the characteristic patterns created by diffraction about each of Thom's stable caustic catastrophes, and compared intricate mathematical calculations with  beautiful experiments~\cite{1977Natur.267...34B, Berry:1977, 1979RSPTA.291..453B, Berry:1980, Berry:2007}. Recently, the need to accurately and efficiently compute similar patterns has arisen in radioastronomy where bright, coherent sources of radio waves like pulsars and fast radio bursts are being detected in rapidly growing numbers~\cite{Spitler:2014, Amiri:2019qbv, Josephy:2019ahz}. These objects are beacons lighting up the universe. They will potentially provide a vast new source of information for astrophysics and cosmology. Typically, they are lensed by diffuse astrophysical plasmas intervening along the line of sight. Since plasma lensing is strongest at long wavelengths, this lensing must be modeled in the full, wave optics regime \cite{Melrose:2006, Cordes:2017, Main:2018}. Although challenging, such modeling will likely be vital to our ability to draw precise inferences from these sources~\cite{Pen:2012}. 

Motivated by this contemporary need, we shall use astrophysical plasma lensing as our main example. However, as should be clear to the reader, the principles involved are far more broadly relevant. The interference patterns created by astrophysical plasma lenses and observed over astronomical or even cosmological distances are governed by exactly the same physics at play in Young's double slit experiment or X-ray crystallography. This is both a striking example of universality in physics and a reminder of how the universe increasingly provides us with a powerful laboratory for studying fundamental physics. 

Spatial variations in the refractive index of astrophysical plasmas can arise due to turbulence in the interstellar medium or other sources of heating~\cite{Rickett:1977, Ishimaru:1978}. Pulsar observations have provided examples where plasma lensing amplifies the brightness of a coherent radio source by factors approaching a hundred~\cite{2014MNRAS.442.3338P,Main:2018,2019ApJ...877..125B}. It has been pointed out that plasma lensing is likely to play an important role in the phenomenology of Fast Radio Bursts (FRBs)~\cite{Cordes:2017,Main:2018}. So-called Extreme Scattering Events (ESEs), where the brightness of radio sources is seen to change by factors of a few, are also thought likely to be due to as yet unexplained plasma lensing~\cite{1987Natur.326..675F,2012MNRAS.421L.132P}. Recently, there has been growing interest in the idea that coherent gravitational wave pulses and trains, emitted from black hole or neutron star mergers, could be gravitationally lensed and thereby magnified. In this situation it will again be important to go beyond geometric optics and include wave diffraction~\cite{2017arXiv170204724D}. In all these examples,  when the line of sight between source and observer passes through a caustic of a lens, at a given frequency, the observed intensity may be enhanced leading to a pulse in frequency, time, or both. These situations have mainly been studied for one-dimensional lenses near fold and cusp caustics \cite{Pen:2012, Er:2018}. Here, we shall explore more complex, two-dimensional examples including the swallowtail, elliptic and hyperbolic umbilic catastrophes which we describe below.

There is already an extensive astrophysical literature on the computation of interference patterns in wave optics~\cite{1995ApOpt..34.2089C,2010ApJ...717.1206C}, but published methods tend to converge slowly~\cite{Grillo:2018}. They are expensive to implement and the results are sometimes inconclusive. In this paper, we shall present faster and more reliable methods. Our approach builds on Picard-Lefschetz theory, a general, exact approach to  multidimensional oscillatory integrals based upon saddle point and steepest descent techniques (for an introduction \cite{2010arXiv1009.6032W}; for applications to quantum cosmology, see \cite{Feldbrugge:2017kzv,Feldbrugge:2017fcc, Feldbrugge:2017mbc, Feldbrugge:2018gin, Feldbrugge:2019} and to relativistic quantum mechanics, see \cite{Feldbrugge:2019Schwinger}). As we shall show, our methods allow for the fast and reliable computation of even very intricate ``diffraction catastrophe'' patterns. The calculations of these patterns by Berry, Nye and collaborators were an analytical {\it tour de force}, but relied heavily on the particularities of Thom's canonical ``normal forms'' of catastrophes, and the mathematical properties of the related special functions, with each case treated separately. Unfortunately, while the normal forms represent the correct universal forms locally, they diverge at large distances. Hence, they are unrealistic as models for natural lenses. Realistic modeling requires a more versatile method which can be efficiently and straightforwardly implemented numerically. We present just such a method here.

Our method applies uniformly, with modest restrictions, to generic lens models. It is simple to implement numerically and computes interference patterns in polynomial time. As far as we have been able to check, our results agree perfectly with those aforementioned. The only requirement of our method is that it should be possible to analytically extend the interference phase into the complexified space of the spatial coordinates over which the integral is taken. Such functions embrace a very large class of lens models including, for example, any rational function, and should be more than sufficient for most modeling purposes. For simplicity, we shall not consider phases which possess branch cuts in the space of complexified coordinates. However, there are physical cases of interest where such phases do occur and an extension of our approach to this more general setting is an interesting problem for the future. 

Instead of using specific properties of special functions and symmetries, our method exploits Cauchy's theorem to exactly transform an integral of an oscillatory phase factor into a sum of absolutely convergent integrals taken over ``Lefschetz thimbles'' in the space of complexified coordinates. These ``thimble" integrals are fast to compute numerically, requiring only polynomial time. They are insensitive to numerical cutoffs and may be performed iteratively in any order with no change to the result. In this paper, we demonstrate the efficacy of our method by computing the interference patterns for one- and two-dimensional thin lenses. We study the most observationally accessible catastrophes, both in their ``normal forms" and in a set of more realistic, localized lens models where these catastrophes appear.  Our one-dimensional numerical code, capable of handling generic one-dimensional lenses, is now publicly available \href{https://github.com/jfeldbrugge/Picard_Lefschetz_Integrator}{online} \footnote{See \url{https://github.com/jfeldbrugge/Picard_Lefschetz_Integrator}.}.

A simple example of the type of integral we are interested in is 
\begin{align}
\Psi(\mu,\alpha,\nu) = \left(\frac{\nu}{\pi}\right)^{1/2} \int^\infty_{-\infty} \mathrm{d}x\, e^{i \phi(x) \nu },  \quad {\rm where}\quad  \phi(x)= (x-\mu)^2 +{\alpha \over 1+x^2}. \label{eq:simplex}
\end{align}
Here, $\Psi(\mu,\alpha,\nu)$ is the amplitude whose square $|\Psi(\mu,\alpha,\nu)|^2$ gives the intensity of light observed at a position, frequency and lens strength controlled by the parameters $\mu$, $\nu$ and $\alpha$. The control parameter $\mu$ is determined by the transverse positions of the observer and the source relative to the lens (See Fig.~\ref{fig:LensSetup} and Eq.~(\ref{eq:amp}) below). The frequency of the waves is proportional to $\nu$ so the spacing of interference fringes shrinks as $\nu$ is increased. The eikonal limit is $\nu\rightarrow \infty$. Finally, $\alpha$ controls the strength of the lens which, in this example, is taken to have a Lorentzian profile. The integral (\ref{eq:simplex}) is analytically intractable. However, it is simple to compute numerically, for reasonable values of $\nu$, $\mu$ and $\alpha$, using the methods we shall describe below. 

In the eikonal limit of large $\nu$, only real saddle point solutions -- real stationary points of the phase $\phi(x)$ -- contribute significantly to the amplitude. Each one corresponds to a particular ray. For $\alpha<1$ the lens is ``weak'' and there is only one real solution of $\partial_x \phi(x)=0$. Hence there is only one contributing ray at each value of $\mu$. For $\alpha>1$, the lens is ``strong:'' for a finite range of $\mu$  values centred on zero, there are three real solutions of $\partial_x \phi(x)=0$ hence three contributing rays. Correspondingly, one finds three images of the source in this range of $\mu$. The values of $\mu$ bounding this range mark a transition from three contributing rays ({\it i.e.} three real saddles in the phase) to one.  At these values of $\mu$, a maximum and a minimum of $\phi(x)$ merge into a {\it cubic} stationary point ({\it i.e.} a point of inflexion), creating the simplest ``fold'' catastrophe. If we now decrease the strength of the lens $\alpha$ towards unity, the two ``fold'' catastrophes approach the point $\mu=0$ where they merge to form a ``cusp'' catastrophe, in which there is a {\it quartic} stationary point in the phase $\phi(x)$. Since the phase (viewed as a function of $x$) is flatter in the vicinity of higher order stationary points, there is less destructive interference. The intensity of light grows more rapidly as $\nu$ is increased as compared to the intensity from a quadratic saddle , so that ``folds'' become increasingly bright compared to the unlensed image and ``cusps''  become even brighter.  While higher order catastrophes are rarer, their brightness makes them easier to detect. This has encouraged the conjecture, yet to be verified~\cite{Pen:2012, Spitler:2014, Cordes:2017, Dai:2017, Grillo:2018}, that the brightest sources seen may be those which happen to be lensed into high order catastrophes.  

In order to emphasize the foundational character of the physics at play and by way of a pedagogical introduction, we show how the standard Fresnel-Kirchhoff integral (see, {\it e.g.}, Ref.~\cite{1999prop.book.....B}, Chapter 8), central to the description of lensing in radioastronomy and in optics\footnote{The integral formula has a fascinating history of successive approximate derivations and subsequent critiques, reviewed in Ref.~\cite{1999prop.book.....B}, Chapter 8. Exact solutions of Maxwell's equations (or their scalar version) representing quasi-realistic interference patterns created by diffraction around physical obstacles of various types are still few in number, and are reviewed in Chapter 11 of the same work. It would be interesting to revisit these solutions and, perhaps to find others, using the ideas we develop here.}, can be derived directly from Feynman's path integral for a massless particle propagating through a refractive medium, {\it i.e.}, one in which the speed of light varies across space. Our main focus in this paper is on dispersive but non-dissipative lensing, in which the lensing phase factor always has modulus unity. However, the methods we use may equally well be applied to dissipative (lossy) lensing, in which the plasma dispersion relation is complex (for a review of dispersion relations, for example in water or in the ionosphere, see, {\it e.g.}, \cite{Jackson} Ch. 7). In this more general circumstance, the ``phase factor" over which the Fresnel-Kirchhoff integral is taken has a varying modulus. 

As an illustration of such a case, as well as to provide a foretaste of the use of our method in describing quantum mechanical interference, in Appendix \ref{ap:Young} we examine Young's famous double-slit experiment. We consider a thin, flat one-dimensional lens which modulates the intensity rather than the phase of the light passing through it. We model the lens with a smooth function which allows very little light through except in two narrow regions comprising the slits. We calculate the resulting interference pattern by deforming the contour onto the relevant Lefschetz thimbles numerically, observing how different real and complex saddle points become relevant and irrelevant, as one moves across the observational screen, through an intricate sequence of Stokes phenomena. Using this smooth lens model, we can also study in detail the emergence of the classical limit as Planck's constant $\hbar$ is taken to zero, so that the de Broglie wavelength becomes small. In this limit we find as expected that only the real, classical saddles contribute and all interference effects disappear. 

Finally, as an aside, we remark that the work presented here represents a step in a larger program, involving two of us~\cite{FT2019} and our collaborators, seeking to rigorously define, calculate and interpret real time (Lorentzian) path integrals, with diverse applications in quantum physics, both nonrelativistic and relativistic, including quantum gravity and cosmology. We expect to report further on this work in the near future. Recently, Dunne, Unsal and collaborators have pursued a very interesting (and closely related) program in quantum field theory and quantum mechanics, based upon Euclidean path integrals \cite{2013JHEP...10..041B, 2015arXiv151003435B, 2015arXiv151105977D, 2016PhRvL.116a1601B, 2018JHEP...06..068B}. See also, the closely related work of \cite{2017JHEP...05..056S}, and earlier work of \cite{Tanizaki:2014}.

The outline of this paper is as follows. In Section \ref{sec:FromFeynmanToFermat} we show how the Fresnel-Kirchhoff integral and Fermat's principle follow from the relativistic path integral for a massless particle, {\it i.e.}, a spinless photon, moving in a medium with a variable speed of light. In Section \ref{sec:Fresnel-Kirchoff_Thin} we discuss the Fresnel-Kirchhoff integral for thin  astrophysical lenses, putting the answer into a canonical dimensionless form. We then discuss the intensity in the geometric optics limit, along with the occurrence of critical points and caustics. We introduce catastrophe theory, describing the ``normal form'' of critical points of increasing complexity and their relation to observable parameters. In Section \ref{sec:1DLensExample} we discuss Picard-Lefschetz theory for a one-dimensional lens -- first in the geometric optics limit and then beyond, to include diffraction. We introduce the key concept of ``flowing'' the integration contour into the complex plane, in order to find the set of relevant Lefschetz thimbles upon which the integral becomes absolutely convergent. We describe a simple and powerful numerical code which implements this idea. In section \ref{sec:catastrophePL} we numerically compute the interference patterns of the seven elementary catastrophes, giving a comprehensive analysis of their ``unfoldings.'' In section \ref{sec:2DlocalLenses} we turn to localized lens models, which are analytically intractable. In section \ref{sec:signatures} we anticipate possible applications to the study of Fast Radio Bursts, which is an exciting current prospect. Section \ref{sec:conclusion} concludes. Appendix \ref{ap:DefiningOscillatoryIntegrals} provides some instructive background on the simplest (Gaussian) oscillatory integrals - both one- and two-dimensional, and Appendix \ref{ap:Young} tackles Young's famous double-slit experiment.

\section{From Feynman to Fermat to Fresnel-Kirchhoff}\label{sec:FromFeynmanToFermat}

Imagine a  bright source emitting coherent electromagnetic waves which traverse an astrophysical plasma on their way to our telescopes on earth. Let us describe the propagation in terms of the elementary {\it quanta} of such waves, considered to be relativistic particles. The Feynman path integral over these particle's trajectories in spacetime yields the quantum mechanical amplitude to propagate from the source to any particular location. The square of the amplitude yields the intensity, determining the interference pattern in position and frequency. As we shall see, one or more classical trajectories dominate the amplitude: these dominant trajectories obey Fermat's ``principle of least time.''  For simplicity, we shall ignore polarization effects, taking the elementary quanta to be spinless. We shall furthermore study only the simplest dispersion relation for astrophysical plasmas, valid in the high frequency regime -- generalizations to more complex and realistic dispersion relations should be straightforward. Our derivation emphasizes the fundamental nature of the physics involved - as we shall show, the Fresnel-Kirchhoff integral (see, {\it e.g.}, Ref. \cite{1999prop.book.....B}, Chapter 8, 8.3.3 (28)) follows directly from the Feynman path integral. We hope the reader will enjoy the directness and economy of this approach compared to more standard (and cumbersome) derivations based on Maxwell's equations, or their scalar counterpart. 

We start from the dispersion relation in a tenuous plasma (see, {\it e.g.}, \cite{Jackson} Section 7.9)
\begin{equation}
 \omega^2=k^2 c^2+\omega_p^2(\bm{x}).
 \label{dispersionrel}
\end{equation}
Here, $\omega$ and $k$ are the angular frequency and wavenumber of the waves, $c$ is the speed of light and $\omega_p(\bm{x})$  
is the plasma frequency at position $\bm{x}$, determined by the local density of electrons, assumed to vary across space on scales much larger than the wavelength of the electromagnetic waves.  Notice that (\ref{dispersionrel}) takes exactly the same form as the dispersion relation for a relativistic particle whose mass varies with spatial position. 

The dispersion relation (\ref{dispersionrel}) yields a phase propagation speed 
\begin{equation}
 v_{p}(\bm{x})\equiv {\omega\over k}  = c \, \sqrt{1+{\omega_p^2(\bm{x})\over k^2 c^2}},
 \label{phv}
\end{equation}
which is greater than the speed of light. This should be no cause for concern, as the analogy with a massive particle assures us, since information only propagates at the group velocity, $ \bm{v}_g\equiv \bm{\nabla_{k}} \omega$ whose magnitude $c_g=c^2/c_p$ is always less than the speed of light.

The quanta of these waves may be described as relativistic particles, following parameterized worldlines in spacetime: $x^\mu(\lambda)=(ct(\lambda), \bm{x}(\lambda))$. Reparameterizations $\lambda \rightarrow \tilde{\lambda}(\lambda)$ are generated by a Hamiltonian, and reparameterization invariance corresponds to the constraint that the Hamiltonian vanishes, ${\cal H}=0$. The correct expression for the Hamiltonian ${\cal H}$ may be read off from the dispersion relation (\ref{dispersionrel}), using the correspondence $p_\mu =(p_0,\bm{p})\leftrightarrow \hat{p}_\mu=-i\hbar \partial_\mu =\hbar(-\omega/c,\bm{k})$:
\begin{equation}
 {\cal H}=-p_0^2 \,c^2+p^2\,c^2+\hbar^2 \,\omega_p^2(\bm{x}).
 \label{hamiltonian}
\end{equation}
The first order (phase space) action, with the initial and final spacetime
locations of the particle held fixed, is: 
\begin{equation}
 S[x;x^\mu(0),x^\mu(1)]=\int_0^1 d\lambda\, \left(p_0\, \dot{x}^0+\bm{p} \cdot \dot{\bm{x}} -\tau(\lambda) {\cal H}\right).
 \label{eq:acti}
\end{equation}
where dots denote derivatives with respect to $\lambda$, taken to run from $0$ to $1$ as the the particle trajectory runs from the initial spacetime point $x^\mu(0)\equiv (c\,t_i,\bm{x}_i)$ 
to the final point $x^\mu(1)\equiv (c\,t_f,\bm{x}_f)$. 
The `einbein' $\tau (\lambda)$ serves as a Lagrange multiplier enforcing the Hamiltonian constraint and ensuring the action is reparameterization invariant (it transforms under reparameterization so that $d\lambda \,\tau(\lambda)$ is invariant). Varying the action with respect to the momenta yields Hamilton's equations for the momenta $p_0 \,c^2=-\dot{x}^0/(2 \tau)$ and $\bm{p} \,c^2=\dot{\bm{x}}/(2 \tau)$. Varying with respect to $\tau$ yields the constraint $ {\cal H}=0$. The energy $E=-p_0 c$ is conserved because the action is invariant under constant translations of the time $x^0$. 

In seeking to derive Fermat's principle, we face a conundrum. If the initial and final times $t_i$ and $t_f$ are held fixed, how can the total time $t_f-t_i$ possibly vary? The resolution is that, for a monochromatic beam, we should fix the initial energy $E$, not the initial time $t_i$. We cannot fix both because of the time-energy uncertainty relation (which follows from the commutator $\left[\hat{p}_0,\hat{x}^0\right]=-i \hbar$).  The action appropriate to fixing the initial energy is obtained by adding a boundary term. The latter must be chosen to ensure that the variation of the action is zero when the initial energy and the final time, as well as the initial and final spatial positions, are held fixed and the equations of motion are satisfied. The initial time is then free to vary, which is how Fermat's principle can arise. The required total action is:
\begin{equation}
 S\left[\bm{x};E,\bm{x}_i,t_f,\bm{x}_f\right]=p_0 x^0(0) +\int_0^1 d\lambda\, \left(p_0\, \dot{x}^0+\bm{p} \,\dot{\bm{x}} -\tau(\lambda) {\cal H}\right),
 \label{eq:actprop}
\end{equation}
with ${\cal H}$ given in (\ref{hamiltonian}).

Since the action (\ref{eq:actprop}) is quadratic in the momenta and linear in $\tau$, we can integrate out those variables. At the relevant saddle, we may use Hamilton's equations for the momenta, and the constraint, to obtain a reduced action expressed purely in terms of reparameterization-invariant quantities:
\begin{equation}
 S_{r}\left[\bm{x}\right]= -E t_i  -\int_{t_{i}}^{t_{f}} dt  {\hbar^2 \omega_p^2(\bm{x}(t)) \over E}.
 \label{eq:actii}
\end{equation}

Writing $t_i=t_f-\int_{t_i}^{t_f} dt $ where the final time $t_f$ is held fixed, we find, up to an irrelevant constant phase,
\begin{equation}
 S_{r}\left[\bm{x}\right]= E \int_{t_{i}}^{t_{f}}  \,dt \, \left(1-{\hbar^2 \omega_p^2(\bm{x}) \over E^2}\right)= E  \int_{\bm{x}_i}^{\bm{x}_f} \, {|d\bm{x}| \over c} \left(1-{\hbar^2 \omega_p^2(\bm{x}) \over E^2}\right)^{1\over 2} = \int_{\bm{x}_i}^{\bm{x}_f} |d \bm{x}|\,|\bm{p}|,
 \label{eq:actiii}
\end{equation}
where, again, we used Hamilton's equations for the momenta and the Hamiltonian constraint. Finally, we  express the result in terms of the phase velocity (\ref{phv}), obtaining
 \begin{equation}
 S_{r}\left[\bm{x}\right]= E \int_{\bm{x}_i}^{\bm{x}_f} {|{\bm{dx}}| \over c_{p}(\bm{x})}.
 \label{fermact}
\end{equation}
Note that, although the phase velocity $c_p({\bf x})$ appearing here is always greater than the speed of light, nowhere in our derivation does any on-shell particle actually travel faster than light. 

The reduced action (\ref{fermact}) embodies Fermat's principle of least time or, more correctly, the principle that the time taken is stationary on dominant classical trajectories. The path integral over all paths, weighted by $e^{i S_{r}[\bm{x}]/\hbar},$ is the Fresnel-Kirchhoff integral we seek. 

As an aside, note that one may, equally well, obtain the result (\ref{fermact}) starting from the square root (Nambu-type) action for a particle with a spatially dependent mass $m(\bm{x})$, by making use of the correspondence $m(\bm{x})\,c^2\leftrightarrow \hbar\omega_p({\bm{x}})$, namely 
\begin{equation} 
 {\cal S}\left[\bm{x}\right]=-\int_{t_i}^{t_f} \, dt \,\hbar\omega_p({\bm{x}}(t))\left(1-{\dot{\bm{x}}(t)^2 \over c^2}\right)^{1\over 2},
 \label{nambuact}
\end{equation}
where the dot now denotes a $t$ derivative. This action is explicitly reparameterization invariant from the start. However, it is the action appropriate to fixing the initial time $t_i$ whereas we need to fix the initial energy $E$. As before, we must supplement the action (\ref{nambuact}) by a boundary term, which turns out to be $+E (t_f-t_i)$. One can easily check that the identity $\partial S_{cl}/\partial t_i=E$ for Hamilton's principal function $S_{cl}$ implies the total action is stationary, provided the desired boundary conditions and the equations of motion are fulfilled.  Using $E =\hbar\omega_p/\left(1-\dot{\bm{x}}^2/c^2\right)^{1\over 2}$, the total action reduces (again, up to a constant phase) to (\ref{eq:actiii}) as before.

\section{Evaluating the Fresnel-Kirchhoff integral}\label{sec:Fresnel-Kirchoff_Thin}

\begin{figure}
\centering
\resizebox {0.6\textwidth}{!}{
\begin{tikzpicture}
\draw[-] (0,0) -- (4,0) node[right] {observer plane};
\draw[-] (0,3) -- (4,3) node[right] {lens plane};
\draw[-] (0,5) -- (4,5) node[right] {source plane};

\draw[<->] (0,0.1) -- (0,2.9);
\draw[<->] (0,3.1) -- (0,4.9);
\draw[<->] (-0.25,0.1) -- (-0.25,4.9);

\node at (0,1.5) [right] {$d_{lo}$};
\node at (0,4)   [right] {$d_{sl}$};
\node at (-0.25,2.5)   [left] {$d_{so}$};

\draw[-,red] (2.5,5) -- (2.,3) -- (2.3,0);
\draw[dotted,red] (2.5,5) -- (1.,3) -- (2.3,0);
\draw[dotted,red] (2.5,5) -- (1.5,3) -- (2.3,0);
\draw[dotted,red] (2.5,5) -- (2.5,3) -- (2.3,0);
\draw[dotted,red] (2.5,5) -- (3.,3) -- (2.3,0);

\filldraw[black] (2.5,5)         circle (1pt) node[above] {$\bm{x}_s$};
\filldraw[black] (2.,3)     circle (1pt) node[above right] {$\bm{x}$};
\filldraw[black] (1.,3)     circle (0.75pt);
\filldraw[black] (1.5,3)         circle (0.75pt);
\filldraw[black] (2.5,3)             circle (0.75pt);
\filldraw[black] (3.,3)       circle (0.75pt);
\filldraw[black] (2.3,0)         circle (1pt) node[below] {$\bm{x}_{obs}$};
\end{tikzpicture}
}
\caption{The geometry of interfering paths passing through a thin lens.}
\label{fig:LensSetup}
\end{figure}
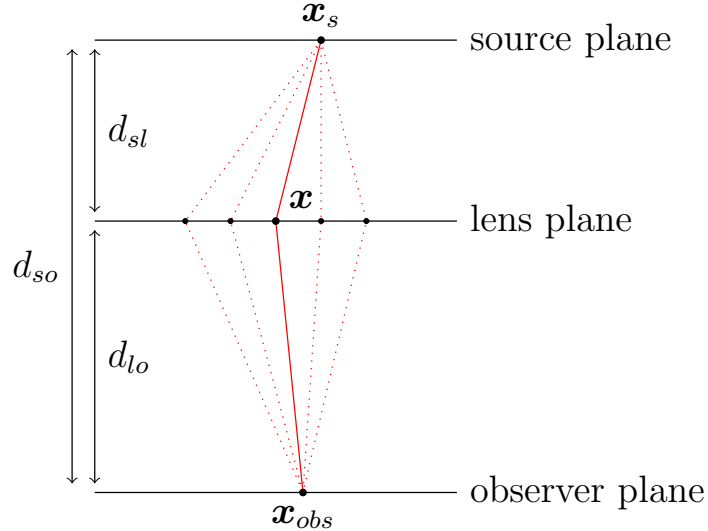

Consider now a radio wave quantum, as described above, traversing an astrophysical plasma from its initial position at the source $\bm{x}_i=\bm{x}_s$ to its final position at the observer $\bm{x}_f=\bm{x}_{obs}$. For simplicity we assume the plasma takes the form of a thin, flat lens, with the phase velocity $c_{p}(\bm{x})=c$, the speed of light {\it in vacuo},  everywhere except on the lens (see Fig. \ref{fig:LensSetup}). Let us redefine the spatial coordinates $\bm{x}\rightarrow (\bm{x},z)$ to separate out the coordinates in the lens plane $\bm{x}$ from the normal coordinate $z$. The real classical paths are piecewise linear, with a possible bend at the lens, and the integral over these paths reduces to an ordinary integral over the lens plane~\cite{Feynman:1965}. The path integral amplitude for a (spinless) photon is obtained by integrating over all paths weighted by the phase factor $e^{i S_{r}\left[\bm{x}\right]/\hbar}$ obtained from (\ref{fermact}):
 \begin{equation}
 \Psi(\bm{x}_{obs},\bm{x}_s)= \int d \bm{x} \,\exp\left[i \omega \int_{\bm{x}_s}^{\bm{x}_{obs}}|d \bm{x}| {n(\bm{x})\over c}\right].
 \label{fermactpi}
\end{equation}
where we replaced $E$ with $\hbar \omega$, $\omega$ is the angular frequency of the light, and the phase velocity $c_p(\bm{x})$ with $ c/n(\bm{x})$ where $c$ is the speed of light {\it in vacuo} and $n(\bm{x})$ is the refractive index. For an astrophysical plasma, as mentioned above, at high frequencu we have $n(\bm{x})\approx 1-{\omega_p^2(\bm{x})/ \omega^2}$ where $\omega_p$ is the plasma frequency, given by $\omega_p^2(\bm{x})\approx n_e(\bm{x}) e^2/(m_e \epsilon_0)$, with $n_e(\bm{x})$, $e$, and $m_e$ respectively the local electron density and the charge and mass of the electron in SI units (see {\it e.g.} \cite{Main:2018}). We explicitly exhibit the $\bm{x}$-dependence since it governs the structure of the lens. 

In the thin lens approximation, variations in the phase arise in part geometrically, from variations in the length of the straight line segments on  either side of the lens, and in part from the passage through the lens. The former are straightforwardly computed using the Pythagorean theorem in the approximation that the relative horizontal displacements in Fig. \ref{fig:LensSetup}, $|{\bf x}_{obs}-{\bf x}|$ and $|{\bf x}_s-{\bf x}|$ are much smaller than the vertical distances $d_{lo}$ and $d_{sl}$. The latter are likewise computed approximately, noting that, to lowest order, the paths pass vertically through the lens so we may replace $\int n_e(\bm{x},z))dz$ with  $\Sigma_e(\bm{x})$, the electron surface density. 

The path integral amplitude then becomes
\begin{equation}
\Psi(\bm{x}_{obs},\bm{x}_s;\nu)  \propto  \int d\bm{x} \, \exp \left[i {\omega\over 2 c} \left({(\bm{x}-\bm{\mu})^2\over \overline{d}}-{\Sigma_e(\bm{ x}) e^2\over m_e \epsilon_0\omega^2} \right)\right],
\label{eq:pathfinal}
\end{equation}
where $\overline{d}=d_{sl}d_{lo}/(d_{sl}+d_{lo})$ is the reduced distance and $\bm{\mu}= ({\bf x}_s d_{lo} +{\bf x}_{obs} d_{sl})/d_{so}$ is a weighted average of the transverse displacements of the source and the observer. Notice that $\bm{x}-\bm{\mu}$ depends only on the {\it relative} displacements of the source, the lens and the observer, so that the answer is independent of the choice of origin for the transverse coordinates. 

It is convenient to normalize the amplitude by dividing it by the amplitude obtained with the same geometry but no lens present. We may then write the resulting normalized amplitude as a dimensionless integral. Redefining $\bm{x}\rightarrow a \bm{x}$,  $\bm{\mu}\rightarrow a \bm{\mu}$ where $a$ is some convenient physical scale associated with the lens, we set $\nu =\omega a^2 /(2 c \overline{d})= a^2/(2 R_F^2)$ where $R_F=(\lambda \overline{d})^{1\over 2}$ is the Fresnel scale~\cite{Berry:1980}. Notice that, because lensing alters the {\it angle} of propagation, the fringe spacing grows with the distance. Hence, it is the Fresnel scale -- the geometric mean of the distance and the wavelength -- rather than the wavelength which should be compared with the source dimensions to determine whether the interference pattern is observed in the heavily diffracted (low $\nu$) or eikonal (high $\nu$) regime. Finally, we define $\varphi({\bf x}) = -\Sigma_e(\bm{ x}) e^2 \overline{d}/( m_e \epsilon_0 a^2\omega^2)$ to obtain the normalized, dimensionless amplitude,
\begin{equation}
\Psi(\bm{\mu};\nu) = \left(\frac{\nu}{\pi}\right)^{N/2} \int_{\mathbb{R}^N} \mathrm{d}\bm{x} \, \exp \left[i  \phi(\bm{x};\bm{\mu})\nu\right]\,,  \quad {\rm with}\quad 
\phi(\bm{x};\bm{\mu}) =(\bm{x}-\bm{\mu})^2 + \varphi(\bm{x})\,,
\label{eq:amp}
\end{equation}
for an $N$-dimensional lens. Since $\nu \propto \omega$, we see that the eikonal limit is high frequency limit. However, the strength of the lens is controlled by $\varphi$ which is proportional to $\omega^{-2}$. Therefore the lens becomes stronger at lower frequencies where, of course, diffraction becomes important. The highest magnifications attained involve a playoff between strong lensing, creating effects like caustics and catastrophes, and diffraction which tends to smear out intensity peaks. Hence, to model the most interesting regime for astrophysical plasma lenses, one must go beyond geometrical optics and include diffractive effects.

The intensity corresponding to the amplitude (\ref{eq:amp}) is proportional to the probability for a photon to be detected at $\bm{\mu}$:
\begin{align}
I (\bm{\mu};\nu) \propto |\Psi(\bm{\mu};\nu)|^2\,.
\label{eq:intensity}
\end{align}
The observed intensity should be normalized to the energy flux received by the detector, at each frequency, integrated over all observed $\bm{\mu}$. For a more detailed analysis see \cite{Brooker:2003}.

Except in special cases, the Fresnel-Kirchhoff integral \eqref{eq:amp} is not possible to evaluate analytically. At large $\nu$ (and with the dimensionless form of the lens $\varphi(\bm{x})$ held fixed) and in the  geometric optics limit, one can easily model the intensity, as we shall explain. However, the most interesting regime for astrophysical plasma lenses occurs in the intermediate regime, where focusing and caustic catastrophes generate bright features whose peak intensity is controlled by diffraction. In this regime, there are characteristic patterns in the intensity,  controlled by the topological character of the lens. In this intermediate-$\nu$ regime, conventional integration techniques typically fail, and it is hard to capture the complex, oscillatory interference pattern numerically. For example, G. Grillo and J. M. Cordes \cite{Grillo:2018} implemented a procedure based on Fourier methods but found this technique to generate numerical artifacts. Here, motivated by our earlier work on Picard-Lefschetz theory, we instead employ analytic continuation and Cauchy's theorem to unambiguously define and to evaluate the relevant oscillatory integrals. We have developed a custom numerical scheme (made available \href{https://github.com/jfeldbrugge/Picard_Lefschetz_Integrator}{online} \footnote{See \url{https://github.com/jfeldbrugge/Picard_Lefschetz_Integrator}.}.) which is fast and accurate, and applicable to a generic one dimensional oscillatory integral. A two dimensional version will be made available shortly. A nice feature of our method is that it typically becomes more efficient, {\it i.e.}, its convergence is improved, as the integrand becomes more oscillatory and difficult to handle via conventional techniques.
\subsection{The geometric optics limit}\label{sec:geometricOptics}

In the limit of large $\nu$, the Fresnel-Kirchhoff integral is dominated by real stationary points of the phase function $\phi$ which, except at special  values of $\mu$, are well-approximated by Gaussians. Furthermore, any interference between different stationary points leads to oscillations in the intensity which, in the limit $\nu\rightarrow \infty$, become increasingly rapid. In the geometric optics approximation, one averages these oscillations away. Physically this averaging occurs through the incoherence of any realistic extended source, as we explain later. Although this paper is devoted to the study of interference phenomena, it proves useful to begin by studying the geometric optics limit. 

In the large $\nu$ (eikonal) limit, we focus on real critical points of the exponent, \textit{i.e.}, those values of $\bm{x}$ for which 
\begin{align}
\nabla_{\bm{x}} \phi(\bm{x};\bm{\mu}) = 0\,,
\end{align}
considered as a function of the parameter $\bm{\mu}$. The critical points are generally smooth complex-valued functions of $\bm{\mu}$. In the eikonal limit, only the real critical points contribute because contributions from complex saddle points are exponentially suppressed. The critical points can be described in terms of the Lagrangian map $\bm{\xi}:X \to M$, mapping the points in the base space $\bm{x}\in X = \mathbb{R}^N$ to points in the parameter space $\bm{\mu} \in M$ according to the critical point condition
\begin{align}
\nabla_{\bm{x}} \phi(\bm{x};\bm{\mu})|_{\bm{\mu} = \bm{\xi}(\bm{x})}  = 0\,.
\end{align}
The Lagrangian map is determined by the gradient of the phase of the lens:
\begin{align}
\nabla_{\bm{x}} \phi(\bm{x};\bm{\mu}) = 2 (\bm{x}-\bm{\mu}) + \nabla \varphi(\bm{x})\quad \implies \quad \bm{\mu} = \bm{\xi}(\bm{x}) = \bm{x} + \frac{1}{2}\nabla \varphi(\bm{x})\,.
\end{align}

\begin{figure}
\centering
\includegraphics[width=0.9\textwidth]{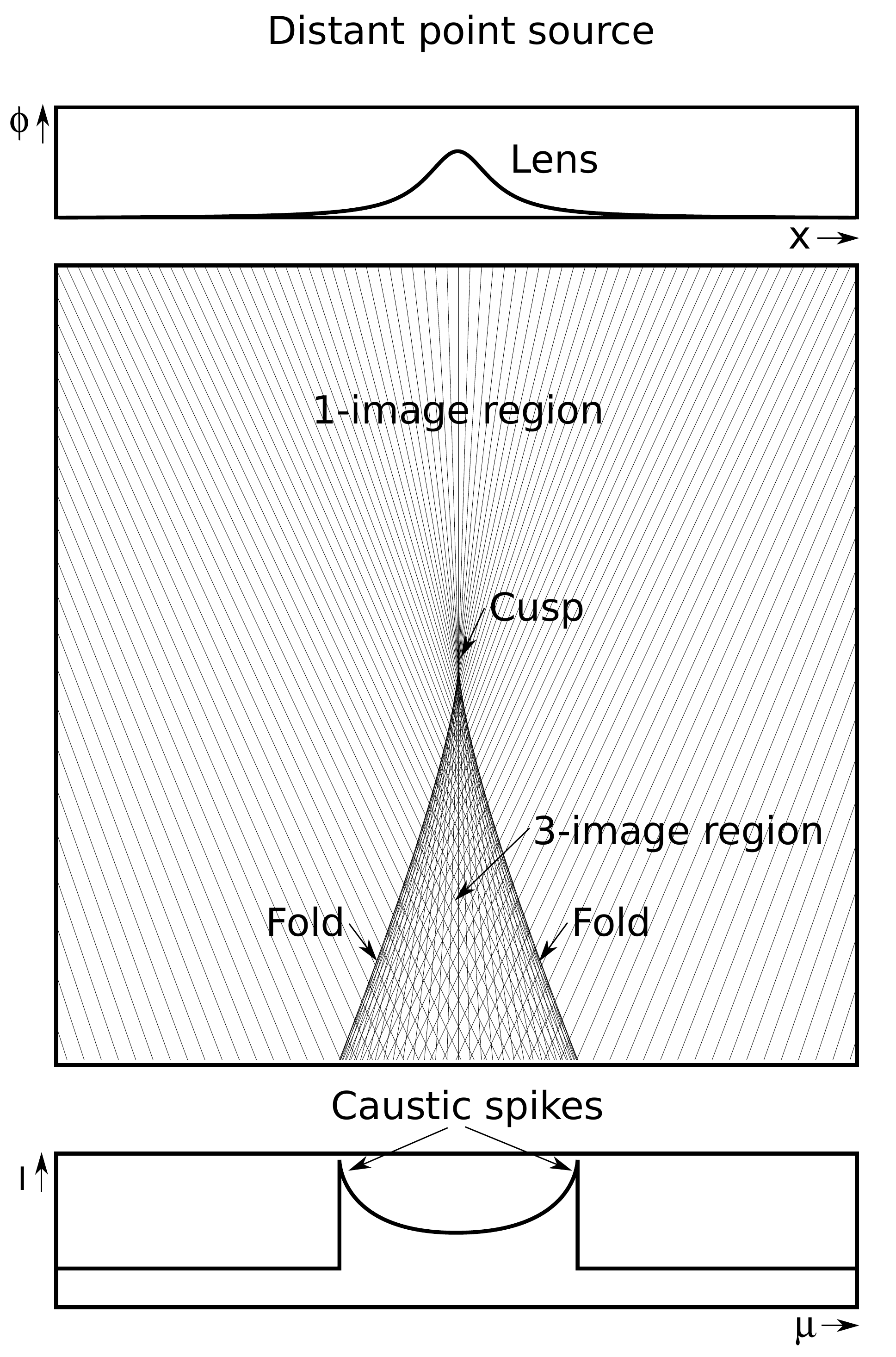}
\caption[The Lagrangian map in geometric optics.]{The Lagrangian map in geometrical optics. The image consists of two single and one triple image regions separated by a fold caustic at which the normalized intensity spikes.}\label{fig:LagrangianMap}
\end{figure}

The Lagrangian map $\bm{\xi}$ determines the optical rays, giving a purely geometric description of the lens. Every point $\bm{x}$ is mapped to a point $\bm{\mu}$ in the space of observational parameters. In general, a point $\bm{\mu} \in M$ might be obtained from several points in $\bm{x}\in X$, \textit{i.e.}, the Lagrangian map can be many-to-one. The regions in $\bm{\mu}$ where each point is obtained from $n$ points in $X$ are known as $n$-image regions. In multi-image regions, one adds the intensities due to each of the contributing paths: performing the relevant Gaussian integrals one finds for the normalized intensity 
\begin{align}
I(\bm{\mu}; \infty) = \sum_{\bm{x} \in \xi^{-1}(\bm{\mu})} \frac{2}{|\lambda_1(\bm{x})|\dots |\lambda_N(\bm{x})|}\,,\label{eq:IntensityGeometric}
\end{align}
with $\bm{\xi}^{-1}$ the pre-image of the Lagrangian map and $\lambda_1,\dots,\lambda_N$ the eigenvalues of the deformation tensor,
\begin{align}
\mathcal{M}_{ij}(\bm{x})= \frac{\partial^2 \phi(\bm{x};\bm{\mu})}{\partial x_i \partial x_j} = 2 \frac{\partial \xi_i(\bm{x})}{\partial x_j}\,,
\end{align}
evaluated at the relevant critical points $\bm{x}=(x_1,x_2,\dots,x_N)$. Below, we shall study these multi-image regions in detail, at finite $\nu$, where they exhibit intricate interference patterns.

At the boundaries between regions with a different number of images, at least one of the eigenvalue fields $\lambda_i$ must vanish. At infinite $\nu$ this leads to an infinite spike in the normalized intensity map, signalling a caustic. See Fig.~\ref{fig:LagrangianMap} for an illustration of the Lagrangian map corresponding to a one-dimensional lens with a single- and a triple-image region. The triple-image region is separated from the single-image regions by two fold caustics. At the fold caustic the normalized intensity profile diverges.

Formally, the Lagrangian map $\bm{\xi}$ forms a caustic at $\bm{x}_c\in X$ when the deformation tensor becomes singular, \textit{i.e.}, its determinant vanishes. However, the $X$ space is generally not observed. In the space $M$ of observable parameters, the caustic at $\bm{x}_c$ appears at the point $\bm{\mu}_c = \xi(\bm{x}_c)$. For one-dimensional lenses caustics occur at isolated points. For higher-dimensional cases, the determinant of the deformation tensor vanishes on a manifold $X_c =\{ \bm{x} \in X| |\mathcal{M}(\bm{x})| = 0\}$
which is mapped to a caustic set $M_c = \bm{\xi}(X_c)$ in the parameter space. Note that the set $M_c$ is generally {\it not} a manifold, as it includes higher order caustics, such as cusps and swallowtails, at which the variety is non-differentiable and therefore $M_c$  is not locally homeomorphic to Euclidean space. We shall discuss examples of this kind later, but note here that they are exactly the points at which the lensing integral exhibits the most interesting behaviour. 

The geometric optics limit is attained in two stages: at short wavelengths, each real stationary point corresponds to a distinct image.  As the wavelength is increased, each image itself forms an interference pattern, as illustrated in the Young's double slit experiment examined in Appendix \ref{ap:Young}. The limit of short wavelengths, in which phase coherence is maintained, is often called the eikonal approximation. However, when phase coherence is lost - for example, when the source size becomes larger than the spacing of its fringes, interference effects disappear altogether.  This assumption of loss of coherence is implicit in the geometric optics limit. However, objects smaller than the Fresnel scale are still seen to scintillate, as a result of coherent interference effects on unresolved scales. This is reflected in the expression {\it stars twinkle, planets don't}. Interstellar scintillation typically occurs for sources smaller than about a micro arcsecond, corresponding to the Fresnel angle $\theta_F = \sqrt{\lambda/d}$ (with $\lambda$ the wavelength and $d$ the distance from the lens) on the sky. This condition is true for most FRBs and pulsars.  Interplanetary scintillation due to the solar wind is commonly seen for many compact extragalactic radio sources at low frequencies \cite{1978SSRv...21..411C}. In this case, the characteristic Fresnel angle for wavelengths of a few meters and distances of an astronomical unit is a fraction of an arcsecond. Ionospheric scintillation is strongest at the lowest frequencies, and is commonly seen at solar maximum or at equatorial locations near sunrise or sunset~\cite{7775820}, and causes loss of lock in GPS.  The Fresnel angular scale for a screen at a distance of 200km at wavelengths of a meter is 8 arc minutes, causing all celestial sources except the sun and the moon to scintillate.

\subsection{Catastrophe theory}\label{sec:catastropheTheoryOptics}

Catastrophe theory is the mathematical classification of stable critical points. Caustics are classified by Lagrangian catastrophe theory \cite{Arnold:1972, Arnold:1976}, which is a special application of the general theory. Given the definition of the Lagrangian map $\bm{\xi}$, the connection between caustics in optical systems and critical points is not surprising. For one-dimensional functions, the classification consists only of minima and maxima. The local minima and maxima of a one-dimensional function are stable, \textit{i.e.}, the addition of a small perturbation merely leads to a displacement of the critical point. Degenerate critical points are not included, as they are not stable in one dimension. For example, a cubic critical point  decomposes into a minimum and a maximum, or no critical point at all, when perturbed.

In the catastrophe theory of higher-dimensional functions, degenerate critical points are included because they are stable. Ren\'e Thom (1972)  proved \cite{Thom:2018} that the stable critical points with co-dimension\footnote{The co-dimension of a caustic is roughly the dimensionality of the singularity. The stable critical points of a $n$-dimensional function are completely classified by the caustics with co-dimension smaller or equal to $n$.} $K$ less than or equal to $4$ are classified by the seven ``elementary catastrophes.'' These seven singularities suffice to classify the full range of caustics emerging in three-dimensional lenses. Thom named the seven catastrophes: \textit{the fold, cusp, swallowtail, butterfly, and the elliptic, hyperbolic and parabolic umbilic}. The caustics were in the subsequent years connected and labeled by the Coxeter reflection groups (Arnol'd \cite{Arnold:1973, Arnold:1975}). The theory was subsequently applied to optical interference patterns by Berry and collaborators, and beautiful experiments were performed~\cite{Berry:1980}. For a more recent theoretical investigation of catastrophe theory and caustics with applications to large-scale structure formation see \cite{Feldbrugge:2018JCAP...05..027F}. Here we briefly review catastrophe theory and its application to oscillatory integrals.

\begin{table}
\center
\begin{tabular}{lcccc}
\hline
Name & Symbol & $K$ & $N$ & $\phi(\bm{x};\bm{\mu})$\\
\hline
Maximum/minimum        &    $A_1^\pm$    &    $0$    &    $1$    &    $\pm x^2$\\
Fold                &    $A_2$        &    $1$    &    $1$    &    $x^3/3+ \mu x$             \\
Cusp                &    $A_3$        &    $2$    &    $1$    &    $x^4/4 + \mu_2 x^2/2 + \mu_1 x$    \\
Swallowtail            &    $A_4$        &    $3$    &    $1$    &    $x^5/5 + \mu_3 x^3/3 + \mu_2 x^2 /2 + \mu_1 x$\\
Butterfly            &    $A_5$        &    $4$    &    $1$    &    $x^6 / 6 + \mu_4 x^4/4 + \mu_3 x^3/3+ \mu_2 x^2/2 + \mu_1 x$\\
Elliptic umbilic         &    $D_4^-$    &    $3$    &    $2$    &    $x_1^3 - 3 x_1 x_2^2 - \mu_3(x_1^2 + x_2^2) - \mu_2 x_2 -\mu_1 x_1$\\
Hyperbolic umbilic    &    $D_4^+$    &    $3$    &    $2$    &    $x_1^3 + x_2^3 - \mu_3 x_1 x_2 - \mu_2 x_2 - \mu_1 x_1$\\
Parabolic umbilic        &    $D_5$        &    $4$    &    $2$    &    $x_1^4 + x_1 x_2^2 + \mu_4 x_2^2 + \mu_3 x_1^2 + \mu_2 x_2 + \mu_1 x_1$\\
\hline
\end{tabular}
\caption[The unfoldings of the seven elementary catastrophes]{The unfoldings of the seven elementary catastrophes with codimension $K \leq 4$, with $\bm{x}=(x_1,x_2,\dots,x_N)$ and $\bm{\mu}=(\mu_1,\mu_2,\dots,\mu_K)$. The normal forms are defined as the unfolding at parameter $\bm{\mu}=\bm{0}$, \textit{i.e.}, $\phi(\bm{x};\bm{0})$.}
\label{tab:unfoldings}
\end{table}

Table \ref{tab:unfoldings} lists the seven ``elementary catastrophes'' and their unfoldings $\phi(\bm{x};\bm{\mu})$. The unfolding $\phi(\bm{x};\bm{\mu})$ evaluated at $\bm{\mu}=0$ is the normal form of the catastrophe, representing the archetypical form of the critical point near $\bm{x}=0$. We observe that the fold and the cusp respectively correspond to a cubic and quartic critical point of $x$. The parameter $\bm{\mu}$ represents the ways in which the caustic can decompose into lower-order caustics. In the case of the fold, we see that a linear perturbation decomposes the fold into a minimum and a maximum for $\mu <0$ and no critical point at all for $\mu>0$. The seven catastrophes belong to two families, classified by their co-rank \footnote{The co-rank is the number vanishing eigenvalues of the Hessian matrix.}. The $A$-family is of co-rank $N=1$, while the $D$-family is of co-rank $N=2$. Critical points with higher co-rank have a co-dimension higher than $4$, and for this reason are not included here. The co-rank $N$ and the co-dimension $K$ characterize the critical point. It generally takes $N$ variables to describe the critical point, and it takes $K$ parameters to describe its unfolding. In more prosaic terms, $N$ is the dimension of the space of $\bm{x}$'s and $K$ is the dimension of the space of $\bm{\mu}$'s.\\

\begin{table}
\center
\begin{tabular}{lcccc}
\hline
Catastrophe & Symbol & $I_0$ & $\beta$ & $\sigma_j$\\
\hline
Fold                &    $A_2$     & $1.584$ &        $1/6$        &    $\sigma_1 = 2/3$                                \\
Cusp                &    $A_3$     & $2.092$ &    $1/4$        &    $\sigma_1=3/4, \sigma_2 = 1/2$                    \\
Swallowtail         &    $A_4$     & $1.848$ &    $3/10$       &    $\sigma_1=4/5, \sigma_2 =3/5, \sigma_3 = 2/5$           \\
Butterfly           &    $A_5$     & $1.991$ &    $1/3$        &    $\sigma_1=5/6,\sigma_2=2/3,\sigma_3=1/2,\sigma_4=1/3$   \\
Elliptic umbilic    &    $D_4^-$ 	& $1.096$ &    $1/3$        &    $\sigma_1=2/3, \sigma_2=2/3,\sigma_3=1/3$          \\
Hyperbolic umbilic  &    $D_4^+$   & $0.580$ &    $1/3$        &    $\sigma_1=2/3, \sigma_2=2/3,\sigma_3=1/3$            \\
Parabolic umbilic   &    $D_5$     & $2.258$ &    $3/8$        &    $\sigma_1=5/8,\sigma_2=3/4,\sigma_3=1/2,\sigma_4=1/4$  \\
\hline
\end{tabular}
\caption[The scaling laws of the elementary catastrophes.]{The intensity and fringe separation scaling relations for the catastrophes shown shown in Table \ref{tab:unfoldings}. At large $\nu$ the maximum intensity \eqref{eq:intensity} is given by $I_0 \nu^{2\beta}$ (see the discussion following Eq.~\eqref{eq:inten}) and the fringe scaling exponents are defined in \eqref{eq:scaling}.}
\label{tab:exponents}
\end{table}

For each of the normal forms listed in Table \ref{tab:unfoldings}, the normalized amplitude 
\begin{align}
\Psi(\bm{\mu};\nu) = \left(\frac{\nu}{\pi}\right)^{N/2} \int e^{i \phi(\bm{x};\bm{\mu}) \nu }\mathrm{d}\bm{x}\,,\label{eq:ffffffffff}
\end{align}
forms a caustic at the critical point $\bm{\mu}=0$. For a detailed analysis including  illustrations of the intensities obtained in each case, see chapter 36 of \cite{Thompson:2011}. As $\nu$ is increased to large values, the normalized intensity $I(\bm{\mu};\nu)=|\Psi(\bm{\mu};\nu)|^2$ diverges and the scale of the associated diffraction fringes shrinks to zero according to scaling laws which are specific for each catastrophe. At large $\nu$, the maximum of the intensity is attained near $\bm{\mu}=0$ as illustrated, for example, by the fold singularity shown in Fig.~\ref{fig:Airy}. The maximum intensity obeys the following scaling law at large $\nu$:
\begin{align}
I(\bm{0},\nu) = I_0 \nu^{2 \beta}\,.
\label{eq:inten}
\end{align}
The constant $\beta$, termed the \textit{singularity index} by Arnold (Arnold \cite{Arnold:1975} and Varchenko \cite{Varchenko:1976}), is universal, being invariant under diffeomorphisms and depending only on the topological class of the catastrophe. It is given, for each case, in the fourth column of  Table \ref{tab:exponents}. The scaling with $\nu$ is easily seen by examining the corresponding normal form. Setting the unfolding parameter to zero, \textit{i.e.}, $\bm \mu=\bm 0$, in the phases listed in Table \ref{tab:unfoldings}, one can render the phase of the integrand independent of $\nu$ by rescaling the integration variables $\bm{x}$. For example, for $A_2$ we set $x=\nu^{-{1\over 3}}y$. Taking into account the $\nu$-dependence arising from the Jacobian in the integration measure as well from the prefactor in (\ref{eq:ffffffffff}), one infers that the amplitude at the caustic scales as $\nu^{1\over 6}$ for $A_2$ and hence that $\beta={1\over 6} $ for this case. For the two dimensional lenses, one has to rescale both $x_1$ and $x_2$ in order to remove $\nu$ from the exponent but the argument is otherwise the same. 

For each of the normal forms of the phase listed in Table \ref{tab:unfoldings}, one may also analytically compute the constant $I_0$, and we provide its numerical value in Table \ref{tab:exponents}. When considering a class of lens models for modeling purposes (such as the localised models we consider later), it may be helpful to notice an additional scaling property. At large $\nu$, the amplitude is determined by the form of the phase near the critical point. Indeed, this is how universality arises. For any lens model which includes a given catastrophe, the leading terms in the Taylor expansion of the phase about the associated critical point will, after coordinate redefinitions, take the form of one of the ``normal forms" listed in Table \ref{tab:unfoldings}. Coordinate rescalings are one of the simplest such transformations, which have a simple effect on any model and on its Taylor expansion about any critical point. One can easily derive the scaling behavior of the intensity under such transformations of the lens model. For an $A_n$ catastrophe, for example, we may consider a set of lens models whose phase $\phi(x,0) \sim {a \over n+1} x^{n+1}$ near $x=0$, with $a$ a constant. By rescaling $x$ we can remove the $a$ dependence from the phase and hence infer that the intensity scales as $a^{-{2\over n+1}}$. Physically, decreasing $a$ means increasing the size of the lens, so it makes sense that the corresponding intensity grows. For a type $D_n$ catastrophes we may likewise have a phase in the form $a x_1^{n+1}\pm b x_1 x_2^2$. Again, the $a$ and $b$ dependence can be removed from the phase by rescaling $x_1$ and $x_2$. Hence one can infer that the intensity scales as $a^{-{1\over n-1}} b^{-1}$. (As an aside, note that for $D_4^+$ the normal form used here differs, from that given in Table \ref{tab:unfoldings}. The two forms may be shown to be equivalent under a linear coordinate transformation $x_i'=A_{ij} x_j,$ $i,j=1,2$, and hence are in the same equivalence class. In the rest of this paper we always use the normal forms listed in Table \ref{tab:unfoldings}.)

The pattern of fringes may likewise be shown to scale as 
\begin{align}
\Psi(\bm{\mu};\nu) = \nu^\beta \Psi\left(( \nu^{\sigma_1} \mu_1,  \dots,  \nu^{\sigma_K} \mu_K),\nu\right)\,,\label{eq:scaling}
\end{align}
where the fringe exponents $\sigma_i$,  defined by Berry \cite{Berry:1977}, are also listed in Table \ref{tab:exponents}.  The sum of the fringe exponents, $\gamma = \sum_{i=1}^K \sigma_i$, represents the scaling exponent for the $K$-dimensional hypervolume of the diffraction pattern known as the \textit{fringe index}. All of these exponents are invariant under diffeomorphisms, making them topological features.

While these catastrophes provide an exhaustive list, the precise forms in Table \ref{tab:unfoldings} are unlikely to arise, since apart from $A_1^\pm$, they are completely de-localized, with the strength of the lens diverging away from the critical point. Later in this paper we shall consider more realistic, localized lenses which generate catastrophes within them of the listed form. In the vicinity of such a catastrophe, one can expect the behaviors indicated in Table \ref{tab:exponents} to hold. Note, however, that to compute the maximum intensity at such a catastrophe, one must first redefine the coordinate $\bm{x}$ to put the exponent locally into the normal form of the catastrophe listed in Table \ref{tab:unfoldings} (this is guaranteed to be possible by the theorems mentioned above), and take into account the associated Jacobian factor when evaluating the integral.  

\begin{figure}
\center
\begin{tikzpicture}
\node (A1) {$A_1$};
\node [right of=A1] (A2) {$A_2$};
\node [right of=A2] (A3) {$A_3$};
\node [right of=A3] (A4) {$A_4$};
\node [right of=A4] (A5) {$A_5$};
\node [below of=A4] (D4) {$D_4$};
\node [below of=A5] (D5) {$D_5$};
  \draw[<-] (A1) to  (A2);
  \draw[<-] (A2) to  (A3);
  \draw[<-] (A3) to  (A4);
  \draw[<-] (A4) to  (A5);
  \draw[<-] (D4) to  (D5);
  \draw[<-] (A3) to  (D4);
  \draw[<-] (A4) to  (D5);
\end{tikzpicture}
\caption{The unfolding diagram of the seven elementary catastrophes.}\label{fig:unfoldingDiagram}
\end{figure}
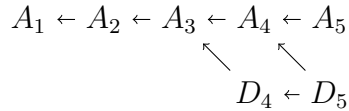

The seven elementary catastrophes listed above form an intricate hierarchy which unfold under perturbations according to the unfolding diagram (see Fig.~\ref{fig:unfoldingDiagram}). As we saw before, the fold caustic ($A_2$) splits under a small perturbation into a maximum and a minimum each corresponding to an $A_1$. Analogously, butterfly caustic ($A_5$) unfolds into a swallowtail caustic ($A_4$), which in its turn unfolds into the cusp ($A_3$) and the fold caustic ($A_2$). The parabolic umbilic caustic ($D_5$) has a more intricate structure as it can unfold into both the elliptic ($D_4^-$), hyperbolic umbilic ($D_4^+$) and the swallowtail caustic ($A_4$). The elliptic ($D_4^-$) and hyperbolic umbilic caustic ($D_4^+$) always unfold into cusp caustics ($A_3$).

\section{Example: a one-dimensional lens}\label{sec:1DLensExample}
Picard-Lefschetz theory, as described for example in \cite{Feldbrugge:2017kzv}, enables us to deform the real integration domain $\mathbb{R}^N$ of the Fresnel-Kirchhoff integral \eqref{eq:amp} into a set of Lefschetz thimbles, {\it i.e.}, steepest descent contours, $\mathcal{J}_i \subset \mathbb{C}^N$ each corresponding to a relevant saddle point $\bar{x}_i$,
\begin{align}
\Psi(\bm{\mu};\nu) = \left(\frac{\nu}{\pi}\right)^{N/2}  \sum_{i} \int_{\mathcal{J}_i} e^{i\phi(\bm{x};\bm{\mu})\nu}\mathrm{d}\bm{x}\,.
\label{eq:thimbles}
\end{align}
The exponent $i\phi(\bm{x};\bm{\mu})\nu$ evaluated along a steepest descent contour $\mathcal{J}_i$   has a constant imaginary part while its real part $h=\text{Re}[i\phi(\bm{x};\bm{\mu})\nu]$ is monotonically decreasing. As a consequence, the conditionally convergent oscillatory integral is transformed into a sum of convex integrals. This is remarkable, as the originally conditionally convergent integral is generally sensitive to regularization and also, if the integral is performed iteratively, to the order in which the partial integrals are taken (see Appendix \ref{ap:DefiningOscillatoryIntegrals} for an instructive example). The integrals over Lefschetz thimbles have none of these ambiguities. It is for this reason that we will interpret the integral over the sum of Lefschetz thimbles as the {\it definition} of the integral taken over the real integration domain. Once we have identified the correct set of thimbles, we can use conventional numerical methods to evaluate the integral on each thimble.

We shall describe two distinct methods to obtain the sum of Lefschetz thimbles corresponding to the Fresnel-Kirchhoff integral for a one-dimensional lens. In the first method, we follow the techniques explained, for example, in Ref.~\cite{Feldbrugge:2017kzv}. We start by computing all the saddle points of the exponent $i \phi(\bm{x};\bm{\mu}) \nu$ and their corresponding steepest descent and ascent contours . We subsequently study the intersections of the steepest ascent contours with the original integration domain to find the relevant saddle points and associated steepest descent contours $\mathcal{J}_i$. This method is well suited to the Picard-Lefschetz analysis of one-dimensional integrals, for which we can plot the steepest descent and ascent contours in the complex plane $\mathbb{C}$.

In the second method, we instead flow the real integration domain along the downward flow of the real part of the exponent,  $h=\text{Re}[i\phi(\bm{x};\bm{\mu})\nu]$. We show that this downward flow terminates on the correct sum of Lefschetz thimbles $\sum_i \mathcal{J}_i$. The relevant saddle points are given by the local maxima of $h$ restricted to this thimble sum $\mathcal{J}$. Note that this second scheme is completely determined by the gradient of the $h$ with respect to the real and imaginary parts of the complexified coordinates $\bm{x}$. We do not need to find all the saddle points nor evaluate the corresponding steepest ascent and descent contours. We moreover do not need to study the intersections of the steepest ascent contours with the original integration domain. Any Stokes transitions are automatically taken care of. This method is furthermore ideally suited to higher-dimensional oscillatory integrals, where the steepest ascent and descent contours are expensive to evaluate and the intersections are computationally difficult to find. 

\subsection{Geometric optics approximation}\label{sec:1DLens}
\begin{figure}
\centering
\includegraphics[width=0.6\textwidth]{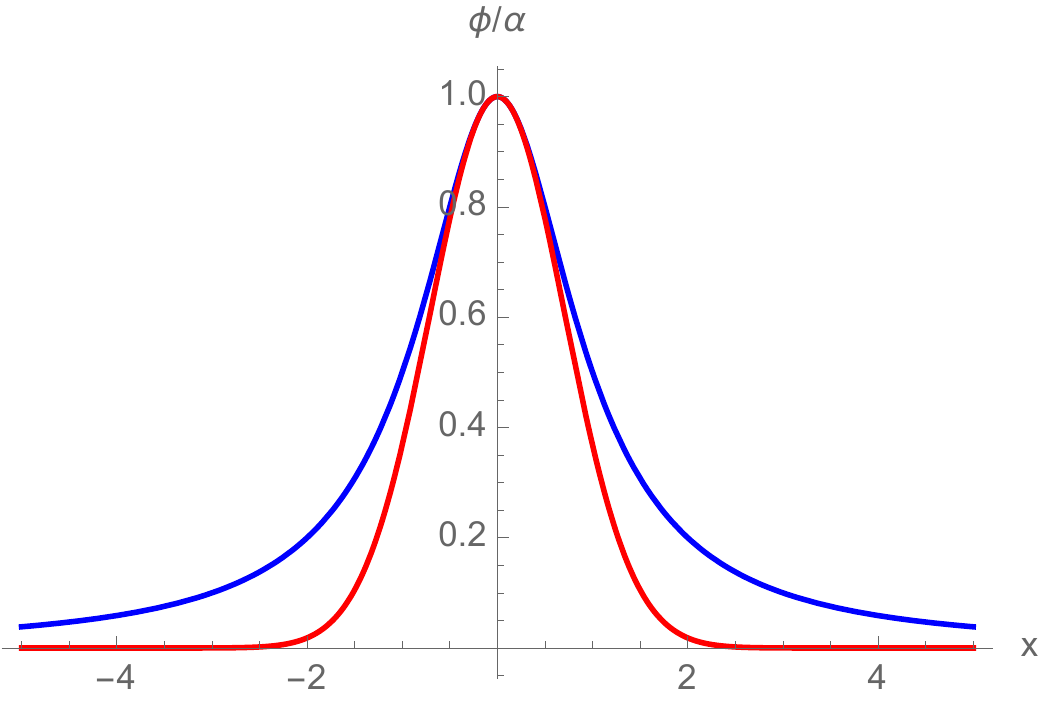}
\caption[A comparison of the Gaussian and the rational lens.]{A comparison of the Gaussian lens (red) and the rational approximation (blue).}
\label{fig:GaussianRational}
\end{figure}
In the introduction to this paper, we discussed a one-parameter family of one-dimensional localized lenses
\begin{align}
\varphi(x) = \frac{\alpha}{1+x^2}\,,
\label{localized1}
\end{align}
with $\alpha \in \mathbb{R}$. For plasma lenses, the parameter $\alpha$ follows the dispersion relation $\alpha \propto \omega^{-2}$ with $\omega$ the angular frequency of the source. The longer the wavelength, the stronger the lens. We restrict our analysis to rational lenses for two reasons:
\begin{enumerate}
\item their analytic continuation into the complex $x$-plane does not contain branch-cuts and consists of only a finite number of poles,
\item the phase $\phi$ has only a finite number of saddle points and corresponding steepest-descent contours. 
\end{enumerate} 
Picard-Lefschetz theory, however, applies to analytic lenses in general. The lens (\ref{localized1}) is a rational approximation to the Gaussian lens
\begin{align}
\varphi(x) = \alpha e^{-x^2}\,,
\end{align}
which has an essential singularity at infinity on the Riemann sphere and an infinite number of saddle points in the complex plane. See Fig.~\ref{fig:GaussianRational} for a comparison between the two lenses. It is a wonderful fact that many real-valued functions with intricate structure in the complex plane, can be well-approximated with a Pad\'e approximation, whose analytic continuation possesses only a finite number of poles.

As we derived in Section \ref{sec:geometricOptics}, the Lagrangian map $\xi$ of the rational lens $\varphi$,
\begin{align}
\xi(x) = x -  \frac{\alpha x}{(1+ x^2)^2}\,,
\end{align}
forms caustics at the real roots of the second order derivative of the exponent
\begin{align}
\frac{\partial^2 \phi(x)}{\partial x^2} = 2 \frac{\partial \xi(x)}{\partial x} = 2 + 2\alpha\frac{3x^2-1}{(1+x^2)^3} = 0\,.
\end{align}
\begin{figure}
\centering
\begin{subfigure}[b]{0.48\textwidth}
\includegraphics[width=\textwidth]{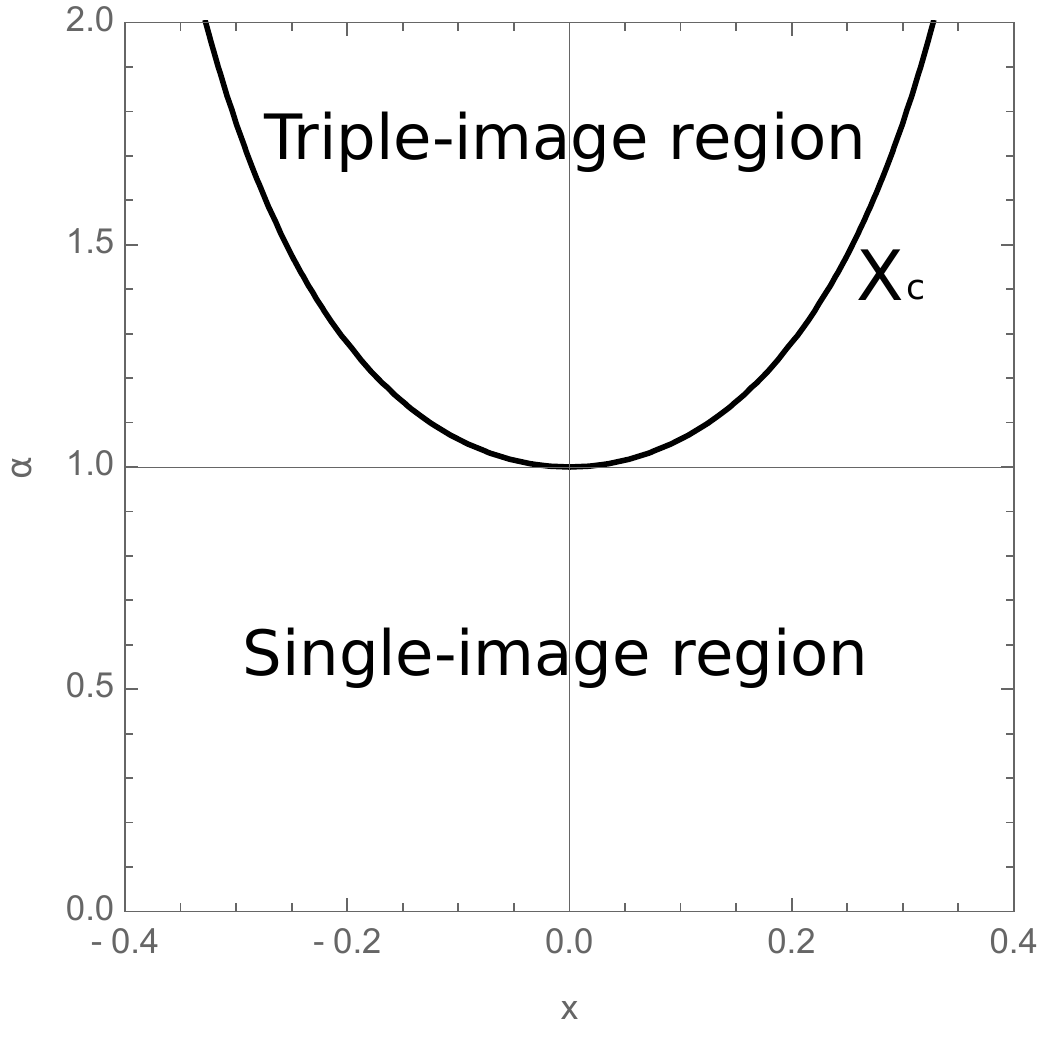}
\caption{Caustic in the $x$-$\alpha$ plane.}
\end{subfigure} ~
\begin{subfigure}[b]{0.48\textwidth}
\includegraphics[width=\textwidth]{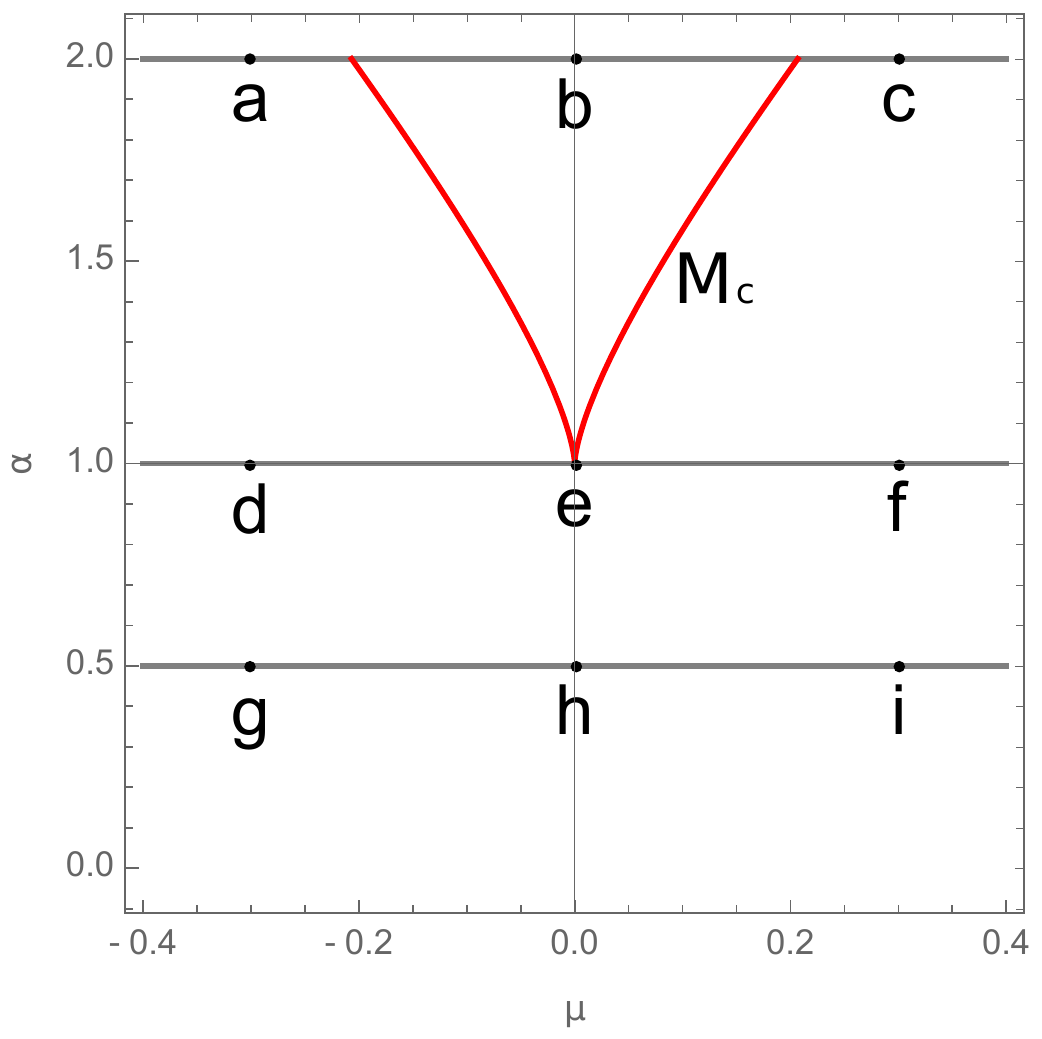}
\caption{Caustic in the $\mu$-$\alpha$ plane.}
\end{subfigure}
\caption[The caustics of the one-dimensional rational lens.]{The caustics of the one-dimensional rational lens in the $x$-$\alpha$ and the $\mu$-$\alpha$ plane. The points in the \textit{left panel} correspond to the Picard-Lefschetz diagrams in Fig.~\ref{fig:1DThimblesFold}. The lines correspond to the panels in Fig.~\ref{fig:1DIntensityAlpha2}. }\label{fig:1DCausticPlane}
\end{figure}

See Fig.~\ref{fig:1DCausticPlane} for the caustic surface in the $x$-$\alpha$ and the $\mu$-$\alpha$ planes.
For $\alpha <1$ no such real root exists. The lensed image consists of a single-image region. For $\alpha = 1$ there is a single real-valued root at $x_c = 0$ with multiplicity two and the corresponding point $\mu_c = \xi(x_c)=0$ in the parameter space $M$. In the $\mu$-$\alpha$ plane, this point is non-differentiable on the caustic set. This is an example of a cusp caustic. For $\alpha > 1$ there are two symmetric real roots. These are examples of fold caustics. For further reference, for $\alpha = 2$, the two caustics are located at
\begin{align}
X_c = \{- 0.327334 \dots, 0.327334 \dots\}\,,
\end{align}
in the base space $X=\mathbb{R}$. In the parameter space $M$ the caustic appears at
\begin{align}
M_c = \xi(X_c) = \{+ 0.206751\dots, - 0.206751\dots\}\,.
\end{align}

\begin{figure}
\centering
\begin{subfigure}[b]{0.32\textwidth}
\includegraphics[width=\textwidth]{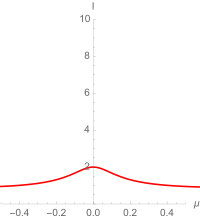}
\caption{$\alpha=1/2$}
\end{subfigure} 
\begin{subfigure}[b]{0.32\textwidth}
\includegraphics[width=\textwidth]{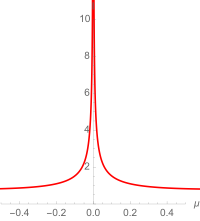}
\caption{$\alpha=1$}
\end{subfigure} 
\begin{subfigure}[b]{0.32\textwidth}
\includegraphics[width=\textwidth]{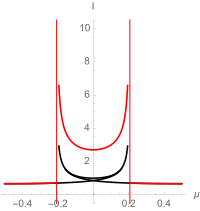}
\caption{$\alpha=2$}
\end{subfigure} 
\caption[The intensity map in geometric optics of the one-dimensional lens \ref{localized1}]{The normalized intensity in geometric optics, $I(\mu;\infty)$ is plotted as a function of $\mu$.}\label{fig:IGeometricOptics}
\end{figure}

The relative normalized intensity of the lens in the geometric optics limit (see Section \ref{sec:geometricOptics}) is plotted in figure \ref{fig:IGeometricOptics}. For $\alpha=1/2$, the lens does not form a caustic. The normalized intensity map is finite. For $\alpha=1$, we see a cusp caustic at $\mu=0$. For $\alpha=2$, we observe two fold caustics at $M_c$ enclosing a triple-image region. The black curves in the triple-image region are the three contributions corresponding to the three images. The red curve is the sum over the multi-image regions.

\subsection{Finding the thimbles}

We now turn to evaluating the full expression (\ref{eq:thimbles}). First, we need to determine which Lefschetz thimbles contribute. We shall describe two distinct methods, the second of which is more efficient for numerical purposes.

\subsubsection{Method 1: following steepest ascent contours}
The exponent 
\begin{align}
i \phi(x;\mu) \nu = i \left[(x-\mu)^2 + \frac{\alpha}{1+x^2}\right]\nu
\end{align}
is imaginary for real $\mu$ and $x$. Its analytic continuation has two poles at $x = \pm i$, and five saddle points in the complex $x$-plane, satisfying 
\begin{align}
\frac{\partial \phi(x;\mu)}{\partial x} = 2(x-\mu) - \frac{2\alpha x}{(1+x^2)^2} =0\,.
\end{align}

For the Picard-Lefschetz analysis we start by writing the analytic continuation of the exponent in terms of its real and imaginary part
\begin{align}
i \phi(\bm{x};\bm{\mu}) \nu = h(\bm{u}+i\bm{v};\mu) + i H(\bm{u}+i\bm{v};\mu)\,,
\end{align}
with the complex expansion $\bm{x}=\bm{u}+i\bm{v}$ and the real-valued functions $h,H$. For generality, we describe the flow of the integration contour in $N$ dimensions. The real part $h$ is, in the Picard-Lefschetz analysis, known as the $h$-function. The downward flow of the $h$-function $\gamma_\lambda:\mathbb{C}^N\to \mathbb{C}^N$ is defined by
\begin{align}
\frac{\partial \gamma_\lambda(\bm{z})}{\partial \lambda} = -\nabla_{\bm{u}+i\bm{v}} h[\gamma_\lambda(\bm{z})]
\end{align}
with the boundary condition $\gamma_0(\bm{z})=\bm{z}\in \mathbb{C}^N$, the parameter $\lambda$ in a subset of $\mathbb{R}$, and the complex gradient defined as
\begin{align}
\nabla_{\bm{u}+i\bm{v}} h = \nabla_{\bm{u}} h + i \nabla_{\bm{v}} h\,.
\end{align}
Note that in defining the gradient, we have assumed a corresponding metric on the space $\mathbb{C}^N$. In this paper we will always associate $\mathbb{C}^N$ with $\mathbb{R}^{2N}$ and us the corresponding Euclidean metric. We are of course free to consider different metrics. Given the saddle points we can compute the steepest ascent and descent contours and intersect the ascent contours with the real axis, to obtain the relevant saddle points.

\begin{figure}
\centering
\begin{subfigure}[b]{0.31\textwidth}
\includegraphics[width=\textwidth]{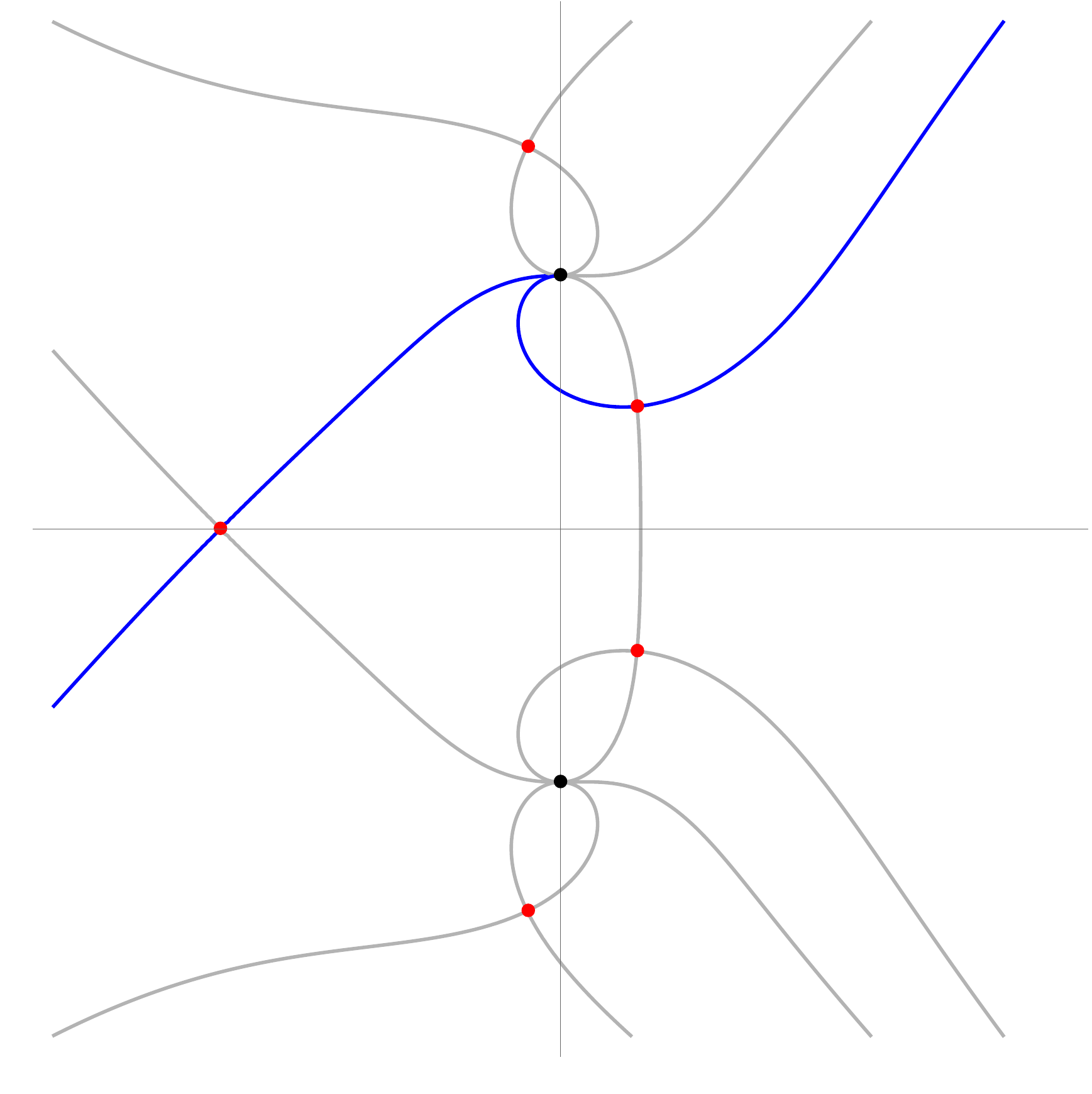}
\caption{$\alpha =2,\mu < -\mu_c$}
\end{subfigure} ~
\begin{subfigure}[b]{0.31\textwidth}
\includegraphics[width=\textwidth]{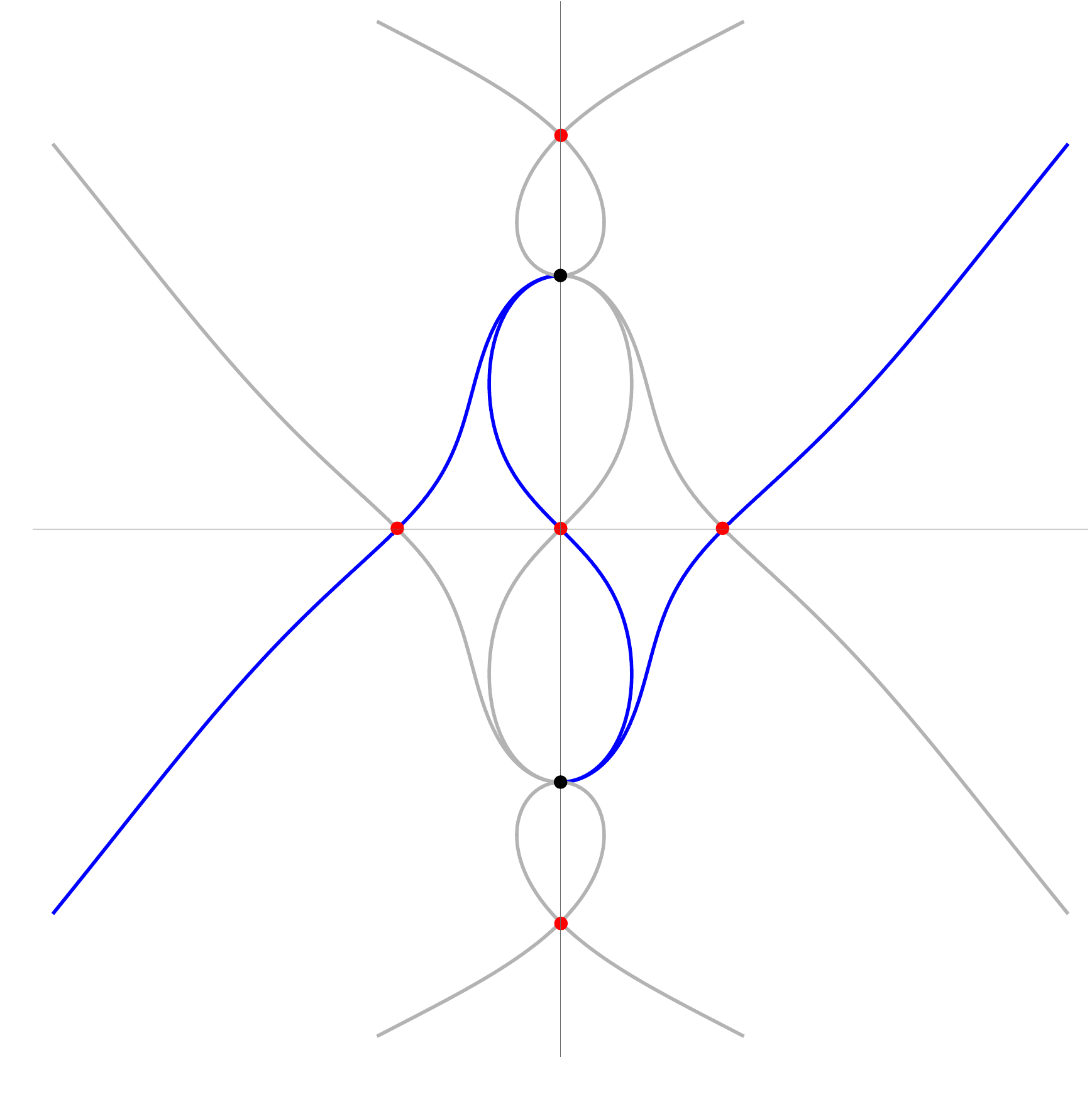}
\caption{$\alpha =2,-\mu_c < \mu < \mu_c$}
\end{subfigure} ~
\begin{subfigure}[b]{0.31\textwidth}
\includegraphics[width=\textwidth]{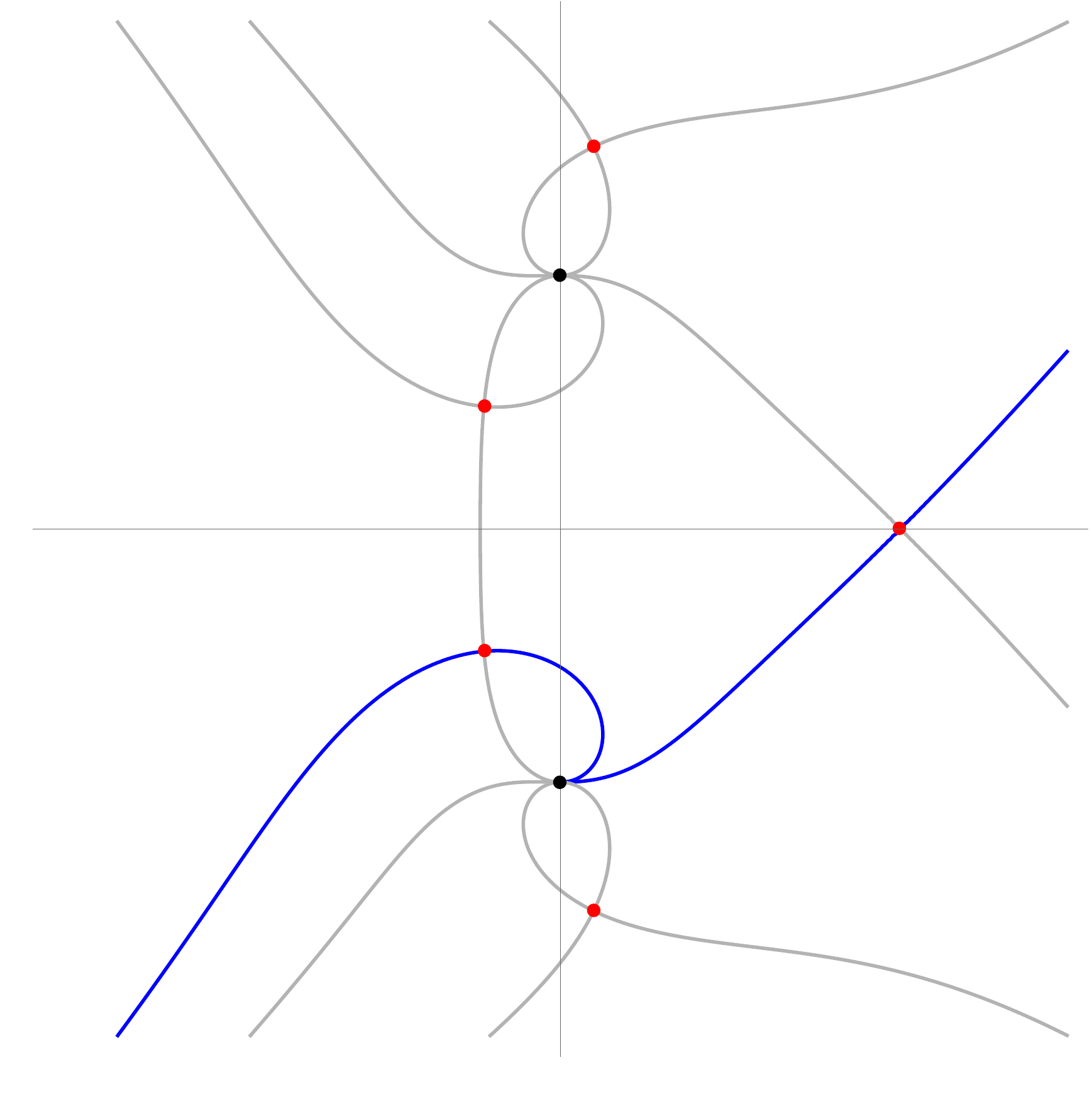}
\caption{$\alpha =2,\mu > \mu_c$}
\end{subfigure}
\begin{subfigure}[b]{0.31\textwidth}
\includegraphics[width=\textwidth]{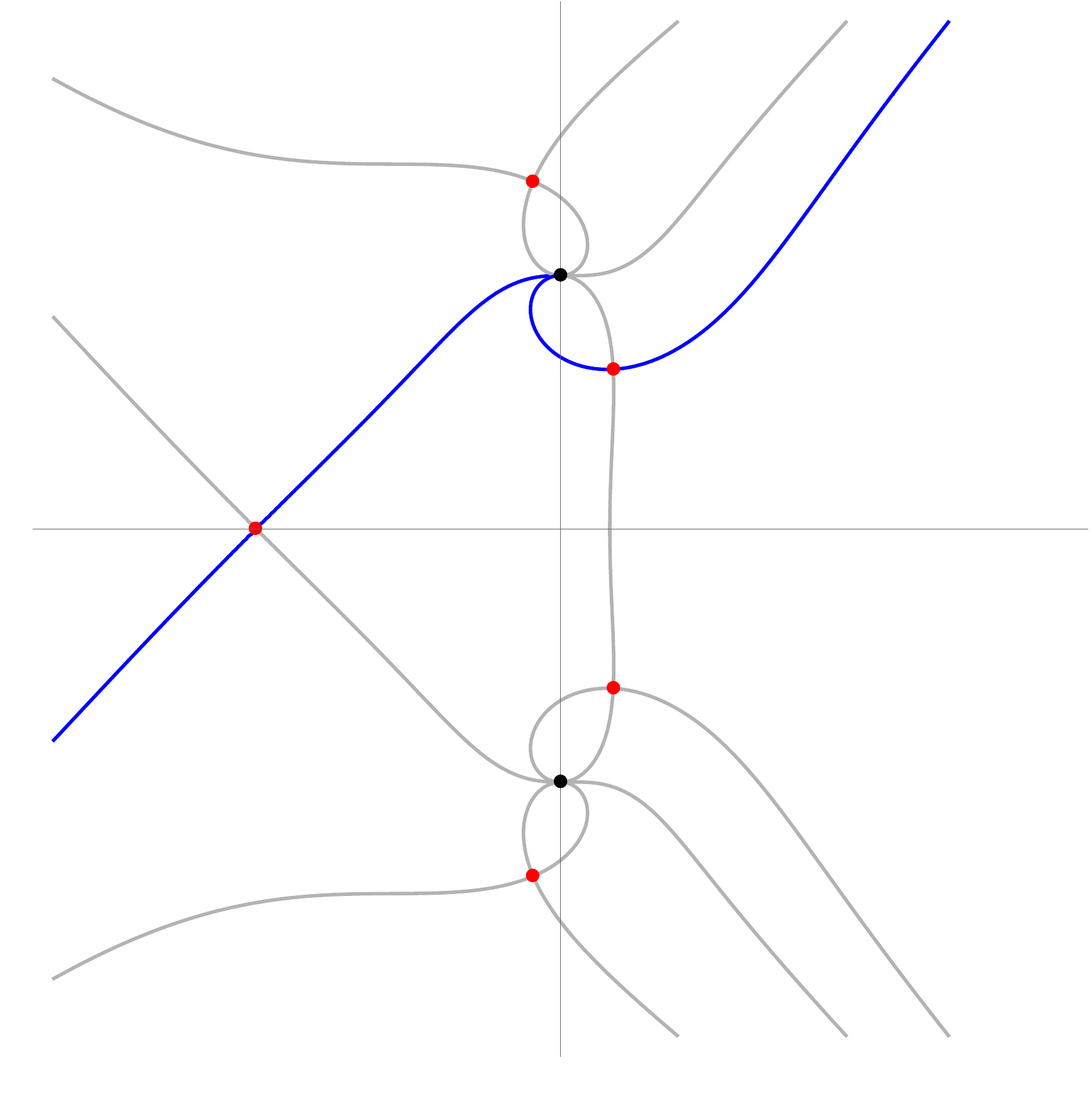}
\caption{$\alpha =1,\mu < -\mu_c$}
\end{subfigure} ~
\begin{subfigure}[b]{0.31\textwidth}
\includegraphics[width=\textwidth]{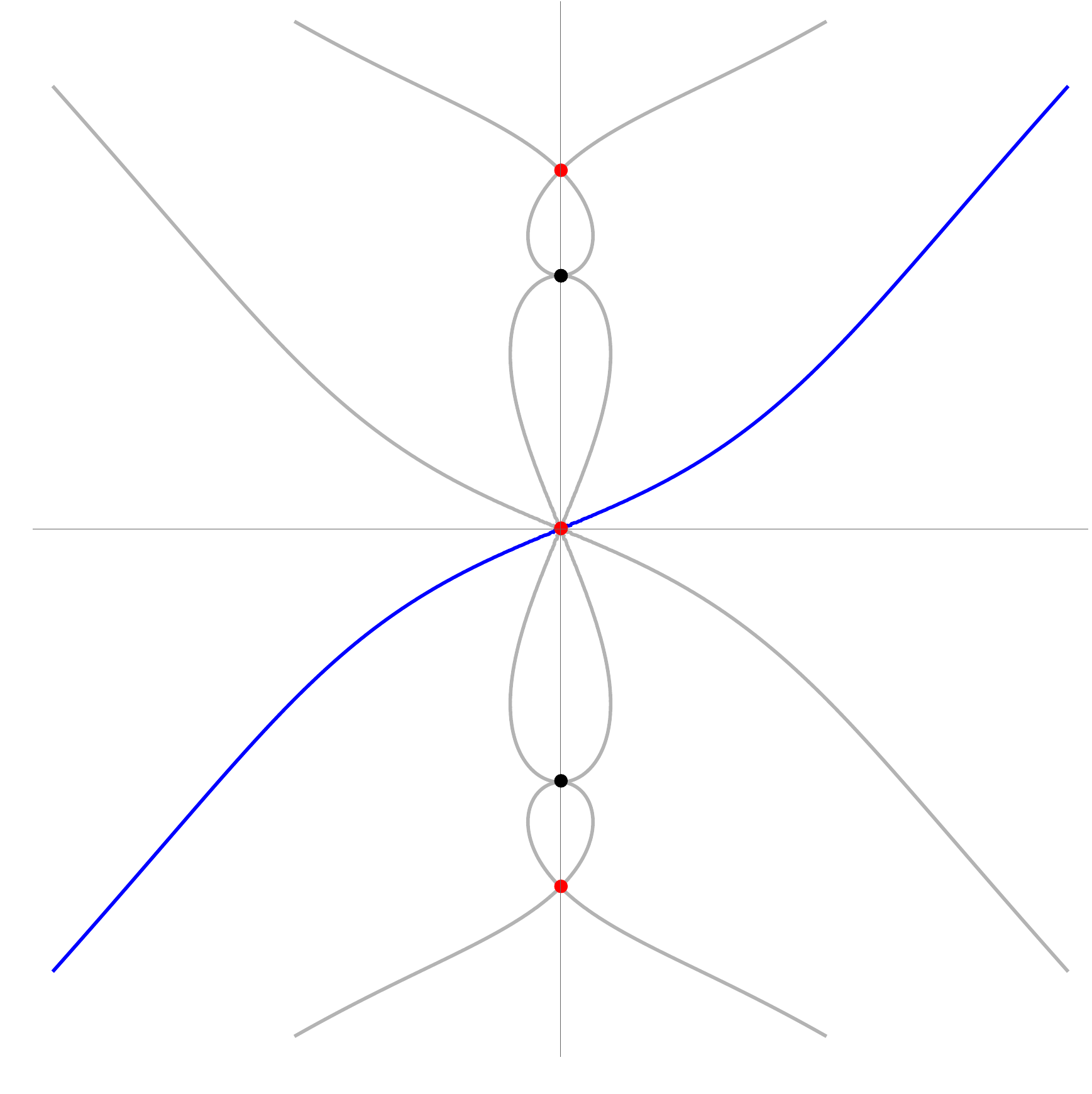}
\caption{$\alpha =1,\mu= \mu_c$}
\end{subfigure} ~
\begin{subfigure}[b]{0.31\textwidth}
\includegraphics[width=\textwidth]{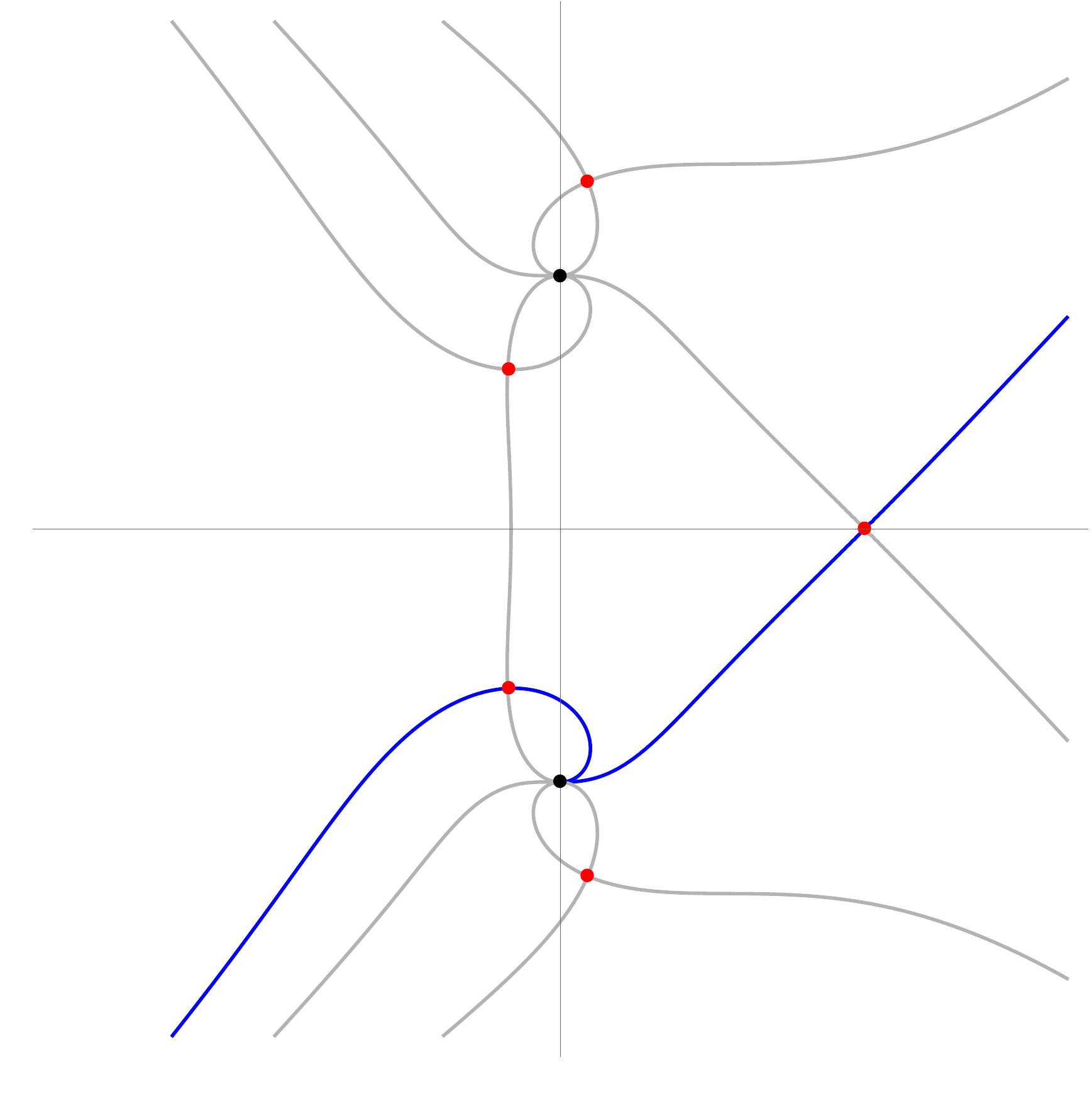}
\caption{$\alpha =1,\mu > \mu_c$}
\end{subfigure}
\begin{subfigure}[b]{0.31\textwidth}
\includegraphics[width=\textwidth]{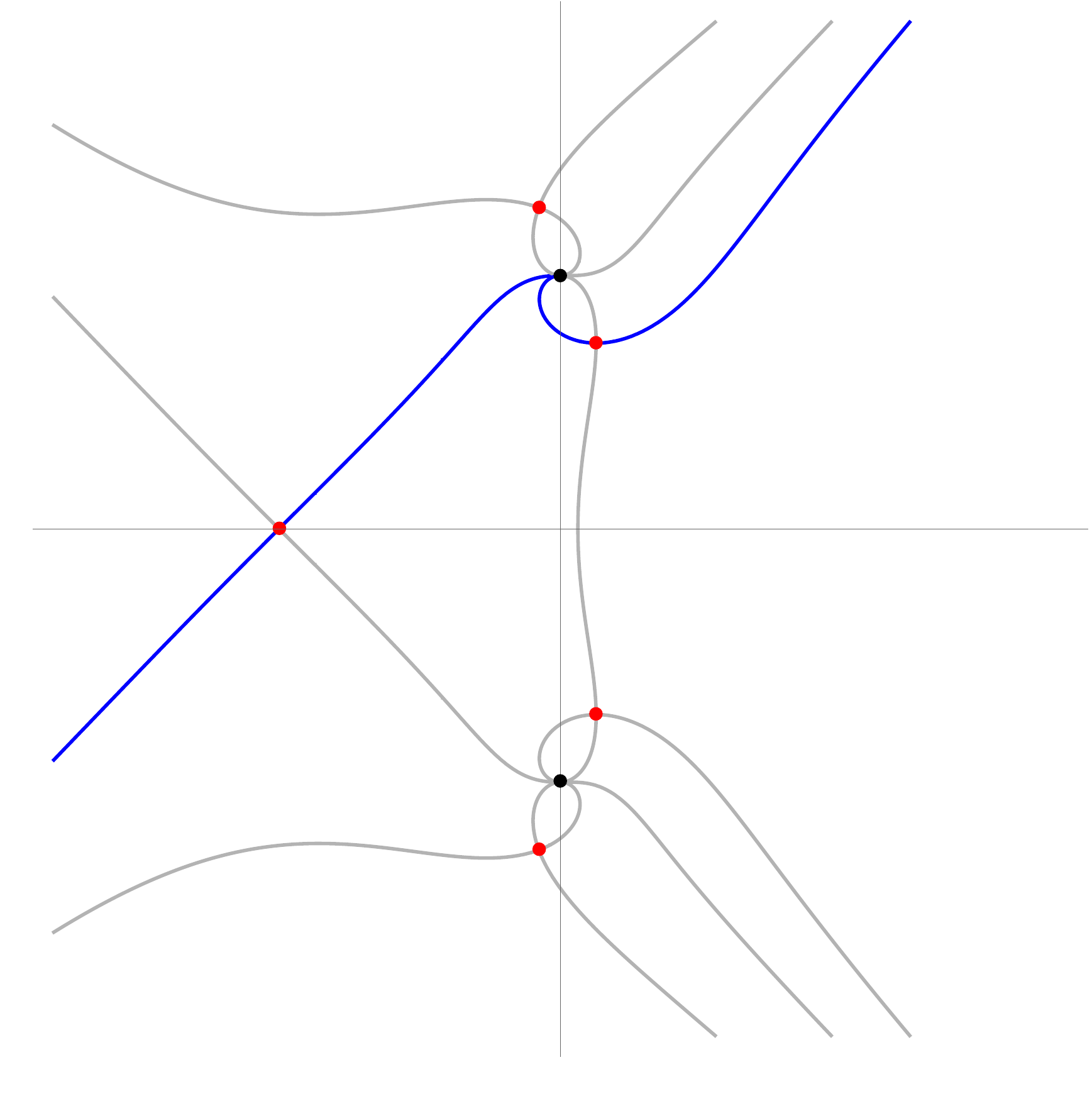}
\caption{$\alpha =1/2,\mu < 0$}
\end{subfigure} ~
\begin{subfigure}[b]{0.31\textwidth}
\includegraphics[width=\textwidth]{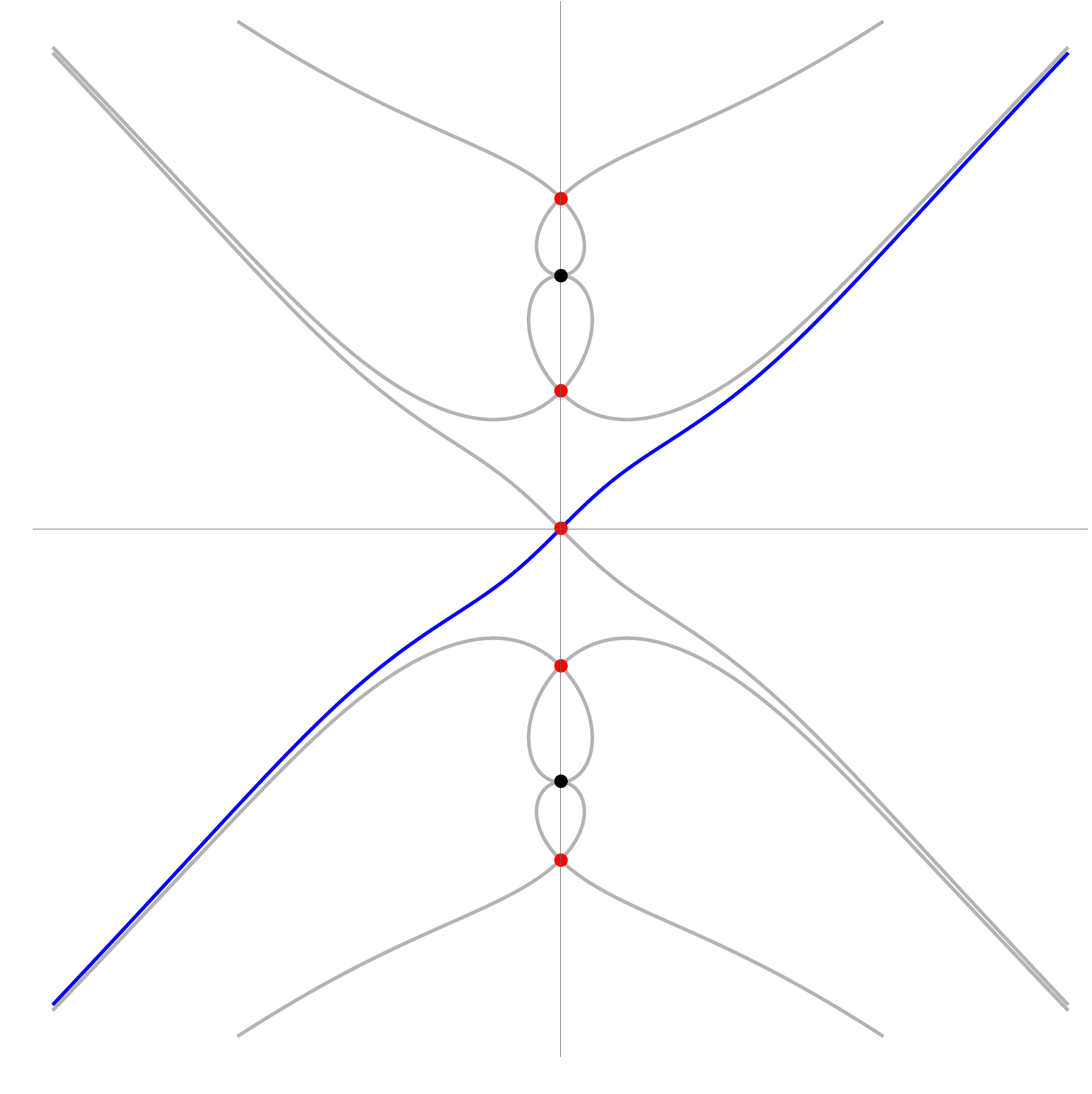}
\caption{$\alpha =1/2,\mu= 0$}
\end{subfigure} ~
\begin{subfigure}[b]{0.31\textwidth}
\includegraphics[width=\textwidth]{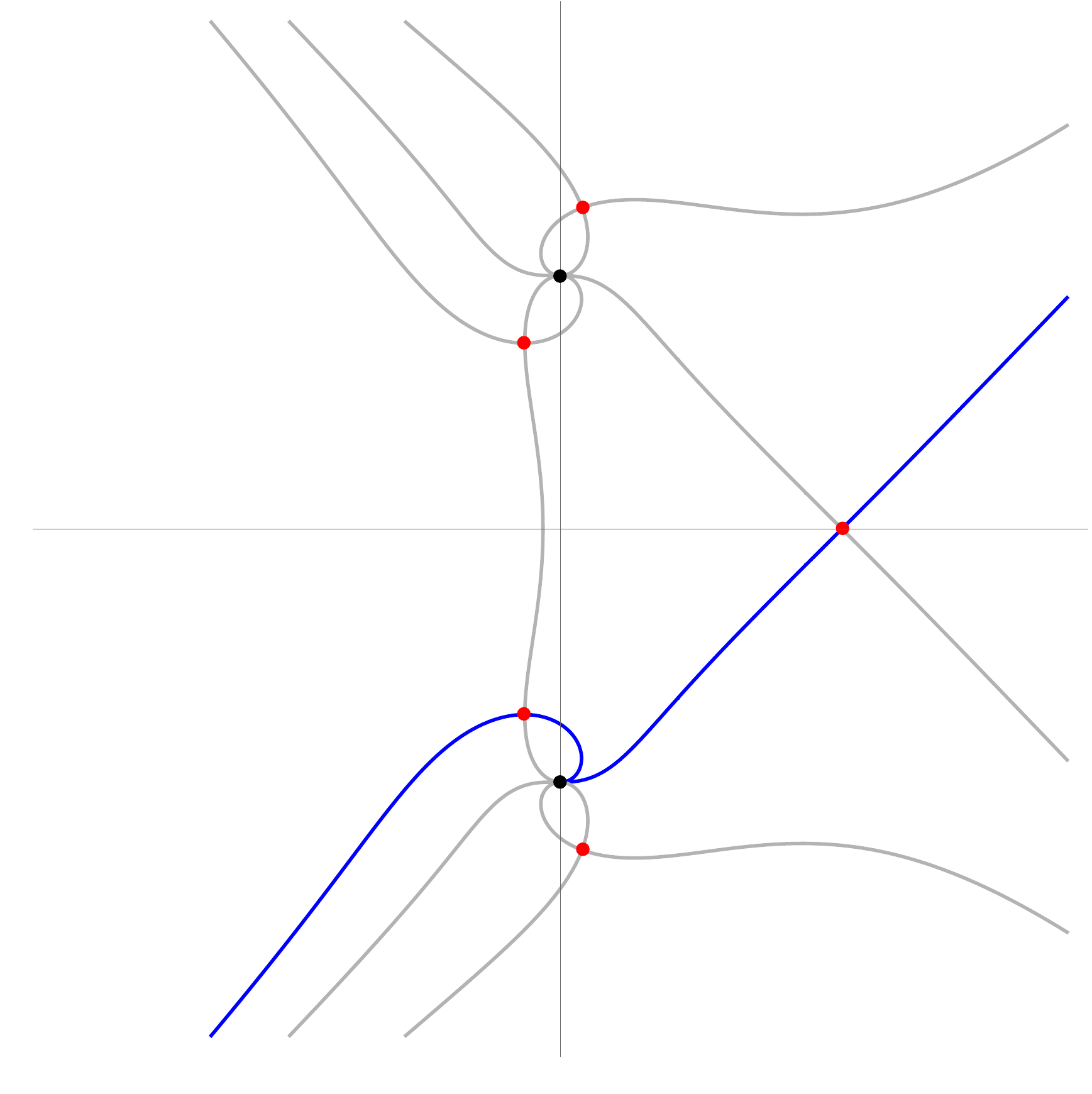}
\caption{$\alpha =1/2,\mu > 0$}
\end{subfigure}
\caption[The Picard-Lefschetz thimbles for the one-dimensional lens.]{The Picard-Lefschetz thimbles for $\alpha=1/2,1,2$ as a function of $\mu$. The red and black points are the saddle points and poles. The curves are paths of steepest descent and ascent. The blue ones are relevant, while the grey ones are irrelevant.}\label{fig:1DThimblesFold}
\end{figure}

Depending on $\mu$ and $\alpha$ either one or three of the saddle points are real-valued. The lens thus has both single- and triple-image regions. See Fig.~\ref{fig:1DThimblesFold} for the five saddle points $\bar{x}_i$ and the corresponding steepest ascent and descent contours. By intersecting the steepest ascent contours with the real line, we obtain the Lefschetz thimble (plotted in blue). The thimbles run from $x=- \infty$ to $x=+\infty$, while passing through the poles at $x = \pm i$.

From the caustic structure in Fig.~\ref{fig:1DCausticPlane} we can distinguish three regimes:
\begin{itemize}
\item
In the regime $\alpha < 1$, the lens forms a single image. The corresponding Picard-Lefschetz analysis yields a single real-valued saddle point. For large $|\mu|$ there is, in addition, a relevant complex saddle point. When $|\mu|$ decreases to $0$, the complex saddle point becomes irrelevant due to a Stokes transition. This phenomenon is discussed in detail in the next section. For $\mu=0$, only the real saddle point is relevant. Note that the thimble can for all $\mu$ be deformed to the original integration domain $\mathbb{R}$. See the lower panels of Fig.~\ref{fig:1DCausticPlane}.
\item 
For $\alpha =1$, the lens contains a cusp caustic at $ \mu_c = 0$. For $\mu \neq \mu_c$, the Picard-Lefschetz analysis is similar to the $\alpha <1$ regime. The thimble passes through one real-valued and one complex-valued saddle point. At the caustic $\mu=\mu_c$, three non-degenerate saddle points merge forming a degenerate saddle point. This is the signature of the cusp caustic, whose normal form is the quartic function $x^4$. See the middle panels of Fig.~\ref{fig:1DCausticPlane}.
\item 
In the regime $\alpha >1$, the Picard-Lefschetz analysis splits into three intervals. In the single-image region, \textit{i.e.}, $\mu$ in $(-\infty, -\mu_c)$ or $(\mu_c,\infty)$, the Picard-Lefschetz analysis consists of two relevant saddle points; one real and one complex. At the caustic, the complex saddle point approaches the real line and merges with its complex conjugate saddle point. This is the signature of the fold caustic. In the triple-image region, \textit{i.e.}, $\mu \in (-\mu_c,\mu_c)$, the analysis consists of three real-valued relevant saddle points. See the upper panels of Fig.~\ref{fig:1DCausticPlane}.
\end{itemize}

\subsubsection{Method 2: flowing the integration domain}
We can alternatively obtain the Lefschetz thimble $\mathcal{J}$ by flowing the original integration domain $\mathbb{R}$ along the downward flow of the real part $h$. 

Given the downward flow $\gamma_\lambda$ for general points $z\in \mathbb{C}$, we flow the original integration domain $X$ to
\begin{align}
X_\lambda = \gamma_\lambda(X) \subset \mathbb{C}\,.
\end{align}
The steepest descent contours $\mathcal{J}_i$ corresponding to the saddle points $\bar{x}_i$ are the fixed points of the flow, \textit{i.e.},
\begin{align}
\gamma_\lambda(\mathcal{J}_i) = \mathcal{J}_i
\end{align}
for all $\lambda$. When the $h$-function has saddle point in the complex plane, it follows from Morse-Smale theory \cite{Morse:1925, Milnor:1963} that the flowed contour $X_\lambda$ will converge to a set of steepest descent contours $\mathcal{J}_i$ as $\lambda \to \infty$. Since $X_\lambda$ is a continuous deformation of the original integration domain $X$, it follows that $X_\lambda$ converges to the Lefschetz contour, \textit{i.e.},
\begin{align}
\lim_{\lambda \to \infty} X_\lambda = \mathcal{J}\,.
\end{align}

\begin{figure}
\centering
\begin{subfigure}[b]{0.32\textwidth}
\includegraphics[width=\textwidth]{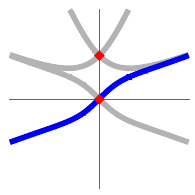}
\end{subfigure} 
\begin{subfigure}[b]{0.32\textwidth}
\includegraphics[width=\textwidth]{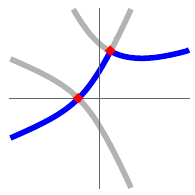}
\end{subfigure} 
\begin{subfigure}[b]{0.32\textwidth}
\includegraphics[width=\textwidth]{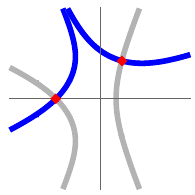}
\end{subfigure}
\caption[Illustration of the Stokes phenomenon.]{The Stokes phenomenon at which the steepest descent contours coincide and the relevant saddle points change. In the \textit{left} panel, we have one real relevant saddle point. In the \textit{central} panel, we see a Stokes phenomenon. The Lefschetz thimble passes to both a real and a complex saddle point. In the \textit{right} pane, we have one real and one complex relevant saddle point. The Stokes phenomenon occurs when the steepest descent contour of a saddle point connects with another saddle point.}\label{fig:StokesPhenomenon}
\end{figure}

When we perform the flow of the original integration domain as a function of the parameter $\mu$, we obtain a family of thimbles. The thimble generally changes smoothly as a function of $\mu$. There are however two ways in which the Picard-Lefschetz structure of the integral can abruptly change its geometry:
\begin{enumerate}
\item If for some $\bm{\mu}$, a few non-degenerate saddle points merge to form a higher order saddle point, the number of relevant critical points will change. At these points, the integral $\Psi(\bm{\mu};\nu)$ forms a caustic. This phenomenon can be described by catastrophe theory (see sections \ref{sec:catastropheTheoryOptics} and \ref{sec:catastrophePL}).
\item When the imaginary part $H$ evaluated in two saddle points coincides for some parameter $\bm{\mu}_s$, the two corresponding steepest-descent contours can coincide. At such a parameter $\bm{\mu}_s$, the Lefschetz thimbles flip changing the number of relevant saddle points (see Fig.~\ref{fig:StokesPhenomenon} for an illustration). This is known as a Stokes transition. The parameters $\bm{\mu}_s$ for which this happens form so-called  Stokes lines.
\end{enumerate}
We study both phenomena in detail in the Section \ref{sec:catastrophePL}.

We numerically evaluate the flow $X_\lambda$ by approximating $X$ by a set of line-segments and flowing the endpoints. Since the real part of the analytic continuation of an analytic function does not have local extrema (this follows from the Cauchy-Riemann equation), all points $z\in \mathbb{C}$ flow to poles as $\lambda \to \infty$. The limit $\lim_{\lambda \to \infty} X_\lambda $ should not be interpreted as a pointwise limit. We, for this reason, trace the length of the line-segments and add points when neighboring points move too far apart. We moreover remove line-segments in the neighborhoods of the poles of the $h$-function. The contour $X_\lambda$ has converged to the thimble when the imaginary part $H$ is approximately constant along the line-segments. 

This idea is implemented by the algorithm:

\begin{algorithm}
\begin{algorithmic}
\REQUIRE Represent a subset $[a,b]$ of the original integration domain $X=\mathbb{R}$ by the regular lattice $p_i = a + i \Delta x$ with $\Delta x = \frac{b-a}{n}$ for some $n\in \mathbb{Z}_{>0}$, and the line-segments $(p_0,p_1), (p_1,p_2),\dots, (p_{n-1},p_n)$.
\WHILE{the variance of the imaginary part $H$ on the points $p_i$ exceeds threshold $T_1$}
\STATE flow the points: $p_i \mapsto p_i - \nabla h(p_i) \Delta t$

\IF{the $h$-function evaluated in the point $p_i$ is smaller than the threshold $T_2$}
\STATE remove the corresponding line segments
\ENDIF

\IF{the length of the line-segments $(p_i,p_{i+1})$ exceeds the threshold $T_3$}
\STATE split the line segment into the two lines $\left(p_i,\frac{p_i + p_{i+1}}{2}\right), \left(\frac{p_i + p_{i+1}}{2}, p_{i+1}\right)$.
\ENDIF

\ENDWHILE
\end{algorithmic}
\caption{The flow of the contour of one-dimensional oscillatory integrals.}
\end{algorithm}

with the parameters $a,b,T_1,T_2,T_3 \in \mathbb{R}$, and $n \in \mathbb{Z}_{>0}$.

\begin{figure}
\centering
\includegraphics[width=0.6\textwidth]{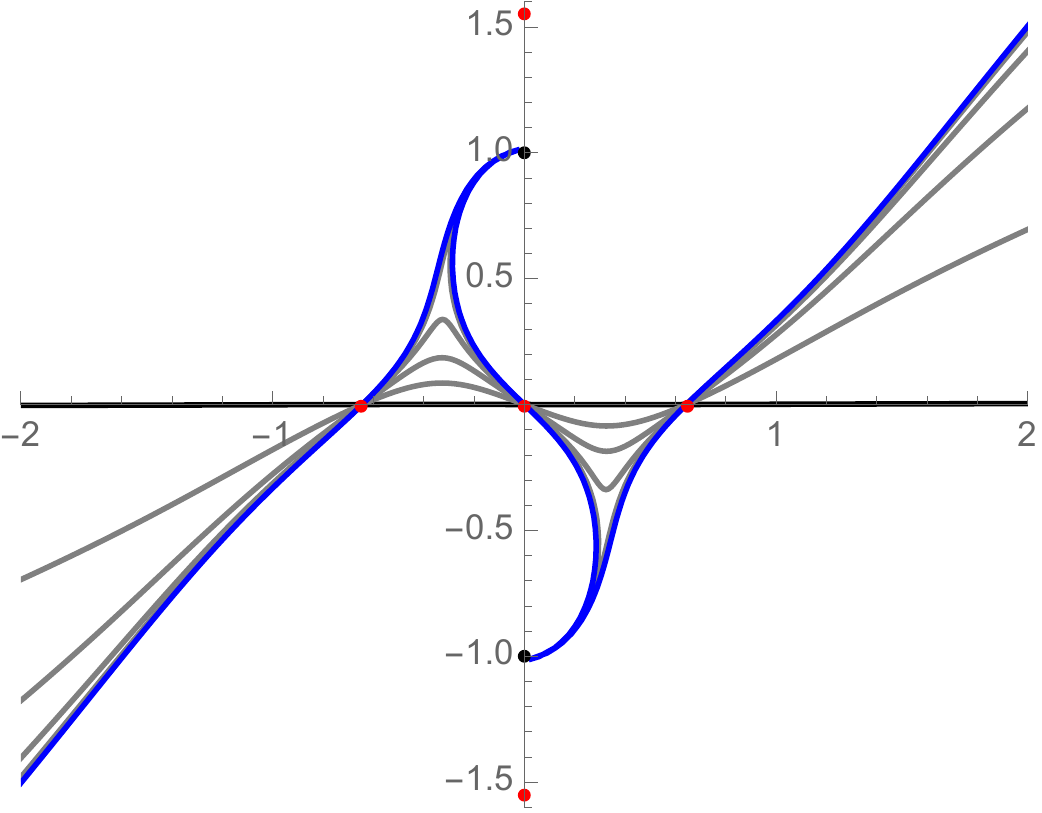}
\caption[The downward flow of the integration domain.]{The downward flow of the integration domain. The contour $X_\lambda$ for $\lambda =0,0.2,0.4,0.6,0.8,1$ is plotted by the black, grey and blue curves. The five saddle points are plotted in red and the poles are plotted in black.}\label{fig:FlowThimble}
\end{figure}

See Fig.~\ref{fig:FlowThimble} for the flow of the original integration domain corresponding the rational lens for $\alpha =2$ and $\mu = 0$. For $\lambda=0$ the contour $X_\lambda$ coincides with the real line. As $\lambda$ is increased to $1$, the original integration domain smoothly flows to the Lefschetz thimble $\mathcal{J}$ consisting of three steepest descent contours $\mathcal{J}_i$ corresponding to three relevant saddle points $\bar{x}_i$. By evaluating the flow for varying $\alpha$ and $\mu$, we obtain the Picard-Lefschetz analysis of the lens.

\begin{figure}
\centering
\begin{subfigure}[b]{0.49\textwidth}
\includegraphics[width=\textwidth]{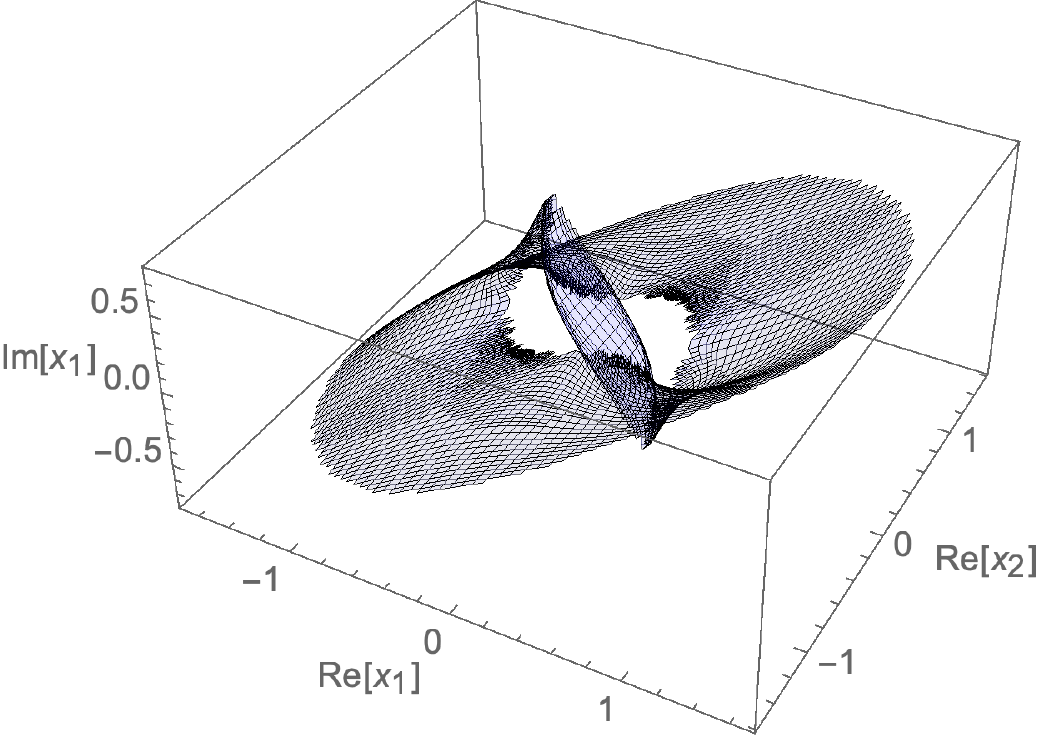}
\end{subfigure} 
\begin{subfigure}[b]{0.49\textwidth}
\includegraphics[width=\textwidth]{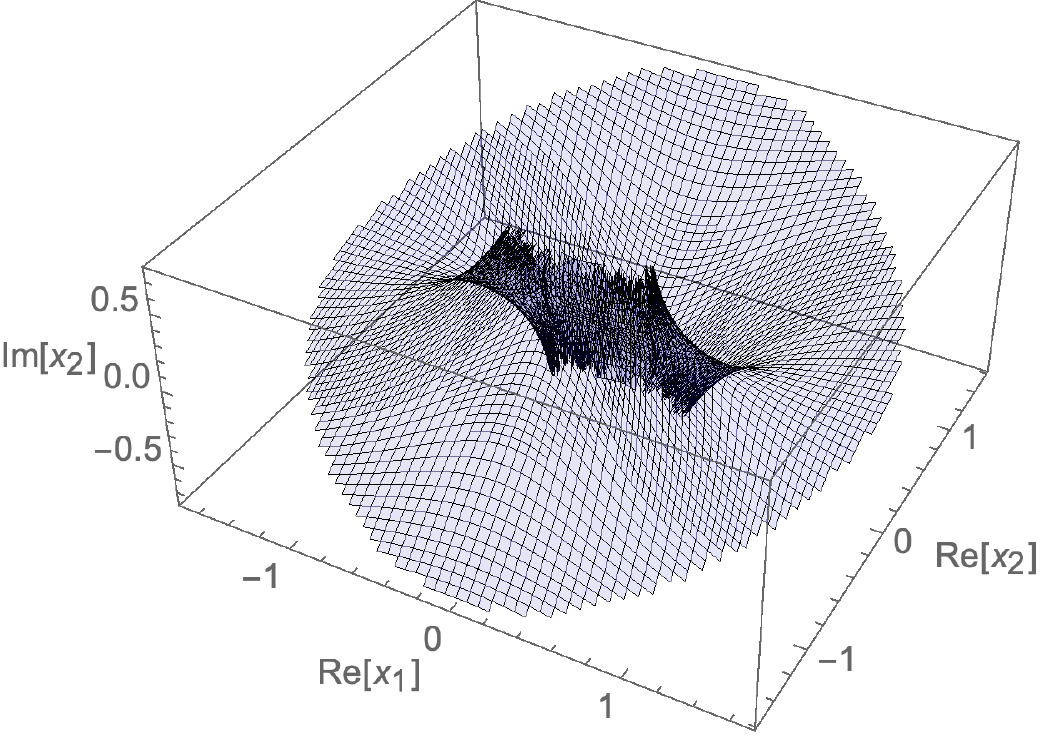}
\end{subfigure} 
\caption{Two projections of the numerically obtained two-dimensional thimble $\mathcal{J}$ in $\mathbb{C}^2$ for a two-dimensional oscillatory integral.}\label{fig:2Dflow}
\end{figure}

For multi-dimensional oscillatory integrals, the flow algorithm can be generalized by flowing the cells of a tessellation of the original integration domain. In this paper, we start our calculations with the tessellation of a rectilinear lattice. For an two-dimensional illustration see Fig.~\ref{fig:2Dflow}.

\begin{algorithm}
\begin{algorithmic}
\REQUIRE Represent a subset of the original integration domain $X$ with a regular tessellation consisting of cells $V_i$ spanned by the points $\bm{p}_{i,1},\bm{p}_{i,2},\dots$.
\WHILE{the variance of the imaginary part $H$ on the points $p_{i,j}$ exceeds threshold $T_1$\\}
\STATE flow the points: $\bm{p}_{i,j} \mapsto \bm{p}_{i,j} - \nabla h(\bm{p}_{i,j}) \Delta t$

\IF{the $h$-function evaluated in the point $p_i$ is smaller than the threshold $T_2$ }
\STATE remove the corresponding cells
\ENDIF

\IF{the volume of a cell $V_i$ exceeds the threshold $T_3$}
\STATE subdivide the cell into smaller cells
\ENDIF

\ENDWHILE
\end{algorithmic}
\caption{The flow of the contour of multi-dimensional oscillatory integrals.}
\end{algorithm}


There are various possible implementations of this algorithm. However, it follows from Cauchy's theorem that the integral is insensitive to the details of the tesselation employed. For all reasonable tesselations, the algorithm terminates in a polynomial number of steps as it scales roughly linearly with the number of simplices. Remarkably, this cost scaling is no worse than that required by the geometric optics approximation.

\subsection{Integrating along the thimbles}
Given a Lefschetz thimble $\mathcal{J}$ for a range of $\alpha$ and $\mu$, obtained with either one of the above-described methods, we perform the resulting integral along the thimble with the trapezium rule. Given a thimble $\mathcal{J}$ represented as a set of line-segments $l_i=(p_{i,1},p_{i,2})$, the integral is approximated by
\begin{align}
\Psi(\mu;\nu) \approx \sum_{i} \frac{e^{i \phi(p_{i,1};\mu)\nu} +e^{i \phi(p_{i,2};\mu)\nu}}{2} (p_{i,2}-p_{i,1}) 
\end{align}
summed over the line segments. For multi-dimensional oscillatory integrals, we evaluate the integral on a linear approximation of the integrand on the tessellation of the thimble. Naively, one might expect to have to compute the Lefschetz contour $\mathcal{J}$ for every $\mu$ for which one wishes to perform the integral. However, since the thimble is a smooth function of $\mu$, it suffices to compute the thimble for a range of $\mu$. When integrating, we instead evaluate the integral on the thimble corresponding to the closest $\mu$ for which we have evaluated the thimble. Finally, it should be noted that for increasing $\nu$, the support of the integral is increasingly concentrated around the relevant saddle points. As a consequence we can, for large $\nu$, restrict the integral to the line segments close to the saddle points. It follows from this that the numerical evaluation of the integral along the thimble becomes more and more efficient as the frequency is increased. This is in sharp contrast with conventional integration techniques which need to trace many oscillations of the integrand along the real line.

See Fig.~\ref{fig:1DIntensityAlpha2} for the normalized intensity profiles of the lens evaluated along the thimble for frequencies $\nu=50, 100, 500$. We observe the following properties of the normalized intensity profiles:

\begin{figure}
\centering
\begin{subfigure}[b]{0.32\textwidth}
\includegraphics[width=\textwidth]{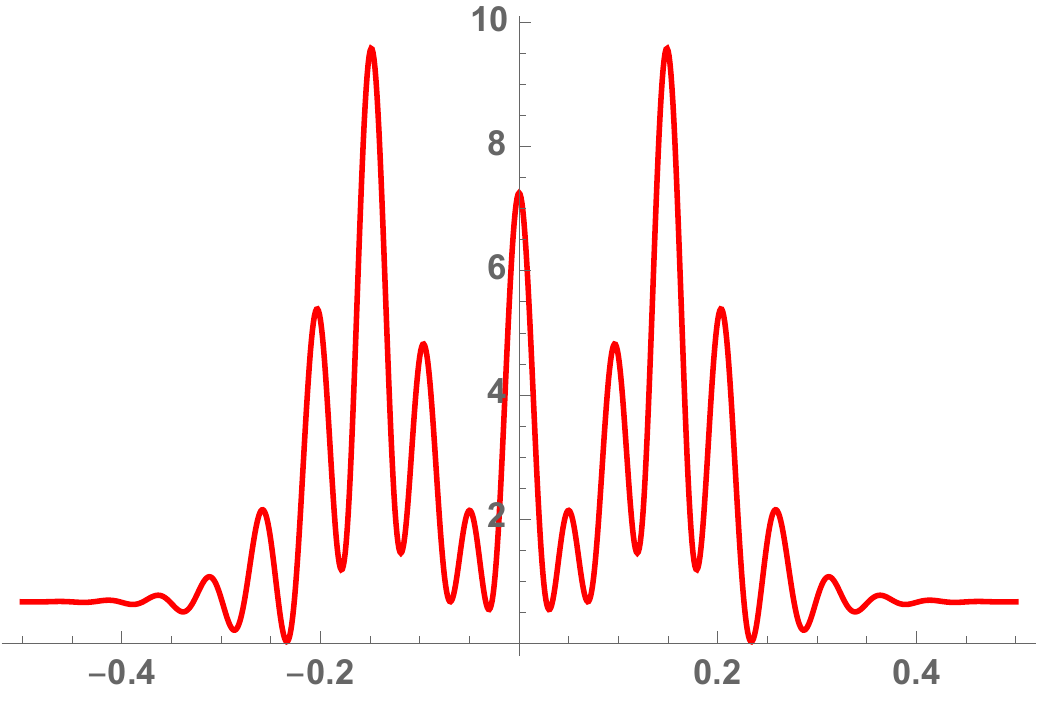}
\caption{$\alpha=2,\nu =50$}
\end{subfigure} 
\begin{subfigure}[b]{0.32\textwidth}
\includegraphics[width=\textwidth]{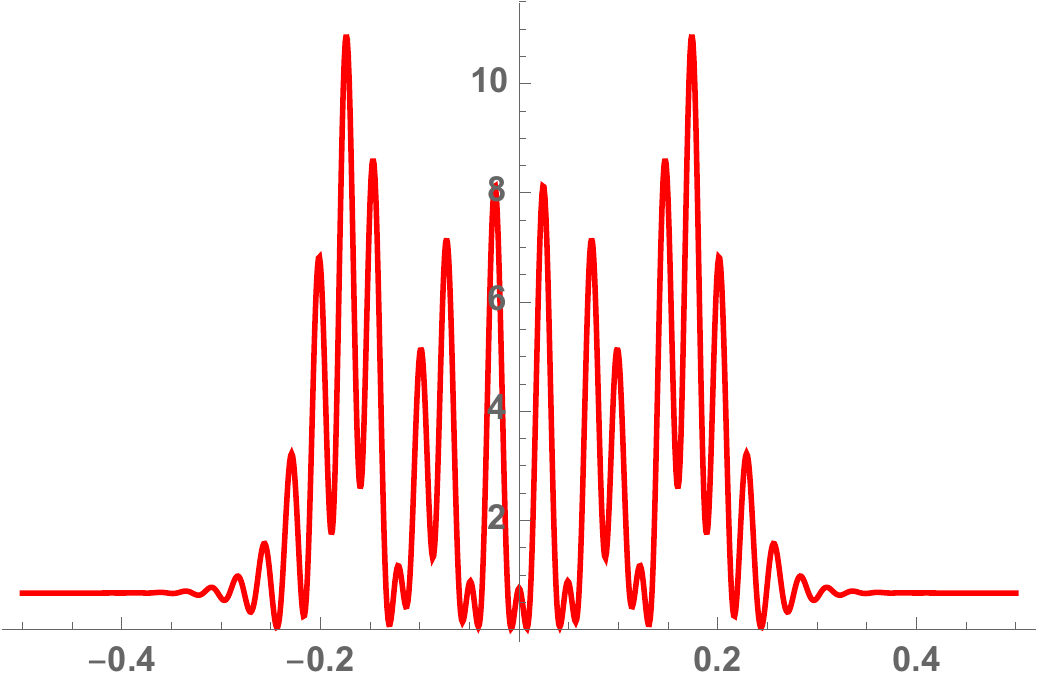}
\caption{$\alpha=2,\nu = 100$}
\end{subfigure} 
\begin{subfigure}[b]{0.32\textwidth}
\includegraphics[width=\textwidth]{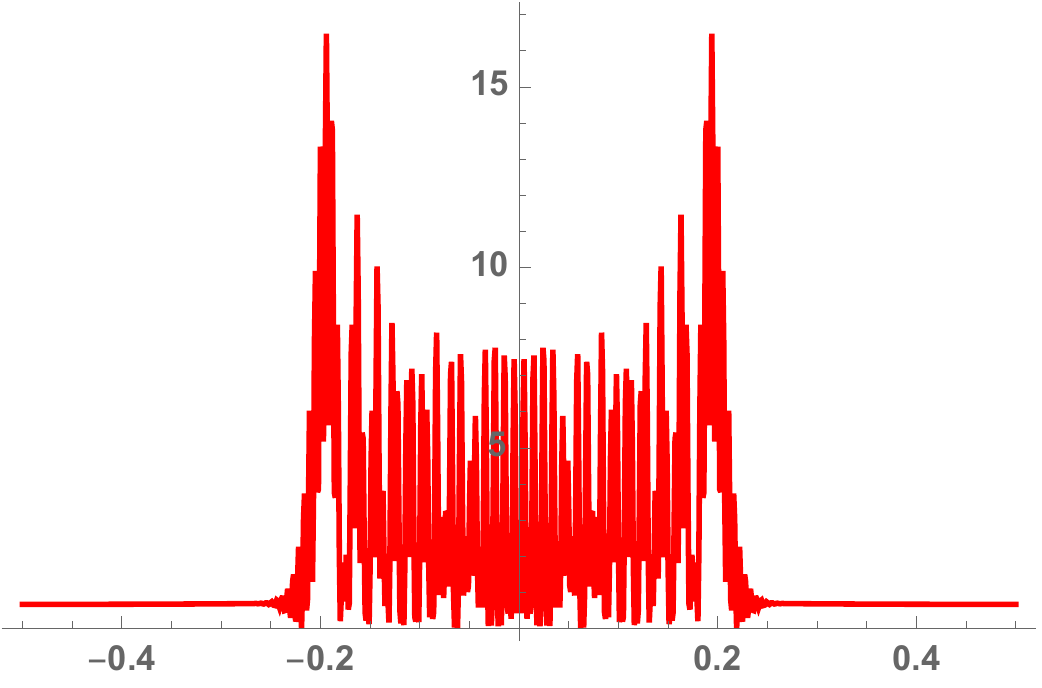}
\caption{$\alpha=2,\nu = 500$}
\end{subfigure}\\
\begin{subfigure}[b]{0.32\textwidth}
\includegraphics[width=\textwidth]{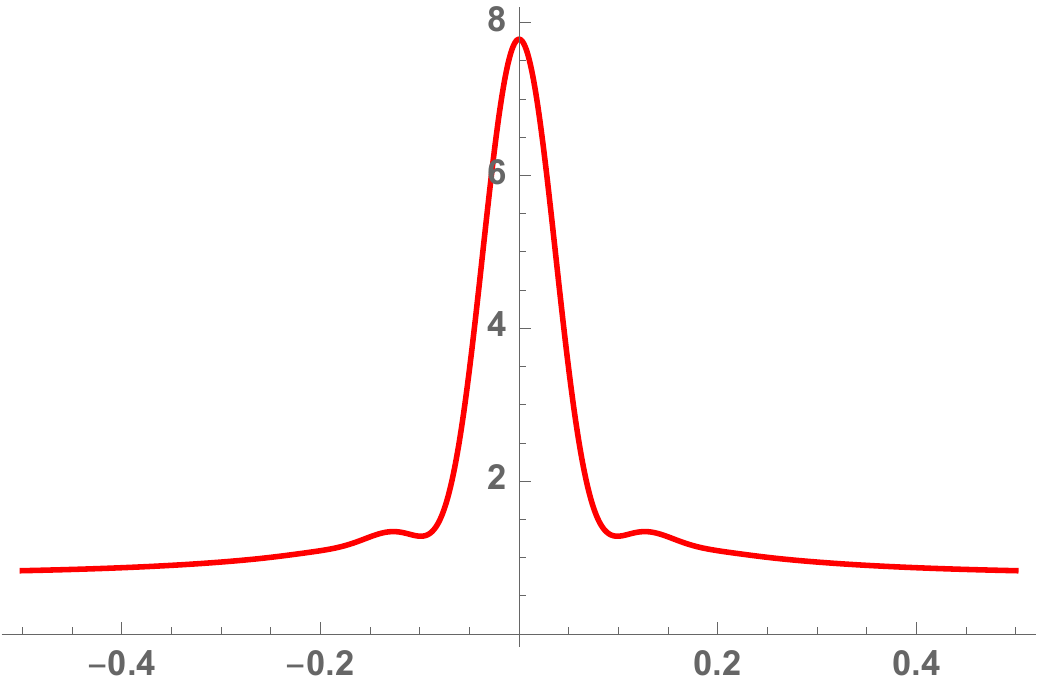}
\caption{$\alpha=1,\nu =50$}
\end{subfigure} 
\begin{subfigure}[b]{0.32\textwidth}
\includegraphics[width=\textwidth]{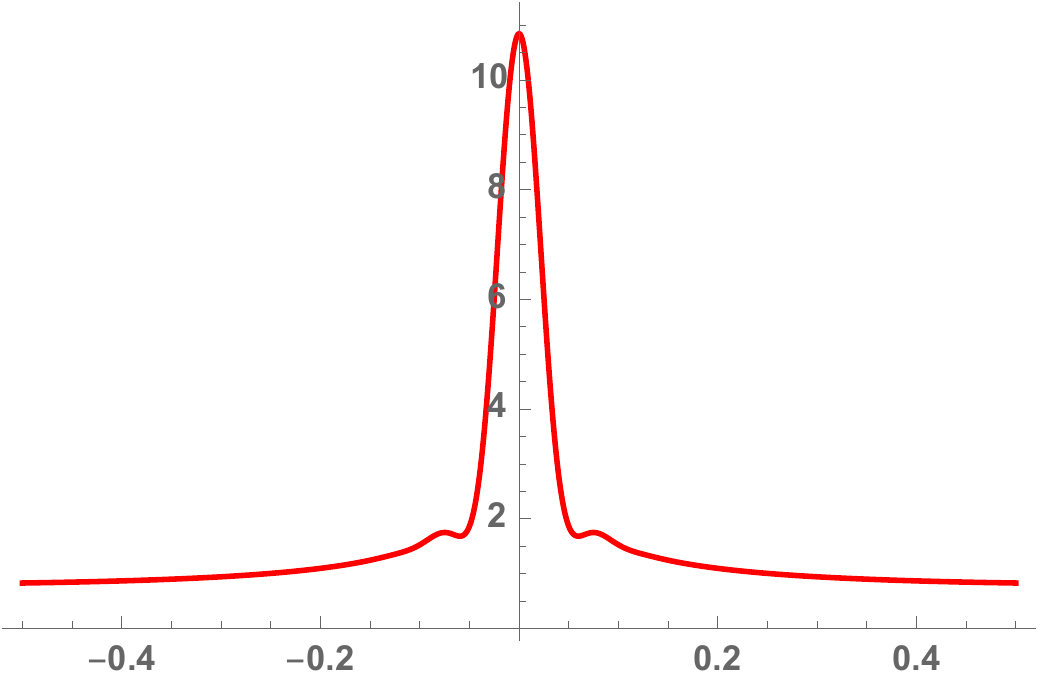}
\caption{$\alpha=1,\nu = 100$}
\end{subfigure} 
\begin{subfigure}[b]{0.32\textwidth}
\includegraphics[width=\textwidth]{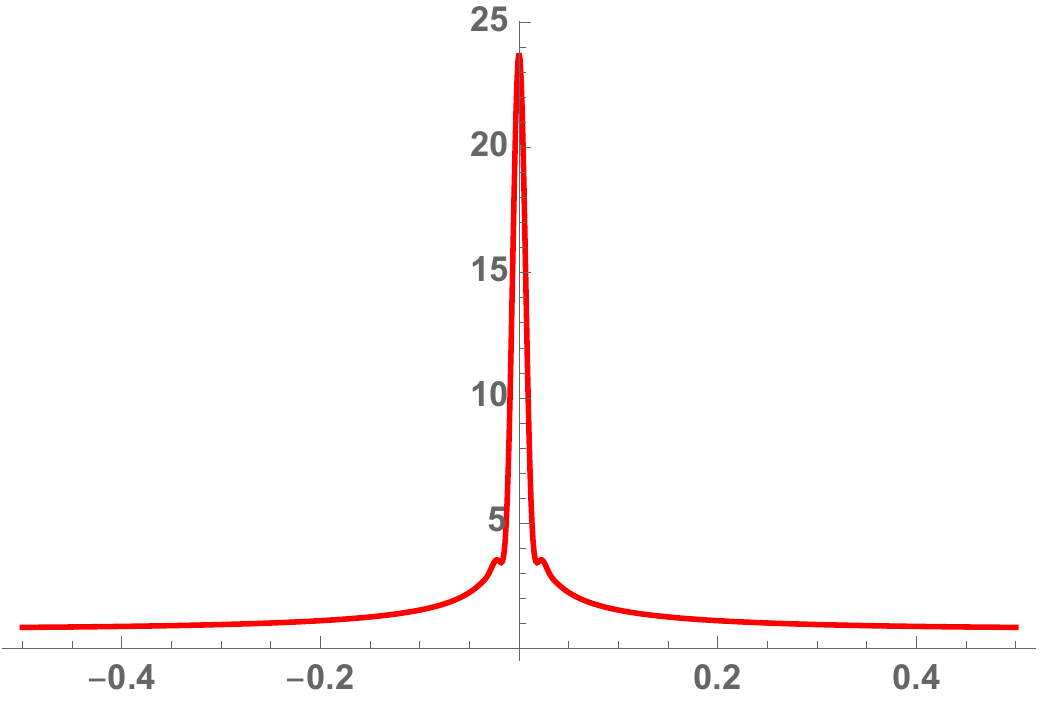}
\caption{$\alpha=1,\nu = 500$}
\end{subfigure}\\
\begin{subfigure}[b]{0.32\textwidth}
\includegraphics[width=\textwidth]{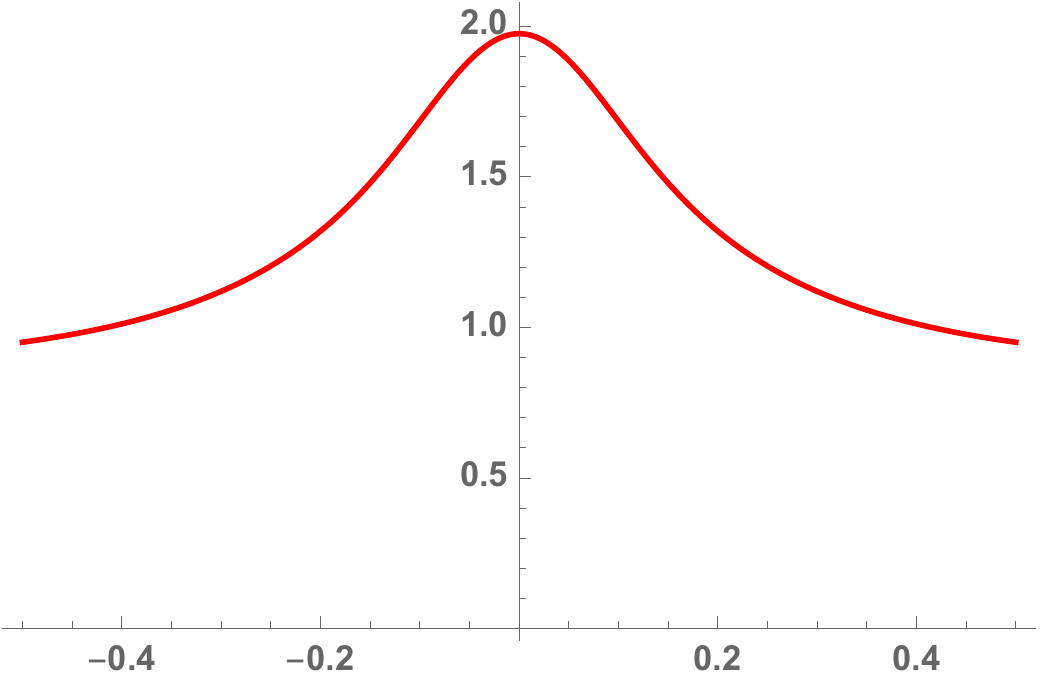}
\caption{$\alpha=1/2,\nu =50$}
\end{subfigure} 
\begin{subfigure}[b]{0.32\textwidth}
\includegraphics[width=\textwidth]{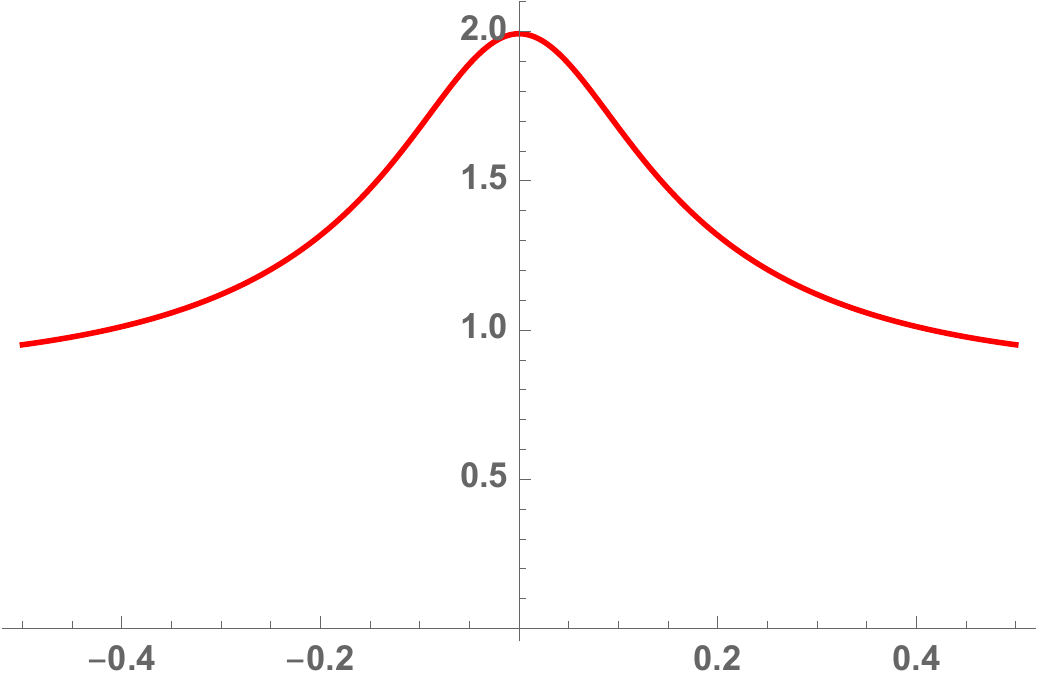}
\caption{$\alpha=1/2,\nu = 100$}
\end{subfigure} 
\begin{subfigure}[b]{0.32\textwidth}
\includegraphics[width=\textwidth]{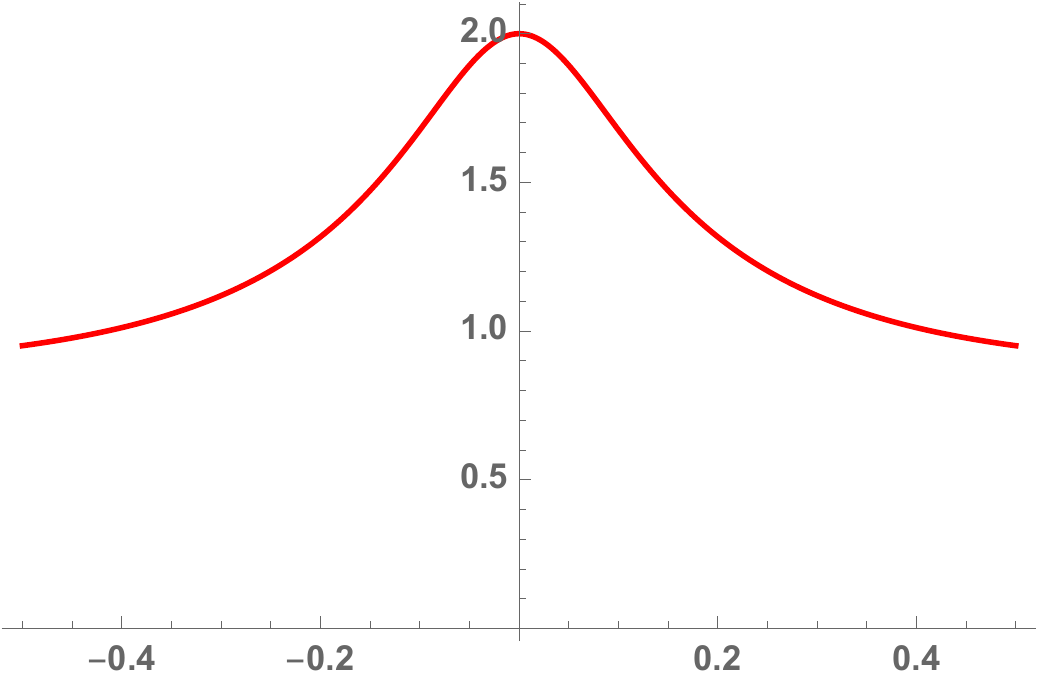}
\caption{$\alpha=1/2,\nu = 500$}
\end{subfigure}
\caption[The intensity map of the one-dimensional lens]{The normalized intensity $I(\mu;\nu)= |\Psi(\mu;\nu)|^2$ for $\alpha =1/2,1,2$ as a function of $\mu$ for $\nu=50,100,500$.}\label{fig:1DIntensityAlpha2}
\end{figure}

\begin{itemize}
\item
In the regime $\alpha <1$, the lens leads to a single-image region. The normalized intensity profile does not oscillate and is moreover independent of the frequency $\nu$.  See the lower panels of Fig.~\ref{fig:1DIntensityAlpha2}.
\item 
For $\alpha =1$, the lens forms a cusp caustic. The caustic corresponds to the peak at $\mu_c=0$. For increasing frequency, $\nu$, the peak is enhanced and becomes increasingly narrow. In the eikonal limit $\nu\to\infty$, the normalized normalized intensity diverges as $\nu^{1/2}$ at the caustic $\mu_c$ (see the scaling relations in Table \ref{tab:exponents}). See the middle panels of Fig.~\ref{fig:1DIntensityAlpha2}.
\item In the regime $\alpha >1$, the lens forms a triple-image region which is bounded by two-fold caustics. We see that the triple-image region $(-\mu_c,\mu_c)$, with $\mu_c=0.206751\dots$ for $\alpha =2$, consists of an interference pattern bounded by two peaks at $\mu=\pm \mu_c$. The interference pattern in the triple-image region is the result of the three real saddle points. The oscillations in the single-image region result from the interplay between the relevant real and the complex saddle point. For increasing $\nu$, the fringes of the interference pattern shrink and spikes corresponding to the fold get sharper and are increasingly enhanced.  For the relevant scalings see Table \ref{tab:exponents}. See the upper panels of Fig.~\ref{fig:1DIntensityAlpha2}.
\end{itemize}
Note that the normalized intensity in the cusp exceeds the normalized intensity in the fold caustic. This related to the co-dimension of the caustic as described in Section \ref{sec:catastropheTheoryOptics}. Moreover remark that the cusp caustic only exists at a single $\alpha$ for the one-dimensional lens, while the fold caustic appears for a range of $\alpha$. Table \ref{tab:exponents} shows the frequency dependence of the pattern. Furthermore, note that the normalized intensity profiles at frequency $\nu=500$, for $\alpha=1/2,1,$ and $2$, are close to the normalized intensity maps predicted by geometric optics (see Fig.~\ref{fig:IGeometricOptics}).

In the context of astronomical radio sources, the signal is dramatically enhanced when the relative position of the observer and the source move through the fold or the cusp caustic of the lens. One would in this context interpret the $\mu$ axis as the line traced by the source on the sky, \text{i.e.} $\mu=vt + \mu_0$ with $\mu_0$ the initial position, $v$ the speed of the source in parameter space and $t$ the time. This amplification of the signal may be relevant as an selection effect for the recently observed Fast Radio Bursts. Note that if the observed FRBs are indeed the result of caustics in plasma lenses, we expect the peaks to evolve in a characteristic way and satisfy specific scaling relations in frequency space. See Section \ref{sec:signatures} for a more detailed discussion.

\section{The elementary catastrophes}\label{sec:catastrophePL}
The unfoldings of the seven \textit{elementary singularities} (see Table \ref{tab:unfoldings}), form a local description of lenses near the caustics. We here study the Picard-Lefschetz analysis of the elementary catastrophes appearing in two-dimensional lenses and evaluate the corresponding normalized intensity maps using the flow algorithm described above. This analysis is complementary to the asymptotic analysis described in chapter 36 of \cite{Thompson:2011}.

\subsection{The fold $A_2$}
\begin{figure}[ht]
\centering
\begin{subfigure}[b]{0.3\textwidth}
\includegraphics[width=\textwidth]{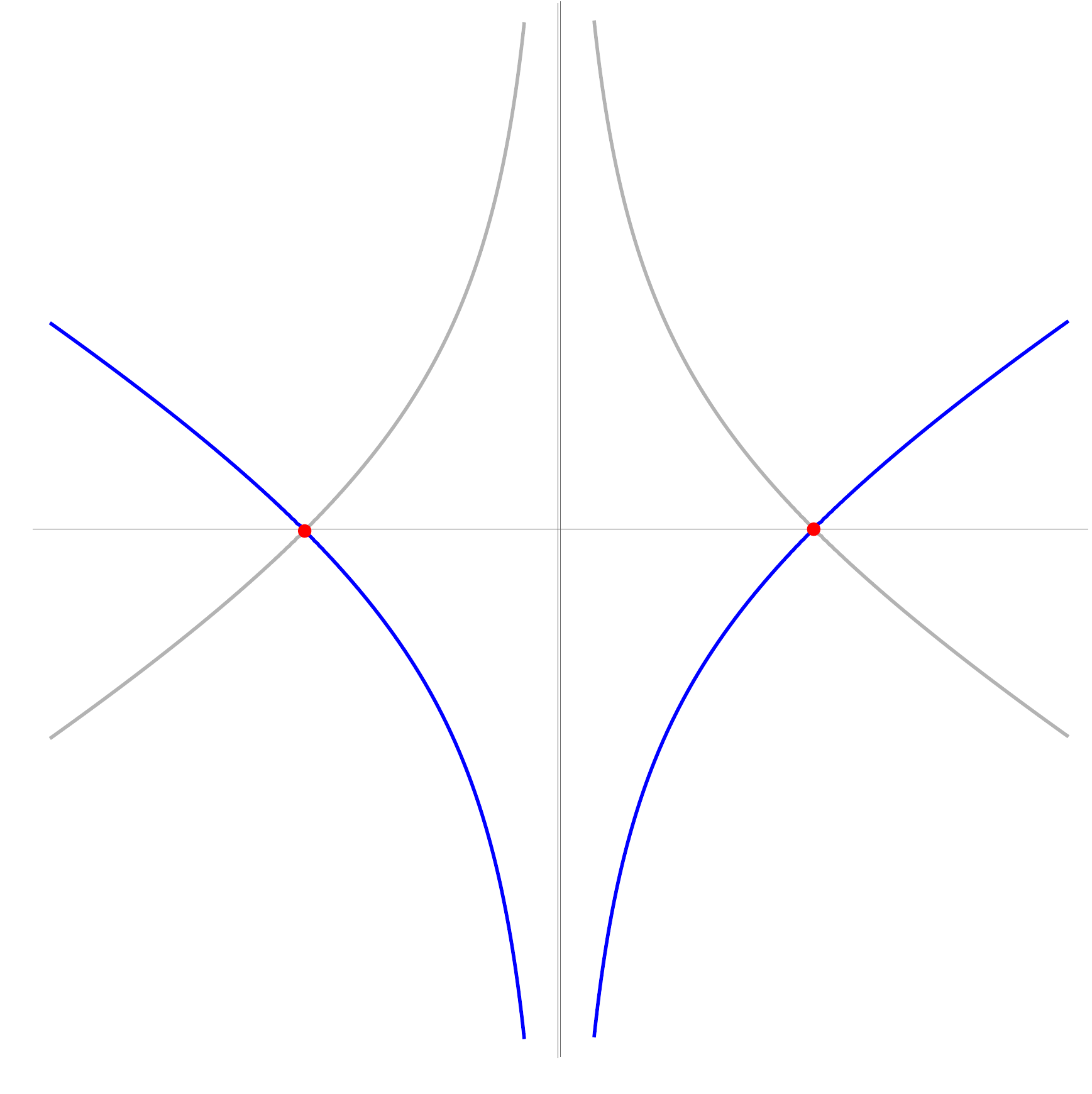}
\caption{$\mu <0$}\label{fig:A2m1-}
\end{subfigure} ~
\begin{subfigure}[b]{0.3\textwidth}
\includegraphics[width=\textwidth]{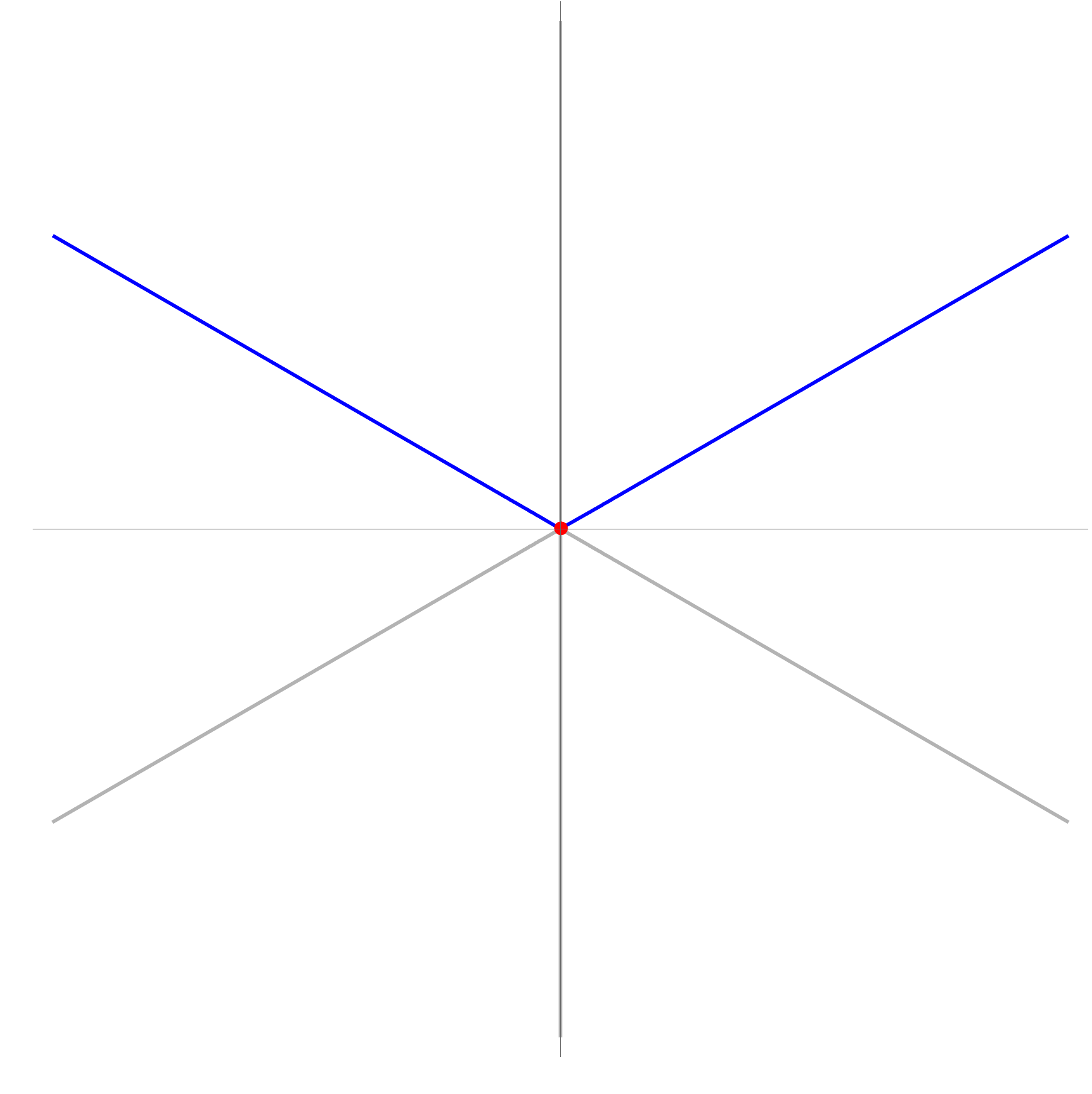}
\caption{$\mu = 0$}\label{fig:A2m}
\end{subfigure} ~
\begin{subfigure}[b]{0.3\textwidth}
\includegraphics[width=\textwidth]{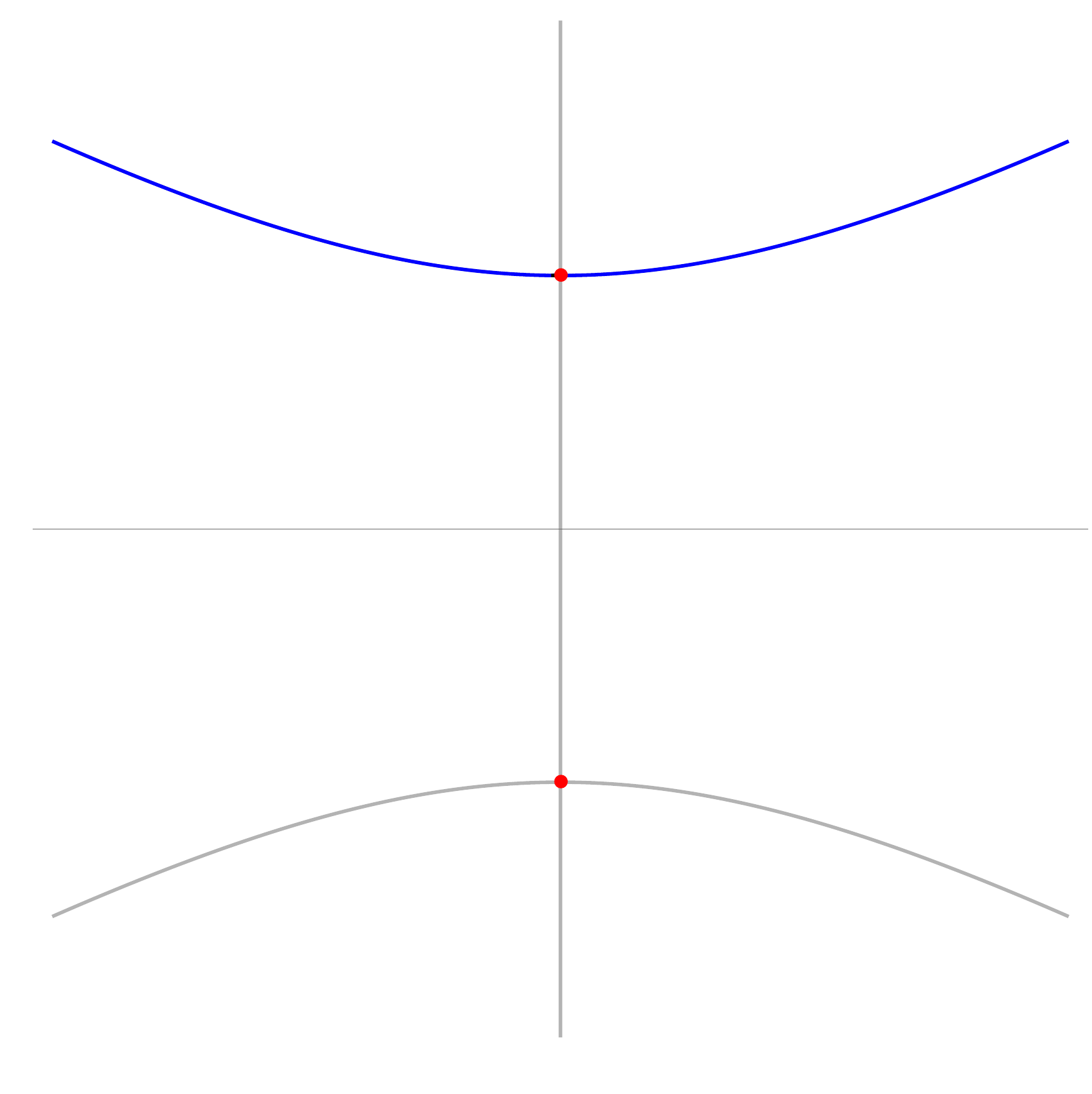}
\caption{$\mu > 0$}\label{fig:A2m1+}
\end{subfigure}
\caption[The Picard-Lefschetz analysis of the fold singularity.]{The saddle points and the Lefschetz thimbles in the complex plane $\mathbb{C}$ of the unfolding of the fold singularity $A_2$.}\label{fig:foldUnfoldingA2}
\end{figure}

The fold singularity is the simplest degenerate critical point and can be viewed as the superposition of two non-degenerate saddle points. The Picard-Lefschetz analysis of the unfolding of the fold singularity is illustrated in Fig.~\ref{fig:foldUnfoldingA2}. For negative $\mu$, there are two relevant real saddle points (see Fig.~\ref{fig:A2m1-}). As $\mu$ approaches the caustic at $\mu_c=0$, the two saddle points merge and form the fold singularity (see Fig.~\ref{fig:A2m}). Note that the fold saddle point emanates three steepest ascent and three descent curves. The thimble is non-differentiable at the degenerate saddle point. When $\mu$ is increased passed the caustic $\mu_c$, the two saddle points move off the real axis and into the complex plane (see Fig.~\ref{fig:A2m1+}). In this regime only one of them remains relevant. 

\begin{figure}
\centering
\begin{subfigure}[b]{0.3\textwidth}
\includegraphics[width=\textwidth]{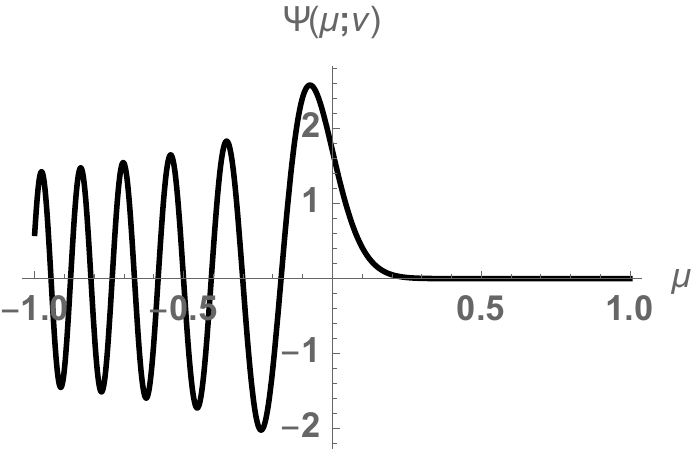}
\caption{$\nu =50$}
\end{subfigure} ~
\begin{subfigure}[b]{0.3\textwidth}
\includegraphics[width=\textwidth]{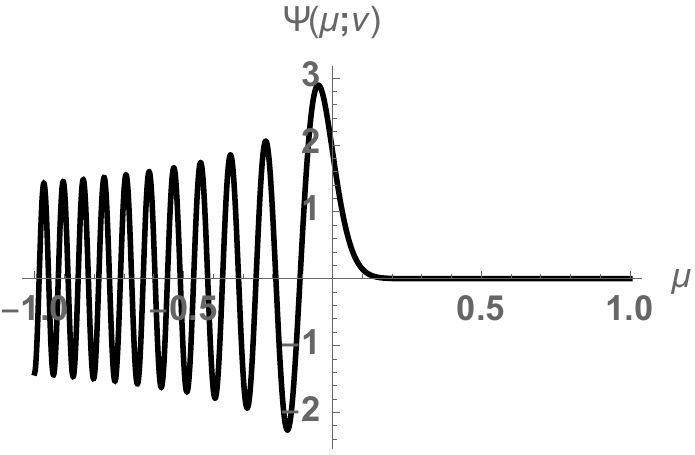}
\caption{$\nu = 100$}
\end{subfigure} ~
\begin{subfigure}[b]{0.3\textwidth}
\includegraphics[width=\textwidth]{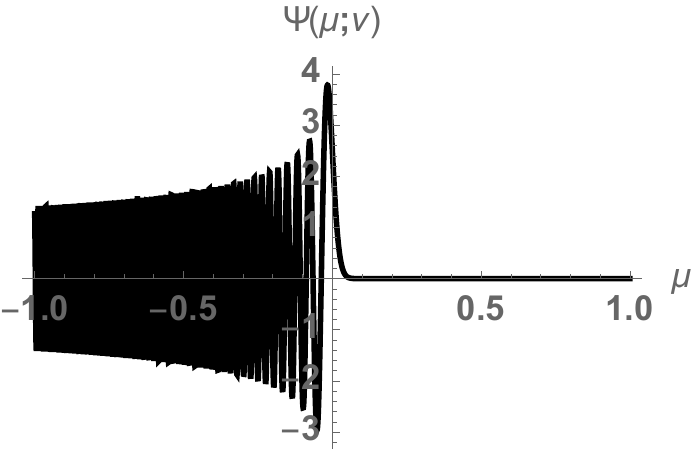}
\caption{$\nu = 500$}
\end{subfigure}
\caption[The integral $\Psi(\mu,\nu)$ for the fold singularity.]{The integral $\Psi(\mu,\nu)$ for the fold singularity as a function of $\mu$ for $\nu=50,100,500$.}\label{fig:Airy}
\end{figure}

The Fresnel-Kirchhoff integral for the fold singularity can be related to the Airy function 
\begin{align}
\Psi(\mu;\nu) =\sqrt{\frac{\nu}{\pi}}\int_{-\infty}^\infty e^{i\left( \frac{x^3}{3}+\mu x\right)\nu}\mathrm{d}x  = 2\sqrt{ \pi} \nu^{1/6} \text{Ai}[\nu^{2/3}\mu]\,.
\end{align}
Note the appearance of the singularity and fringe indices $1/6$ and $2/3$ as listed in Table \ref{tab:exponents}. It straightforward to derive the scaling of the amplitude and the fringes, with the change of coordinates $z=\nu^{1/3}x$. The other scaling relations are derived analogously. The Airy function is a good illustration of the interference phenomenon present in multi-image regions (seen in Fig.~\ref{fig:Airy}). The range $\mu <0$, for which the two relevant saddle points reside on the real line corresponds to a double-image region, where two saddle points lead to an interference pattern. The range $\mu>0$ corresponds to a zero-image region in which the amplitude asymptotes to zero as $\nu \to \infty$. 

\subsubsection{Asymptotics}
Using the Picard-Lefschetz diagrams (Fig.~\ref{fig:foldUnfoldingA2}) we can derive asymptotics for the integral $\Psi(\mu;\nu)$. For $\mu >0$ the Picard-Lefschetz analysis consists of a single relevant saddle point located at $i\sqrt{\mu}$. The exponent can be approximated around the saddle point by
\begin{align}
\phi(x;\mu) = i\frac{2}{3} \mu^{3/2} + i \sqrt{\mu} (x-i\sqrt{\mu})^2+\mathcal{O}\left((x-i\sqrt{\mu})^3\right)\,.
\end{align}
The saddle point approximation for this point gives an exponential falloff
\begin{align}
\Psi(\mu;\nu) \approx \frac{e^{-\frac{2}{3}\mu^{3/2}\nu}}{\sqrt{2}\mu^{1/4}}\,.
\end{align}
This matches the the behaviour in Fig.~\ref{fig:Airy}.

For $\mu <0$, the Picard-Lefschetz analysis consists of two real relevant saddle points located at $x= \pm \sqrt{|\mu|}$. A saddle point approximation around these points gives us the oscillatory behaviour 
\begin{align}
\Psi(\mu;\nu) \approx \frac{e^{-\frac{2 i}{3}\mu^{3/2}\nu}+ i e^{\frac{2 i}{3}\mu^{3/2}\nu}}{\mu^{1/4}}\,,
\end{align}
seen in Fig.~\ref{fig:Airy}. Observe that wave function becomes increasingly oscillatory and falls off as a power law $\Psi(\mu;\nu) \propto \frac{1}{|\mu|^{1/4}}$ in the geometric limit $\nu \to \infty$.

\subsection{The cusp $A_3$}
\begin{figure}
\centering
\includegraphics[width= \textwidth]{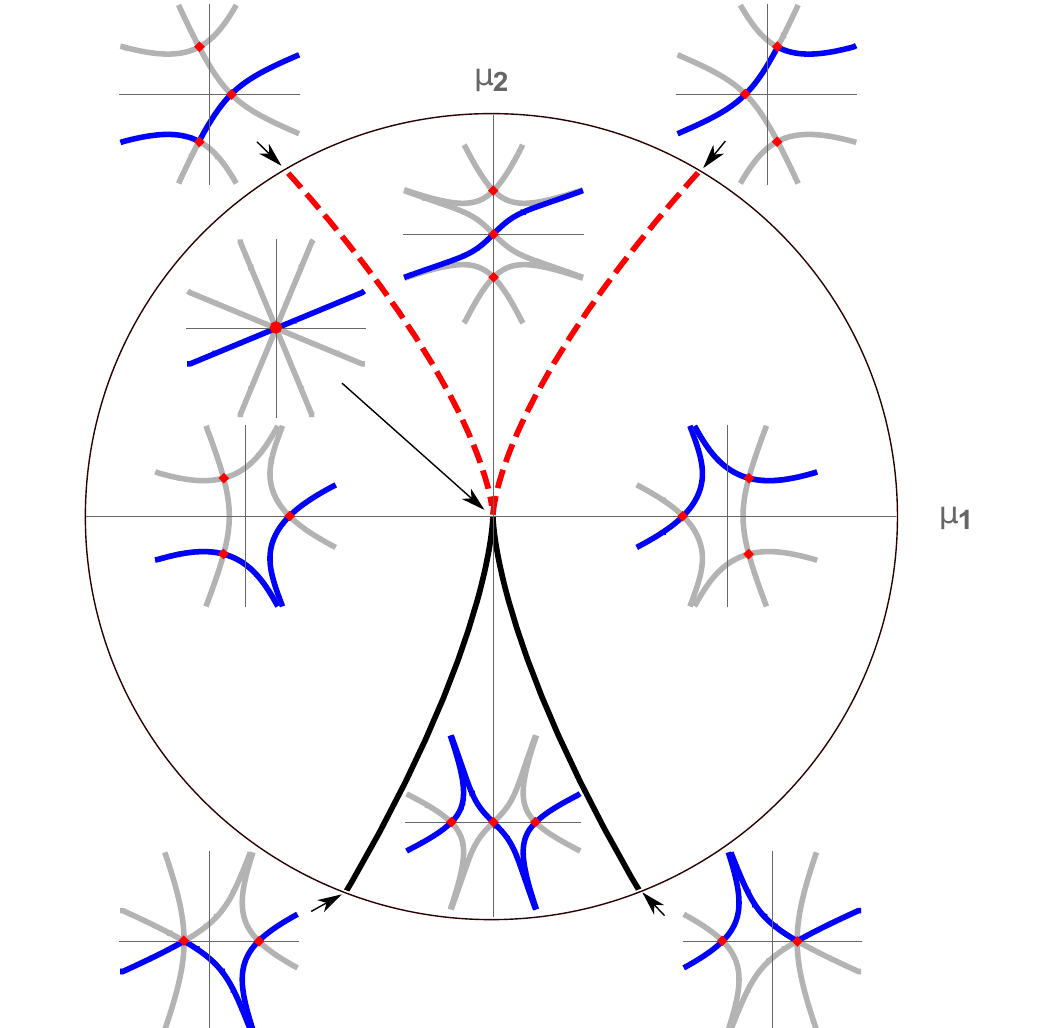}
\caption[The cusp singularity in the unfolding plane.]{The cusp singularity in the unfolding $\bm{\mu}$-plane. The black curve is the fold line separating the single-image region (upper region) from the triple-image region (lower region). The dashed red line is the Stokes line. We observe that the Stokes transition corresponds to one complex saddle point becoming (ir)relevant, and that the caustics correspond to the superposition of real non-degenerate saddle points. The upper region is a single-image region since there is only one real relevant saddle point. The lower region is a triple-image region since there are three real relevant saddle points.}\label{fig:Cusp}
\end{figure}

The cusp singularity consists of the superposition of three non-degenerate saddle points. The singularity is of co-dimension $K=2$ and has two unfolding parameters $\mu_1$ and $\mu_2$, \textit{i.e.}, 
\begin{align}
\Psi(\bm{\mu};\nu) =\sqrt{\frac{\nu}{\pi}}\int_{-\infty}^\infty e^{i\left(\frac{x^4}{4} + \mu_2 \frac{x^2}{2} + \mu_1 x\right)\nu}\mathrm{d}x \,.
\end{align}
See Fig.~\ref{fig:Cusp} for an illustration of unfolding of the cusp caustic and the Picard-Lefschetz analysis in the $(\mu_1,\mu_2)$-plane.  \\

The the exponent $\phi(x;\bm{\mu})$ has three saddle points $\bar{x}_i$,  given by the roots of the cubic equation
\begin{align}
x^3+ \mu_2 x +\mu_1=0\,.
\end{align}
Depending on $\mu$, either one or three of the saddle points are real-valued. The complex-valued saddle points always come in conjugate pairs since $\phi(x;\bm{\mu})$ is a real-valued function, \textit{i.e.} real for real $x$.

Geometric optics applied to this integral shows that the cusp caustic at $(\mu_1,\mu_2)=(0,0)$ emanates two fold-lines $A_2 \subset M$, given by cubic root
\begin{align}
\mu_2 = - \frac{3}{2^{2/3}}|\mu_1|^{2/3}\,.
\end{align}
The fold lines are non-differentiable at the cusp singularity $(\mu_1,\mu_2)=(0,0)$.

In the triple-image region enclosed by the two fold-lines, the thimble passes through three real-valued saddle points. When approaching one of the fold lines, we see that two of the real saddle points merge and move in the complex plane. Only one of the two complex saddle points remains relevant to the integral. This is analogous to the behavior observed in the analysis of the fold caustic. At the cusp saddle point at $(\mu_1,\mu_2)=(0,0)$ all three saddle points merge at the origin. Finally, note that the single-image region consists of three subregions, for which the Picard-Lefschetz analysis either consists of one or two relevant saddle points. These subregions are separated by two Stokes lines (red dashed lines in Fig.~\ref{fig:Cusp}). Along these lines, the Lefschetz thimbles flip while the saddle points remain separated. The Stokes lines can be found by equating the imaginary parts of the exponents evaluated at the saddle points, \textit{i.e.},
\begin{align}
\text{Im}[i\phi(\bar{x}_i;\bm{\mu})\nu] = \text{Im}[i\phi(\bar{x}_j;\bm{\mu})\nu]
\end{align}
for $i \neq j$. For the unfolding of the cusp, we see that the Stokes lines are described by
\begin{align}
\mu_2 = 3 \sqrt[3]{\frac{3\sqrt{3}-5}{2}}|\mu_1|^{2/3}\,,
\end{align}
for $\mu_1<0$ and $\mu_2$. Note that the amplitude across a Stokes line is smooth, even though the saddle point structure changes abruptly. The Stokes lines can be interpreted as the points for which the saddle point approximation of the integral fails. 

\subsubsection{Numerics}
\begin{figure}
\centering
\begin{subfigure}[b]{0.32\textwidth}
\includegraphics[width=\textwidth]{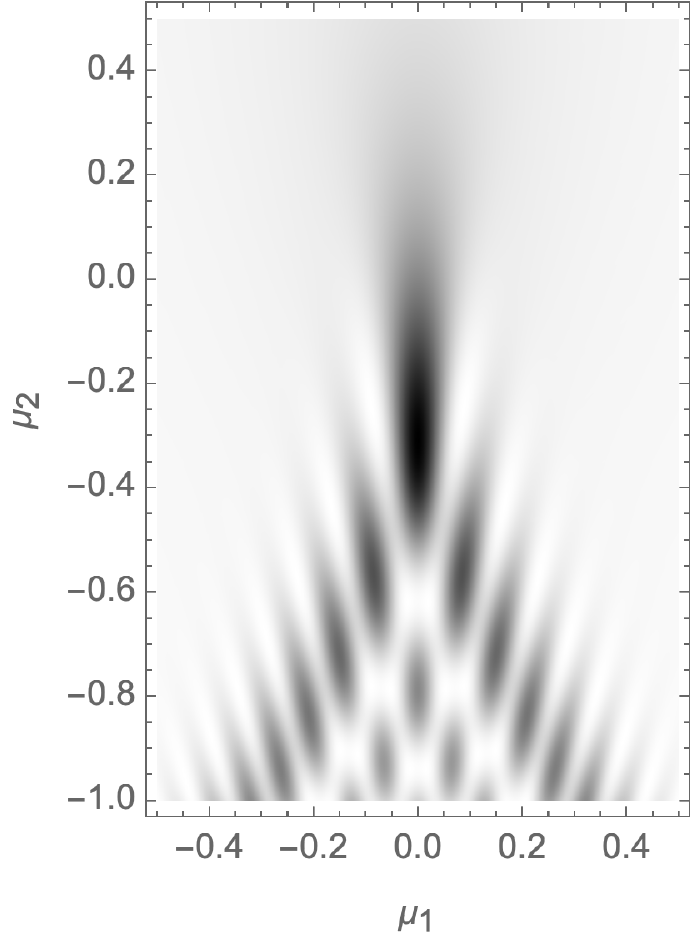}
\caption{$\nu=50$}
\end{subfigure} 
\begin{subfigure}[b]{0.32\textwidth}
\includegraphics[width=\textwidth]{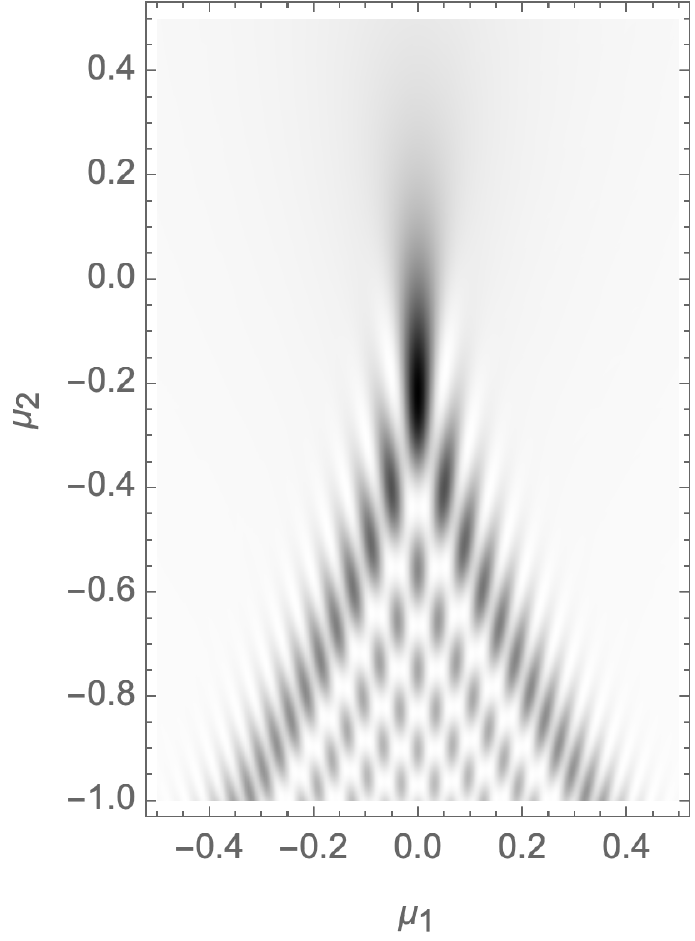}
\caption{$\nu=100$}
\end{subfigure} 
\begin{subfigure}[b]{0.32\textwidth}
\includegraphics[width=\textwidth]{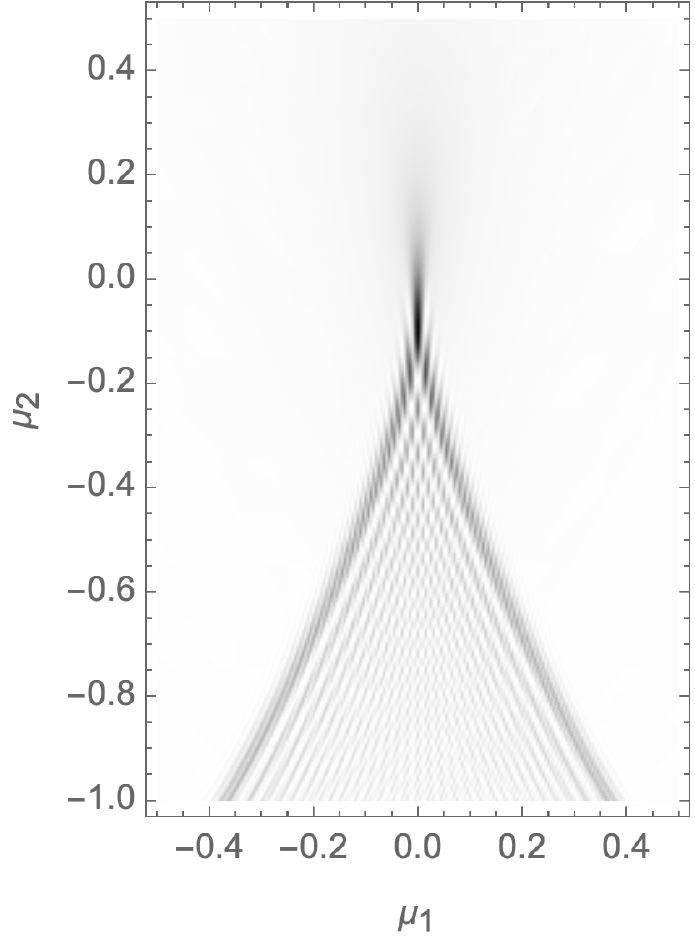}
\caption{$\nu=500$}
\end{subfigure}
\caption[The normalized intensity map for the cusp caustic.]{The normalized intensity $I(\bm{\mu};\nu)$ for the cusp caustic for $\nu=50,100$ and $500$.}\label{fig:CuspNumerical}
\end{figure}

Given the Lefschetz thimble, we can numerically evaluate the amplitude (see Fig.~\ref{fig:CuspNumerical}). In the eikonal limit $\nu \to \infty$ we observe the emergence of a fold-line ($A_2$) with a sharp exponential falloff in most of the single-image region and the power-law falloff in the triple-image region. We also see the emergence of a cusp caustic at the origin with a power-law falloff along the line $\{\mu_1=0\}$.

\subsubsection{Asymptotics}
The Picard-Lefschetz diagrams (Fig.~\ref{fig:Cusp}) allow us to derive limiting behaviour for $\Psi(\bm{\mu};\nu)$. From the functional form of $\phi(x;\bm{\mu})$ along the line $\{\mu_1=0\}$, \textit{i.e.},
\begin{align}
\phi(x;\bm{\mu}) = \frac{\mu_2 x^2}{2} + \frac{x^4}{4} \,,
\end{align}
we observe that one of the relevant saddle points is located at the origin $x=0$. For $\mu_2 >0$ this is the only relevant saddle point, whereas for $\mu_2<0$ it is one of three real relevant saddle points. 

In the case $\mu_2>0$, we find that the single saddle leads to a power-law
\begin{align}
\Psi(x;\bm{\mu}) \approx \sqrt{\frac{\nu}{\pi}} \int_{-\infty}^{\infty} e^{i \frac{\mu_2 \nu}{2}x^2}\mathrm{d}x = \sqrt{2} (-i \mu_2)^{-1/2}
\end{align} 
which in the normalized intensity corresponds to the falloff
\begin{align}
I(x;\mu) = |\Psi(x;\bm{\mu})|^2 \approx \frac{2}{\mu_2}\,,
\end{align}
independent of the frequency. This feature is absent in the unfolding of the fold caustic.

For $\mu_2<0$, the Picard-Lefschetz analysis consists of three real relevant saddle points located at $\pm \sqrt{\mu_2}$ and $0$. The exponent at the saddle point $\pm \sqrt{-\mu_2}$ can be approximated by 
\begin{align}
\phi(x;\bm{\mu}) = -\frac{\mu_2^2}{4} - \mu_2 (x\pm\sqrt{-\mu_2})^2 + \mathcal{O}\left((x \pm \sqrt{-\mu_2})^3\right)\,.
\end{align}
In the saddle point approximation,
\begin{align}
\Psi(\bm{\mu};\nu) \approx \sqrt{2}\frac{-(-1)^{3/4} + (1+i) e^{-\frac{i}{4}\mu_2^2 \nu}}{\sqrt{-\mu_2}}\,.
\end{align}
The normalized intensity $I(\mu;\nu)$ thus oscillates in $\mu_2$ with increasing frequency with power-law suppression 
\begin{align}
I(\bm{\mu};\nu) \propto \frac{2}{\mu_2}\,.
\end{align}

Along the line $\{\mu_2=0\}$, the Picard-Lefschetz analysis consists of two relevant saddle points, one real and one complex. The real relevant saddle point is located at $\bar{x} =- \mu_1^{1/3}$, giving the oscillatory behaviour 
\begin{align}
\Psi(\bm{\mu};\nu) \approx \sqrt{2}\frac{(-1)^{1/4} e^{-i\mu_1^{4/3}\nu}}{-\sqrt{3}\mu_1^{1/3}}\,,
\end{align}
so that the normalized intensity again falls off as a power-law
\begin{align}
I(\bm{\mu};\nu) = |\Psi(\bm{\mu};\nu)|^2 \approx 2 \mu_1^{-2/3}\,.
\end{align}

\subsection{The swallowtail $A_4$}
\begin{figure}
\centering
\includegraphics[width=0.9 \textwidth]{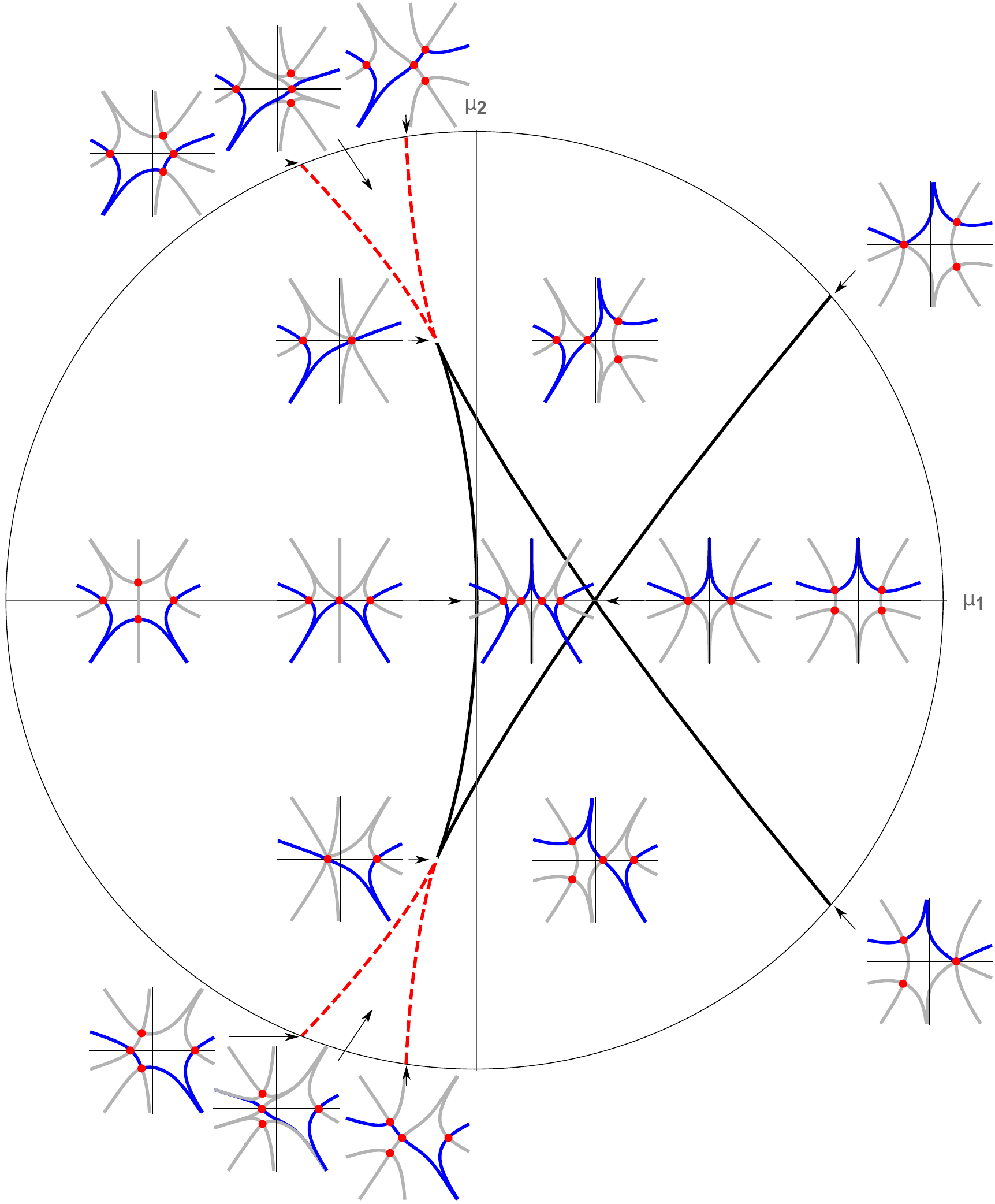}
\caption[The Picard-Lefschetz analysis of the unfolding of the swallowtail.]{The Picard-Lefschetz analysis of the unfolding of the swallowtail ($A_4$) singularity at $\mu_3=-1$. The number of real saddles gives the number of images in geometric optics.}\label{fig:ST_mu3=-1}
\end{figure}

\begin{figure}
\centering
\begin{subfigure}[b]{0.49\textwidth}
\includegraphics[width= \textwidth]{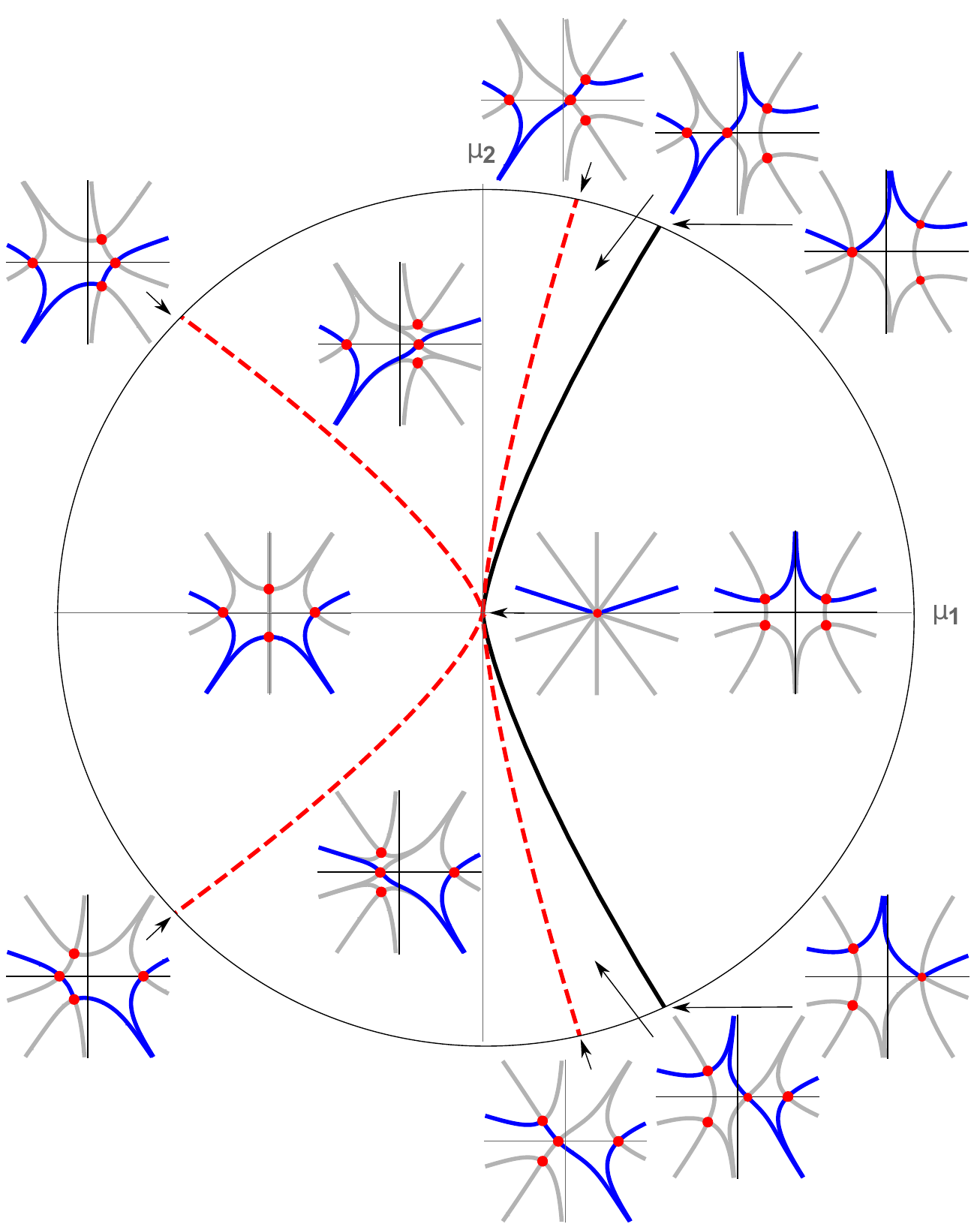}
\caption{$\mu_3=0$}\label{fig:ST_mu3=1_0}
\end{subfigure}
\begin{subfigure}[b]{0.49\textwidth}
\includegraphics[width= \textwidth]{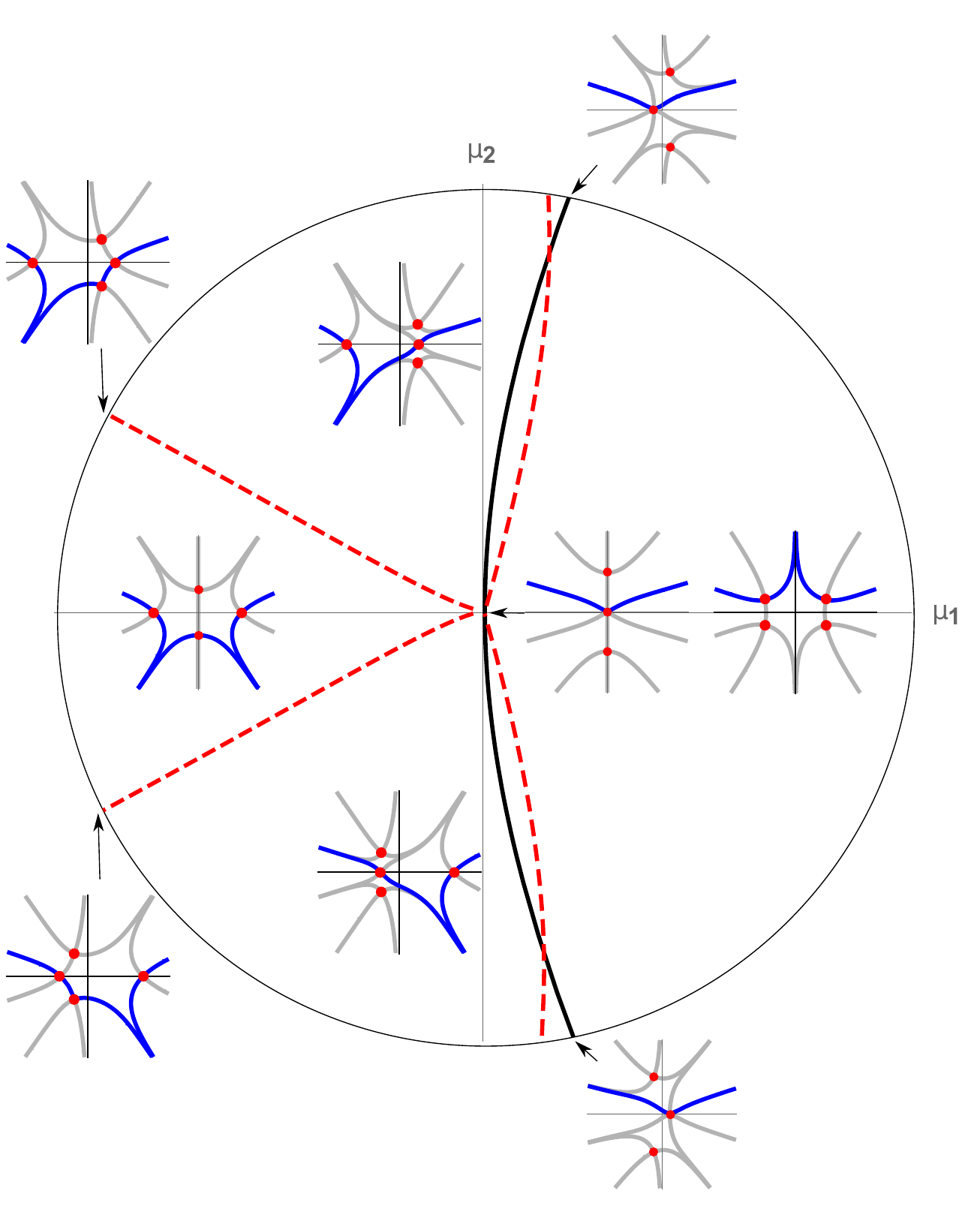}
\caption{$\mu_3=+1$}\label{fig:ST_mu3=1_1}
\end{subfigure}
\caption[The Picard-Lefschetz analysis of the unfolding of the swallowtail.]{The Picard-Lefschetz analysis of the unfolding of the swallowtail ($A_4$) singularity at $\mu_3=0$ and $+1$. The number of real saddles gives the number of images in the geometric optics approximation.}\label{fig:ST_mu3=1}
\end{figure}

The swallowtail singularity is more complicated, as it consists of the superposition of four non-degenerate saddle points. The singularity is of co-dimension $K=3$ and has three unfolding parameters $\mu_1,\mu_2$ and $\mu_3$, \textit{i.e.}, 
\begin{align}
\Psi(\bm{\mu};\nu) =\sqrt{\frac{\nu}{\pi}}\int_{-\infty}^\infty e^{i\left(\frac{x^5}{5} + \mu_3 \frac{x^3}{3} + \mu_2 \frac{x^2}{2} + \mu_1 x\right)\nu}\mathrm{d}x \,.
\end{align}
See figures \ref{fig:ST_mu3=-1} and \ref{fig:ST_mu3=1} for an lustration of unfolding of the swallowtail caustic and the Picard-Lefschetz analysis in the $(\mu_1,\mu_2,\mu_3)$-space.  \\

The analytic continuation of the exponent $i \phi(x;\bm{\mu}) \nu$ has four saddle points $\bar{x}_i$, given by the roots of the quartic equation
\begin{align}
x^4+ \mu_3x^2 +\mu_2 x +\mu_1=0\,.
\end{align}
Depending on $\bm{\mu}$, either zero, two or four of the saddle points are real-valued. The complex-valued saddle points always come in conjugate pairs since $\phi(x;\bm{\mu})$ is a real-valued function.

\begin{figure}[h]
\centering
\includegraphics[width=0.5 \textwidth]{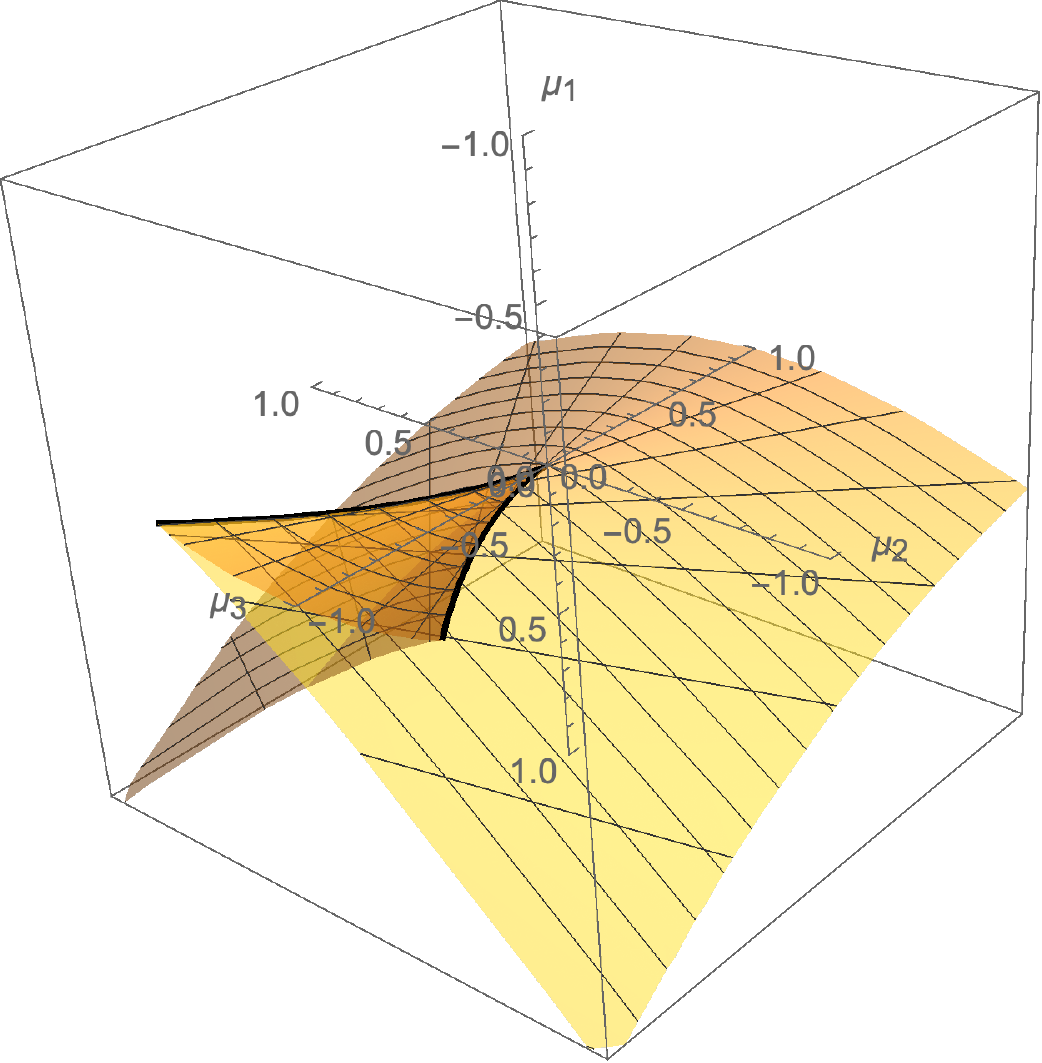}
\caption[The swallowtail singularity in the unfolding space.]{The swallowtail singularity in the unfolding space $(\mu_1,\mu_2,\mu_3)$. The yellow surface is the fold surface separating the single- double- and triple-image regions. The black line is the cusp line, along which we find the cusp saddle points.}\label{fig:SwallowTail}
\end{figure}

Geometric optics applied to this integral shows that the swallowtail caustic at $\bm{\mu}=(0,0,0)$ emanates a cusp-line and a fold-surface (see Fig.~\ref{fig:SwallowTail}). The fold-surface (the yellow surface in Fig.~\ref{fig:SwallowTail}) is given by
\begin{align}
A_2 = \left\{(3 u^4 + u^2 v, -4 u^3 -2 u v, v)| (u,v)\in \mathbb{R}^2 \right\} \subset M
\end{align}
satisfying the two constraints
\begin{align}
\frac{\mathrm{d} \phi(x;\bm{\mu})}{\mathrm{d}x} =0\,, \quad \frac{\mathrm{d}^2 \phi(x;\bm{\mu})}{\mathrm{d}x^2} =0\,.
\end{align}
The cusp-line (the black curve in Fig.~\ref{fig:SwallowTail}) lays on the fold-surface and is given by
\begin{align}
A_3=\{(-3 t^4, 8 t^3, -6 t^2)|t \in \mathbb{R}\} \subset M
\end{align}
satisfying the three constraints
\begin{align}
\frac{\mathrm{d} \phi(x;\bm{\mu})}{\mathrm{d}x} =0\,, \quad \frac{\mathrm{d}^2 \phi(x;\bm{\mu})}{\mathrm{d}x^2} =0\,, \quad \frac{\mathrm{d}^3 \phi(x;\bm{\mu})}{\mathrm{d}x^3} =0\,.
\end{align}
Note that the caustics are symmetric in the $(\mu_1,\mu_3)$-plane and that caustics only appear for negative $\mu_1$. This aids our analysis, since we can consider the three-dimensional swallowtail unfolding as a one parameter family of unfoldings in the $\{\mu_3=\text{const}\}$ planes.\\

In figures \ref{fig:ST_mu3=-1} and \ref{fig:ST_mu3=1} we plot three slices of the fold-surface and cusp-line for $\mu_3=-1,0$ and $+1$. For $\mu_3=-1$ we obtain the characteristic swallowtail shape in the fold-surface with the cusp-line intersecting at the tips, which gives the singularity its name. For $\mu_3=0$ we see the actual swallow caustic. The slice $\mu_3=+1$, is simpler as it does not contain intersections with the cusp-line and only consists of the fold-surface separating two regions. 

Given the caustics of geometric optics, we can evaluate the Lefschetz thimble. It again suffices to study the three cases $\mu_3 <0, \mu_3=0$ and $\mu_3>0$:

\begin{itemize}
\item 
We start by analysing the saddle points in the $\mu_3=-1$ plane (Fig.~\ref{fig:ST_mu3=-1}). The Picard-Lefschetz analysis for the enclosed region in the middle of the circle consists of four relevant real saddle points. This is a quadruple-image region. Note that multiple-image regions for localized lenses always consist of an odd number of images. In such lenses, the swallowtail will in practice always appear near another caustic such as a fold. 

Starting from the quadruple-image region and moving through the fold-line on the left, we observe that the two central saddle points merge to form a fold saddle point. The two saddle points subsequently move in the complex plane, one remaining relevant. Since this region corresponds to two real saddle points it is a double-image region.

Again, starting from the quadruple-image region and moving in the vertical direction, we observe that two of the outer saddle points merge to form a fold saddle point and subsequently move into the complex plane. The resulting Picard-Lefschetz analysis again consists of three relevant saddle points; two real and one complex. This again is a double-image region. If we, however, move from this double-image region to the double-image region on the left of the quadruple-image region, we pass through two Stokes lines, at which the complex saddle point switches from relevant to irrelevant. The Stokes lines are defined by 
\begin{align}
\text{Im}[i\phi(\bar{x}_i;\bm{\mu})\nu] = \text{Im}[i\phi(\bar{x}_j;\bm{\mu})\nu]
\end{align}
for $i \neq j$. Note that the Stokes lines can be associated with the cusp caustic at the tips of the fold-line. Note that the three relevant saddle points merge at these tips, to form a cusp saddle point.

Finally, if we move from the quadruple-image region along the line $\mu_2=0$ to the right, we pass through the intersection of the fold lines. At this point, both the left and right two real saddle points merge to form a `double' fold caustic. After passing this point, the four saddle points move in the complex plane. The Picard-Lefschetz analysis consists here of two relevant complex saddle points. This is a zero-image region (which will not be realized in localized lenses). If we pass from the double-image region to the zero-image region, we again observe a fold caustic in which two relevant real saddle points merge and move in the complex plane. This completes the analysis of the unfolding of the swallowtail caustic at $\mu_3=-1$.

\item
For $\mu_3 =0$, the geometry of the fold-line is simpler as the quadruple-image region has merged into the swallowtail caustic at the origin (see Fig.~\ref{fig:ST_mu3=1_0}). The Picard-Lefschetz analysis of this slice is largely similar to the one at $\mu_3=-1$. The double-image region (including the Stokes lines) has been deformed but is otherwise the same. The zero-image region is also unchanged. However, the intersection of the two fold-lines is replaced by the swallowtail saddle point at the origin of in the $(\mu_1,\mu_2)$-plane. Since this saddle point is the superposition of four non-degenerate saddle points, the amplitude integral is enhanced.

\item
For $\mu_3=+1$, the geometry of the caustics is depicted in Fig.~\ref{fig:ST_mu3=1_1}. The fold-line separates the zero-image region on the right from the double-image region on the left. Since the Picard-Lefschetz diagram in the zero-image region consists of four complex saddle points -- two of them being relevant -- there exist two distinct ways in which we can pass to the double-image region; either by merging the two saddle points on the left or on the right (see upper and lower diagram). The transition between these to takes place at the origin, where the four saddle points are located on the imaginary axis. The double-image region consists of three subregions. The rightmost Stokes lines at $\mu_3=0$ (see Fig.~\ref{fig:ST_mu3=1_0}) have partly moved into the zero-image region.
\end{itemize}

By patching the Picard-Lefschetz analysis at $\mu_3=-1,0$ and $+1$ together, we obtain a complete description of the unfolding of the swallowtail singularity in the $(\mu_1,\mu_2,\mu_3)$-space. Note that the Stokes lines obtained in figures \ref{fig:ST_mu3=-1} and \ref{fig:ST_mu3=1} are intersections of Stokes-surfaces, which together with the fold-surface partition the $\mu$-space.

\subsubsection{Numerics}
\begin{figure}
\centering
\begin{subfigure}[b]{0.25\textwidth}
\includegraphics[width= \textwidth]{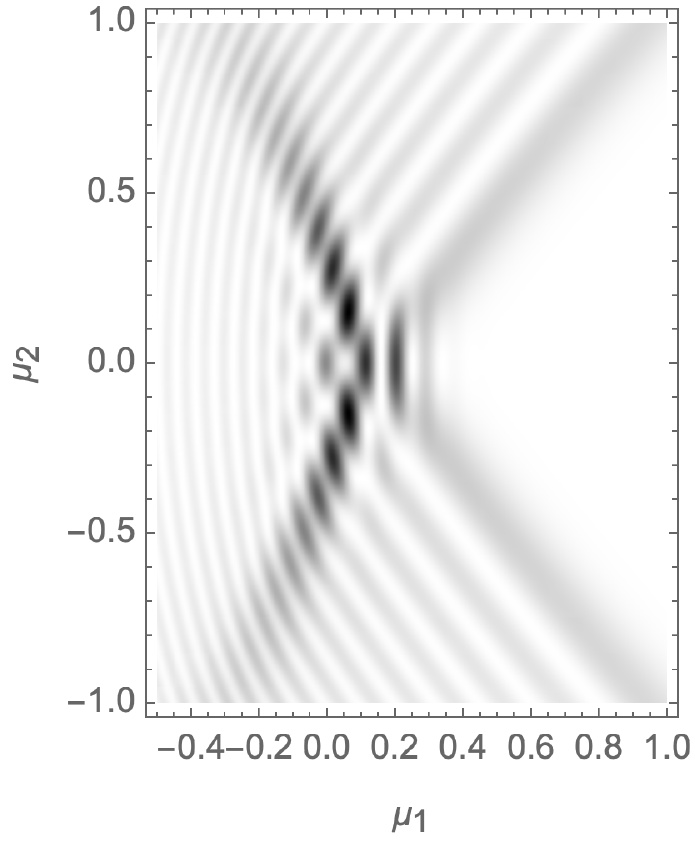}
\caption{$\mu_3=-1,\nu=50$}\label{fig:mu3=-1_nu=50}
\end{subfigure}
\begin{subfigure}[b]{0.25\textwidth}
\includegraphics[width= \textwidth]{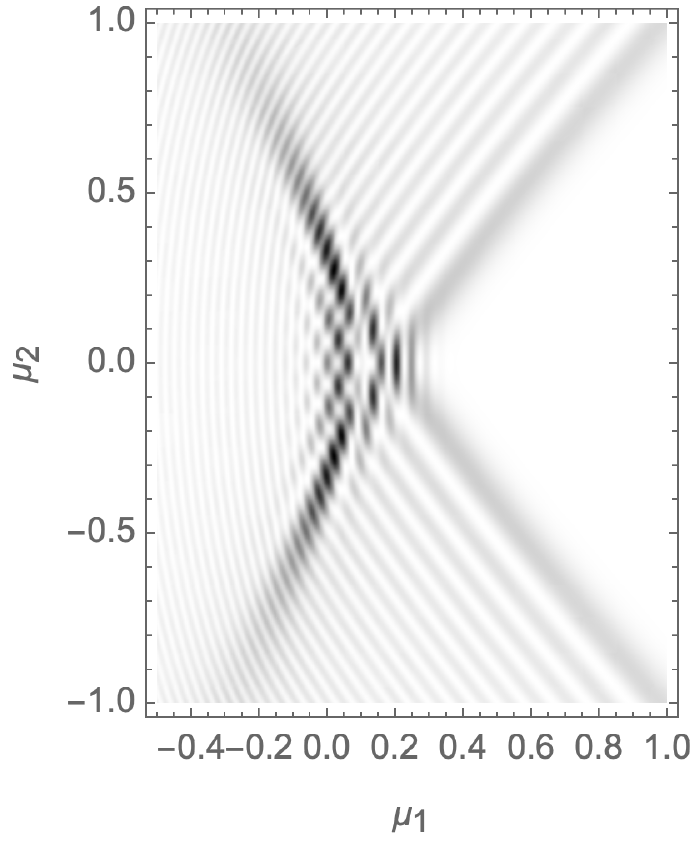}
\caption{$\mu_3=-1,\nu=100$}
\end{subfigure}
\begin{subfigure}[b]{0.25\textwidth}
\includegraphics[width= \textwidth]{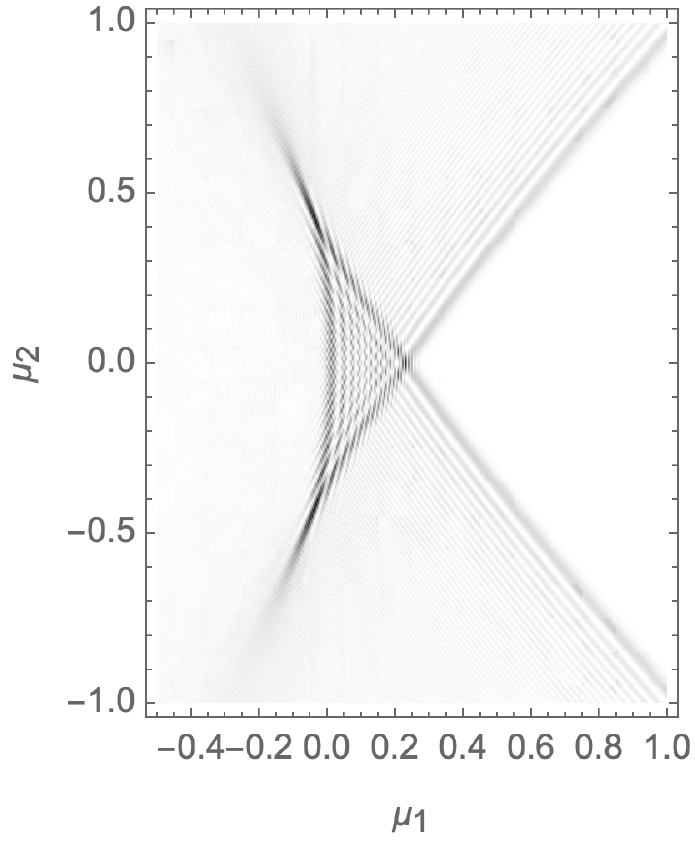}
\caption{$\mu_3=-1,\nu=500$}
\end{subfigure}\\
\begin{subfigure}[b]{0.25\textwidth}
\includegraphics[width= \textwidth]{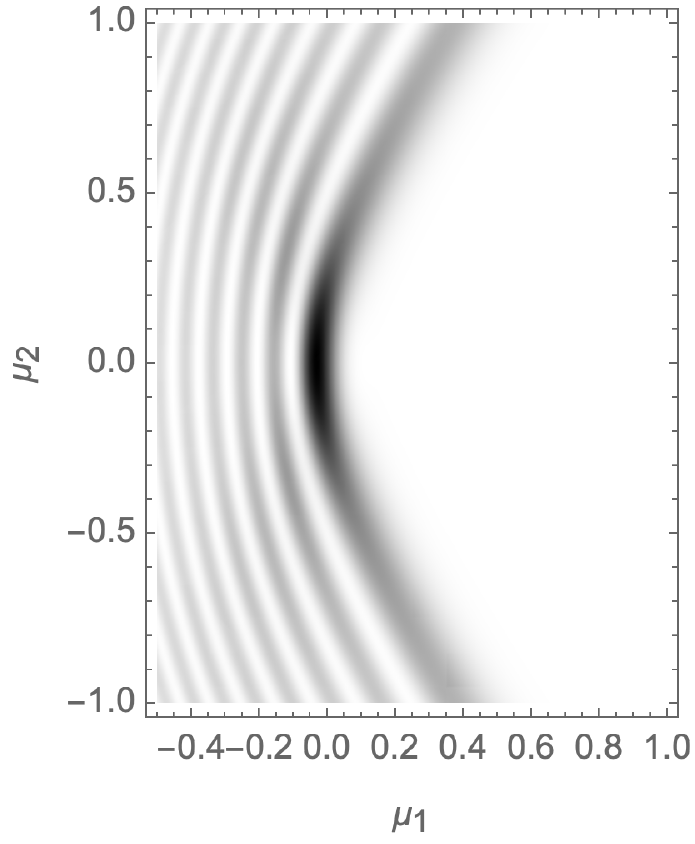}
\caption{$\mu_3=0,\nu=50$}
\end{subfigure}
\begin{subfigure}[b]{0.25\textwidth}
\includegraphics[width= \textwidth]{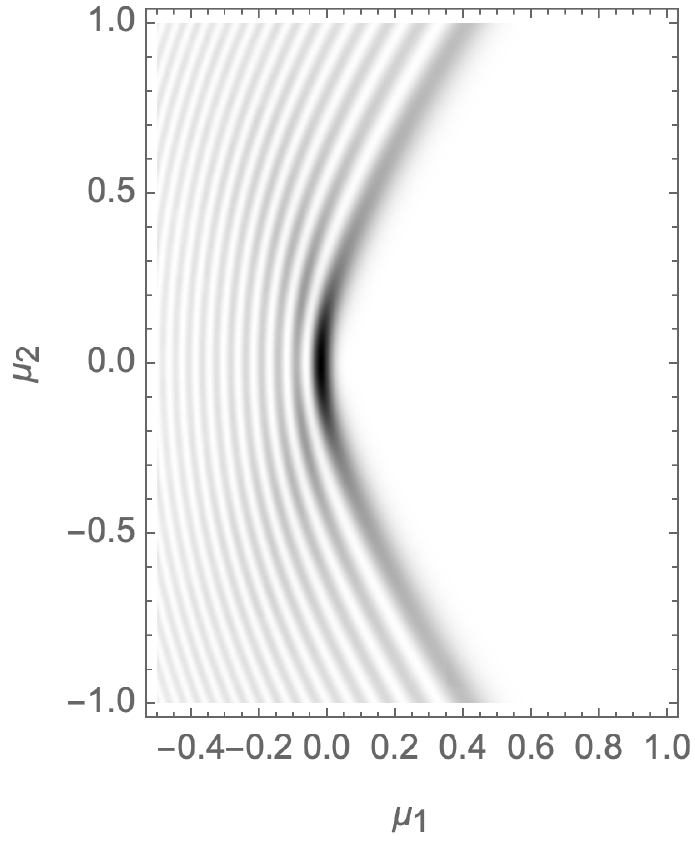}
\caption{$\mu_3=0,\nu=100$}
\end{subfigure}
\begin{subfigure}[b]{0.25\textwidth}
\includegraphics[width= \textwidth]{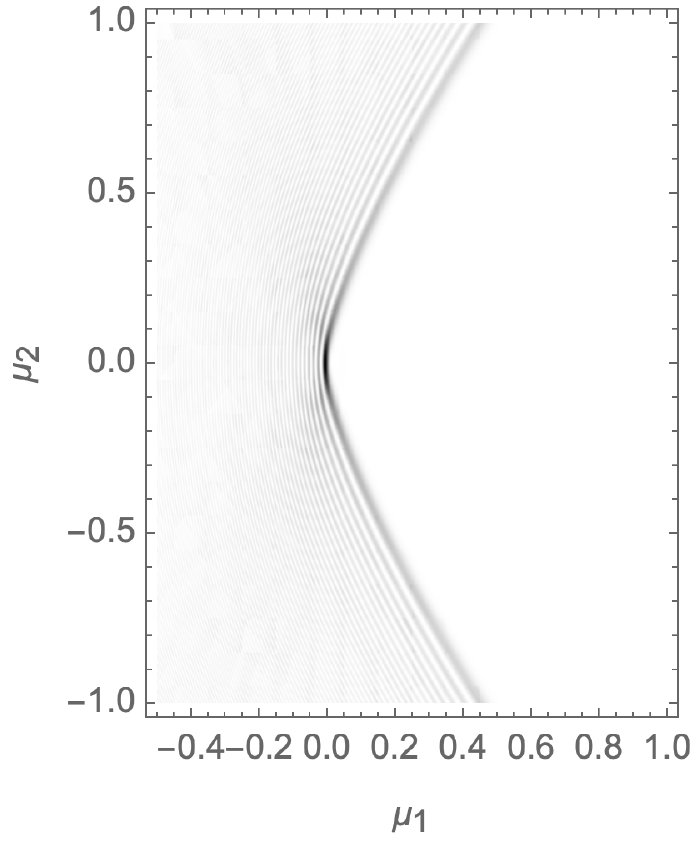}
\caption{$\mu_3=0,\nu=500$}
\end{subfigure}\\
\begin{subfigure}[b]{0.25\textwidth}
\includegraphics[width= \textwidth]{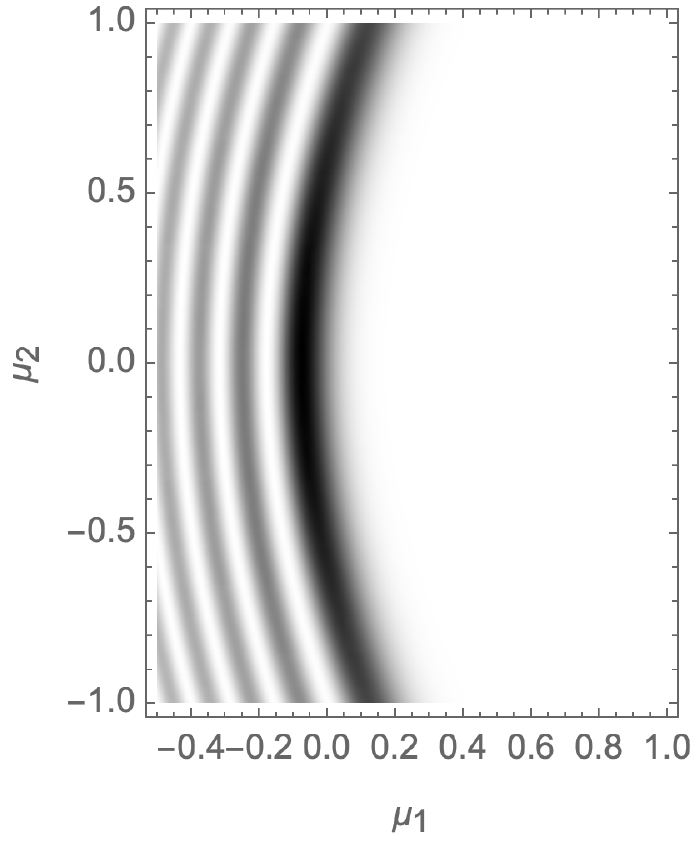}
\caption{$\mu_3=+1,\nu=50$}
\end{subfigure}
\begin{subfigure}[b]{0.25\textwidth}
\includegraphics[width= \textwidth]{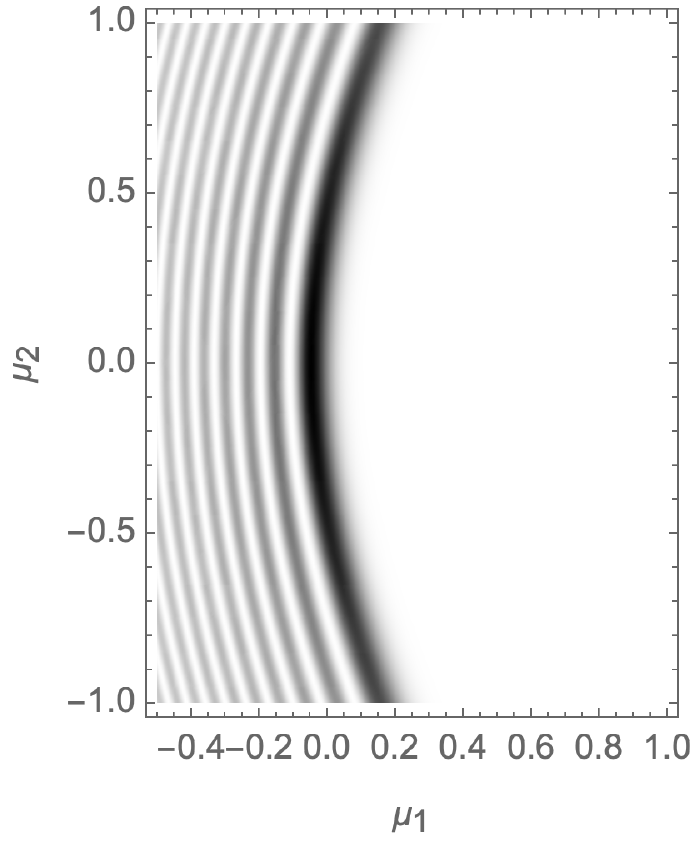}
\caption{$\mu_3=+1,\nu=100$}
\end{subfigure}
\begin{subfigure}[b]{0.25\textwidth}
\includegraphics[width= \textwidth]{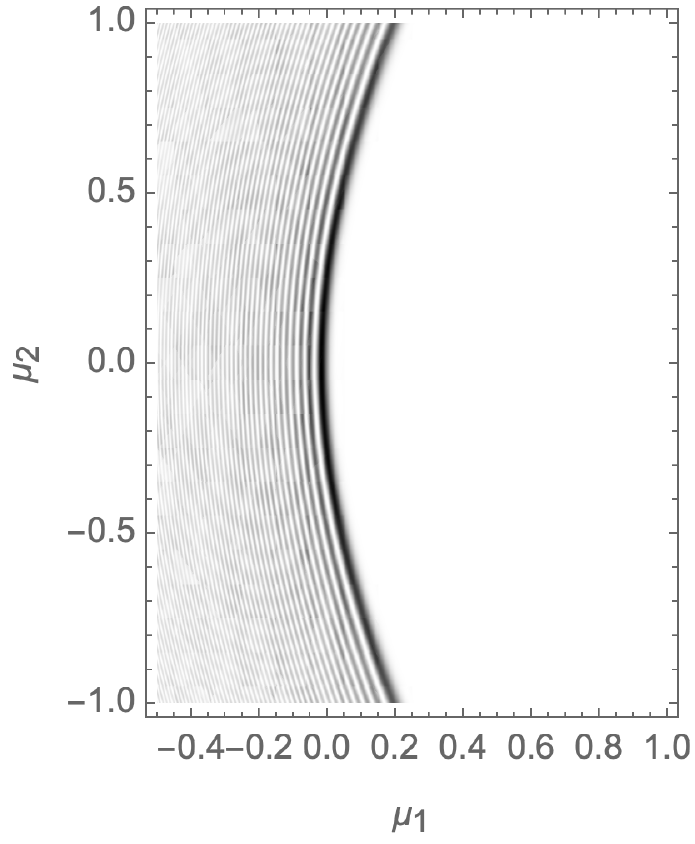}
\caption{$\mu_3=+1,\nu=500$}
\end{subfigure}
\caption[The normalized intensity of the unfolding of the swallowtail caustic.]{The normalized intensity, $I(\bm{\mu};\nu)$,  of the unfolding of the swallowtail caustic ($A_4$) sliced by the surfaces $\{\mu_3=-1\}, \{\mu_3=0\}, \{\mu_3=+1\}$ (respectively the left, central and right panels) for the frequencies $\nu=50,100$ and $500$ (respectively the upper, the middle and lower panels).}\label{fig:SwallowtailNumerics}
\end{figure}

Given the Lefschetz thimble, we can numerically compute the normalized intensity map of the lens (see Fig.~\ref{fig:SwallowtailNumerics}). The left, central and right panels depict the normalized intensity $I(\bm{\mu};\nu)$ for $\mu_3=-1,0$ and $+1$. The upper, middle and lower panels depict the different frequencies $\nu=50,100$ and $500$.

We observe that for $\nu=50$, interference is a dominant feature of the geometry of the caustic. The images are blurry and the geometry of the swallowtail is not resolved (Fig.~\ref{fig:mu3=-1_nu=50}). We do observe the power-law falloff associated with the cusp singularities, which contrasts with the exponential falloff of the fold singularities.

In the eikonal limit $\nu \to \infty$ we observe the emergence of a fold-line ($A_2$) with cusps ($A_3$). For $\nu=500$ the swallowtail structure at $\mu_3=-1$ is fully resolved. Note the difference in normalized intensity between the double- and quadruple-image regions. As the frequency $\nu$ is increased we observe that the enhanced flares, in the double-image regions, corresponding to the cusp caustics get thinner. However, note that they are independent of the frequency $\nu$.

\subsection{The elliptic umbilic $D_4^-$}
The caustics described above were part of the $A$-family. They are of co-rank $1$ and can be described by a one-dimensional integral. This should be contrasted by the $D$ family which is of co-rank $2$ and can only be studied in two-dimensional integrals.

The elliptic umbilic $D_4^-$ is a singularity with co-rank $2$ and co-dimension $K=3$. The unfolding is described in terms of the three unfolding parameters $(\mu_1,\mu_2,\mu_3)$. We consider the interference pattern emerging from the integral
\begin{align}
\Psi(\bm{\mu};\nu) =\frac{\nu}{\pi}\int_{\mathbb{R}^2} e^{i\left(x_1^3 - 3 x_1 x_2^2 - \mu_3(x_1^2 + x_2^2) - \mu_2 x_2 -\mu_1 x_1\right)\nu}\mathrm{d}x_1\mathrm{d}x_2 \,.
\end{align}

The analytic continuation of the exponent $i \phi(\bm{x};\bm{\mu}) \nu$ has four saddle points $\bar{x}_i$, given by the roots of the two quadratic equations
\begin{align}
3 x_1^2 - 3 x_2^2 - 2 \mu_3 x_1  - \mu_1 &= 0\\
-6 x_1 x_2 -   2 \mu_3x_2  - \mu_2 & = 0\,.
\end{align}
Depending on $\bm{\mu}$, either two or four of the saddle points are real-valued. The complex-valued saddle points always come in conjugate pairs since $\phi(\bm{x};\bm{\mu})$ is real-valued for real $x$. Solving this set of equations for $\mu_1$ and $\mu_2$ we obtain the Lagrangian map as a function of $\mu_3$,
\begin{align}
\xi_{\mu_3}(x_1,x_2)  = 
(3 x_1^2 - 3 x_2^2 - 2 x_1 \mu_3, -2 x_2 (3 x_1 + \mu_3), \mu_3)\,.
\end{align}

In the geometric limit, we form a fold-surface and three cusp lines. The fold-surface in base space $X=\mathbb{R}^2$ is given by
\begin{align}
A_2^X(\mu_3) = \left\{\left(\frac{\mu_3}{3}\cos \theta,\frac{\mu_3}{3}\sin \theta\right)|\theta \in [0,2\pi) \right\}
\end{align}
which is a cylinder with radius $\frac{\mu_3}{3}$, satisfying the equation
\begin{align}
|\mathcal{M}| = 0\,,
\end{align}
where the deformation tensor is given by
\begin{align}
\mathcal{M} &= \left[\frac{\partial^2 \phi(\bm{x};\bm{\mu})}{\partial x_i \partial x_j}\right]_{i,j=1,2}\\
&=
\begin{pmatrix}
6 x_1 - 2 \mu_3 & -6 x_2\\
-6 x_2 & -6 x_1 - 2 \mu_3
\end{pmatrix}
\,.
\end{align}
The three cusp-lines are straight lines and lay on the fold-surface,
\begin{align}
A_3^X(\mu_3)=\left\{ \left( \frac{\mu_3}{3}, 0\right), 
\left(\frac{\mu_3}{3}\cos \frac{2\pi}{3},\frac{\mu_3}{3}\sin \frac{2\pi}{3}\right),
\left(\frac{\mu_3}{3}\cos \frac{4\pi}{3},\frac{\mu_3}{3}\sin \frac{4\pi}{3}\right) \right\}
\end{align}
in the $X$ space.

\begin{figure}
\centering
\begin{subfigure}[b]{0.49\textwidth}
\includegraphics[width= \textwidth]{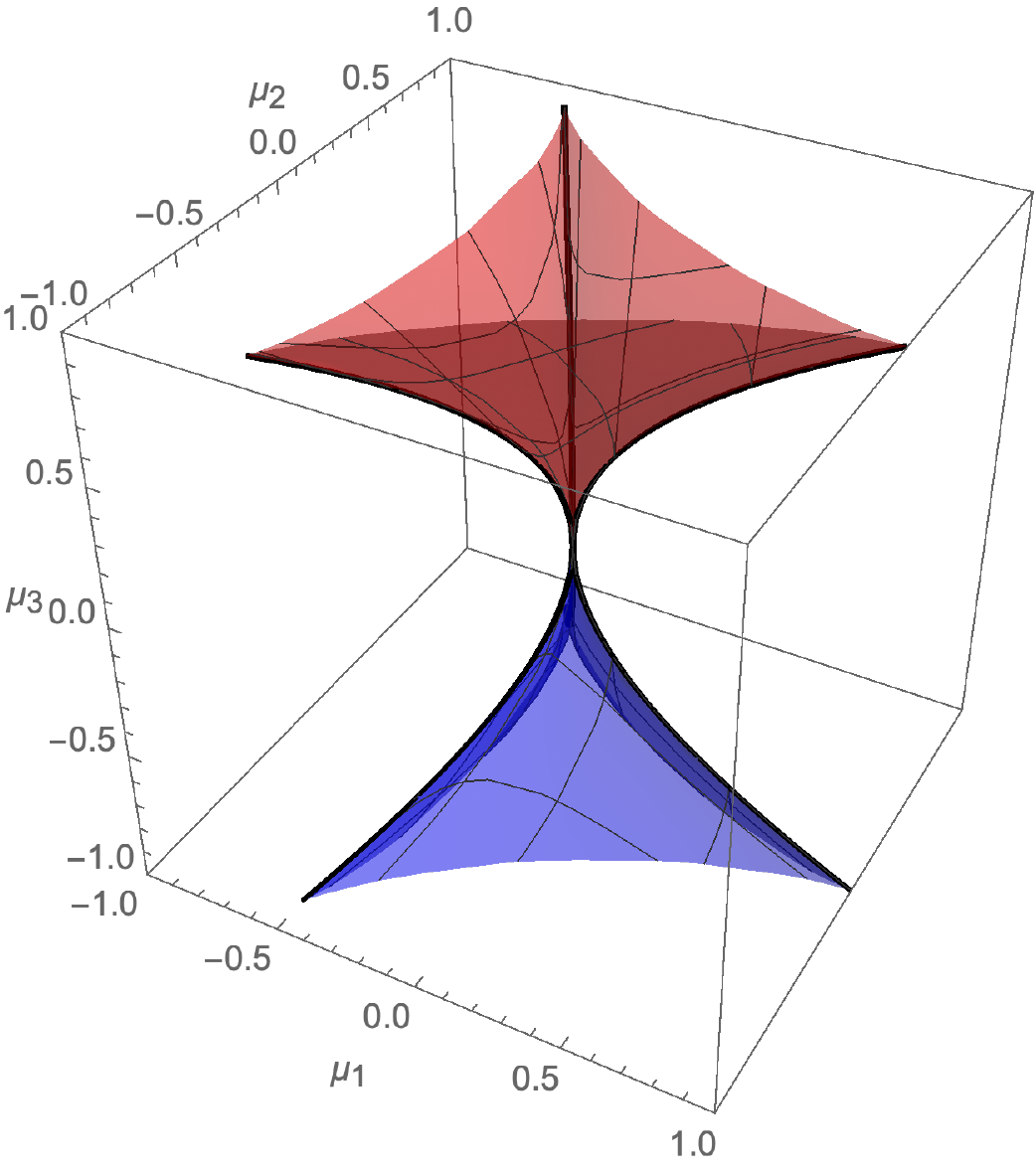}
\caption{Elliptic umbilic $D_4^-$}\label{fig:EllipticUmbilic}
\end{subfigure}
\begin{subfigure}[b]{0.49\textwidth}
\includegraphics[width= \textwidth]{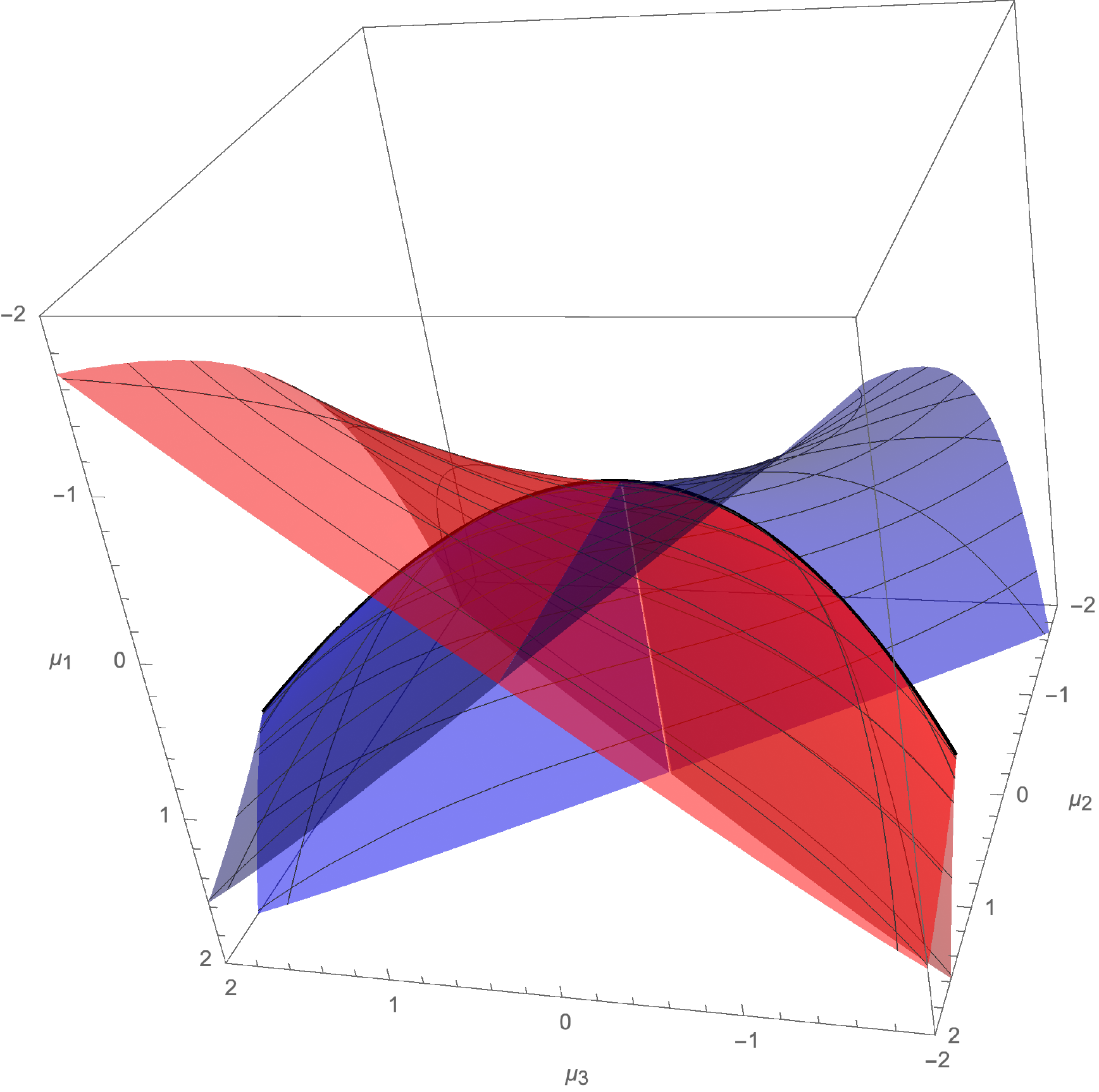}
\caption{Hyperbolic umbilic $D_4^+$}\label{fig:HyperbolicUmbilic}
\end{subfigure}
\caption[The elliptic and hyperbolic umbilic singularity in the unfolding space.]{The elliptic ($D_4^-$) and hyperbolic umbilic ($D_4^+$) singularity in the unfolding space $(\mu_1,\mu_2,\mu_3)$. The red and blue surfaces are the fold-surfaces corresponding to the eigenvlaue field $\lambda_1$ and $\lambda_2$ separating the single- double- and triple-image regions. The black lines are the cusp-lines, along which we find the cusp saddle points.}\label{fig:Umbilic}
\end{figure}

In $M$ space, after being mapped by $\xi_{\mu_3}$, the elliptic umbilic point is located at the origin. The fold-surface is given by
\begin{align}
A_2 = \left\{\left(\frac{ \mu_3^2}{3} ( \mp 2 \cos\theta + \cos(2 \theta), -\frac{2\mu_3^2}{3}(\pm 1 + \cos(\theta) )\sin(\theta), \pm \mu_3\right)\bigg| \theta \in [0,2\pi), \mu_3 \in \mathbb{R} \right\}
\end{align}
where the two branches corresponding to $\pm$ correspond to two disconnected pieces corresponding to the two eigenvalue fields of $\mathcal{M}$. The cusp lines are given by
\begin{align}
A_3= \left\{ (t^2, 0, t),( - t^2/2,\sqrt{3}t^2/2, t) , ( - t^2/2,-\sqrt{3}t^2/2, t) |t \in \mathbb{R}\right\}\,.
\end{align}
The fold-surface and cusp lines are illustrated in Fig.~\ref{fig:EllipticUmbilic}. The red and the blue surfaces denote the fold surfaces corresponding to the eigenvalue fields $\lambda_1$ and $\lambda_2$. The fold surface has a harp edge at the cusp lines (in black).

\begin{figure}
\centering
\begin{subfigure}[b]{0.49\textwidth}
\includegraphics[width= \textwidth]{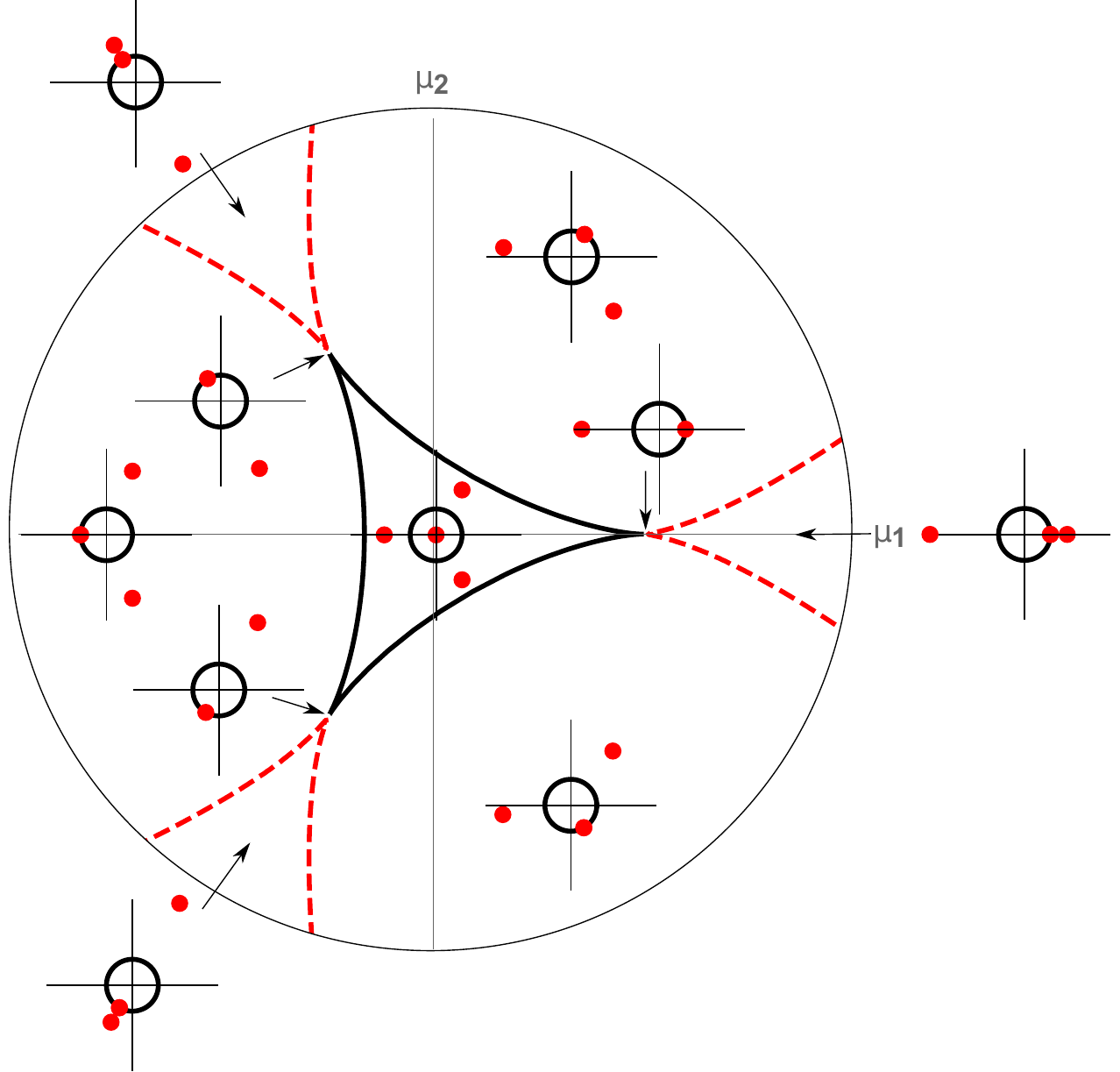}
\caption{$\mu_3=\pm 1$}\label{fig:EllipticUnfolding_1}
\end{subfigure}
\begin{subfigure}[b]{0.49\textwidth}
\includegraphics[width= \textwidth]{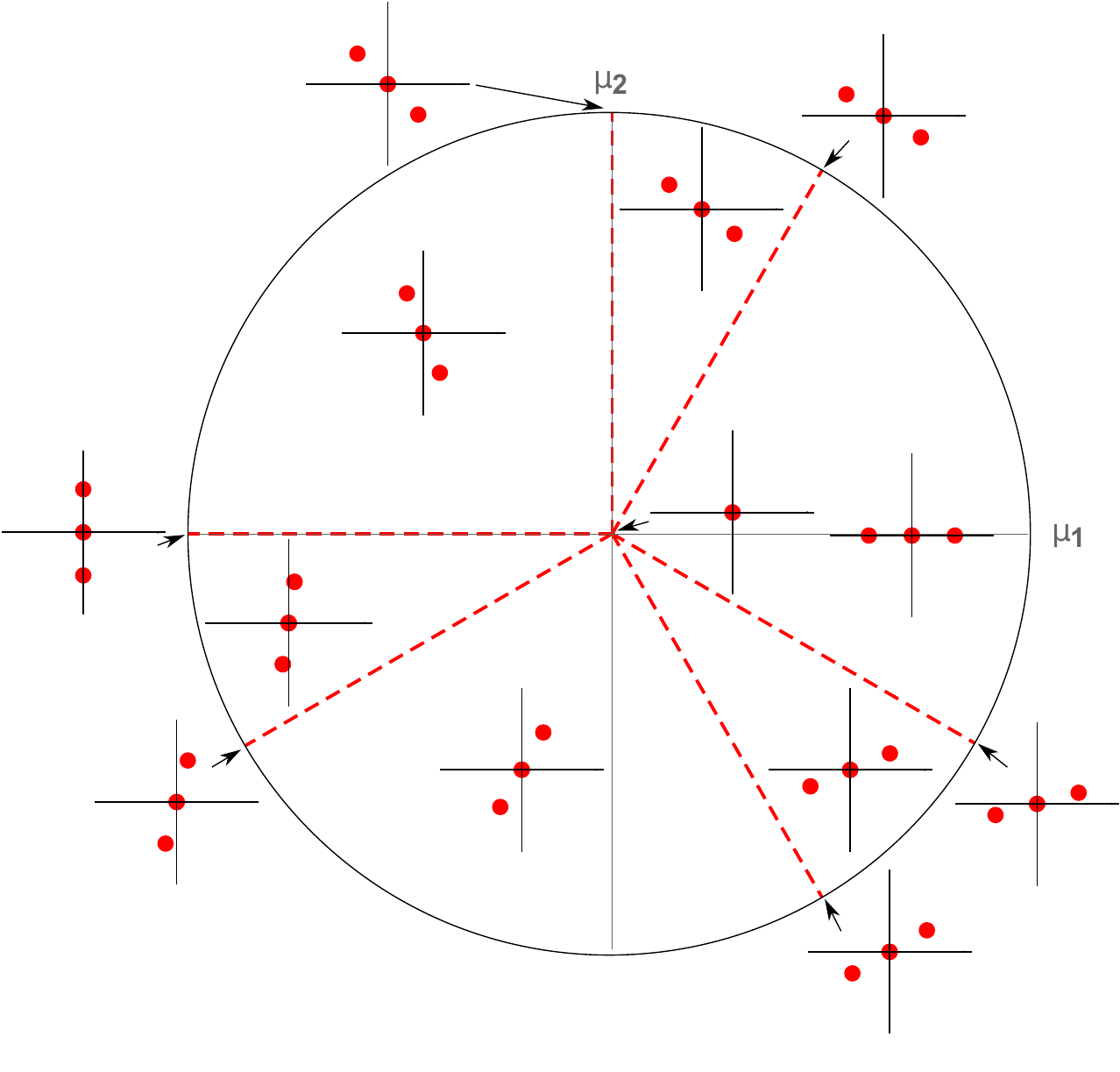}
\caption{$\mu_3= 0$}\label{fig:EllipticUnfolding_0}
\end{subfigure}
\caption[The Picard-Lefschetz analysis of the elliptic umbilic $D_4^-$ by the surface.]{Intersection of the elliptic umbilic $D_4^-$ by the surface $\{\mu_3=-1\}$. The black line is the fold-line and the red line is the Stokes line.  The number of real saddles gives the number of images in the geometric optics approximation.}\label{fig:EllipticUnfolding}
\end{figure}

Note the symmetry of the triangular singularity and point symmetry of the caustic. By performing the Picard-Lefschetz analysis for the two slices $\mu_3= \pm 1$ and $\mu_3=0$ we can obtain the Picard-Lefschetz diagram of the unfolding of the singularity. See Fig.~\ref{fig:EllipticUnfolding} for the Picard-Lefschetz analysis of the two slices. The small diagrams are the real parts of the four saddle points in the $(x_1,x_2)$-plane. The black circle is the caustic in the base space at the corresponding $\mu_3$.

\begin{itemize}
\item
We first consider the case $\mu_3 \neq 0$. At the origin, the four saddle points are real-valued (see Fig.~\ref{fig:EllipticUnfolding_1}). As a consequence, we conclude that they are all relevant. This is a quadruple-image region. One of the four saddle points is located inside the black circle. The other three are symmetrically distributed around the circle.  

When crossing the fold-line, the saddle point in the circle merges with one of the outer saddle points on the circle. After passing the fold-line, the two saddle points become complex. The saddle point with the smallest real part of the exponent $i \phi(\bar{\bm{x}};\bm{\mu})$ will remain relevant whereas the other saddle point becomes irrelevant. The outside of the triangle is a double-image region. Note that the real parts of the two complex saddle points always coincides with the black circle. 

When approaching one of the three the cusp points, three of the four saddle points merge at a single point on the circle. Note that the four saddle points are collinear in the cusps. 

Finally, note that the double-image region consists of six subregions divided by six Stokes lines. In the regions on the left, the upper right and the lower right, the Lefschetz thimble passes through two real and one complex saddle point. In the regions to the right, upper left and lower right, the Picard-Lefschetz analysis consist of only two real saddle points (the ones outside the circle. The Stokes lines are again associated with the three cusps.

\item
In the case, $\mu_3=0$, the central region is replaced by the elliptic umbilic saddle point (see Fig.~\ref{fig:EllipticUnfolding_0}). The rest of the $\mu_1$-$\mu_2$-plane is divided into six distinct regions by the six Stokes lines. The upper left, lower left, and the upper right regions consist of two relevant real saddle points. These regions correspond to the upper left, lower left, and right region in Fig.~\ref{fig:EllipticUnfolding_1}. The three remaining regions consist of two real and one complex relevant saddle points.
\end{itemize}

These slices form a complete description of the Lefschetz thimble of the unfolding of the elliptic umbilic in the $\mu$-space.
 
\subsubsection{Numerics}
\begin{figure}
\centering
\begin{subfigure}[b]{0.32\textwidth}
\includegraphics[width= \textwidth]{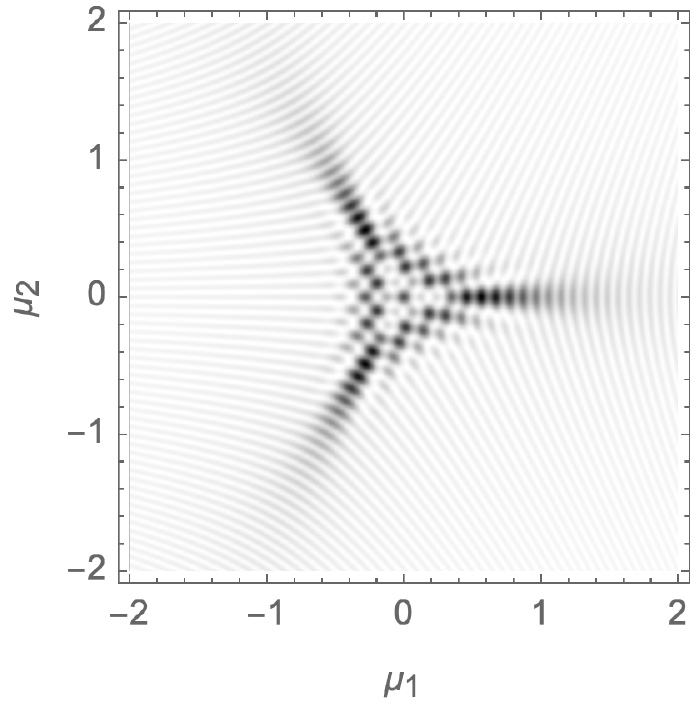}
\caption{$\mu_3=\pm1,\nu=50$}
\end{subfigure}
\begin{subfigure}[b]{0.32\textwidth}
\includegraphics[width= \textwidth]{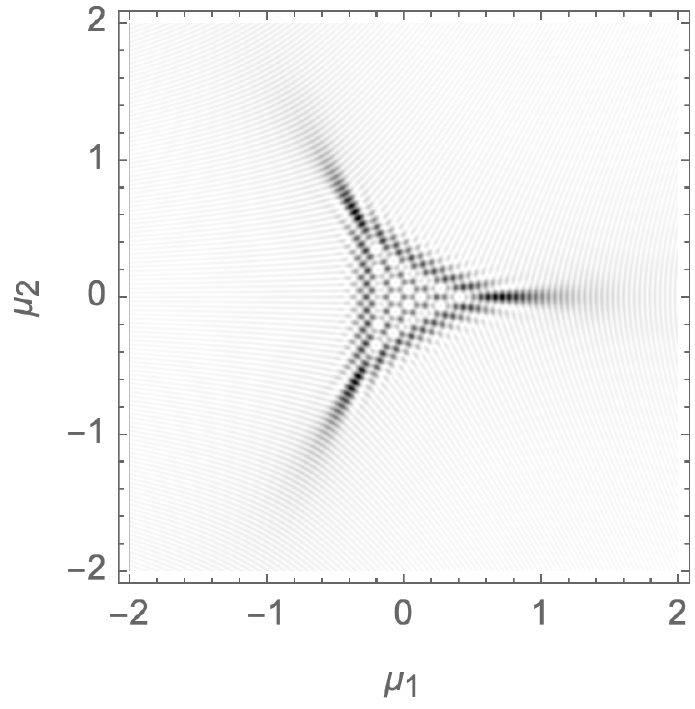}
\caption{$\mu_3=\pm 1,\nu=100$}
\end{subfigure}
\begin{subfigure}[b]{0.32\textwidth}
\includegraphics[width= \textwidth]{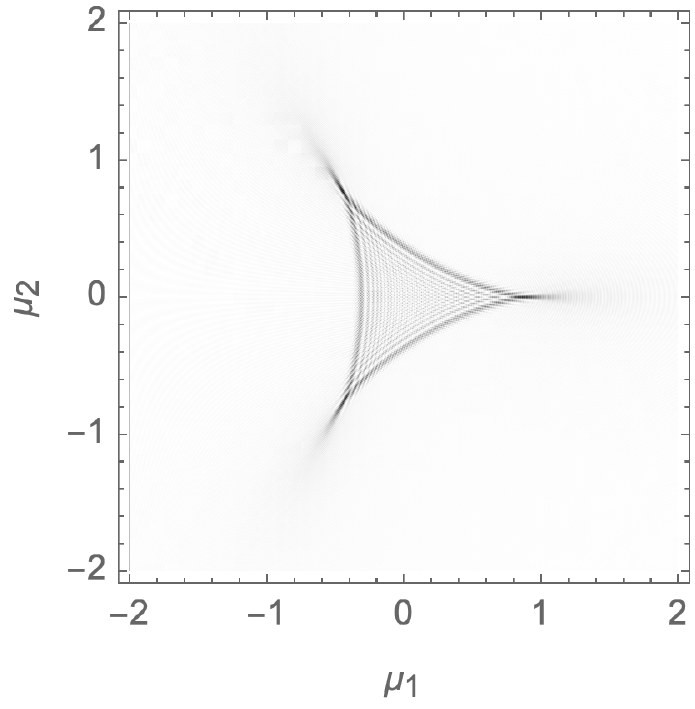}
\caption{$\mu_3=\pm1,\nu=500$}
\end{subfigure}\\
\begin{subfigure}[b]{0.32\textwidth}
\includegraphics[width= \textwidth]{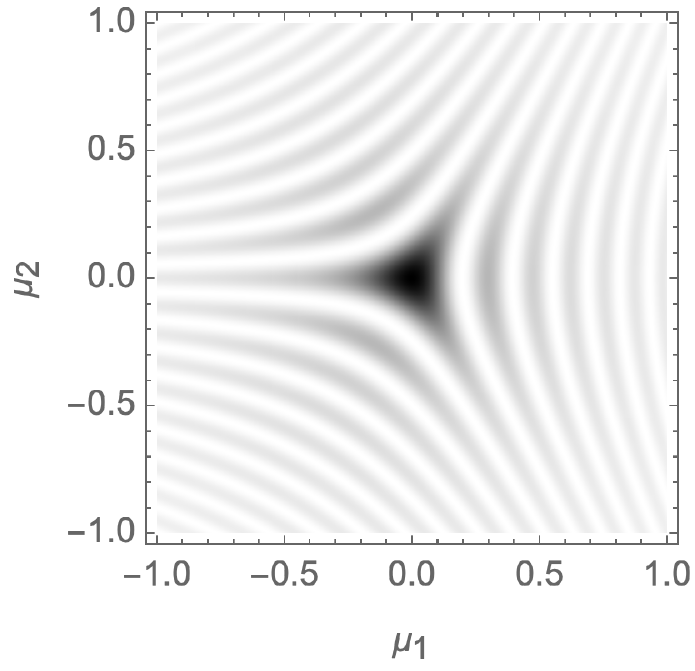}
\caption{$\mu_3=0,\nu=50$}
\end{subfigure}
\begin{subfigure}[b]{0.32\textwidth}
\includegraphics[width= \textwidth]{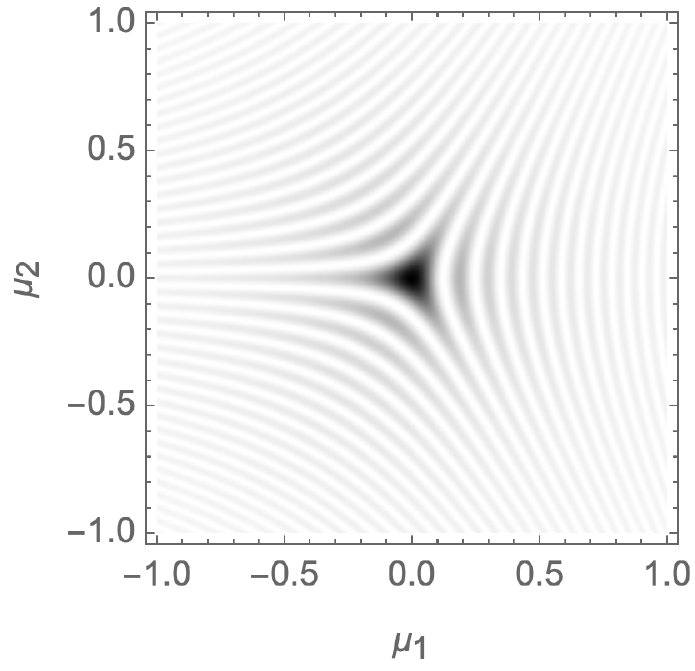}
\caption{$\mu_3=0,\nu=100$}
\end{subfigure}
\begin{subfigure}[b]{0.32\textwidth}
\includegraphics[width= \textwidth]{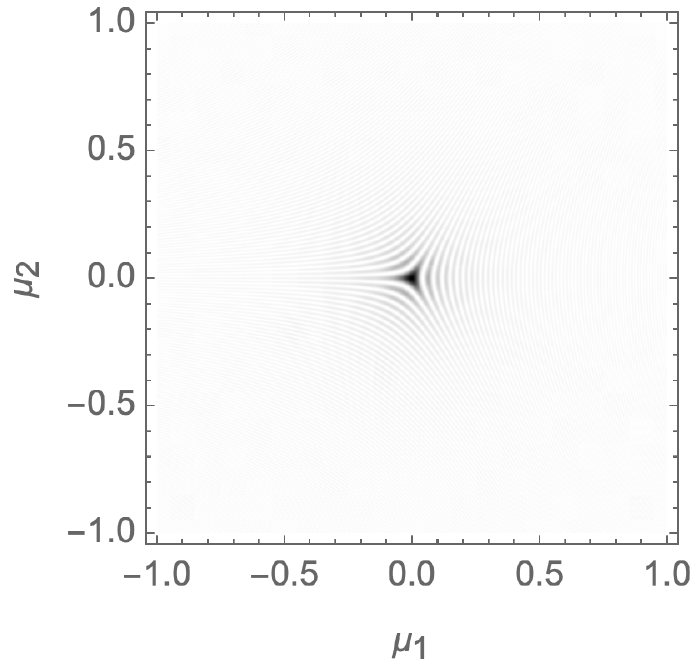}
\caption{$\mu_3=0,\nu=500$}
\end{subfigure}
\caption[The intensity map of the unfolding of the elliptic umbilic caustic ($D_4^-$).]{The normalized intensity, $I(\bm{\mu};\nu)$, of the unfolding of the elliptic umbilic caustic ($D_4^-$) sliced by the surfaces $\{\mu_3=\pm 1\}, \{\mu_3=0\}$ (respectively the upper and lower panels) for the frequencies $\nu=50,100$ and $500$ (respectively the left, the centre and right panels).}\label{fig:EllipticNumerics}
\end{figure}

Given the Lefschetz thimble, we can numerically evaluate the normalized intensity map of the lens (Fig.~\ref{fig:EllipticNumerics}). The upper and lower panels depict the normalized intensity $I(\bm{\mu};\nu)$ for $\mu_3=\pm 1$ and $0$. The left, central and right panels depict the frequencies $\nu=50,100$ and $500$.

The normalized intensity map corresponding to the unfolding of the elliptic umbilic ($D_4^-$) has a triangular symmetry. As the frequency increases, the normalized intensity profile steepest and increases in amplitude. In the plane $\mu_3 = \pm 1$ we observe a fold-line in a triangular configuration with three cusp caustics at the corners. For the frequency, $\nu=50$ the fold-line is relatively blurry. We again observe outward stripes emanating from the cusp caustics. These again follow a power-law falloff independent of the frequency. As the frequency is raised to $\nu=100$ and $\nu=500$ we observe that the fold lines become sharper and the fringes in the quadruple image region shrink. The normalized intensity at frequency $\nu=500$ is very close to the normalized intensity map predicted by geometric optics.

\subsection{The hyperbolic umbilic $D_4^+$}
The hyperbolic umbilic $D_4^+$ completes the set of caustics appearing in two-dimensional lenses. It is again a singularity with co-rank $2$ and co-dimension $K=3$. The unfolding is described in terms of the three unfolding parameters $(\mu_1,\mu_2,\mu_3)$. We consider the integral
\begin{align}
\Psi(\bm{\mu};\nu) =\frac{\nu}{\pi}\int_{\mathbb{R}^2} e^{i\left(x_1^3 + x_2^3 - \mu_3 x_1 x_2 - \mu_2 x_2 - \mu_1 x_1\right)\nu}\mathrm{d}x_1\mathrm{d}x_2 \,.
\end{align}

The analytic continuation of the exponent $i \phi(\bm{x};\bm{\mu}) \nu$ has four saddle points $\bar{x}_i$, given by the roots of the two quadratic equations
\begin{align}
3 x_1^2 - \mu_3 x_2 - \mu_1  &=0\,,\\
3 x_2^2 - \mu_3 x_1 - \mu_2  &= 0\,.
\end{align}
Depending on $\mu$, either zero, two or four of the saddle points are real-valued. The complex-valued saddle points always come in conjugate pairs since $\phi(\bm{x};\bm{\mu})$ is real-valued for real $x$. Solving this set of equations for $\mu_1$ and $\mu_2$ we obtain the Lagrangian map as a function of $\mu_3$,
\begin{align}
\xi_{\mu_3}(x_1,x_2)  = 
(3 x_1^2 - x_2 \mu_3, 3 x_2^2 - x_1 \mu_3)\,.
\end{align}

In the geometric limit, we form a fold-surface and a cusp lines. The fold-surface in $X$ space is given by
\begin{align}
A_2^X(\mu_3) = \left\{\left( \pm \frac{\mu_3^2}{36 t},t \right)|t \in \mathbb{R} \right\}
\end{align}
which is a cylinder with radius $\frac{\mu_3}{3}$, satisfying the equation
\begin{align}
|\mathcal{M}| = 0
\end{align}
where the deformation tensor is given by
\begin{align}
\mathcal{M} &= \left[\frac{\partial^2 \phi(\bm{x};\bm{\mu})}{\partial x_i \partial x_j}\right]_{i,j=1,2}\\
&=
\begin{pmatrix}
6 x_1& -\mu_3 \\
-\mu_3 & 6 x_2
\end{pmatrix}
\,.
\end{align}

The three cusp-lines are linear lines laying on the fold-surface,
\begin{align}
A_3^X(\mu_3)=\left\{(-\mu_3/6, -\mu_3/6 ) \right\}
\end{align}
in the $X=\mathbb{R}^2$ space.

In the parameter space $M$, the elliptic umbilic point is located at the origin. The fold-surface is given by
\begin{align}
A_2 &= \left\{\left(
3 u^4 \pm 6 u v^3, \pm 6 u^3 v+ 3 v^4, \mp 6 u v
\right)| u,t \in \mathbb{R} \right\}\\
A_2 &= \left\{\left(
3 u^4 \mp 6 u v^3, \mp 6 u^3 v+ 3 v^4, \mp 6 u v
\right)| u,t \in \mathbb{R} \right\}
\end{align}
where the two solutions correspond to two disconnected pieces corresponding to the two eigenvalue fields of $\mathcal{M}$. The cusp line in the parameter space is given by
\begin{align}
A_3= \left\{  (t^{2}/4, t^2/4, t)|t \in \mathbb{R}\right\}\,.
\end{align}
The fold-surface and cusp-line are illustrated in Fig.~\ref{fig:HyperbolicUmbilic}. The red and the blue surfaces denote the fold surfaces corresponding to the eigenvalue fields $\lambda_1$ and $\lambda_2$ of the deformation tensor $\mathcal{M}$. The fold surface has a harp edge at the cusp lines (in black).

Note the symmetry of the triangular singularity and point symmetry of the caustic. By performing the Picard-Lefschetz analysis for the two slices $\mu_3= \pm 1$ and $\mu_3=0$ we can obtain an understanding of the relevant saddle points.\\

\begin{figure}[h]
\centering
\begin{subfigure}[b]{0.49\textwidth}
\includegraphics[width= \textwidth]{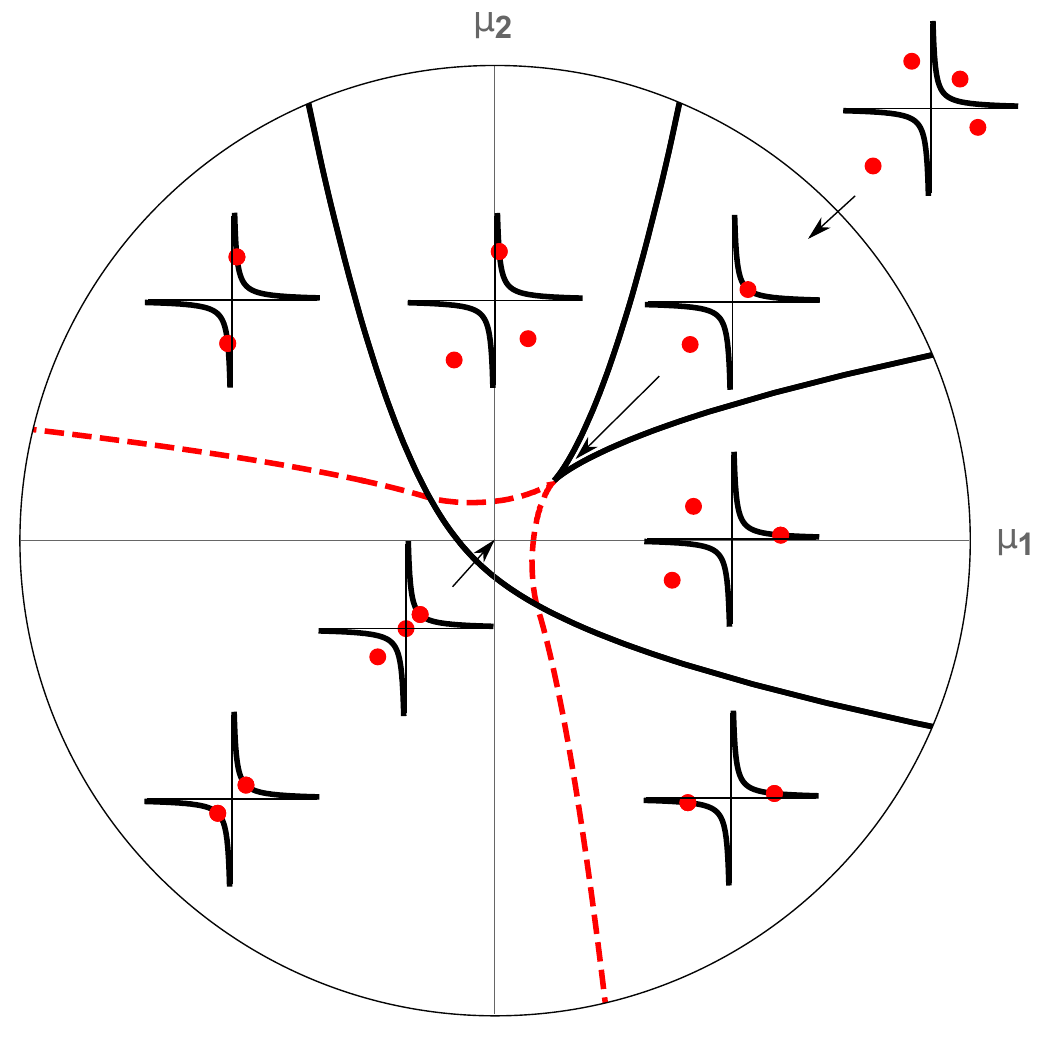}
\caption{$\mu_3=\pm 1$}\label{fig:HyperbolicUnfolding_1}
\end{subfigure}
\begin{subfigure}[b]{0.49\textwidth}
\includegraphics[width= \textwidth]{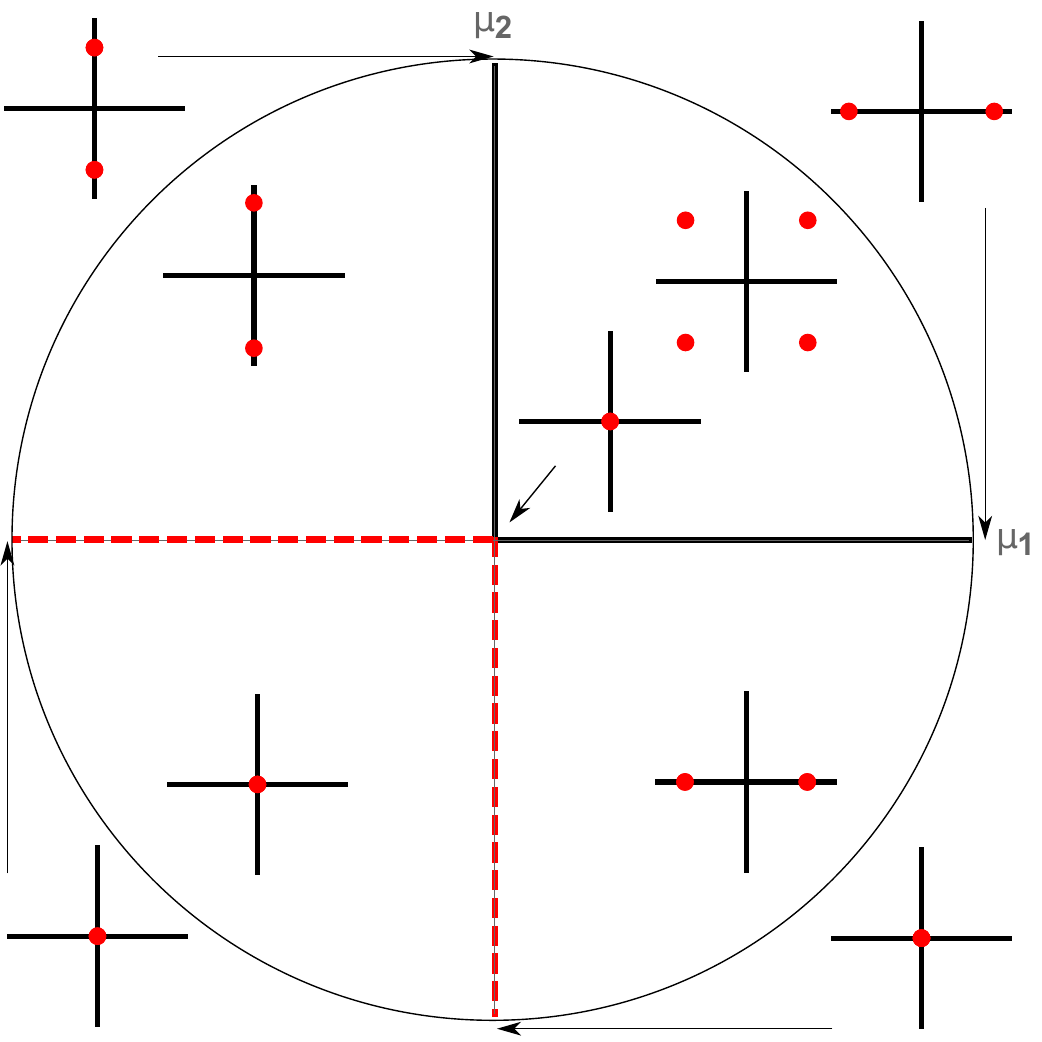}
\caption{$\mu_3= 0$}\label{fig:HyperbolicUnfolding_0}
\end{subfigure}
\caption[The Picard-Lefschetz analysis of the hyperbolic umbilic in parameter space.]{Intersection of the hyperbolic umbilic $D_4^+$ with the surface $\{\mu_3=-1\}$. The black line is the fold-line and the red line is the Stokes line.  The number of real saddles gives the number of images in the geometric optics approximation.}\label{fig:HyperbolicUnfolding}
\end{figure}

See Fig.~\ref{fig:HyperbolicUnfolding} for the Picard-Lefschetz analysis of the two slices. The small diagrams are the real parts of the four saddle points in the $(x_1,x_2)$ plane. The black circle is caustic in $X$ space at $\mu_3=\pm 1$ and $0$. 
\begin{itemize}
\item
Consider the slice $\mu_3 = \pm 1$ (see Fig.~\ref{fig:HyperbolicUnfolding_1}). In the upper right corner, the four saddle points are real-valued. In the corresponding Picard-Lefschetz analysis, they are all relevant. This is a quadruple-image region. When we pass the left or lower fold-line, two of the four saddle points merge at the hyperbola in $X$ space, to form a fold singularity. Afterward, both saddle points become complex. The one with the smallest real part of the exponent $i\phi(\bar{\bm{x}};\bm{\mu})\nu$ remains relevant whereas the other saddle point becomes irrelevant. Just like in the elliptic umbilic, the real part of the complex saddle points remains on the hyperboloid. This is a double-image region. Depending on whether we cross the fold line to the left or below the quadruple region, two different saddle points merge. 

If we move from the quadruple-image region to the cusp, we obtain a singularity due to the merger of three saddle points. After passing through the cusp, only the two real saddle points will be relevant. The two complex saddle points are irrelevant. 

From the double-image region, we can pass the second fold-line. At this fold-line, the two remaining real saddle points merge to form a fold saddle point after which they move in the complex plane. Note that the real parts of these two saddle points remain on the second branch of the hyperbolic. Since the Picard-Lefschetz analysis does not contain any real-valued saddle points after passing the second fold-line, this is a zero-image region. The zero-image region is again subdivided into three subregions. In the upper left and lower right regions, the Picard-Lefschetz analysis consists of two relevant complex saddle points. In the lower left region, the Picard Lefschetz analysis consists of one relevant complex saddle point.\\

\item
In the case, $\mu_3=0$ the analysis is similar to the one obtained for $\mu_3=\pm 1$, since the regions are trivially deformed (see Fig.~\ref{fig:HyperbolicUnfolding_0}). In the upper right region, again four saddle points are real. All of them are relevant. This is still a quadruple-image region. The fold line along the positive $\mu_1$ and $\mu_2$ axis is double fold lines, as the two fold lines at $\mu_3=\pm 1$ have merged. The left and lower right regions are zero-image regions. In the upper left and lower right regions, the Picard-Lefschetz analysis consists of two relevant complex saddle points. In the lower left region, the Picard-Lefschetz analysis again consists of one relevant complex saddle point. This concludes the Picard-Lefschetz analysis.
\end{itemize}

\subsubsection{Numerics}
\begin{figure}
\centering
\begin{subfigure}[b]{0.32\textwidth}
\includegraphics[width= \textwidth]{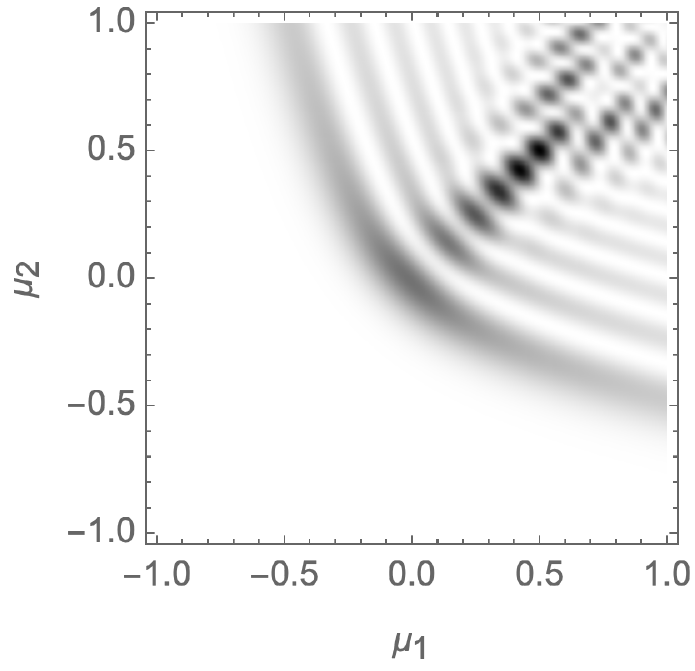}
\includegraphics[width= \textwidth]{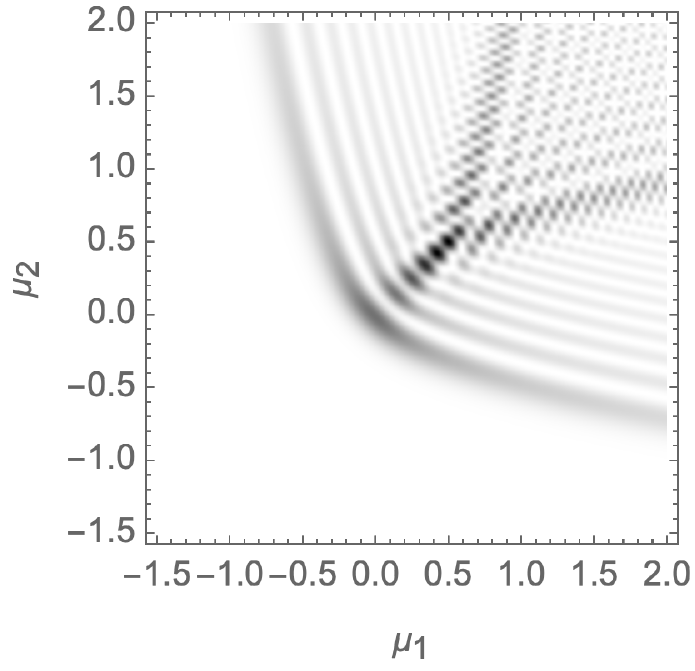}
\caption{$\mu_3=\pm1,\nu=50$}
\end{subfigure}
\begin{subfigure}[b]{0.32\textwidth}
\includegraphics[width= \textwidth]{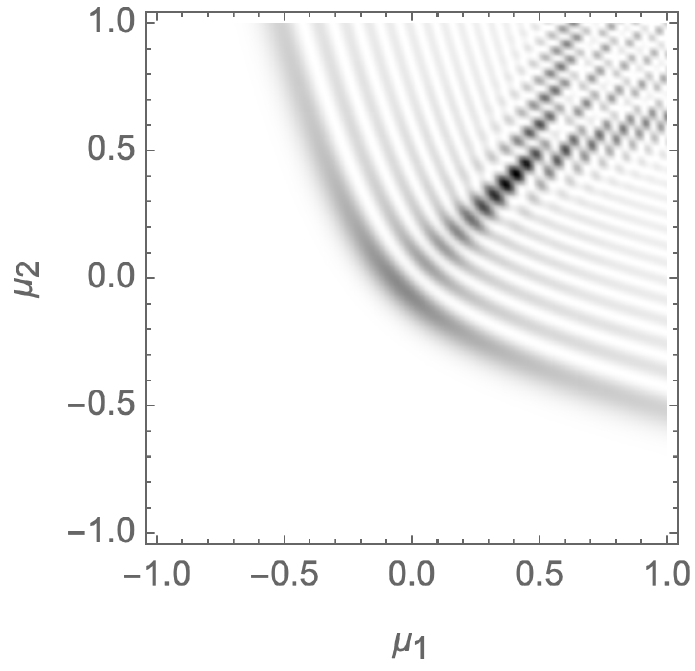}
\includegraphics[width= \textwidth]{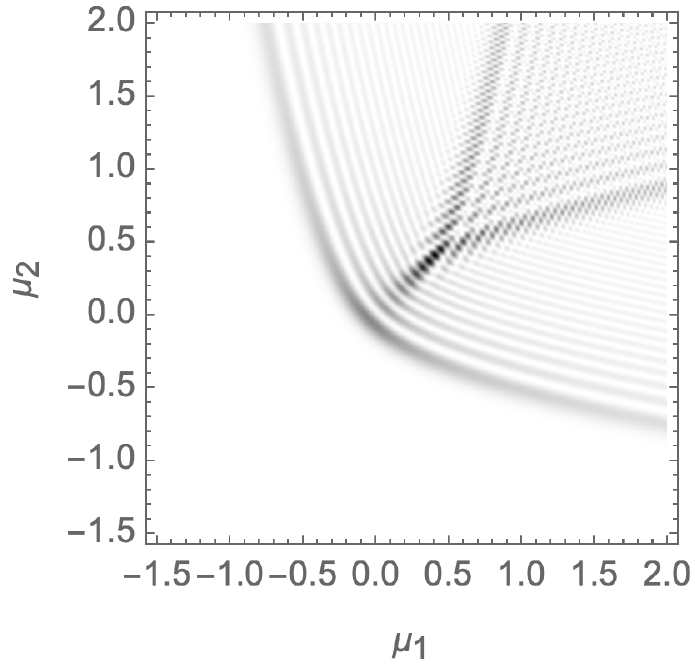}
\caption{$\mu_3=\pm 1,\nu=100$}
\end{subfigure}
\begin{subfigure}[b]{0.32\textwidth}
\includegraphics[width= \textwidth]{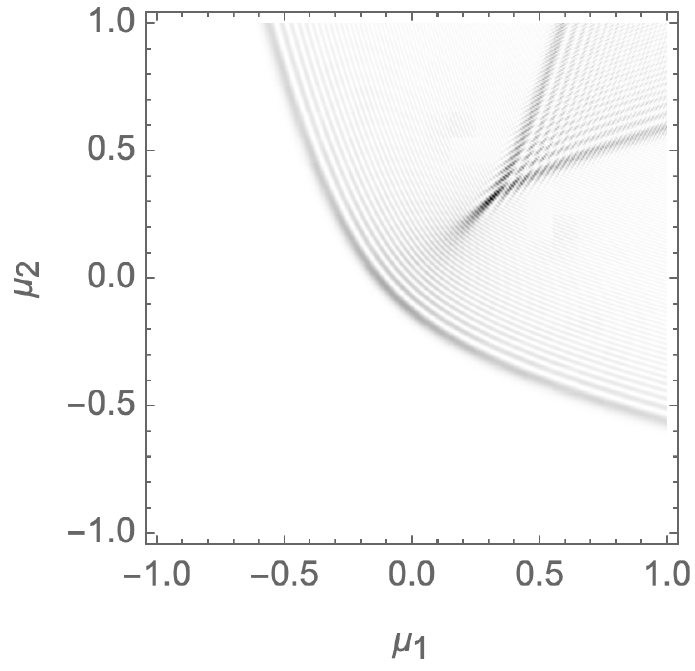}
\includegraphics[width= \textwidth]{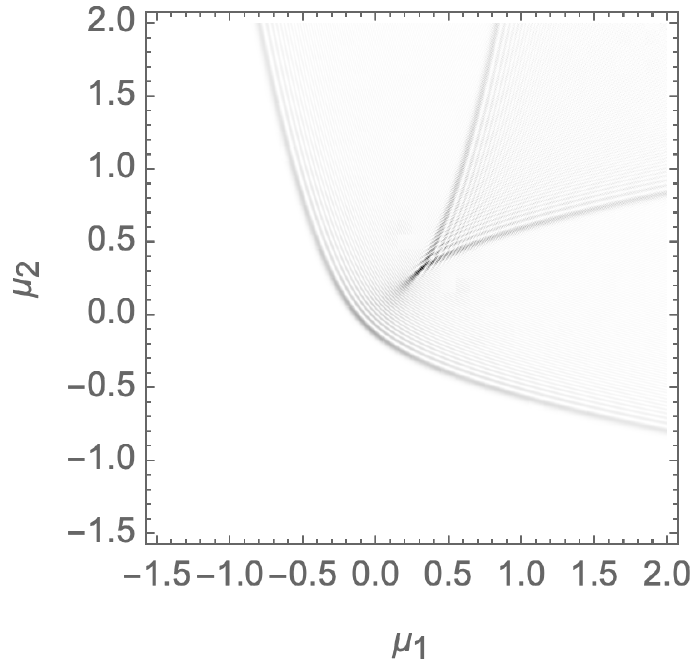}
\caption{$\mu_3=\pm1,\nu=500$}
\end{subfigure}\\
\begin{subfigure}[b]{0.32\textwidth}
\includegraphics[width= \textwidth]{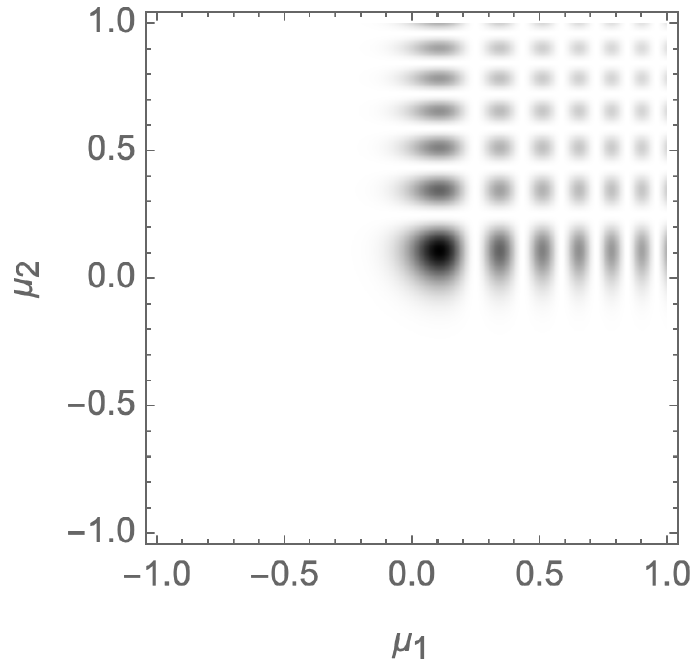}
\caption{$\mu_3=0,\nu=50$}
\end{subfigure}
\begin{subfigure}[b]{0.32\textwidth}
\includegraphics[width= \textwidth]{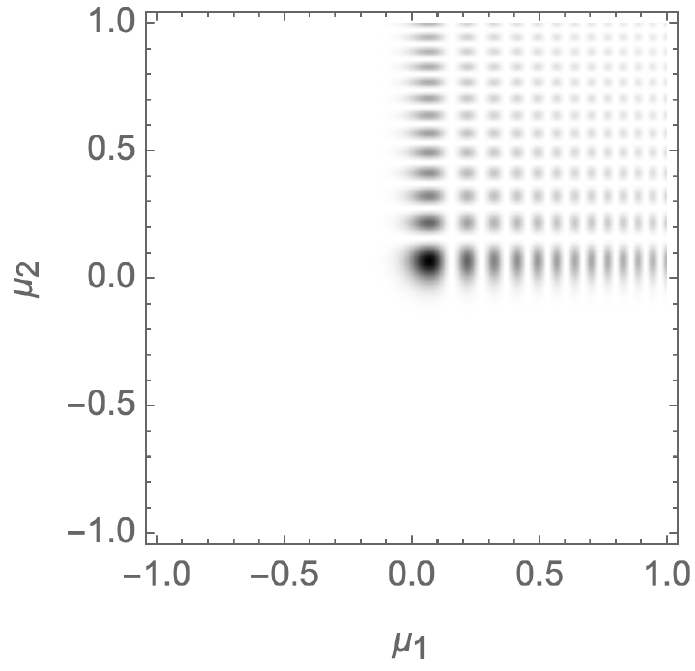}
\caption{$\mu_3=0,\nu=100$}
\end{subfigure}
\begin{subfigure}[b]{0.32\textwidth}
\includegraphics[width= \textwidth]{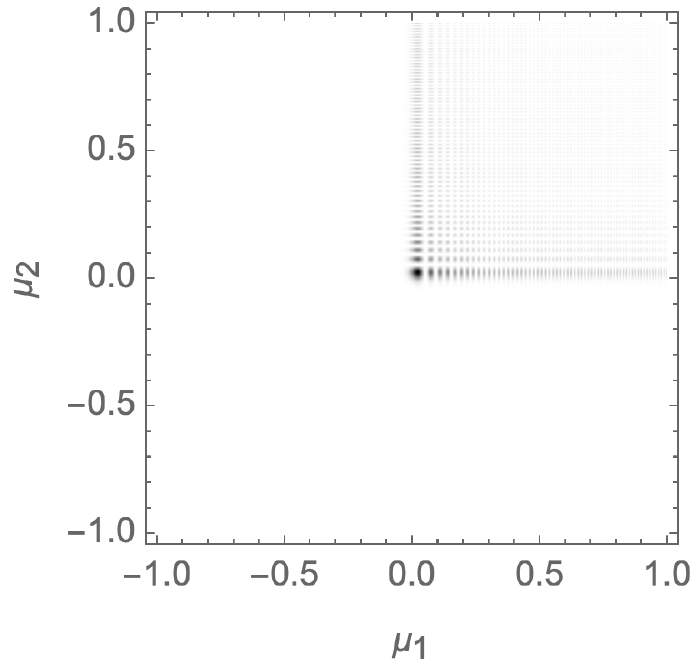}
\caption{$\mu_3=0,\nu=500$}
\end{subfigure}
\caption[The intensity map of the unfolding of the hyperbolic umbilic caustic.]{The normalized intensity, $I(\bm{\mu};\nu)$, of the unfolding of the hyperbolic umbilic caustic ($D_4^+$) sliced by the surfaces $\{\mu_3=\pm 1\}, \{\mu_3=0\}$ (respectively the upper and lower panels) for the frequencies $\nu=50,100$ and $500$ (respectively the left, the centre and right panels).}\label{fig:HyperbolicNumerics}
\end{figure}

Given the Lefschetz thimble, we can numerically compute the normalized intensity map (see Fig.~\ref{fig:HyperbolicNumerics}). The upper and lower panels depict the normalized intensity $I(\bm{\mu};\nu)$ for $\mu_3=\pm 1$ and $0$. The left, central and right panels depict the different frequencies $\nu=50,100$ and $500$.

For both unfoldings at $\mu_3 = \pm 1 $ and $\mu_3 = 0$, the normalized intensity map closely follows the caustics structure represented in Fig.~\ref{fig:HyperbolicUnfolding}. In the zero-image region, the normalized intensity vanishes. In the double-image regions, for $\mu_3 = \pm 1$, the normalized intensity oscillates forming lines of equal normalized intensity as should be expected from the presence of the left fold line. In the quadruple-image regions, the normalized intensity oscillates in two directions, for $\mu_3=\pm 1$ forming the structure we observed for the cusp caustic, and for $\mu_3 =0$ forming an interference pattern with rectangular symmetry.

In the eikonal limit $\nu \to \infty$, the normalized intensity becomes sharper and the caustics become more pronounced. It should, in particular, be noted that the normalized intensity at the hyperbolic umbilic (in the origin in the plots corresponding to $\mu_3=0$), the normalized intensity rises rapidly with $\nu$. This is in correspondence with the scaling relations we found above.

\section{Two dimensional localized lenses}\label{sec:2DlocalLenses}
The seven elementary singularities form a dictionary of the local behavior of the lens integral 
\begin{align}
\Psi(\bm{\mu};\nu) &= \left(\frac{\nu}{\pi}\right)^{N/2} \int_{\mathbb{R}^N} e^{i \phi(\bm{x};\bm{\mu})\nu}\mathrm{d}\bm{x}\,, \\
\phi(\bm{x};\bm{\mu}) &=(\bm{x}-\bm{\mu})^2 + \varphi(\bm{x})\,,
\end{align}
near caustics. Their corresponding normalized intensity map completely describes the local properties of lensed images. However, the global structure of the caustic is in general different. Since the normal forms of the elementary singularities are polynomials, the corresponding phase $\varphi$ has support throughout the base space $X=\mathbb{R}^N$. The catastrophes with an even co-dimension $K$, lead to an image with an even number of images. In contrast, localized lenses lead to $n$-image regions with $n$ an odd integer. We now turn to the study of interference patterns appearing in localized lenses near caustics. We evaluate three two-dimensional lenses, which simulate the behavior of a localized lens and include the five elementary catastrophes appearing in two-dimensional lenses. In the process, we also demonstrate the accuracy of the integration scheme along the Lefschetz thimble.

\subsection{A generic peak}
In general, lensing effects are strongest near the extrema of the variation of the phase $\varphi$. It is for this reason natural to study the effect of an asymmetric peak in the phase variation $\varphi$, with
\begin{align}
\varphi(\bm{x}) = \frac{\alpha}{1+x_1^2+2 x_2^2}\,,
\end{align}
the two-dimensional generalization of the one-dimensional lens studied in Section \ref{sec:1DLensExample}.
For astrophysical plasma lenses, the parameter $\alpha$ scales according to the dispersion relation $\alpha \propto \nu^{-2}$.

The Lagrangian map is given by
\begin{align}
\xi(\bm{x}) &= \bm{x} + \frac{1}{2}\nabla \varphi(\bm{x}) \\
&= \bm{x} - \frac{\alpha}{(1+x_1^2+2 x_2^2)^2} (x_1,2 x_2) \,.
\end{align} 
The map forms a caustic where the deformation tensor
\begin{align}
\mathcal{M}_{ij}=\frac{\partial^2 \phi(\bm{x};\bm{\mu})}{\partial x_i\partial x_j}\,,
\end{align}
with the eigenvalue and eigenvector fields $\lambda_i(x),v_i(x)$, is singular, \textit{i.e.},
\begin{align}
|\mathcal{M}(\bm{x})|=\lambda_1(\bm{x}) \lambda_2(\bm{x}) = 0\,.
\end{align}
For convenience, we order the eigenvalue and eigenvector fields by $\lambda_1(\bm{x}) \leq \lambda_2(\bm{x})$.

\begin{figure}
\centering
\begin{subfigure}[b]{0.3\textwidth}
\includegraphics[width=\textwidth]{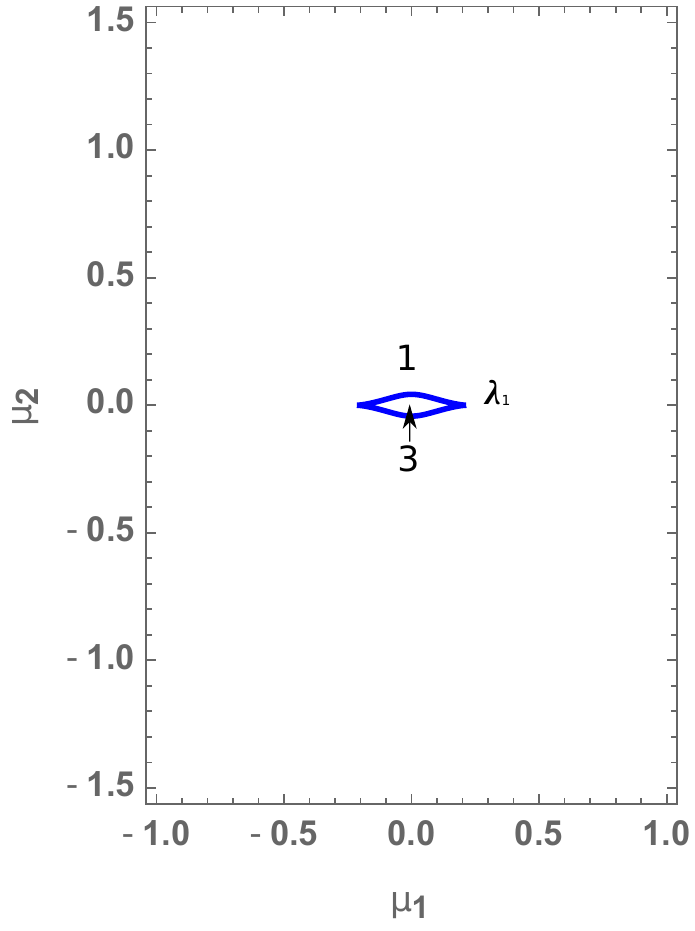}
\caption{$\alpha = 0.7$}
\end{subfigure} ~
\begin{subfigure}[b]{0.3\textwidth}
\includegraphics[width=\textwidth]{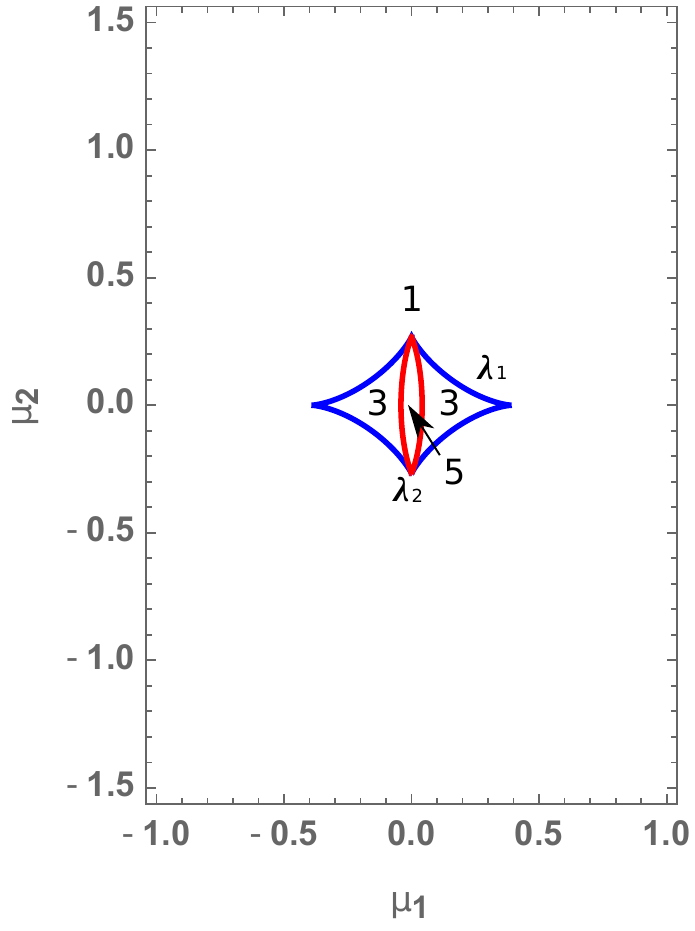}
\caption{$\alpha = 64/49$}
\end{subfigure} ~
\begin{subfigure}[b]{0.3\textwidth}
\includegraphics[width=\textwidth]{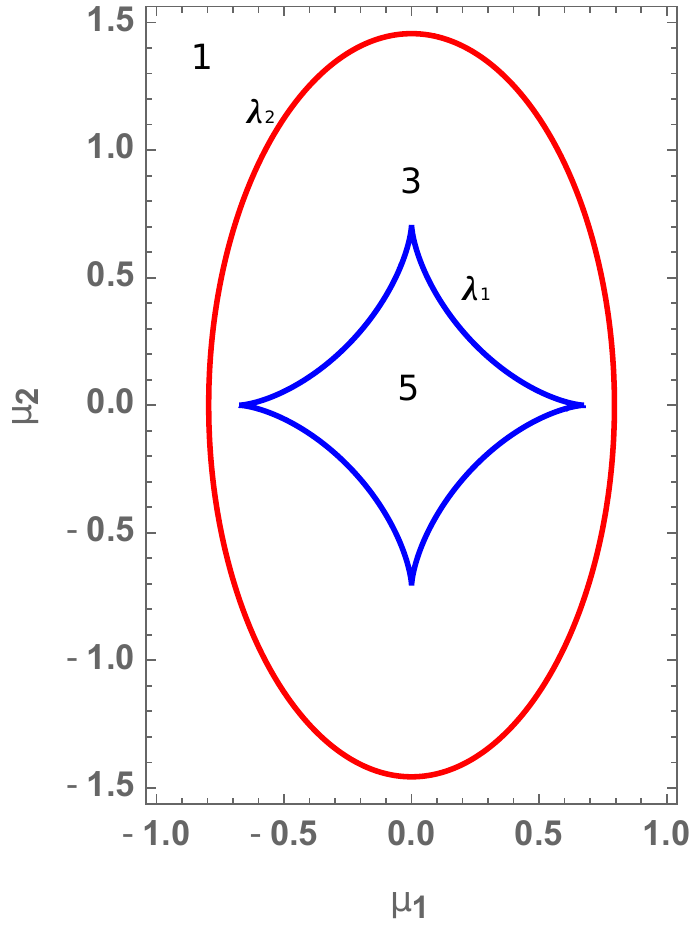}
\caption{$\alpha = 4$}
\end{subfigure}
\caption[The caustics of the Zel'dovich pancake lens.]{The caustics corresponding to $\lambda_1$ (blue) and $\lambda_2$ (red) as a function of $\alpha$. We observe the formation of a triple- and a five-image region.}\label{fig:GeometricOptics}
\end{figure}

The first caustic forms at the origin $(\mu_1,\mu_2)=(0,0)$ for the parameter $\alpha =\frac{1}{2}$ (see Fig.~\ref{fig:GeometricOptics}). This is a cusp singularity. Note that by construction this caustic corresponds to the eigenvalue field $\lambda_1$. For $\frac{1}{2} <\alpha <\frac{64}{49}$ the $A_3$ point forms an outgoing fold-line ($A_2$) with two cusps ($A_3$) on the left and the right. At $\alpha = 1$, a new $A_3$ point is created, this time corresponding to the second eigenvalue field $\lambda_2$. For $1 < \alpha  < \frac{64}{49}$ the $A_3$ point forms a fold-line ($A_2$) with two cusps ($A_3$) at the top and the bottom. At $\alpha=\frac{64}{49}$ the two fold lines merge in a hyperbolic umbilic ($D_4^+$) at $(\mu_1,\mu_2) = (0,\pm 1/\sqrt{14})$. For $\alpha >\frac{64}{49}$ the two fold lines continue to move outwards, where the fold-line corresponding to $\lambda_1$ has four cusps while the fold-line corresponding to $\lambda_2$ does not contain a cusp. Outside the fold-line of the caustics, the image consists of a single-image region. Inside the blue fold line, we find a triple-image and a five-image region enclosed by the red fold line.

The analytic continuation of the exponent, $\phi(\bm{x};\bm{\mu})$, possesses a pole on the two-dimensional surface $x_1^2+2 x_2^2+1 = 0$. Note that poles are never isolated in multi-dimensional complex analysis \cite{Range:2003}. The exponent has nine saddle points $\bar{\bm{x}}_i$. By evaluating the gradient of the $h$-function and flowing the original integration domain, we obtain a numerical representation of the thimble $\mathcal{J} \subset \mathbb{C}^2$. 

\begin{figure}
\centering
\begin{subfigure}[b]{0.3\textwidth}
\includegraphics[width=\textwidth]{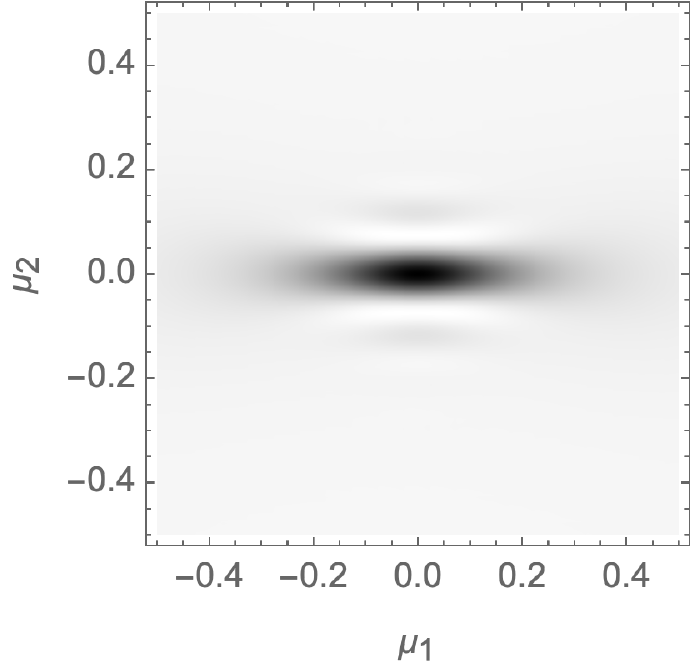}
\caption{$\alpha=0.7, \nu=50$}
\end{subfigure} 
\begin{subfigure}[b]{0.3\textwidth}
\includegraphics[width=\textwidth]{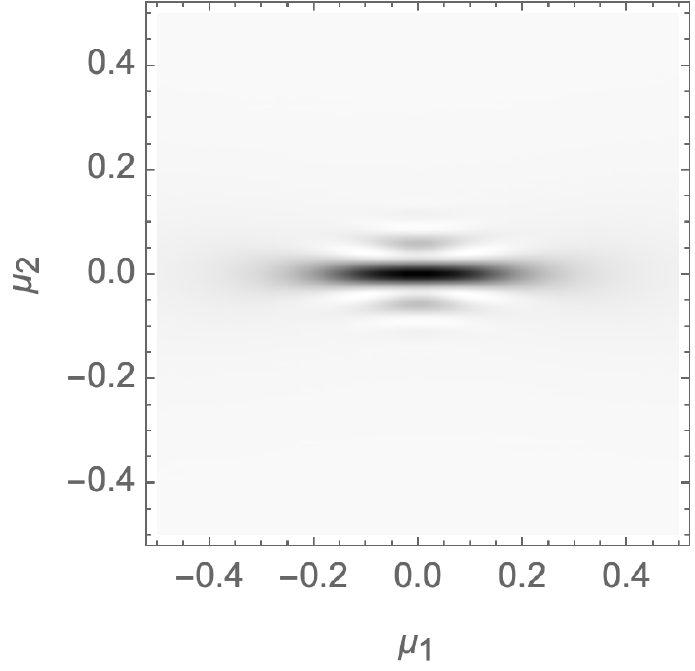}
\caption{$\alpha = 0.7, \nu=100$}
\end{subfigure} 
\begin{subfigure}[b]{0.3\textwidth}
\includegraphics[width=\textwidth]{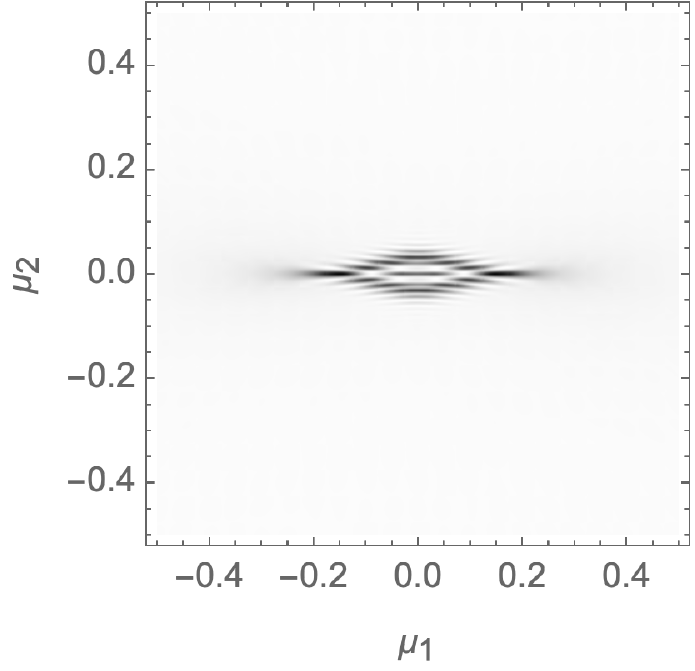}
\caption{$\alpha = 0.7, \nu=500$}
\end{subfigure}\\
\begin{subfigure}[b]{0.3\textwidth}
\includegraphics[width=\textwidth]{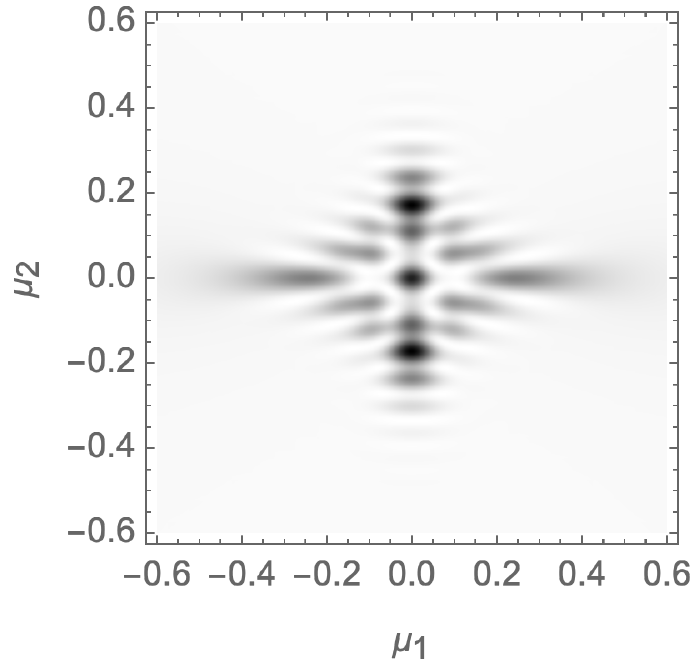}
\caption{$\alpha=64/49, \nu=50$}
\end{subfigure} 
\begin{subfigure}[b]{0.3\textwidth}
\includegraphics[width=\textwidth]{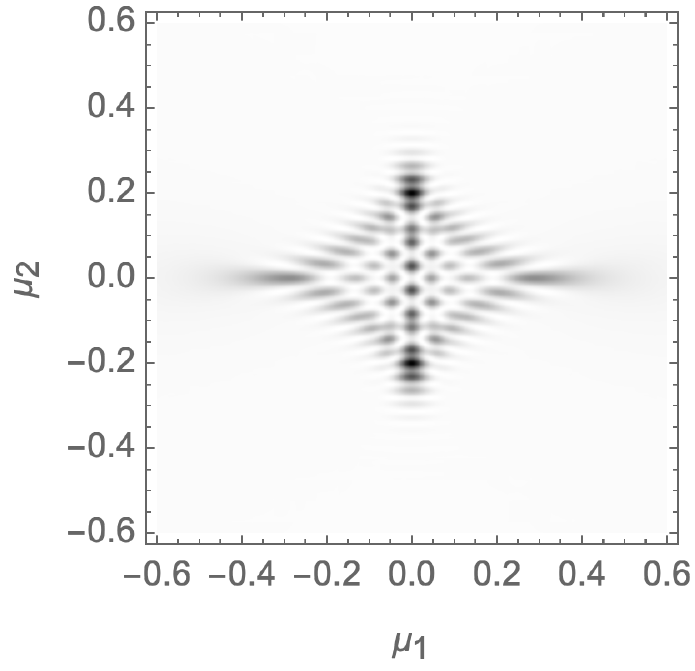}
\caption{$\alpha = 64/49, \nu=100$}
\end{subfigure} 
\begin{subfigure}[b]{0.3\textwidth}
\includegraphics[width=\textwidth]{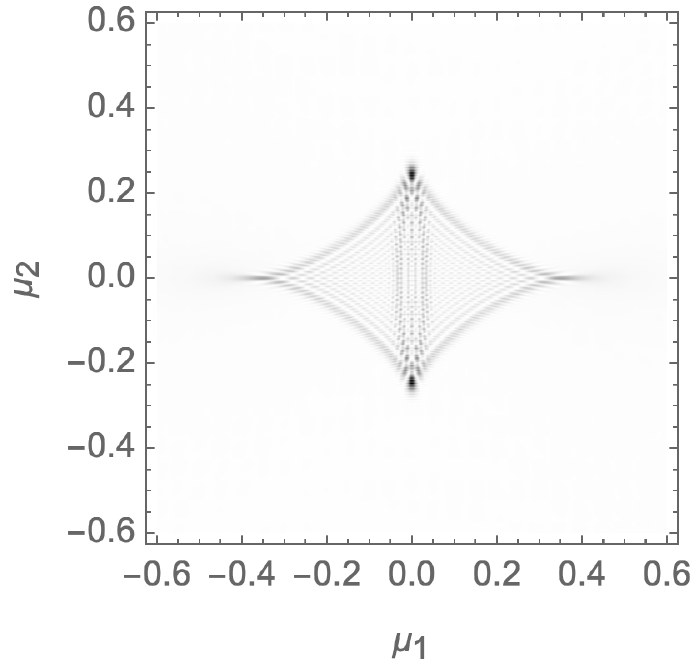}
\caption{$\alpha = 64/49, \nu=500$}
\end{subfigure}\\
\begin{subfigure}[b]{0.3\textwidth}
\includegraphics[width=\textwidth]{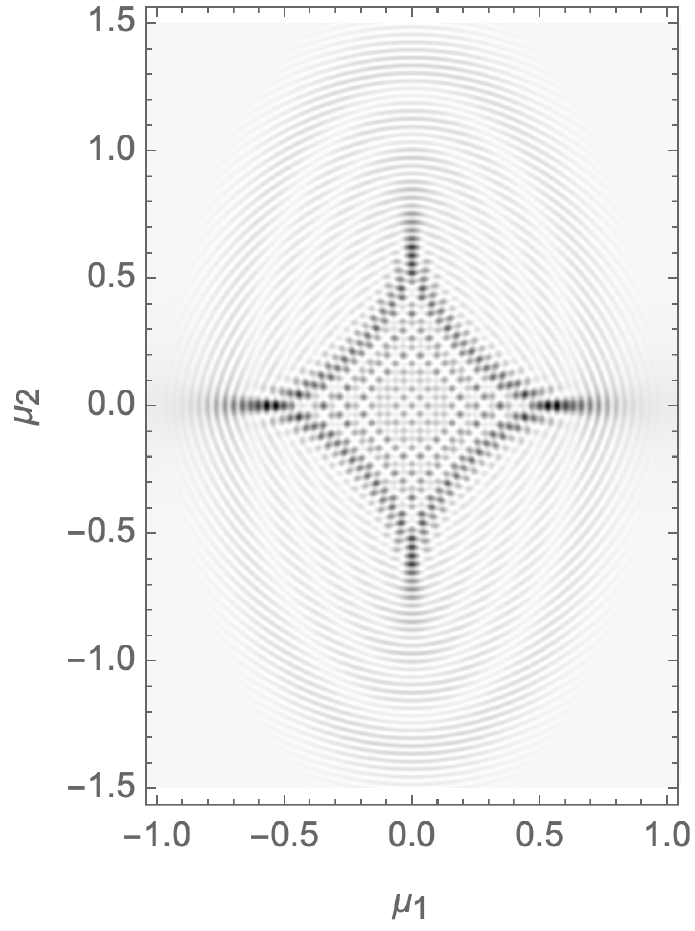}
\caption{$\alpha= 4, \nu=50$}
\end{subfigure} 
\begin{subfigure}[b]{0.3\textwidth}
\includegraphics[width=\textwidth]{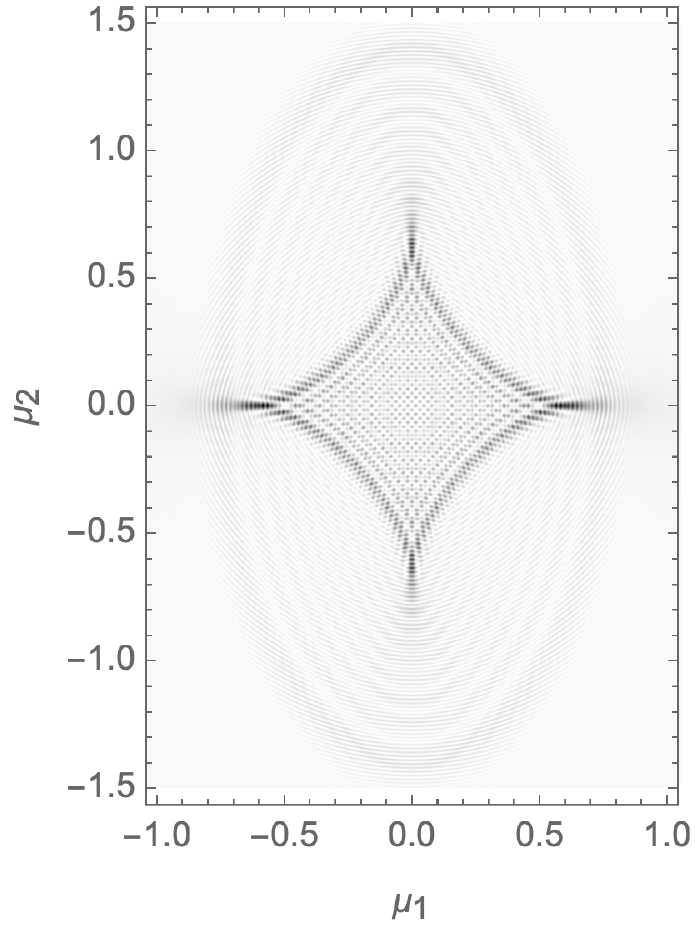}
\caption{$\alpha = 4, \nu=100$}
\end{subfigure} 
\begin{subfigure}[b]{0.3\textwidth}
\includegraphics[width=\textwidth]{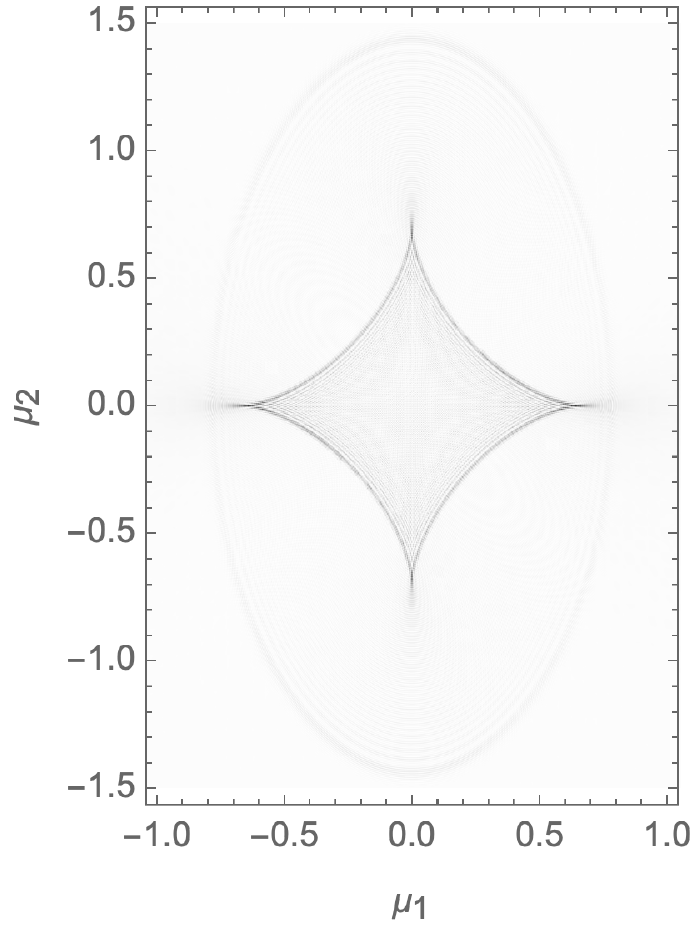}
\caption{$\alpha = 4, \nu=500$}
\end{subfigure}
\caption[The intensity map of the local lens with a non-degenerate peak.]{Intensity $I(\bm{\mu};\nu)$ of the local lens at $\alpha =0.7,64/49,4$ for $\nu=50,100$ and $500$.}\label{fig:Local2DNumerical}
\end{figure}

\begin{figure}
\centering
\begin{subfigure}[b]{0.31\textwidth}
\includegraphics[width=\textwidth]{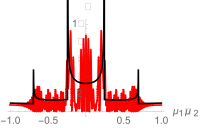}
\caption{$\nu =50$}
\end{subfigure} ~
\begin{subfigure}[b]{0.31\textwidth}
\includegraphics[width=\textwidth]{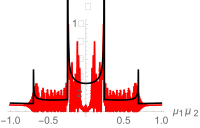}
\caption{$\nu =100$}
\end{subfigure} ~
\begin{subfigure}[b]{0.31\textwidth}
\includegraphics[width=\textwidth]{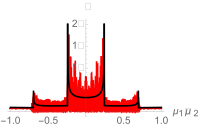}
\caption{$\nu =500$}
\end{subfigure} ~
\caption[The intensity along a cross-section of the lens.]{The normalized intensity evaluate along the diagonal in the $(\mu_1,\mu_2)$-plane for $\alpha = 4$ for $\nu=100$ and $500$. The black curve is the envelope predicted by geometric optics (see equation \eqref{eq:IntensityGeometric}).}\label{fig:Intensity2D}
\end{figure}
Given the two-dimensional thimble $\mathcal{J}$, we numerically evaluate the integral $\Psi(\bm{\mu};\nu)$. In Fig.~\ref{fig:Local2DNumerical}, we plotted the normalized intensity of the sensed signal for $\alpha = 7/10, 64/49$ and $4$ as a function of the frequency. Observe that when the wavelength is comparable to the size of the caustic structure, the normalized intensity is blurred. The caustics emerge when the wavelength becomes shorter. At the frequency $\nu=500$, we accurately recover the image corresponding to geometric optics. Remark the stripes emanating from the cusp singularities. This is the frequency independent power-law falloff we observed in the elementary singularities.

In Fig.~\ref{fig:Intensity2D}, we plot the cross-section of the normalized intensity map along the diagonal $\mu_1=\mu_2$ for the lens with $\alpha =4$ for $\nu=50,100$ and $500$. Observer the four spikes while passing through the fold catastrophe. Note that the spikes increase in magnitude as $\nu$ is raised. In the astronomical context, these spikes correspond to amplification in the light-curve of the lensed source.

\subsection{A degenerate peak}
\begin{figure}
\centering
\begin{subfigure}[b]{0.24\textwidth}
\includegraphics[width=\textwidth]{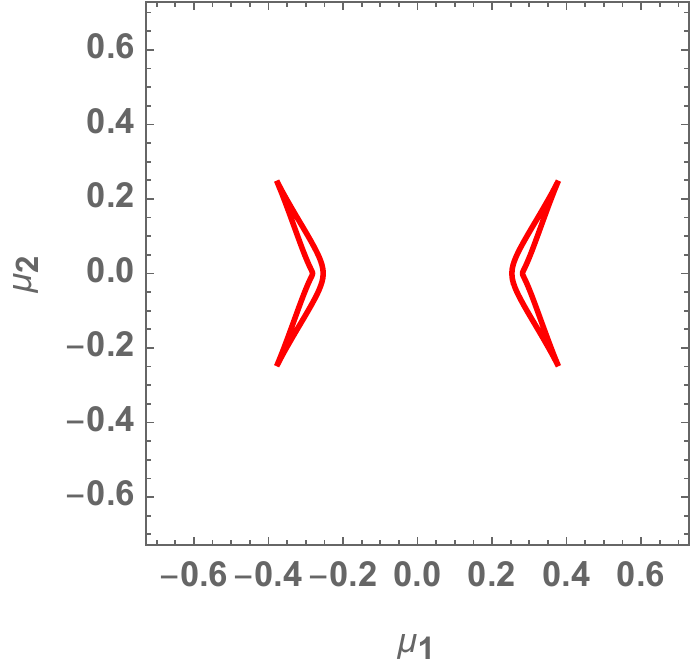}
\caption{$\alpha=1$}
\end{subfigure} 
\begin{subfigure}[b]{0.24\textwidth}
\includegraphics[width=\textwidth]{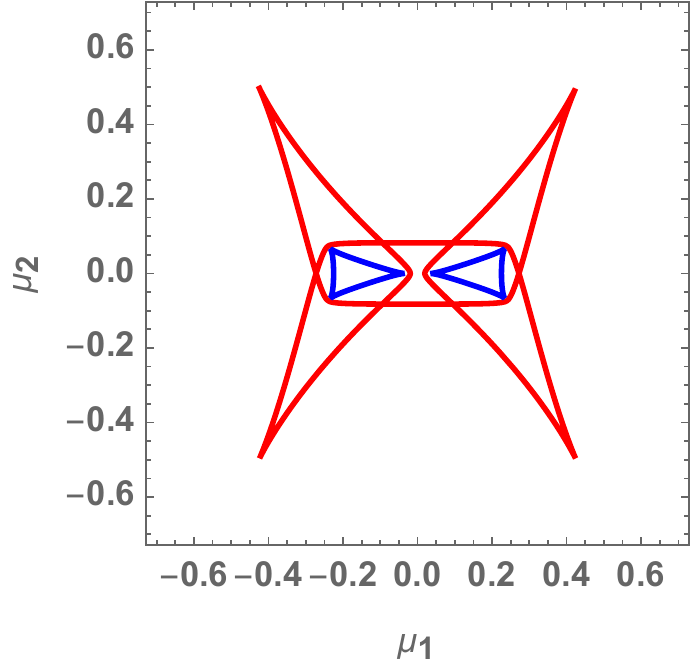}
\caption{$\alpha = 1.5$}
\end{subfigure} 
\begin{subfigure}[b]{0.24\textwidth}
\includegraphics[width=\textwidth]{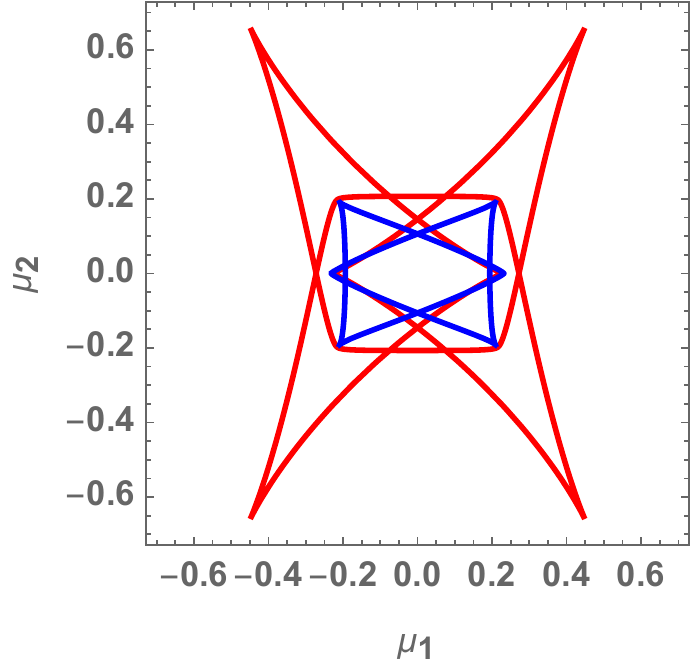}
\caption{$\alpha = 2$}
\end{subfigure}
\begin{subfigure}[b]{0.24\textwidth}
\includegraphics[width=\textwidth]{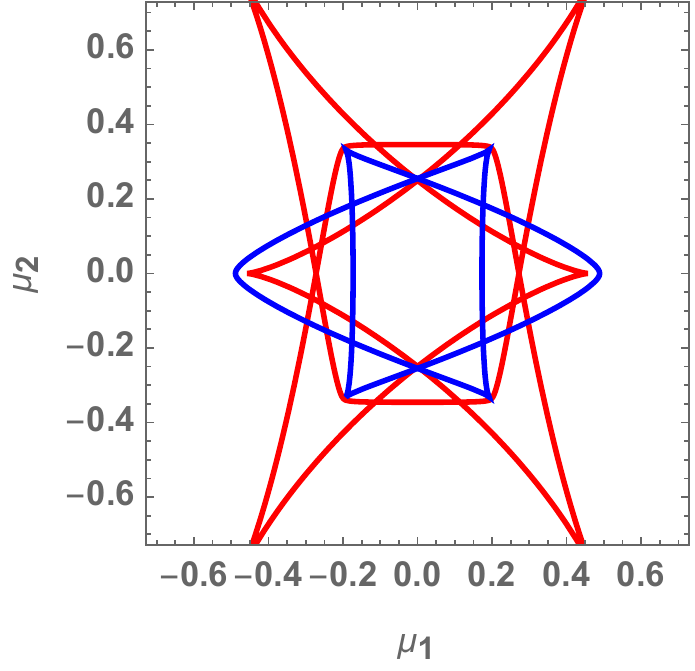}
\caption{$\alpha = 2.5$}
\end{subfigure}
\caption[The caustics in geometric optics for the degenerate peak.]{The caustics of the Lagrangian map for varying $\alpha$. The caustics corresponding to the first and second eigenvalue fields $\lambda_1,\lambda_2$ in red and blue.}\label{fig:Lens2Caustics}
\end{figure}

A more intricate structure arises for the lens corresponding to the degenerate peak in the phase,
\begin{align}
\varphi(\bm{x}) = \frac{\alpha}{1+x_1^4+ x_2^2}\,,
\end{align}
with the Lagrangian map
\begin{align}
\xi(\bm{x}) &= \bm{x} + \frac{1}{2}\nabla \varphi(\bm{x})\\
& = \bm{x} - \frac{\alpha}{(1+x_1^4+x_2^2)^2} (2 x_1^3,x_2)\,.
\end{align} 
The caustics structure of the Lagrangian map for varying $\alpha$ is plotted in Fig.~\ref{fig:Lens2Caustics}. For $\alpha = 1$ we find two disconnected components, which are joined at $\alpha=1.5$ and form an intricate pattern at $\alpha = 2$ and $\alpha = 2.5$. At $\alpha=2$ we again find a hyperbolic umbilic caustic ($D_4^+$) at the two points where the cusps corresponding to the first and second eigenvalue fields $\lambda_1,\lambda_2$ coincide. We thus see that not only the structure at the peak but also the falloff of the variation in the phase $\varphi$ is important in the study of caustics in lensed images. The caustic structure is generally sensitive to the Hessian of the phase $\varphi$, \textit{i.e.}, the second order derivatives.

\begin{figure}
\centering
\begin{subfigure}[b]{0.32\textwidth}
\includegraphics[width=\textwidth]{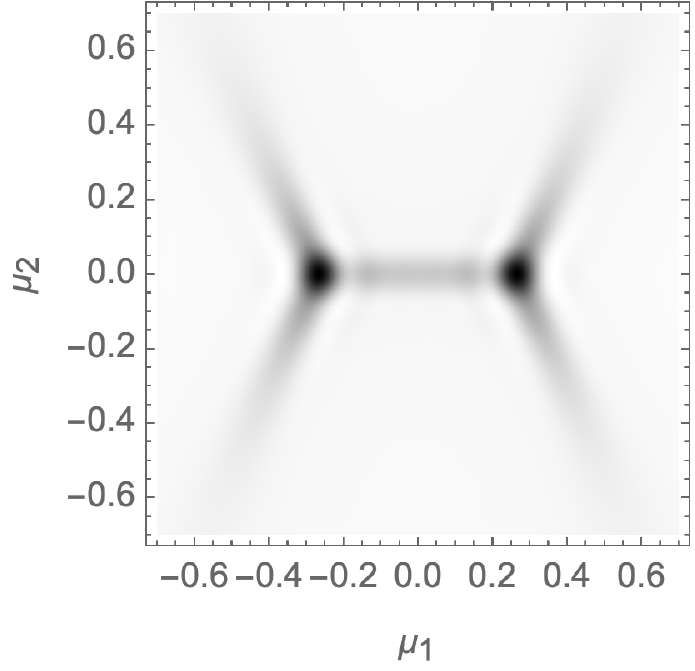}
\caption{$\alpha=1, \nu=50$}
\end{subfigure} 
\begin{subfigure}[b]{0.32\textwidth}
\includegraphics[width=\textwidth]{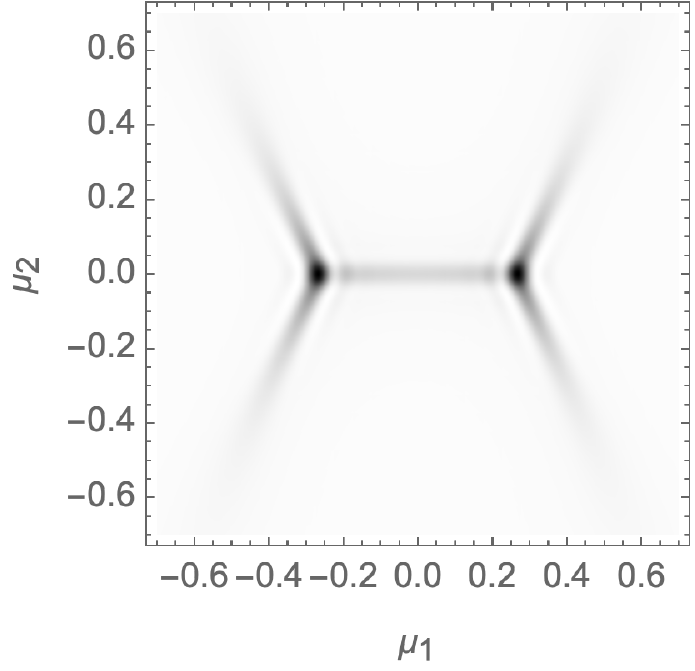}
\caption{$\alpha=1, \nu=100$}
\end{subfigure} 
\begin{subfigure}[b]{0.32\textwidth}
\includegraphics[width=\textwidth]{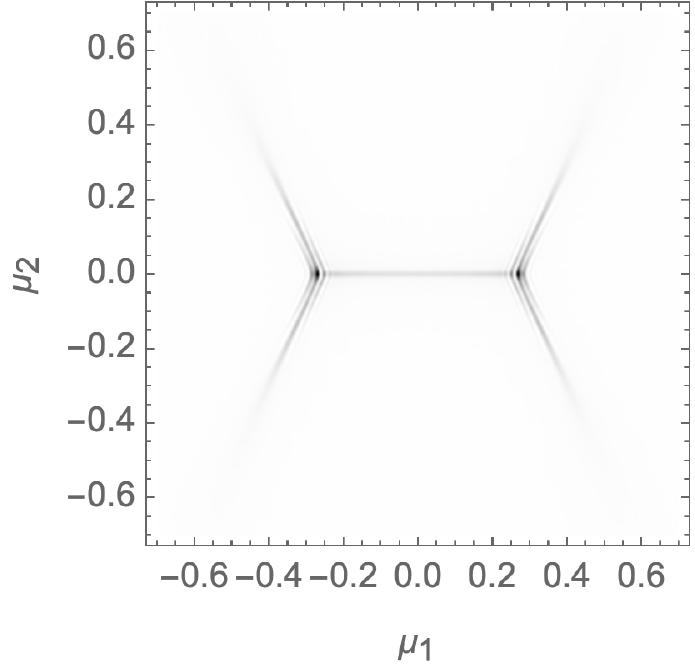}
\caption{$\alpha=1, \nu=500$}
\end{subfigure} \\

\begin{subfigure}[b]{0.32\textwidth}
\includegraphics[width=\textwidth]{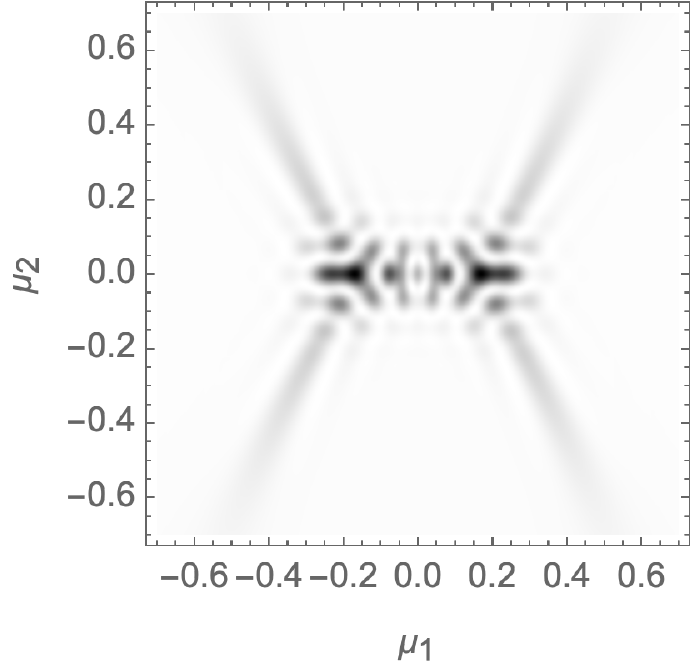}
\caption{$\alpha = 1.5, \nu=50$}
\end{subfigure} 
\begin{subfigure}[b]{0.32\textwidth}
\includegraphics[width=\textwidth]{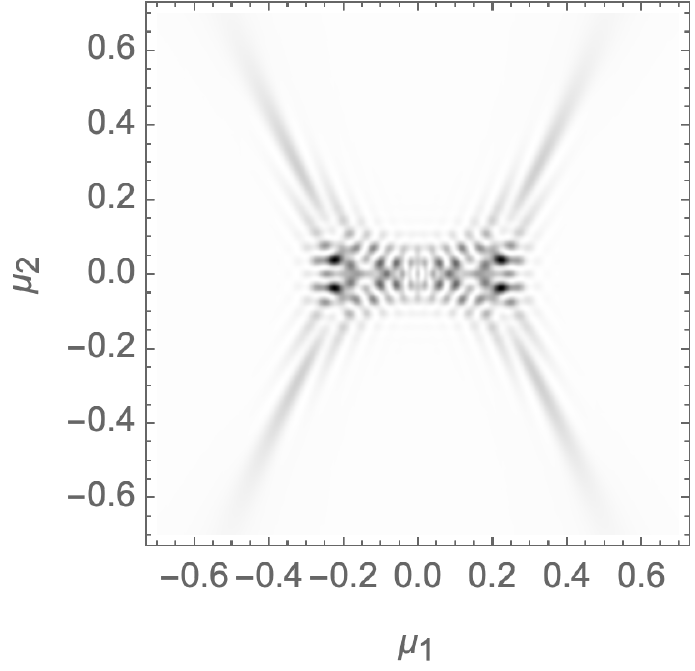}
\caption{$\alpha = 1.5, \nu=100$}
\end{subfigure} 
\begin{subfigure}[b]{0.32\textwidth}
\includegraphics[width=\textwidth]{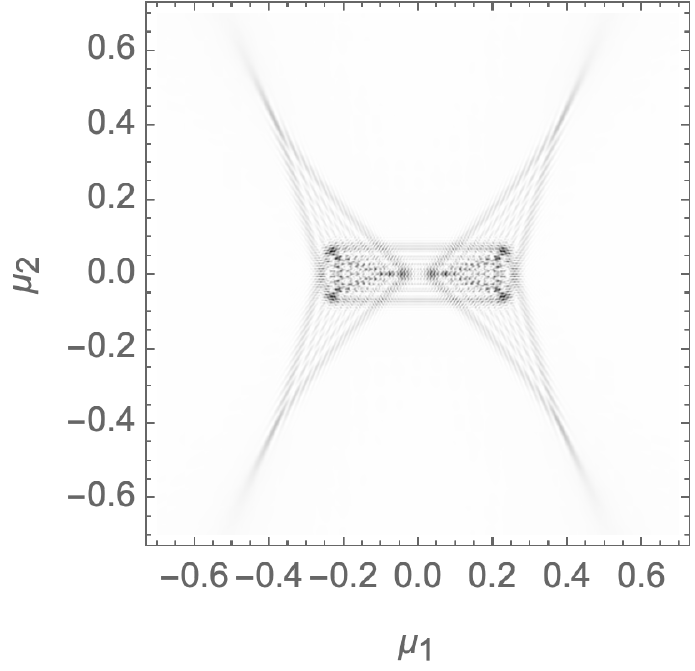}
\caption{$\alpha = 1.5, \nu=500$}
\end{subfigure} \\

\begin{subfigure}[b]{0.32\textwidth}
\includegraphics[width=\textwidth]{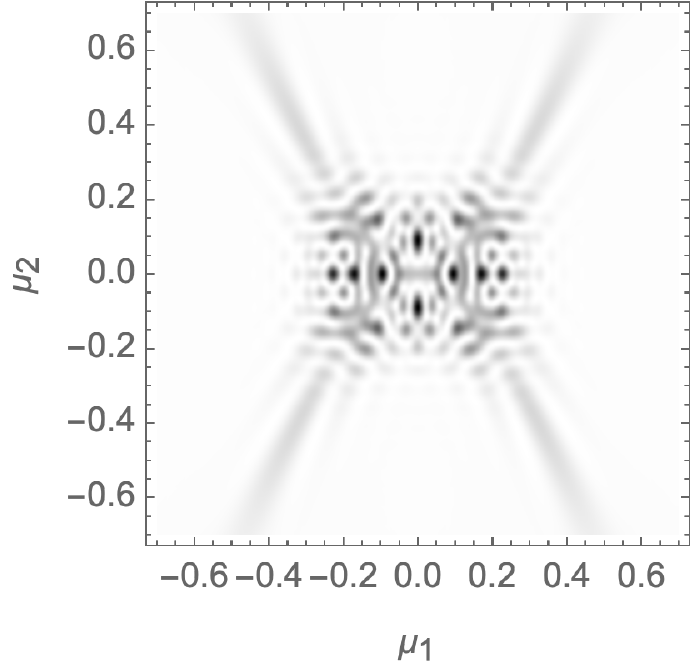}
\caption{$\alpha = 1, \nu=100$}
\end{subfigure}
\begin{subfigure}[b]{0.32\textwidth}
\includegraphics[width=\textwidth]{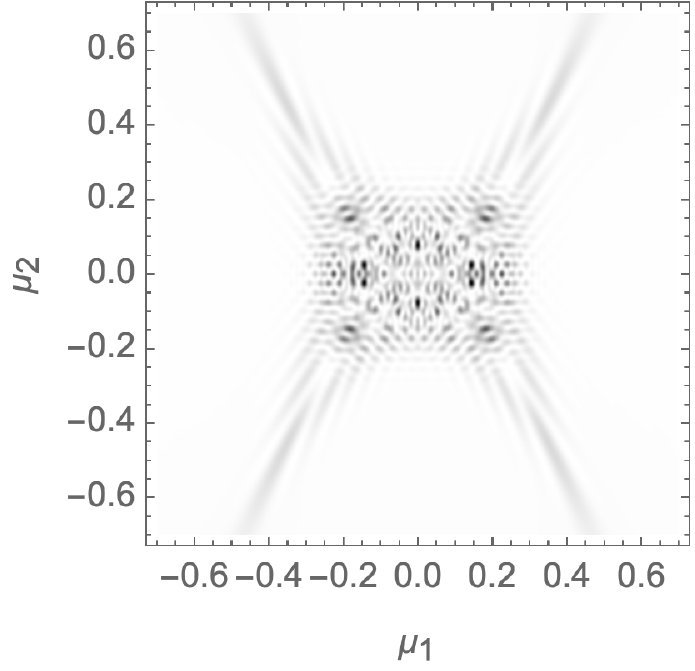}
\caption{$\alpha = 0.7, \nu=500$}
\end{subfigure}
\begin{subfigure}[b]{0.32\textwidth}
\includegraphics[width=\textwidth]{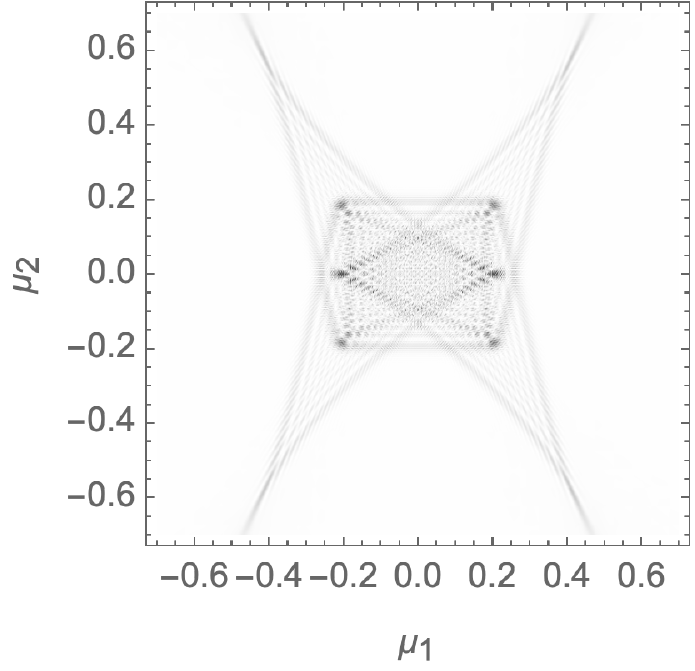}
\caption{$\alpha =2, \nu=500$}
\end{subfigure}
\caption[The intensity map for the degenerate lens.]{The normalized intensity map, $I(\bm{\mu};\nu)$, for different frequencies.}\label{fig:Lens2_1}
\end{figure}

\begin{figure}
\centering
\begin{subfigure}[b]{0.32\textwidth}
\includegraphics[width=\textwidth]{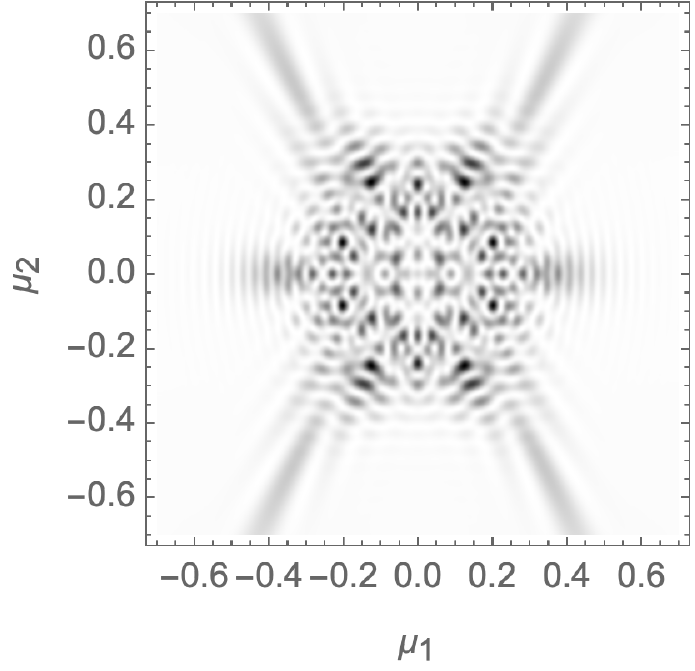}
\caption{$\alpha=2.5,\nu=50$}
\end{subfigure} 
\begin{subfigure}[b]{0.32\textwidth}
\includegraphics[width=\textwidth]{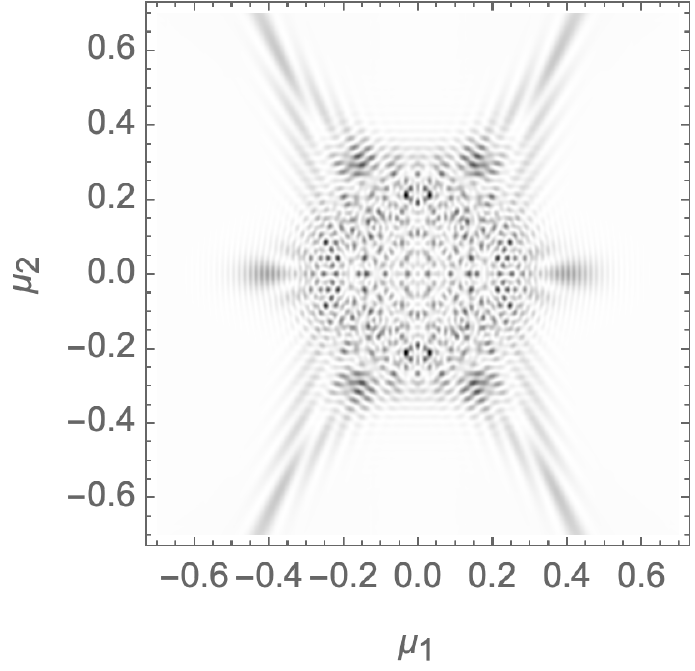}
\caption{$\alpha=2.5,\nu=100$}
\end{subfigure} 
\begin{subfigure}[b]{0.32\textwidth}
\includegraphics[width=\textwidth]{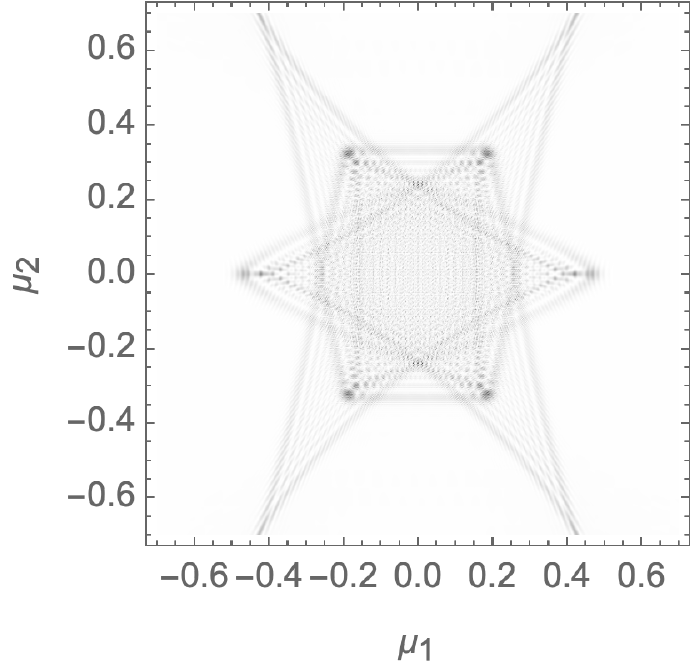}
\caption{$\alpha=2.5,\nu=500$}
\end{subfigure} 
\caption[The intensity map for the degenerate lens.]{The normalized intensity map, $I(\bm{\mu};\nu)$, for different frequencies.}\label{fig:Lens2_2}
\end{figure}

After flowing the original integration contour to the Lefschetz thimble $\mathcal{J}$, we numerically evaluate the amplitude $\Psi(\bm{\mu};\nu)$ and the corresponding normalized intensity $I(\bm{\mu};\nu)$. The resulting normalized intensity maps are plotted in figures \ref{fig:Lens2_1} and \ref{fig:Lens2_2}. For the frequency $\nu=50$, the image is again rather blurry. We can see the general shape, but cannot distinguish the detailed line structure. For the frequency $\nu=100$, the lines are better resolved. However, the length scale of the caustics is comparable to the length scales of the interference patterns in the multi-image regions. For $\nu=500$, we see the complete geometric structure of the caustics. The oscillations in the multi-image regions are now very fine. For this frequency, we are very close to the geometric optics approximation.

\subsection{The swallowtail caustic}
In the previous two examples of lenses corresponding to the simple peaks, we found both fold ($A_2$) and cusp caustics ($A_3$) corresponding to a single eigenvalue field, and the interaction between two eigenvalue fields via the hyperbolic umbilic ($D_4^+$). The two remaining caustics, \textit{i.e.}, the swallowtail ($A_4$) and the elliptic umbilic ($D_4^-$), appear in slightly more involved lenses. For the swallowtail caustic, consider the lens
\begin{align}
\varphi(\bm{x}) = \frac{ \alpha x_1}{1+x_1^4+ x_2^2}\,.
\end{align}
Again, in the astrophysical context, $\alpha$ follows the dispersion relation $\alpha \propto \nu^{-2}$.

\begin{figure}
\centering
\begin{subfigure}[b]{0.24\textwidth}
\includegraphics[width=\textwidth]{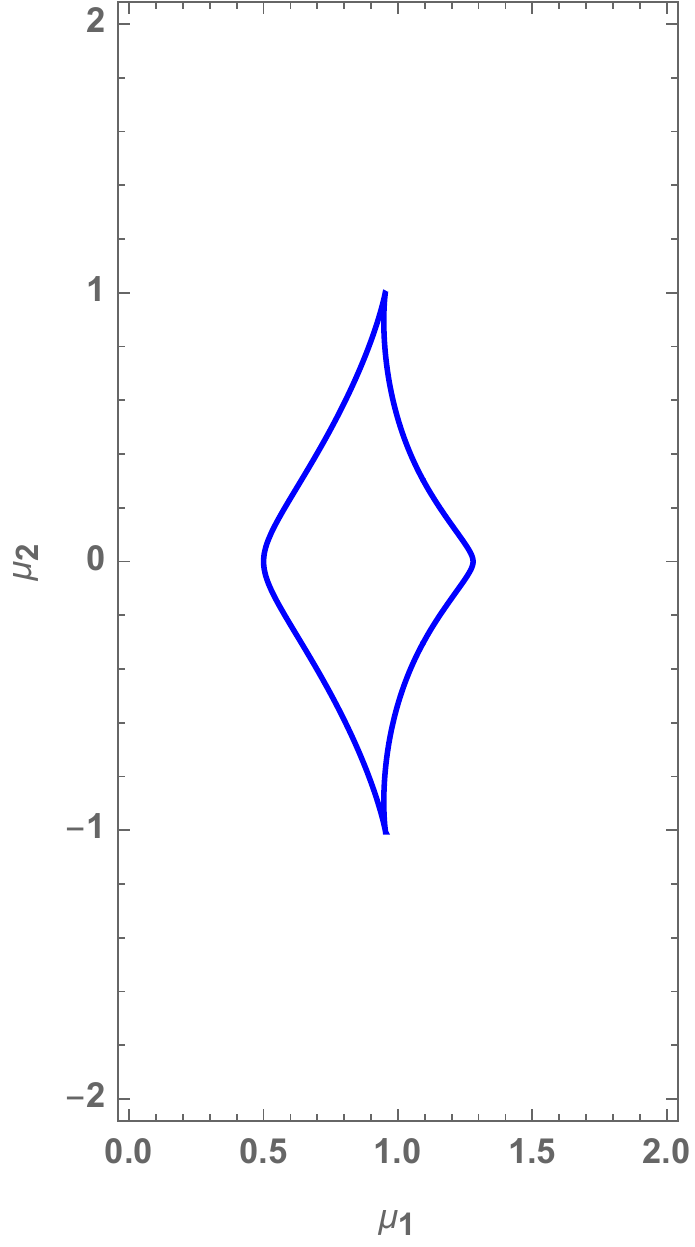}
\caption{$\alpha=2$}
\end{subfigure} 
\begin{subfigure}[b]{0.24\textwidth}
\includegraphics[width=\textwidth]{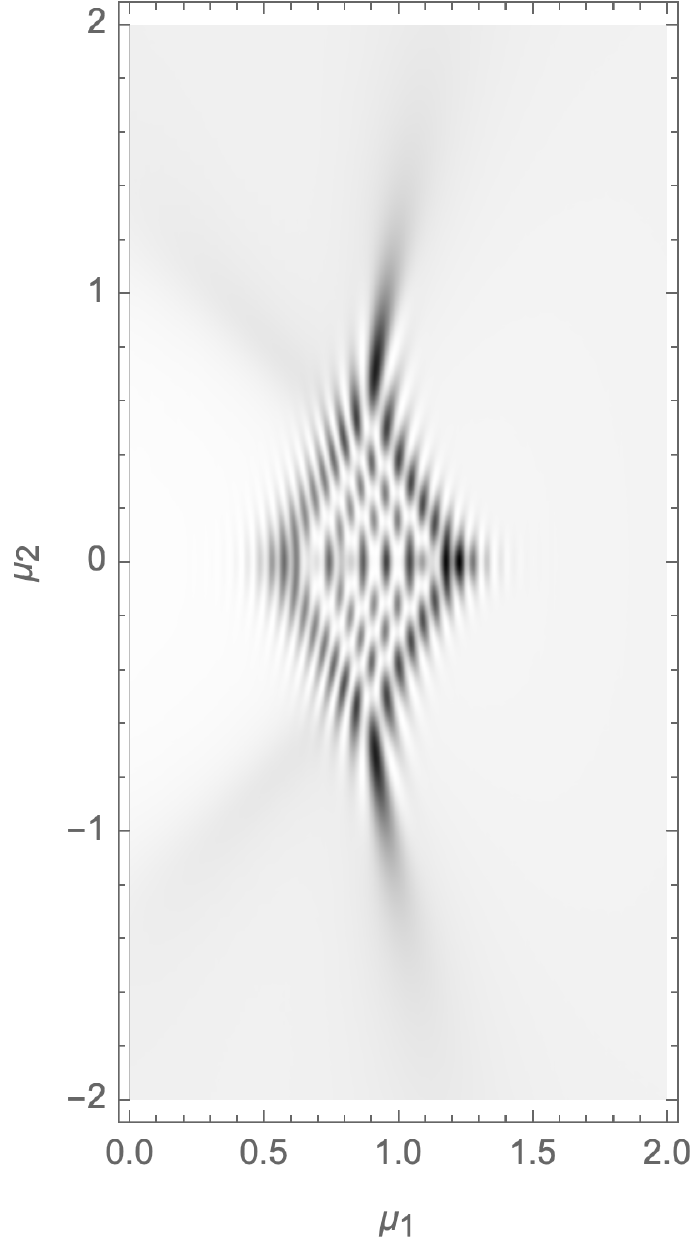}
\caption{$\alpha=2,\nu=50$}
\end{subfigure} 
\begin{subfigure}[b]{0.24\textwidth}
\includegraphics[width=\textwidth]{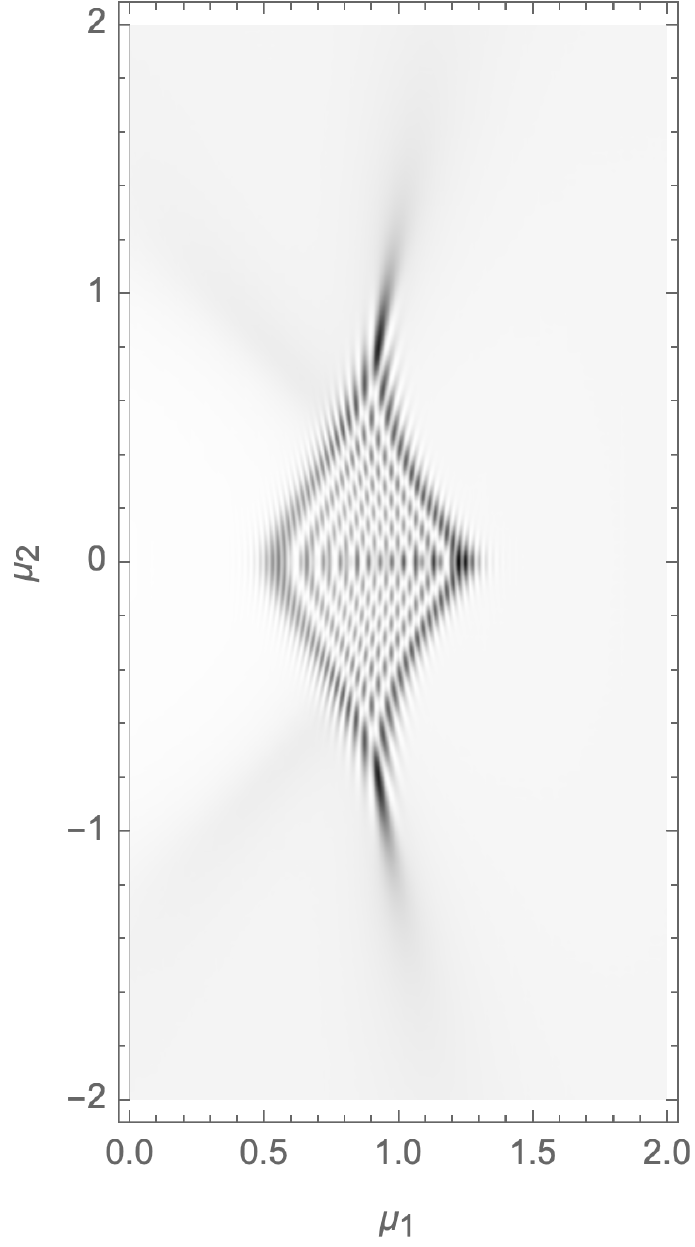}
\caption{$\alpha=2,\nu=100$}
\end{subfigure} 
\begin{subfigure}[b]{0.24\textwidth}
\includegraphics[width=\textwidth]{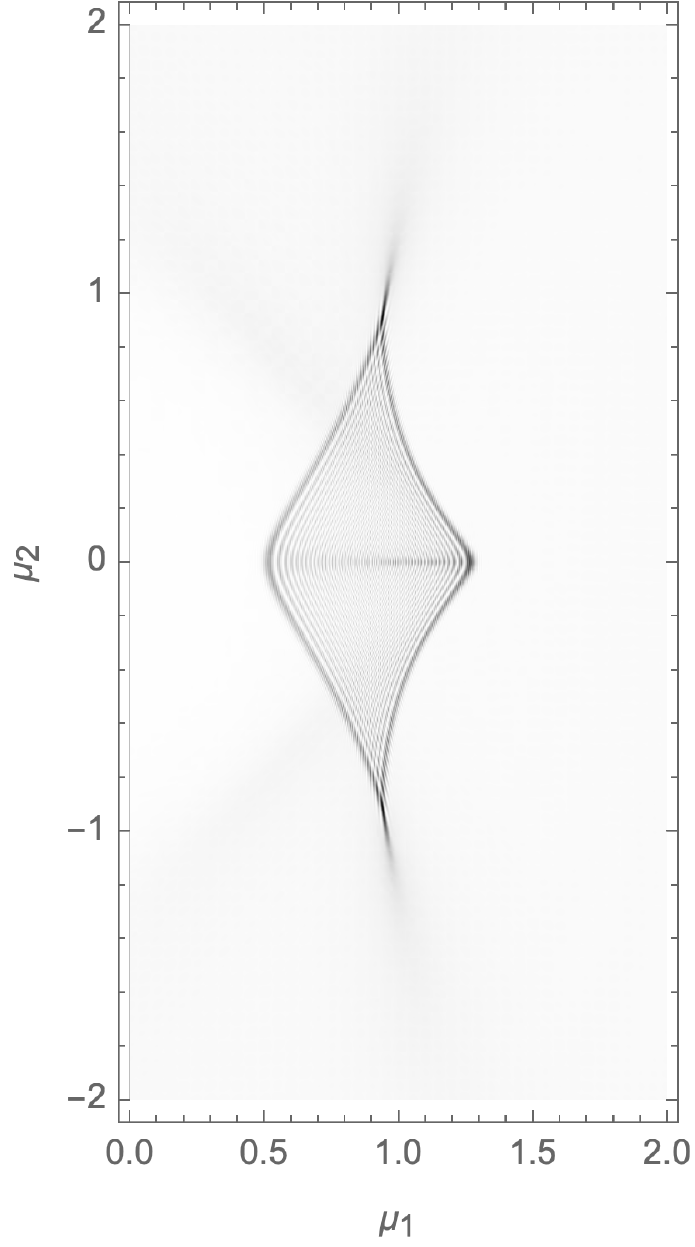}
\caption{$\alpha=2,\nu=500$}
\end{subfigure} \\
\begin{subfigure}[b]{0.24\textwidth}
\includegraphics[width=\textwidth]{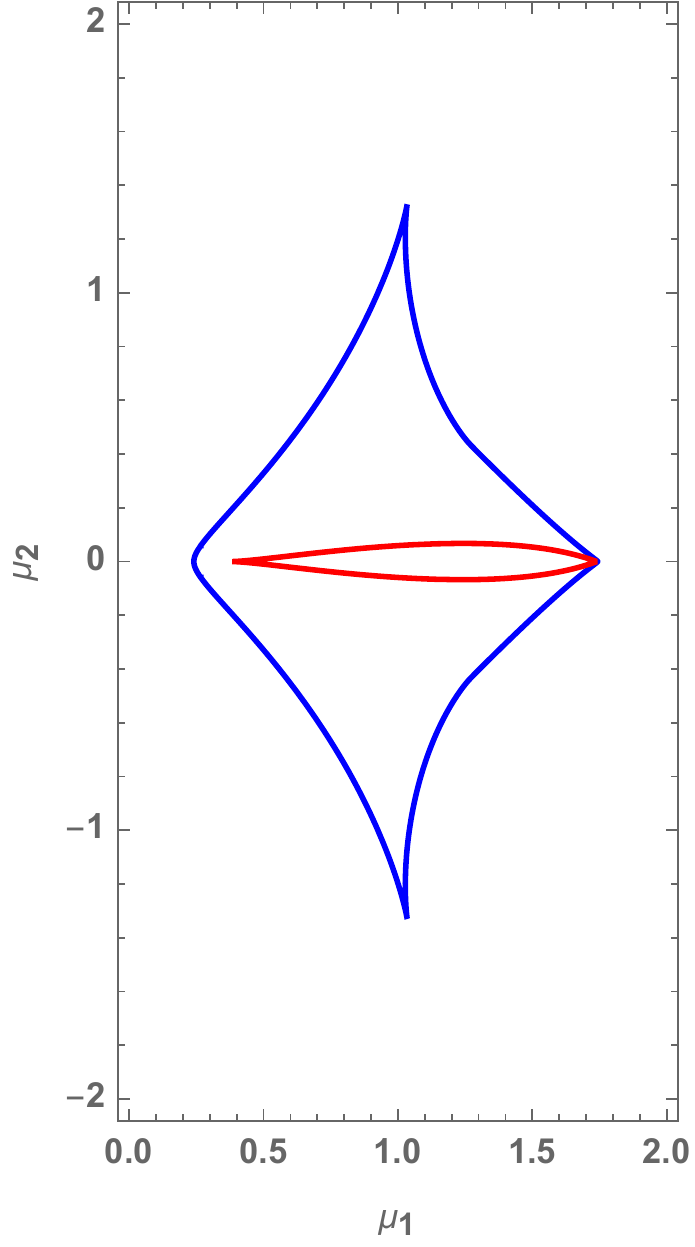}
\caption{$\alpha=3$}
\end{subfigure} 
\begin{subfigure}[b]{0.24\textwidth}
\includegraphics[width=\textwidth]{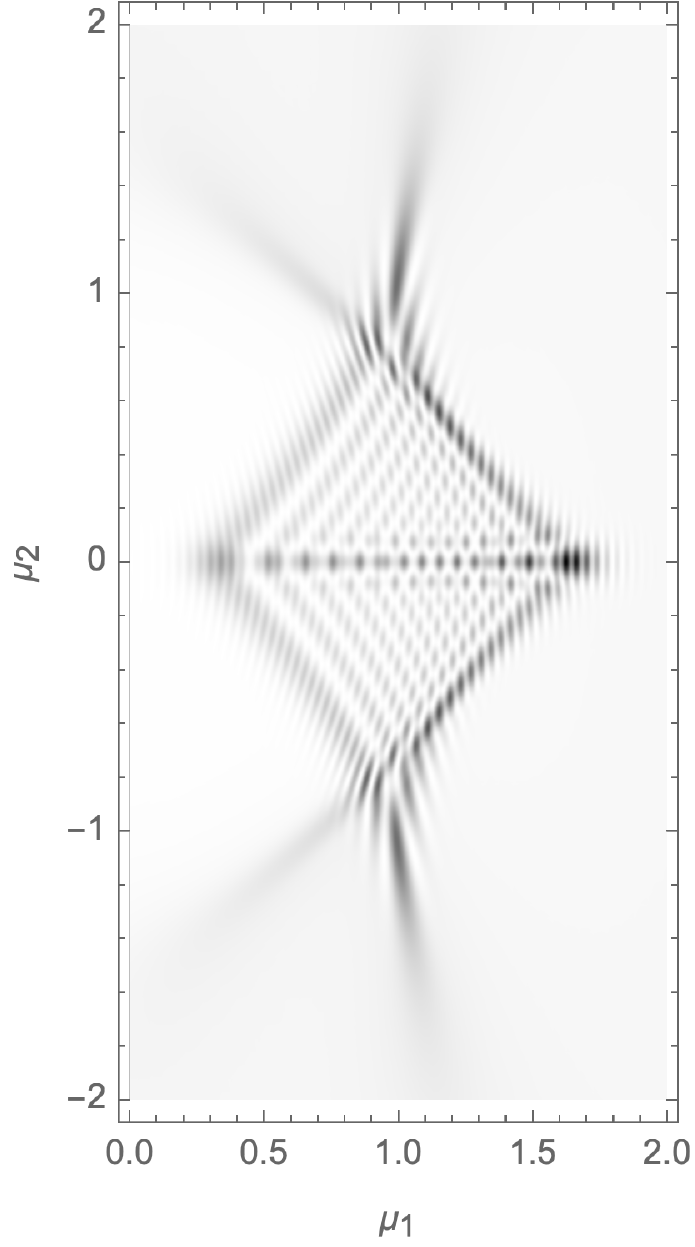}
\caption{$\alpha=3,\nu=50$}
\end{subfigure} 
\begin{subfigure}[b]{0.24\textwidth}
\includegraphics[width=\textwidth]{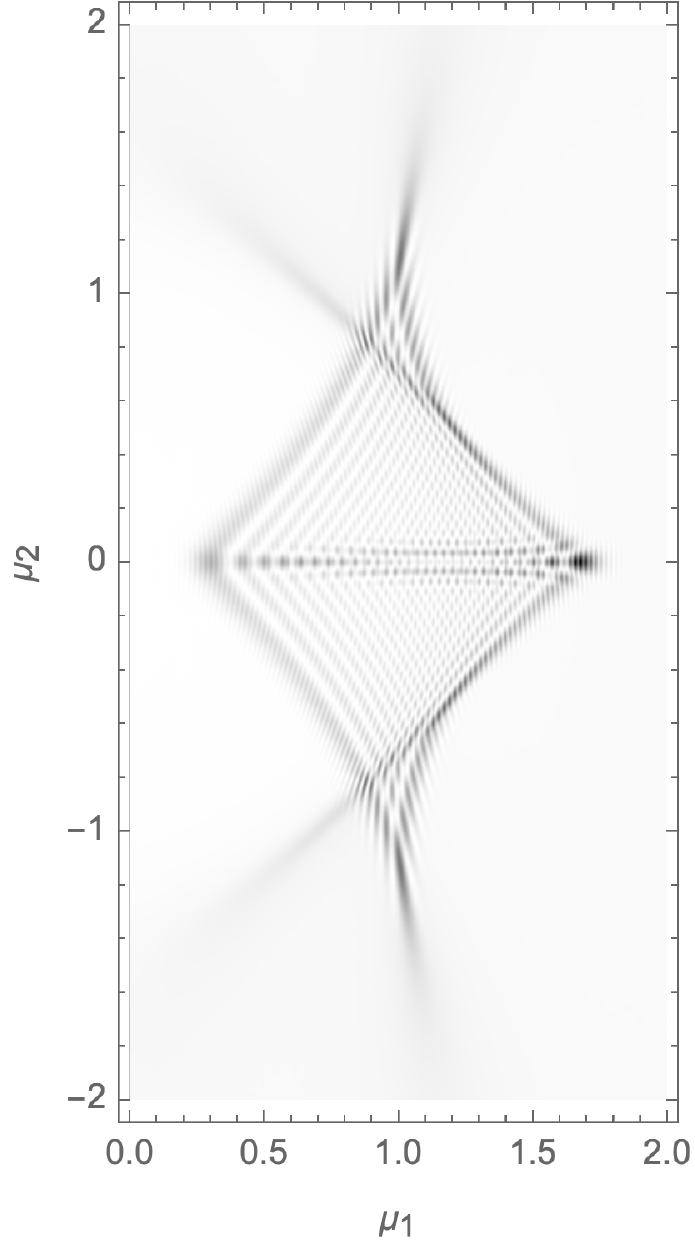}
\caption{$\alpha=3,\nu=100$}
\end{subfigure} 
\begin{subfigure}[b]{0.24\textwidth}
\includegraphics[width=\textwidth]{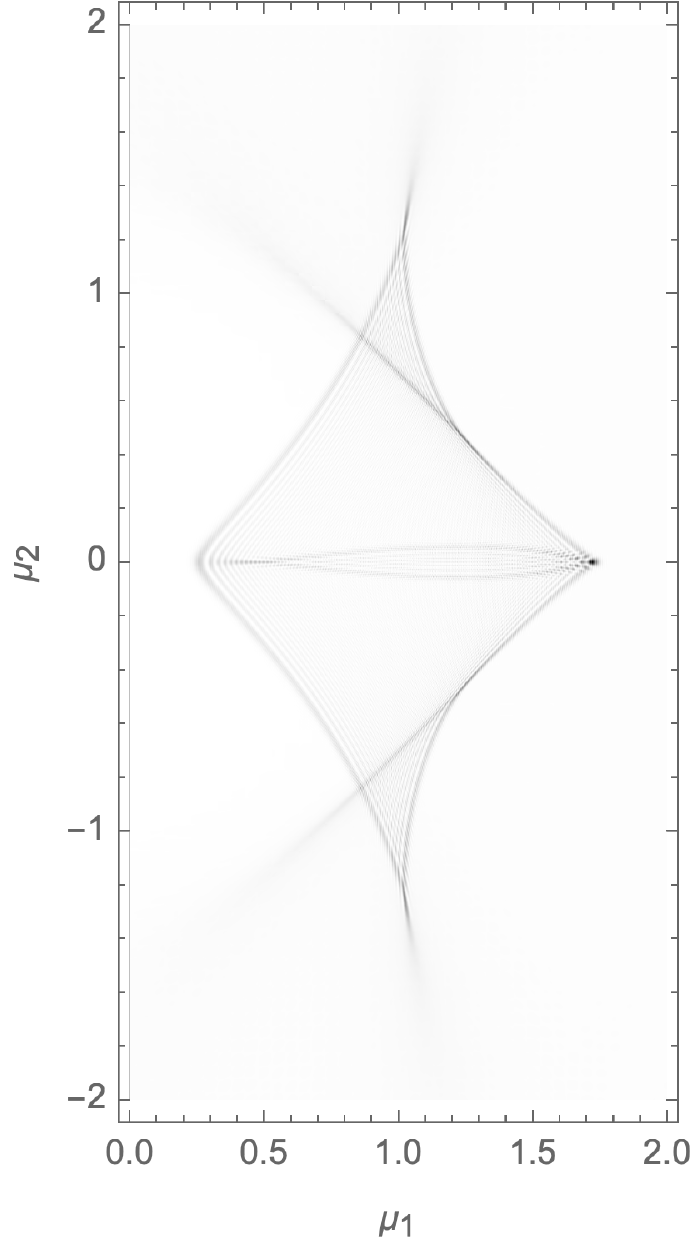}
\caption{$\alpha=3,\nu=500$}
\end{subfigure} 
\caption[The intensity map of the lens with the swallowtail caustic.]{The normalized intensity map, $I(\bm{\mu};\nu)$, for $\alpha=2,3$ and frequencies $\nu=50,100,500$.}\label{fig:Lens3}
\end{figure}

\begin{figure}
\centering
\begin{subfigure}[b]{0.24\textwidth}
\includegraphics[width=\textwidth]{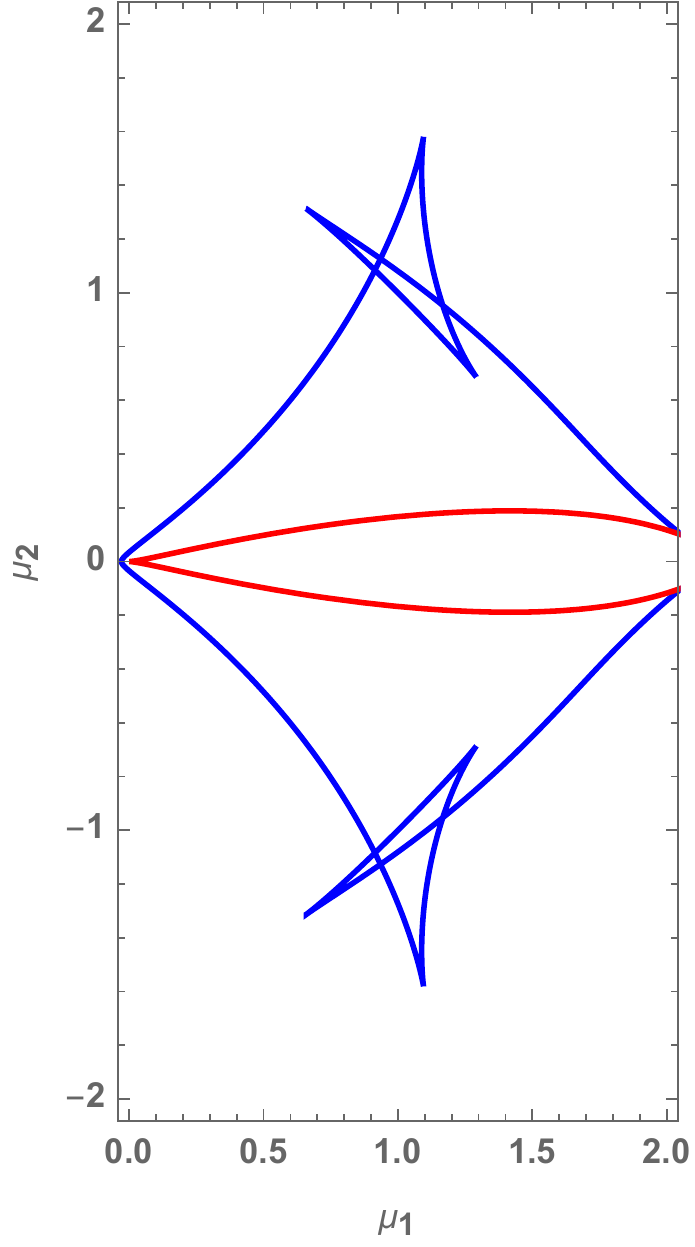}
\caption{$\alpha=4$}
\end{subfigure} 
\begin{subfigure}[b]{0.24\textwidth}
\includegraphics[width=\textwidth]{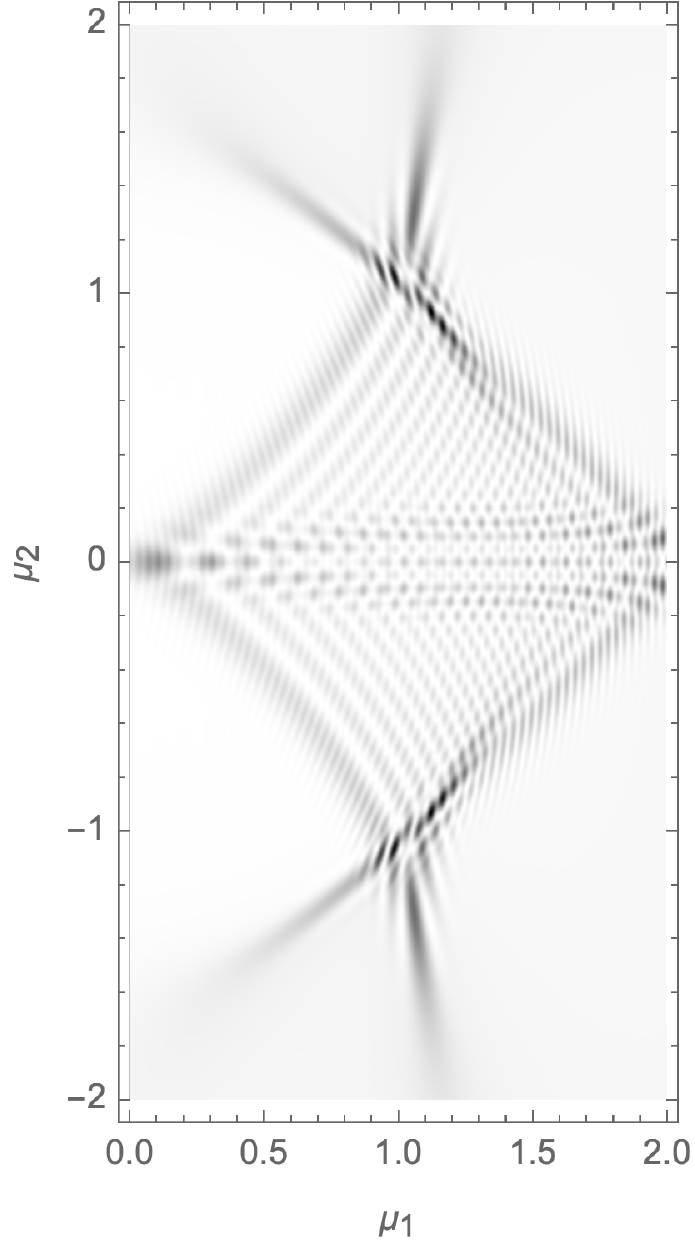}
\caption{$\alpha=4,\nu=50$}
\end{subfigure} 
\begin{subfigure}[b]{0.24\textwidth}
\includegraphics[width=\textwidth]{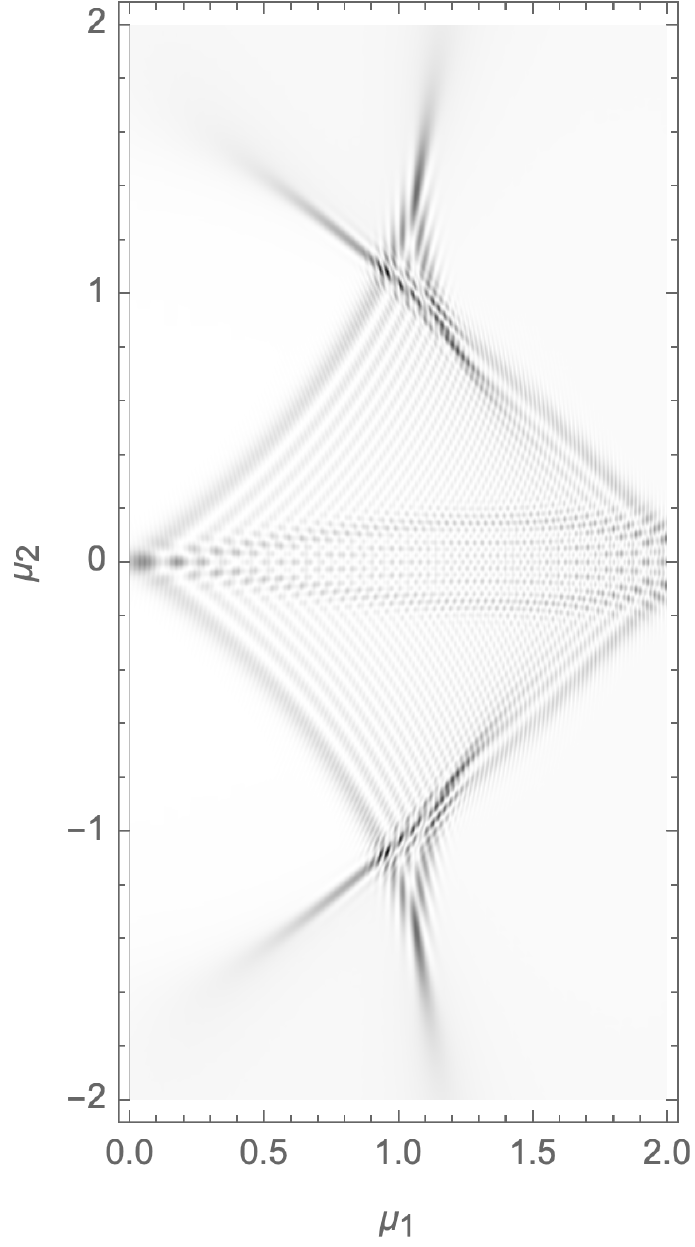}
\caption{$\alpha=4,\nu=100$}
\end{subfigure} 
\begin{subfigure}[b]{0.24\textwidth}
\includegraphics[width=\textwidth]{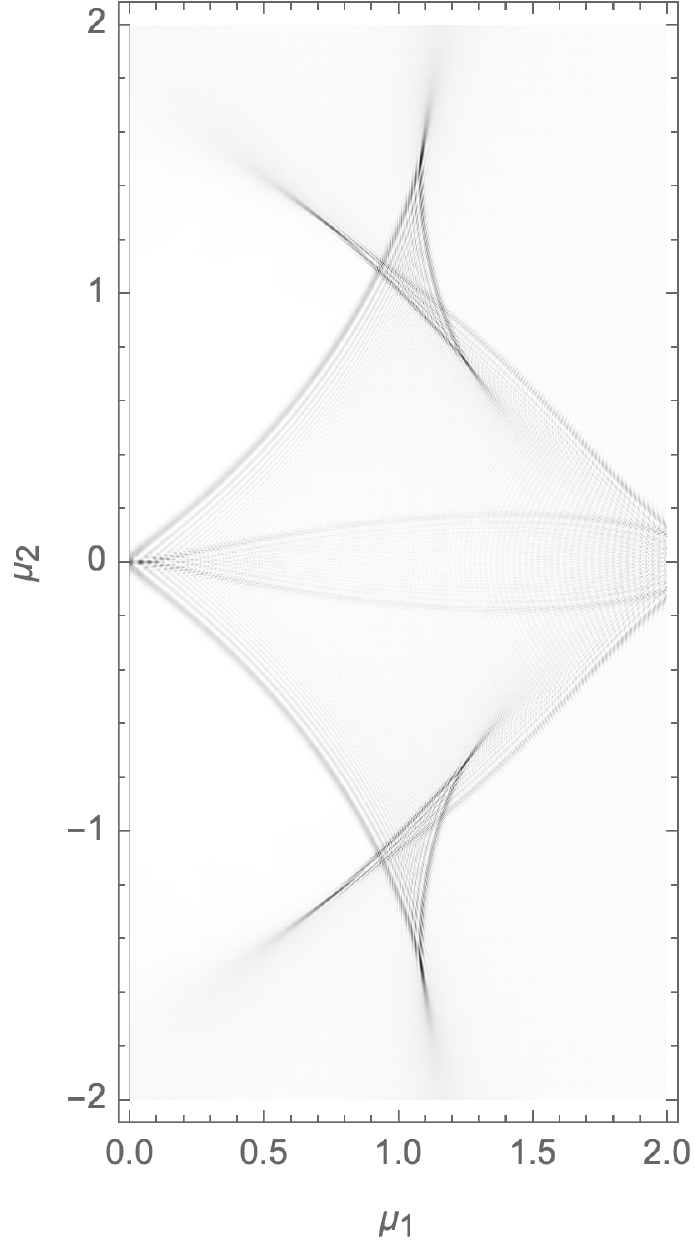}
\caption{$\alpha=4,\nu=500$}
\end{subfigure} 
\caption[The intensity map of the lens with the swallowtail caustic.]{The normalized intensity map, $I(\bm{\mu};\nu)$, for $\alpha=4$ and frequencies $\nu=50,100,500$.}\label{fig:Lens3_2}
\end{figure}

The corresponding integrand $i\phi(\bm{x};\bm{\mu})\nu$, consists of $23$ saddle point in the complex plane. By deforming the integration domain to the thimble, we evaluate the two-dimensional lens integral numerically. See figures \ref{fig:Lens3} and \ref{fig:Lens3_2} for the caustics obtained from geometric optics and the corresponding normalized intensity maps for the frequency $\nu=50,100,500$.
\begin{itemize}
\item
For $\alpha=2$, the lens forms a caustic corresponding to a single eigenvalue field (see the upper panels of figure \ref{fig:Lens3}.). The profile a pancake with two cusps at the tips. In the corresponding normalized intensity field, we see an interference pattern in the triple-image region, two stripes emanating from the cusps and more strikingly two diagonal stripes going to the left in the single image region. These stripes are a precursor of the swallowtail caustic emerging at later $\alpha$.
\end{itemize}


\begin{itemize}
\item
As $\alpha$ is raised to $3$, a second caustic emerges in the triple-image region (see the lower panels of figure \ref{fig:Lens3}). This caustic corresponds to the second eigenvalue field of the deformation tensor. At $\alpha=3$ one of the two cusps of the second fold line merges with the outer fold line and transfers the cusp singularity via an elliptic umbilic caustic ($D_4^-$). For larger $\alpha$, the blue line will thus have three cusps whereas the red line has only one. 

However, more importantly, the lens forms a swallowtail caustic ($A_4$) in the blue line at $\alpha=2$. This phenomenon cannot be observed in the blue fold-line but is apparent in the normalized intensity map. The two stripes already visible for $\alpha=2$ are amplified. At the location where the swallowtail stripe coincides with the fold-line, we see an amplification of the normalized intensity in the swallowtail point. 

In the normalized intensity map, we see that the geometry becomes sharper and sharper as we increase the frequency and approach the geometric optics limit. Note that the normalized intensity of the hyperbolic umbilic ($D_4^+$) outshines the other caustics at frequency $\nu=500$.

\item 
Finally, for $\alpha = 4$, we see that the swallowtail caustic has unfolded into its characteristic shape in the blue fold-line (see Fig.~\ref{fig:Lens3_2}). We see the same structure emerge in the normalized intensity map. However, in addition, we how to obtain a large number of stripes emanating from the cusp caustics.  

We also see that the lens at $\alpha=4$, consists of a second hyperbolic umbilic ($D_4^+$) appearing at the origin, where the blue and the red fold-lines meet. As the frequency is raised, we again see that the normalized intensity spikes for this caustic.
\end{itemize}

\subsection{The elliptic umbilic caustic}
We conclude this section by studying the elliptic umbilici ($D_4^-$) caustic in a localized lens. The elliptic umbilic forms when the deformation tensor is singular due to two eigenvalues vanishing simultaneously. The geometry of the caustic however differs from the hyperbolic umbilic ($D_4^+$), in that it includes the merger of three cusp caustics. We here study the localized lens
\begin{align}
\varphi(\bm{x}) = \frac{ \alpha (x_1^3-3 x_1 x_2^2)}{1+x_1^2+ x_2^2}\,.
\end{align}

\begin{figure}
\centering
\begin{subfigure}[b]{0.24\textwidth}
\includegraphics[width=\textwidth]{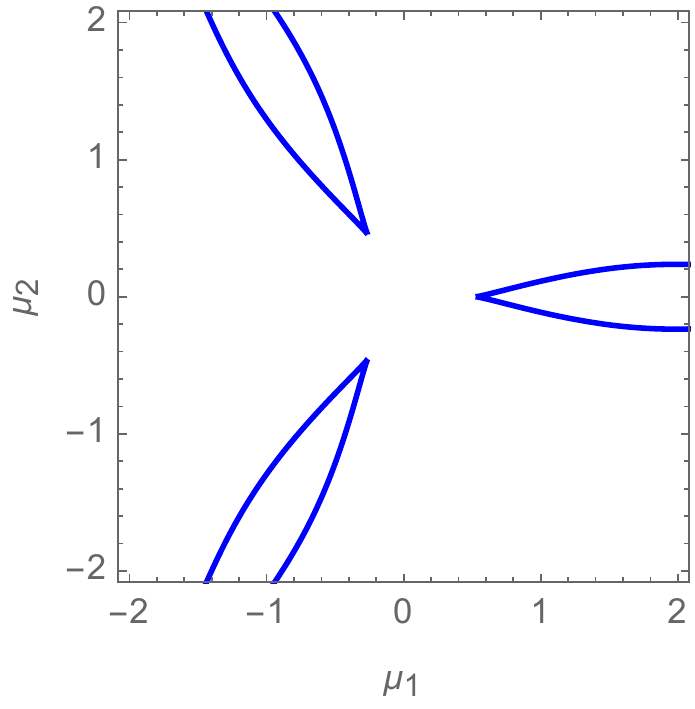}
\caption{$\alpha=1$}
\end{subfigure} 
\begin{subfigure}[b]{0.24\textwidth}
\includegraphics[width=\textwidth]{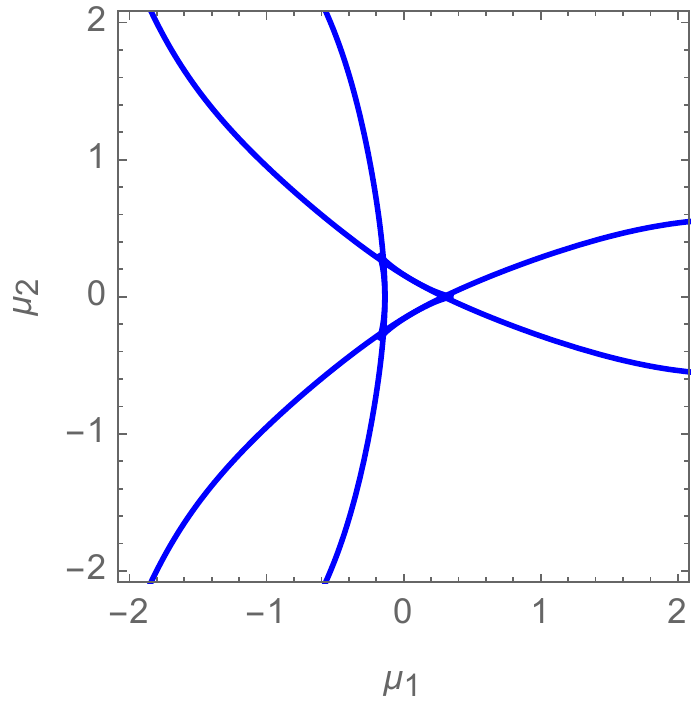}
\caption{$\alpha = 1.4$}
\end{subfigure} 
\begin{subfigure}[b]{0.24\textwidth}
\includegraphics[width=\textwidth]{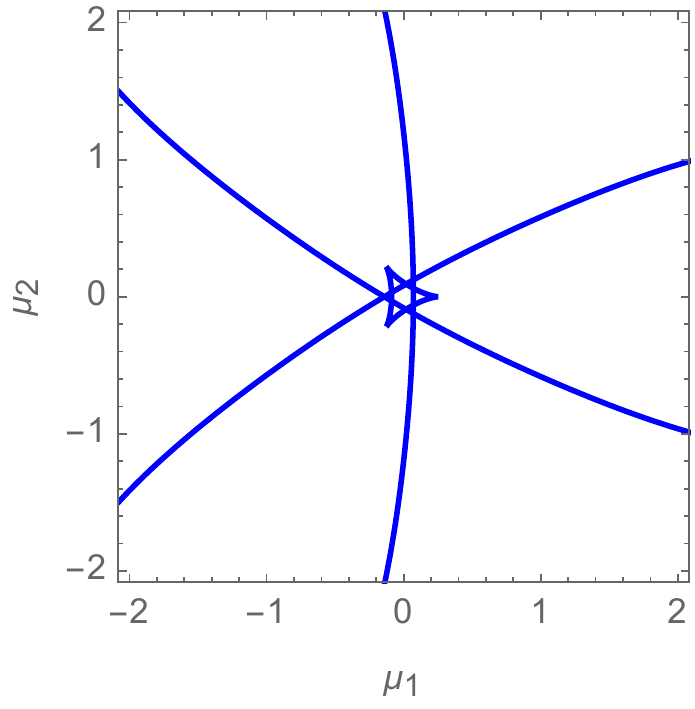}
\caption{$\alpha = 2$}
\end{subfigure}
\begin{subfigure}[b]{0.24\textwidth}
\includegraphics[width=\textwidth]{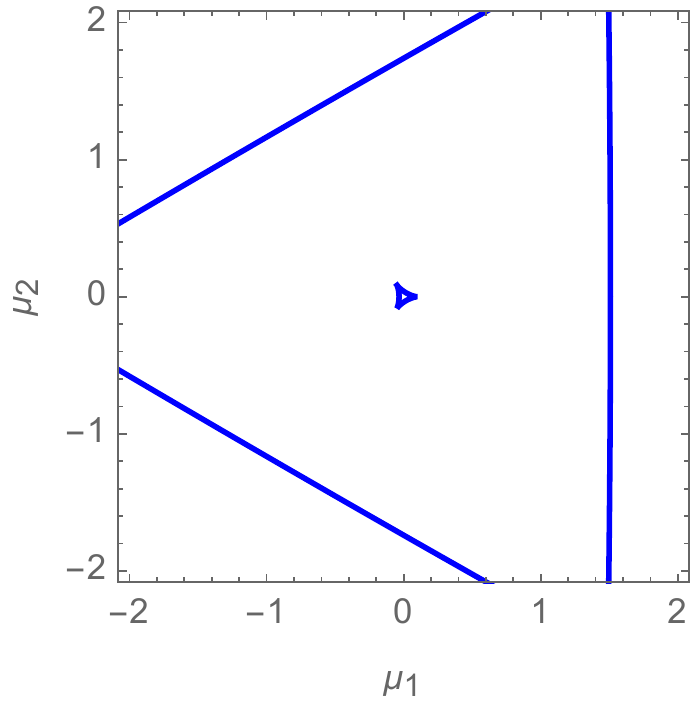}
\caption{$\alpha = 5$}
\end{subfigure}
\caption[The caustics in geometric optics for the local elliptic umbilic singularity.]{The caustics of the Lagrangian map for varying $\alpha$.}\label{fig:Lens6Caustics}
\end{figure}

From geometric optics, we observe the caustic structure of the lens (see Fig.~\ref{fig:Lens6Caustics}). 
\begin{itemize}
\item For small, $\alpha < 1.4$, the lens consists of three Zel'dovich pancakes with a triangular symmetry. Three of the cusp caustic point to the origin of the parameter space.
\item At $\alpha=1.4$, we observe that the three Zel'dovich pancakes are joined by three fold-lines forming a triangular structure. 
\item As $\alpha >1.4$, the triangle decouples from the three Zel'dovich pancakes. The three resulting fold lines move away from the origin and the triangle shrinks to a point. The point a which the triangle is contracted to a point is the elliptic umbilic caustic. The region enclosed by the large triangle is a $5$-image region. The region enclosed by the small triangle is a $7$-image region.
\end{itemize}
Note that since the elliptic umbilic caustic only forms after three cusp caustics have formed a triangular fold line, the caustic will be rare in simple simple lenses. It is nonetheless a stable configuration, as a small deformation of the lens preserves the structure.

\begin{figure}
\centering
\begin{subfigure}[b]{0.32\textwidth}
\includegraphics[width=\textwidth]{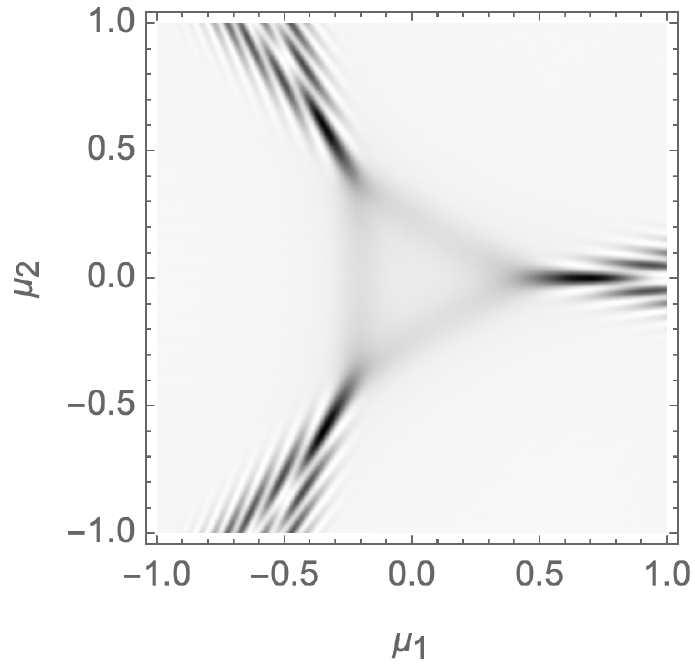}
\caption{$\alpha=1,\nu=50$}
\end{subfigure} 
\begin{subfigure}[b]{0.32\textwidth}
\includegraphics[width=\textwidth]{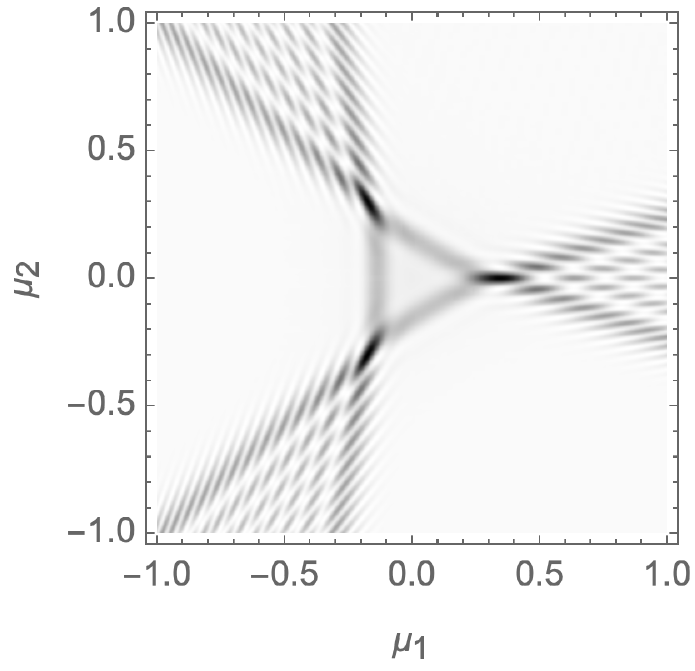}
\caption{$\alpha=1.4,\nu=50$}
\end{subfigure} 
\begin{subfigure}[b]{0.32\textwidth}
\includegraphics[width=\textwidth]{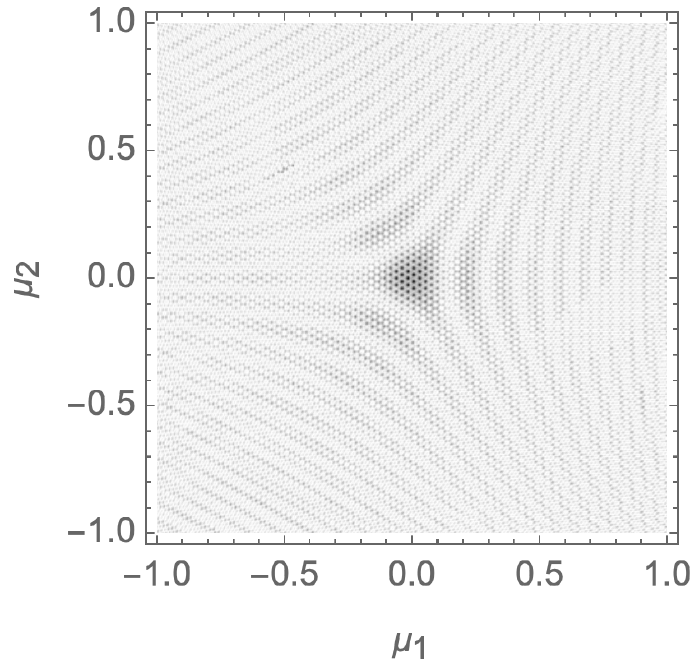}
\caption{$\alpha=5,\nu=50$}
\end{subfigure} \\
\begin{subfigure}[b]{0.32\textwidth}
\includegraphics[width=\textwidth]{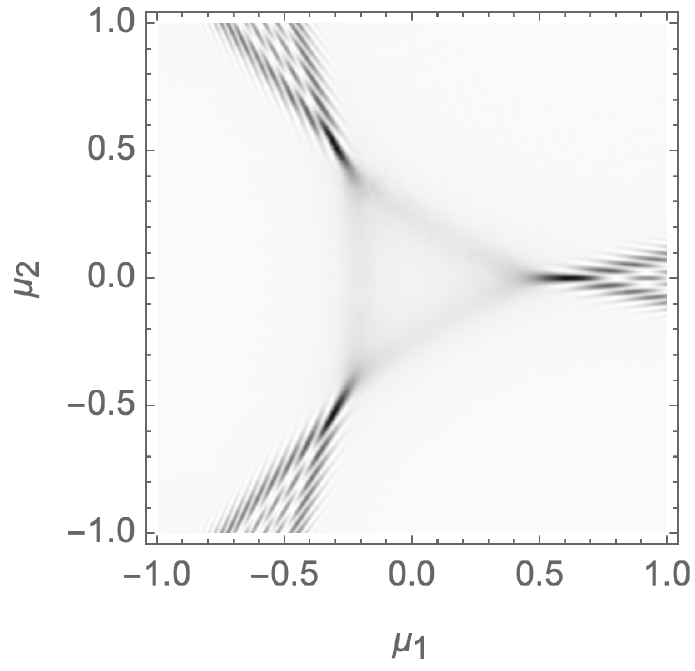}
\caption{$\alpha=1,\nu=100$}
\end{subfigure} 
\begin{subfigure}[b]{0.32\textwidth}
\includegraphics[width=\textwidth]{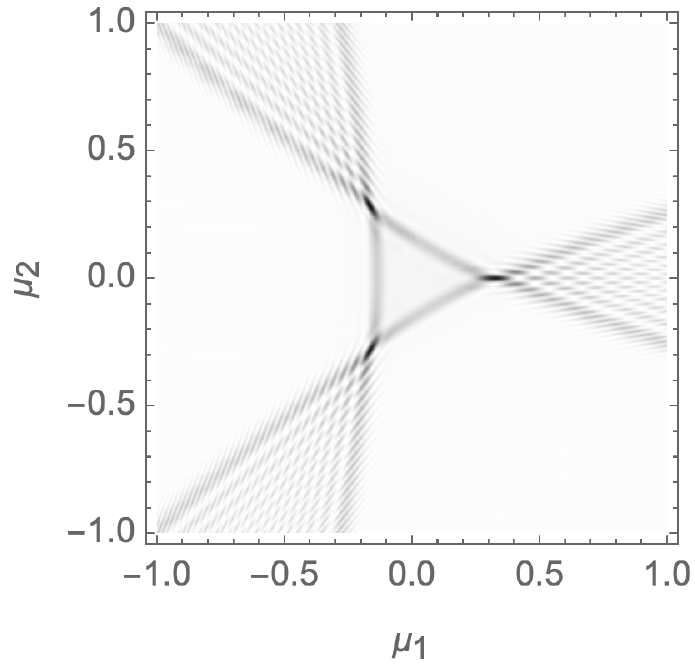}
\caption{$\alpha=1.4,\nu=100$}
\end{subfigure} 
\begin{subfigure}[b]{0.32\textwidth}
\includegraphics[width=\textwidth]{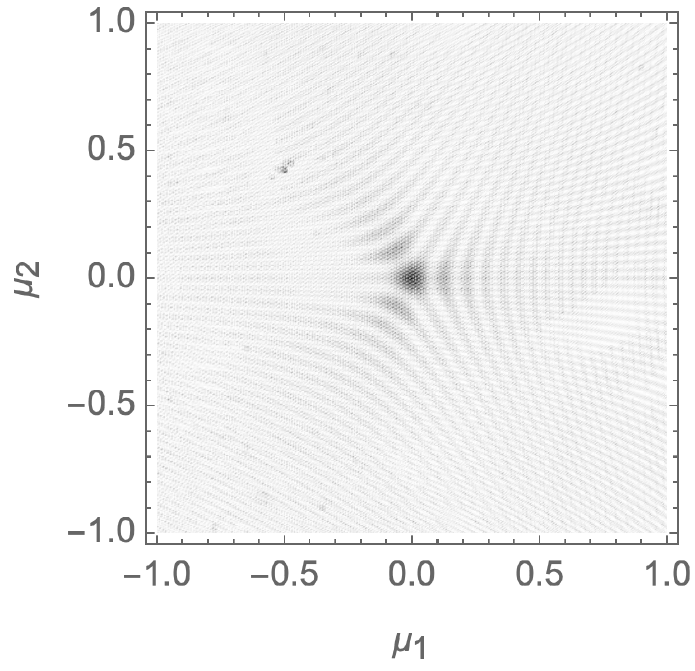}
\caption{$\alpha=5,\nu=100$}
\end{subfigure} 
\caption[The intensity map for the local lens with the elliptic umbilic caustic.]{The normalized intensity map, $I(\bm{\mu};\nu)$, for $\alpha=1,1.4,5$ and frequencies $\nu=50,100$.}\label{fig:Lens6}
\end{figure}

Using the Picard-Lefschetz analysis, we evaluate the normalized intensity map for the configurations $\alpha=1,1.4$, and $5$ for the frequencies $\nu=50,100$ (see Fig.~\ref{fig:Lens6}).
\begin{itemize}
\item For $\alpha=1$, we observe that even though the triangular structure is not yet present in the geometric optics analysis, it is present in the normalized intensity map at finite frequency (see the left panels of Fig.~\ref{fig:Lens6}). That is to say, the normalized intensity is enhanced at the triangle, however as $\nu\to \infty$ the normalized intensity at the triangle will remain finite.
\item At $\alpha=1.4$, the triangle has formed in the geometric optics analysis (see the central panels of Fig.~\ref{fig:Lens6}). In the normalized intensity maps, the triangle is enhanced. The normalized intensity will now diverge in the geometric optics limit.
\item As $\alpha$ is further increased to $\alpha =5$, the triangle shrinks to a point and interference effects between the different fold lines start to appear (see the right panels of Fig.~\ref{fig:Lens6}). At $\alpha=5$, we do no longer observe the fold lines but rather observe a triangular blob at the origin of the parameter space. This closely resembles the normalized intensity map of the elementary elliptic umbilic catastrophe. It is however a bit more intricate as a close inspection demonstrates that caustic structure oscillates at a high frequency due to the interference of the elliptic umbilic with the surrounding multi-image region.
\end{itemize}
Note that there are a few small numerical artefacts present in the normalized intensity map for the lens at $\alpha=5$. The lens outside of the triangle, is a $5$-image region in which some of the real saddle points are located far away from the origin in the lens plane. The inside of the triangular region is a $7$-image region. The Lefschetz thimble has a complicated shape and the tessellation of the thimble can occasionally miss a few points.

\section{Signatures of caustics in fast radio bursts}\label{sec:signatures}
A Fast Radio Burst (FRB) is a millisecond transient radio pulse, caused by some yet to be identified high-energy astrophysical process. The first burst was found by Duncan Lorimer and his student David Narkevic in 2007 while scanning through archival pulsar survey data~\cite{Lorimer:2007}. The burst in question had been detected in 2001 by the Parkes Observatory in Australia. In subsequent years, several other bursts were observed, among which the first repeating source (named FRB 121102)~\cite{Spitler:2014} was detected in 2012 by the Arecibo Observatory in Puerto Rico.   In the last few months, several new detections have been announced by the Canadian Hydrogen Intensity Mapping Experiment (CHIME) collaboration, including the second repeating FRB source (named FRB 180814) \cite{CHIME:2019}. FRBs are now known to be relatively common, with approximately $10,000$ bright fast radio bursts occurring per day over the entire sky. Telescopes capable of detecting a significant fraction of these bursts should become possible in coming decades, an exciting prospect indeed.  

The source of fast radio bursts is yet to be identified. Many different models have been proposed but none is yet compelling. They range from rapidly spinning neutron stars or black holes and regions of very high electromagnetic fields, to more exotic sources~\cite{2018NatAs...2..842P,2018arXiv181005836P}. It seems likely that the bursts are extragalactic in origin, as the first observed repeater, FRB 121102, has been identified with a galaxy at a distance of approximately $3$ billion light years~\cite{Chatterjee:2017, Chatterjee:2017b, Michilli:2018}. As mentioned in the introduction, it is likely that the phenomenology of fast radio bursts is strongly affected by astrophysical plasma lensing. They have a characteristic time-frequency profile, their frequency typically falling during the pulse, or series of pulses. This profile is probably due to the fact that lower frequencies are more strongly lensed and thus follow longer geometrical paths, and also because they propagate more slowly.

The methods and results we have reported here should be helpful in modeling the effects of plasma lensing on observed FRBs. The lensing may take place in a variety of places -- near the source, near the observer or in between. If the line of sight encounters a caustic due to a plasma lens, the FRB may be amplified, enhancing the chances of detection. For reasons we have explained, caustics are likely to be localized in frequency, leading to the observed spectral shape. The ``marching down'' features could also be due to asymmetric structures in the lens, leading to angled caustics. This requires a preferred time asymmetry, which could in turn provide hints about the structure of the lens itself.  In the lensing example of B1957+20 \cite{Main:2018}, the lens is due to a companion wind. In this specimen, the time-frequency caustics march both up and down.  This symmetry could be broken if the wind contained shock waves, which could preferentially move retrograde in the rotating frame.  Quantitative lens modeling can be tested on the pulsar binary system, and then applied to FRB data. This could be the scope of a future paper.

Since the observed radio waves have a relatively long wavelength, the corresponding diffraction catastrophes are likely to fill a significant volume in the parameter space of the normalized intensity maps. Therefore it is important to study the complete interference pattern. It follows from Table \ref{tab:exponents} that the elliptic ($D_4^-$), the hyperbolic ($D_4^+$) umbilic and to a lesser extent the swallowtail ($A_4$) caustic lead to the largest spikes in the normalized intensity map. Of these three caustics, the swallowtail ($A_4$) and the hyperbolic ($D_4^+$) umbilic caustics are most likely to be realized in simple lenses, of which the hyperbolic caustic gives the greatest amplification. However, these caustics will not generically occur in time-frequency data, as they are formed at point in three-dimensional functions. The line of sight, is, however, reasonably likely to pass close to them, as they fill a finite volume of the parameter space. In principle, we do expect to see the cusp ($A_3$) points and the fold ($A_2$) lines caustics, in the data. However, note that these caustics lead to a lesser amplification of the source.

As we observed in the previous sections, caustics due to multi-dimensional lenses never occur as isolated events. The caustics of co-dimension four, \textit{i.e.}, the umbilics $D_4^\pm$, and the swallowtail $A_4$ caustics, are always accompanied by cusp ($A_3$) points and fold ($A_2$) lines. It thus follows that when a fast radio burst is indeed amplified by a lens, that the corresponding peak in time-frequency space will be of characteristic shape. More concretely, after identifying the time and the frequency with the two of the unfolding parameters $\mu$, we expect the peak to resample the normalized intensity map of the corresponding elementary catastrophe computed in Section \ref{sec:catastrophePL}. That is to say, the peak corresponding to elliptic ($D_4^-$) umbilic caustic should exhibit a triangular symmetry and the peak corresponding to the swallow $(A_4)$ caustic will exhibit the characteristic swallowtail geometry in the fold-line and two cusps caustics.

Further investigation is required to estimate the number density of the different caustics for generic two-dimensional lenses and the most likely normalized intensity profiles along the line of sight.

\section{Conclusions}\label{sec:conclusion}
Conditionally convergent oscillatory integrals play a central role in modern physics. However, these integrals are often difficult to define as their definition, in the multi-dimensional case, can depend on the order of integration or the regularization scheme. They are, moreover, generically impossible to evaluate analytically and too expensive to evaluate with conventional numerical methods. In this paper we have brought Picard-Lefschetz theory to bear. We have shown how in a multi-dimensional oscillatory integral, the integrand generically defines a set of relevant Lefschetz thimbles in the complexified integration domain, along which the integral is absolutely convergent. These thimbles can be thought of as an `integrand-dependent Wick rotation'. The integral evaluated along the set of relevant thimbles in fact provides an unambiguous definition of the original integral itself. We moreover have presented a new, efficient numerical scheme both to find the thimbles and to efficiently evaluate the integral along them in polynomial time. The virtue of this new method that the efficiency actually {\it increases} as the integrand becomes more oscillatory.

In particular, we have studied the Lefschetz thimbles for  caustic catastrophes and the Stokes phenomenon occurring in two-dimensional lenses. Given the thimbles, we numerically evaluate the normalized intensity maps over all frequencies study the resulting interference patterns. We have shown that the normalized intensity maps smoothly converge to the caustics predicted by geometric optics, without introducing numerical artifacts. 

Our method renders feasible the calculation of interference patterns in a wide variety of interesting astrophysical contexts, in particular to model the effect of plasma lenses on radio sources. So far, such modeling has been restricted to the simplest examples of  fold and cusp singularities, produced by one dimensional lenses. More realistic, two-dimensional models, including the swallowtail, elliptic umbilic and hyperbolic umbilic caustics are now accessible. We have computed the normalized intensity maps for a few representative examples, and commented briefly on likely observational signatures. A statistical analysis of the normalized intensity profiles for the diffraction catastrophes generated by a realistic plasma lens ensembles will be the subject of further investigations.

Finally, we analyzed a simple model of Young's double slit experiment, representing an initial exploration of the use of these methods for describing interference in quantum mechanics. 

\acknowledgments

We thank Roger Blandford, Claudio Bunster, Neal Dalal, Angelika Fertig, Sterl Finney, Steven Gratton, James Hartle, Estelle Inack, Nick Kaiser, Nynke Niezink, Laura Sberna and Doug Scalapino for interesting comments and encouragement. Research at Perimeter Institute is supported by the Government of Canada through Industry Canada and by the Province of Ontario through the Ministry of Research and Innovation. Ue-Li Pen holds Associate positions at the Dunlap Institute for Astronomy and Astrophysics and at Perimeter Institute. Ue-Li Pen and Neil Turok are Associate Fellows in the Canadian Institute for Advanced Research (CIFAR) Gravity and the Extreme Universe program. The authors also gratefully acknowledge support from the Centre for the Universe at Perimeter Institute. 

\bibliographystyle{utphys}
\bibliography{library}

\appendix
\section{Defining oscillatory integrals} \label{ap:DefiningOscillatoryIntegrals}
Oscillatory integrals, which do not converge absolutely, are sometimes claimed to be ill-defined since the key theorems of measure theory, \textit{e.g.} the dominated convergence theorem and Fubini's theorem, do not apply \cite{Doob:2012}. We here study conditionally convergent oscillatory integrals for the one- and multi-dimensional case and propose a definition using Picard-Lefschetz theory in terms of absolutely convergent ones.

\subsection{One-dimensional integral}
\begin{figure}
\centering
\includegraphics[width=0.5\textwidth]{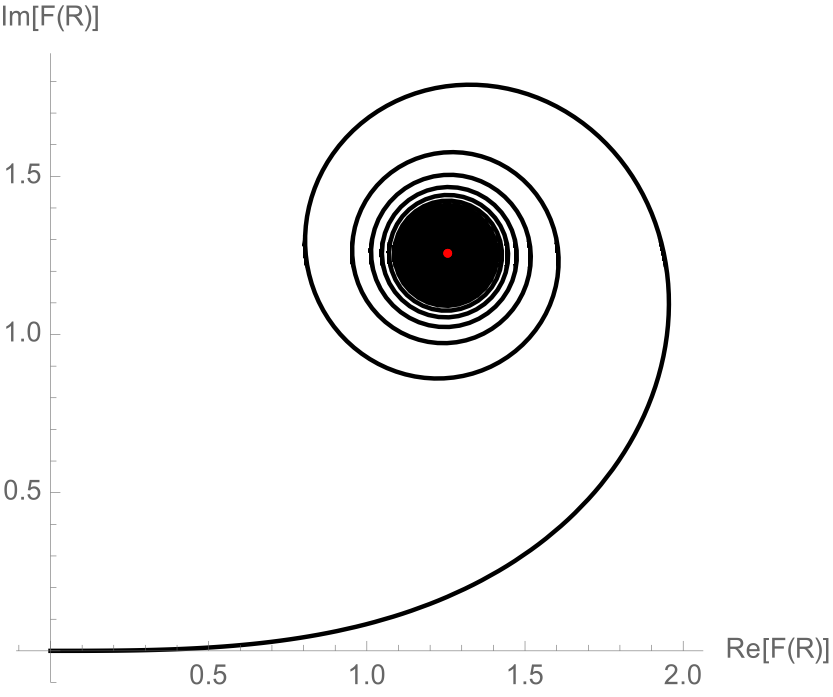}
\caption[The Euler or Cornu spiral.]{The Euler or Cornu spiral. The black line is the real and imaginary part of $F(R)$ as a function of $R$. The black point is the real and imaginary part of the limit $\lim_{R\to\infty}F(R) = (1+i)\sqrt{\frac{\pi}{2}}$.}\label{fig:Cornu}
\end{figure}

The Fresnel integral 
\begin{align}
F(\infty) = \int_{-\infty}^\infty e^{i x^2}\mathrm{d}x = (1 + i) \sqrt{\frac{\pi}{2}}
\end{align}
exists, even though the integral is only conditionally convergent. The integral is usually defined as a limit of the partial integral
\begin{align}
F(R) = \int_{-R}^R e^{i x^2}\mathrm{d}x,
\end{align}
\textit{i.e.}, $\lim_{R\to \infty} F(R) = (1 + i) \sqrt{\frac{\pi}{2}}$ following the Euler or Cornu spiral (see Fig.~\ref{fig:Cornu}). This definition is as important to the integral as the integrand, as different regularization schemes -- which do not approach the real line by adding points incrementally -- lead to different answers. 

The definition of the conditionally convergent integral in terms of the limit $R\to \infty$ is equivalent to the assumption of analyticity, since Cauchy's integral theorem only applies to integrals over $\mathbb{R}$ defined this way. We can alternatively define the integral by deforming the integration contour $\mathbb{R}$ in the complex plane $\mathbb{C}$ to the Lefschetz thimble 
\begin{align}
\mathcal{J}=\{ (1+i)u| u \in \mathbb{R}\}= (1+i)\mathbb{R}\,,
\end{align} 
for which the integrand is convex and the integral is absolutely convergent, \textit{i.e.},
\begin{align}
\int_\mathrm{R} e^{i x^2}\mathrm{d}x &= \int_\mathcal{J} e^{i x^2}\mathrm{d}x\\
&= (1+i) \int_\mathrm{R} e^{-2 u^2}\mathrm{d}u\\
&= (1+i) \sqrt{\frac{\pi}{2}}\,.
\end{align}
Note that this definition does not depend on a limit. The regularization is completely determined by the assumption of analyticity.

\subsection{Multi-dimensional integrals}\
Multi-dimensional conditionally convergent oscillatory integrals such as 
\begin{align}
\int_{\mathbb{R}^N} e^{if(x_1,\dots,x_N)}\mathrm{d}x_1\dots \mathrm{d}x_N
\end{align}
for $N\in \mathbb{N}$ and appropriate functions $f$, play an important role in optics but cannot be uniquely defined using an extension of the regularization scheme described above for the one-dimensional case. To show this, lets consider the two-dimensional analogue of the Fresnel integral
\begin{align}
\int_{\mathbb{R}^2} e^{i (x^2 + y^2)}\mathrm{d}x\mathrm{d}y\,.
\end{align}
Since this integral factorizes, it is reasonable to require the integral to converge to 
\begin{align}
F(\infty)^2 = \left((1+i)\sqrt{\frac{\pi}{2}}\right)^2=i \pi\,.
\end{align}
However, for general $f(x_1,\dots,x_n)$ we are not able to write the integral as a product of one-dimensional integrals. This thus should not be considered as a desirable definition of the integral. 

To see the dependence on the regularization scheme, consider the integral in polar coordinates. We write
\begin{align}
I(R) &= \int_{\mathbb{D}_R} e^{i(x^2+y^2)}\mathrm{d}x \mathrm{d}y\\
&= 2 \pi \int_0^R r e^{i r^2} \mathrm{d}r\\
&= i \pi\left(1- e^{i R^2}\right)\,,
\end{align}
with $\mathbb{D}_R$ the disk of radius $R$ centred at the origin. We thus find that the limit $\lim_{R\to \infty} I(R)$ does not exist! The function $I(R)$ instead circles the `correct answer' $i\pi$ with increasing angular velocity.

It is instead appropriate to define the integral in terms of the Lefschetz thimble
\begin{align}
\mathcal{J} &= \{(1+i)(u,v)| (u,v)\in \mathbb{R}^2\}\\
&=(1+i)\mathbb{R}^2\,.
\end{align}
Along the thimble, the integral is absolutely convergent
\begin{align}
\int_{\mathbb{R}^2} e^{i (x^2 + y^2)}\mathrm{d}x\mathrm{d}y 
&= \int_\mathcal{J} e^{i (x^2 + y^2)}\mathrm{d}x\mathrm{d}y \\
&= (1+i)^2\int_{\mathbb{R}^2} e^{-2 (u^2 + v^2)}\mathrm{d}u\mathrm{d}v \\
&= 2i \int_{-\infty}^\infty\int_{-\infty}^\infty  e^{-2 (u^2 + v^2)}\mathrm{d}u\mathrm{d}v \\
&= i \pi\,.
\end{align}
On the thimble we can safely convert the integral over the real plane $\mathbb{R}^2$ into the iterative integral using Fubini's theorem, since the integral in $u$ and $v$ over $\mathbb{R}^2$ is absolutely convergent. This definition straightforwardly generalizes to general multi-dimensional conditionally convergent integrals.

\section{Young's double-slit experiment}\label{ap:Young}
In this appendix, we generalize our treatment of interference in order to tackle Young's famous double slit experiment. 
In spite of the extreme simplicity of this example, and its centrality to introductory discussions of quantum physics, detailed interference patterns are surprisingly hard to compute. By generalizing our treatment of the Fresnel-Kirchhoff integral we shall be able to efficiently study the pattern created by a pair of smooth, finite size slits in detail. In particular, we shall see how quantum interference effects disappear in the classical limit, as $\hbar$ is taken to zero.

 The generalization required is to make the interference ``phase" complex in order to damp out the amplitude away from two narrow slits. Modeling this complex phase with a simple rational function, our numerical techniques allow us to efficiently find the relevant Lefschetz thimbles and compute the detailed interference pattern at all values of the parameters. 

\begin{figure}
\includegraphics[width=0.5\textwidth]{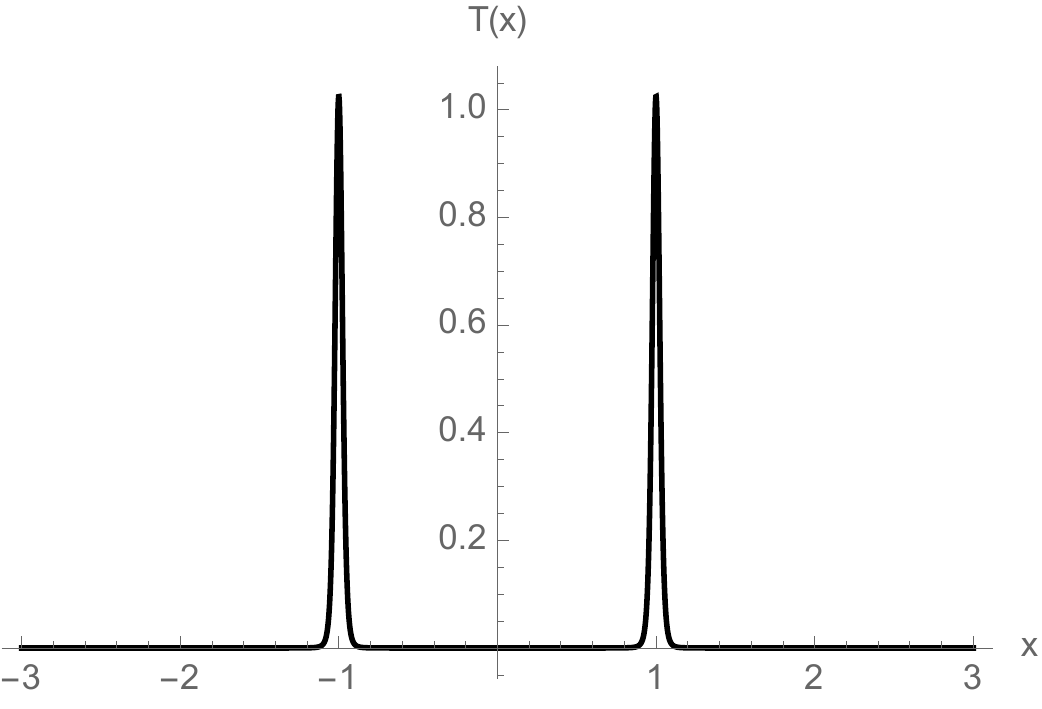}
\caption{The transition amplitude of the wall $T(x)$ with two slits at $s_1=-1$, $s_2=1$ with width $\epsilon=0.1$.}\label{fig:doubleSlit}
\end{figure}

Consider a distant point source emitting particles towards a screen, with a thin barrier separating the screen from the source. The barrier is opaque to the particles except in the neighbourhood of two slits. In dimensionless coordinates (which we shall define below), the transmission amplitude takes the form
\begin{align}
T(x) \propto \exp\left[ \frac{\epsilon}{\epsilon^2 + (x-s_1)^2} + \frac{\epsilon}{\epsilon^2 + (x-s_2)^2} - \frac{1}{\epsilon}\right]\,,
\end{align}
consisting of two peaks each of strength unity, centered respectively at $x = s_1$ and $ x = s_2$. Here, $\epsilon >0$ is a small number representing both the width of the slits (see Fig.~\ref{fig:doubleSlit}) and the opacity of the barrier: away from the slits, the latter is given by  $T\sim \exp(-1/\epsilon)$.

Assuming the incident amplitude for the particles to be coherent and constant across the slits, we may then compute the path integral amplitude just as in Section \ref{sec:FromFeynmanToFermat}. Here, however, we deal with a particle of fixed mass $m$, energy $E$ and momentum $p=\sqrt{2 m E}$. The last formula in (\ref{eq:actiii}), in the same small displacement-approximations made in Eq.~(\ref{eq:pathfinal}) above, yields a Pythagorean contribution to the phase, $ p (x-\mu)^2/(2 d \hbar)$ where $d$ is the distance from the slits to the screen. Setting $x\rightarrow x a$ where $a$ is the characteristic dimension of the slits and $x$ is dimensionless, we take the quantity $\hbar 2 d /(p a^2)$ to be our new, dimensionless $\hbar$. In terms of these dimensionless quantities, the amplitude for the particle to arrive at position $\mu$ on the screen is therefore given by the oscillatory integral
\begin{align}
\Psi(\mu) = \mathcal{N} \int e^{\frac{i}{\hbar} (x-\mu)^2} T(x)\mathrm{d}x\label{eq:waveFunction}
\end{align}
with the normalization constant $\mathcal{N}$, ensuring unitarity $\int |\Psi(\mu)|^2\mathrm{d}\mu = 1$. The probability for the particle to arrive at $\mu$ on the screen is given by the absolute square of the wavefunction
\begin{align}
I(\mu) = |\Psi(\mu)|^2\,.
\end{align}
Note that the dimensionless version of Planck's constant $\hbar$ appears in this nonrelativistic problem, whereas it cancelled out of our earlier formulae for a massless particle, as a result of the latter's scale covariance. 

We evaluate the wavefunction (equation \eqref{eq:waveFunction}), by analytically continuing the exponent
\begin{align}
\phi(x;\mu) = \frac{i}{\hbar} (x-\mu)^2 + \frac{\epsilon}{\epsilon^2 + (x-s_1)^2} + \frac{\epsilon}{\epsilon^2 + (x-s_2)^2} - \frac{1}{\epsilon}\,,\label{eq:YoungExp}
\end{align}
in the complex plane and evaluating the Lefschetz thimble. The exponent, $\phi$, has four poles and nine saddle points. The poles at $x = s_i \pm i \epsilon$ correspond to the slit centereed at $s_i$. The saddle points are roots of a nineth order polynomial. We can associate four saddle points to each slit. The remaining saddle point is shared and moves between the the poles corresponding to the two slits as a function of the position on the screen $\mu$.

\begin{figure}
\includegraphics[width=0.84\textwidth]{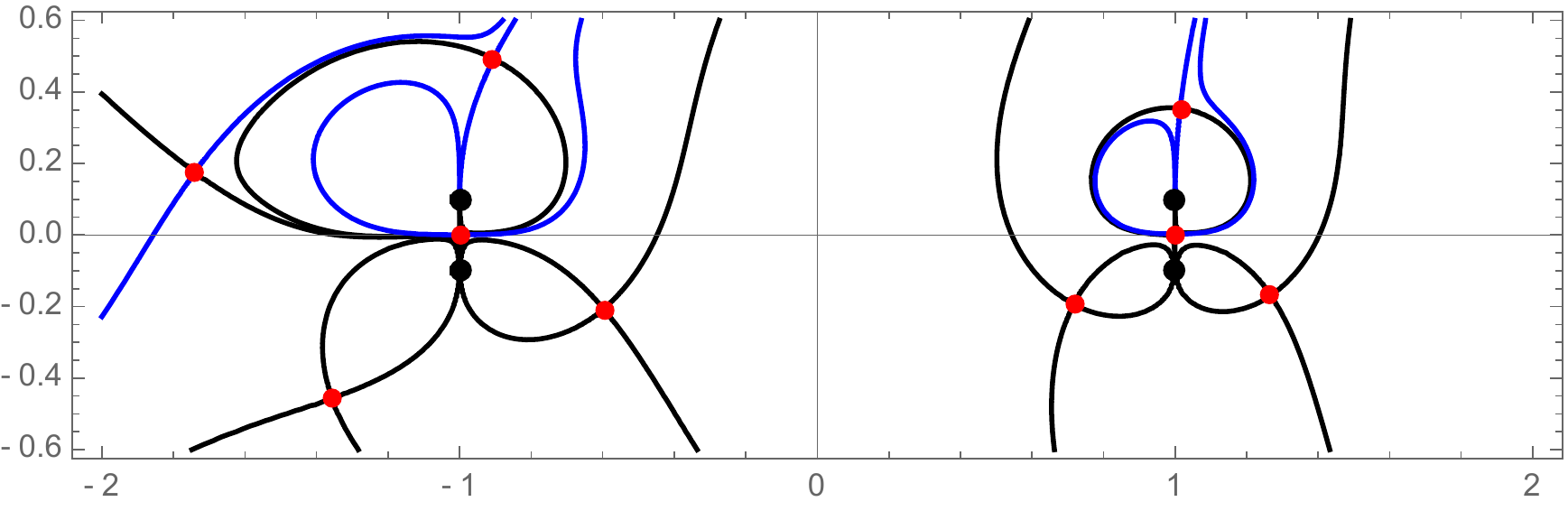}
\includegraphics[width=0.84\textwidth]{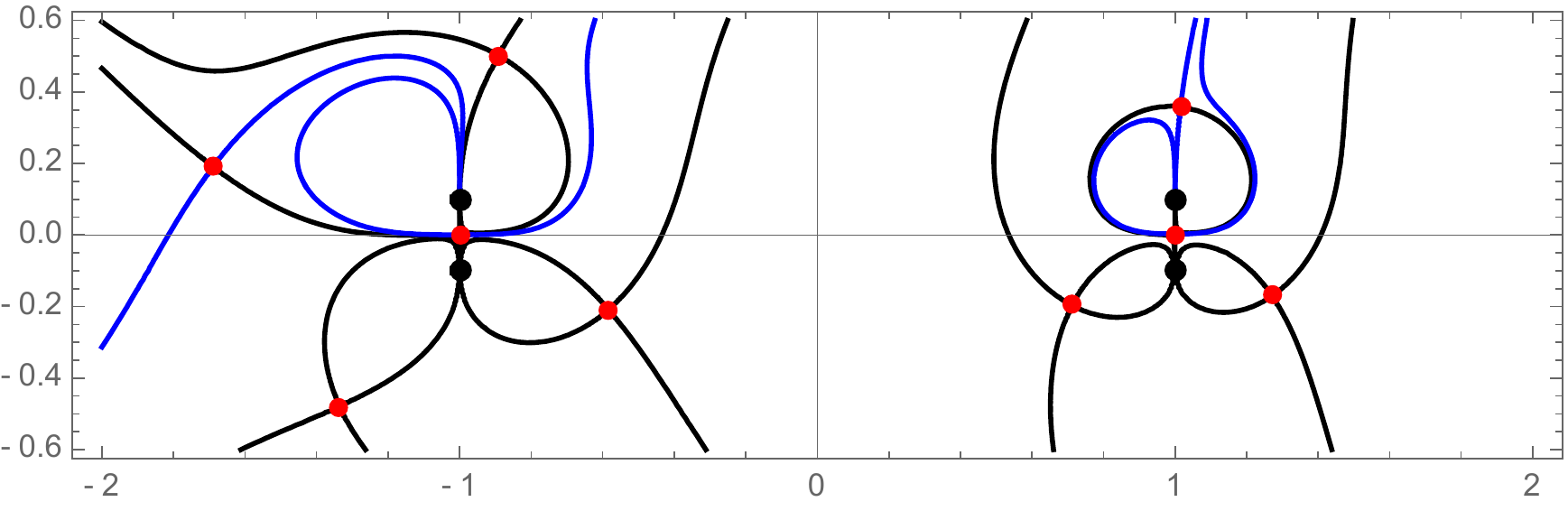}
\includegraphics[width=0.84\textwidth]{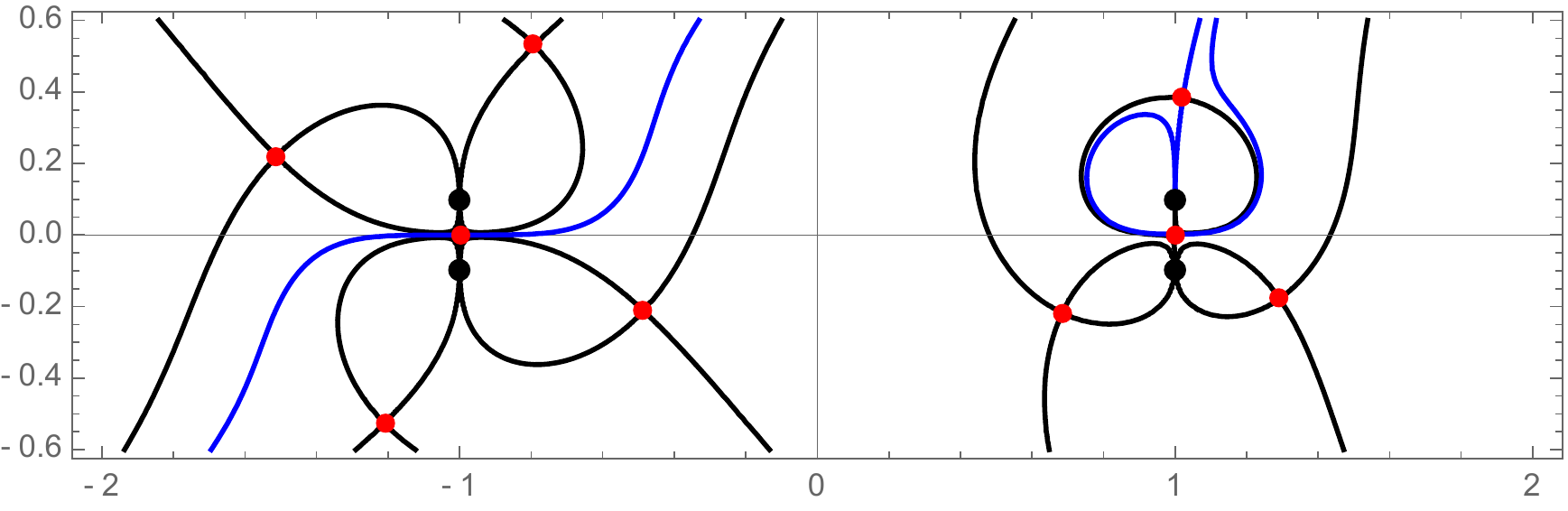}
\includegraphics[width=0.84\textwidth]{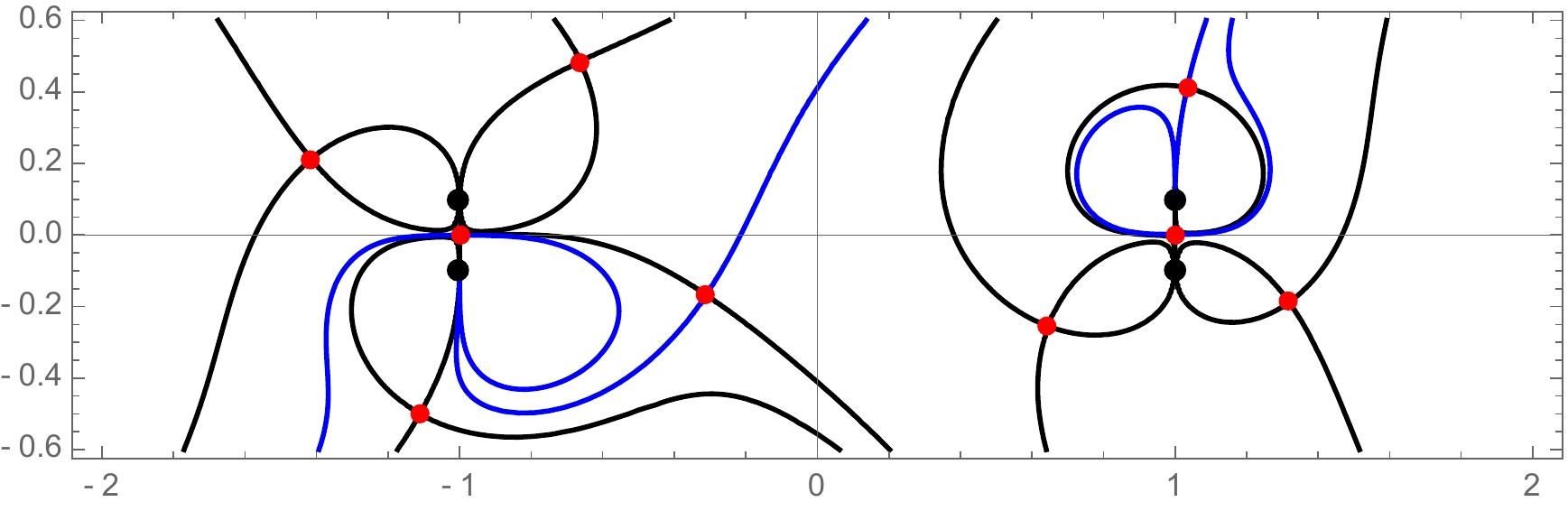}
\includegraphics[width=0.84\textwidth]{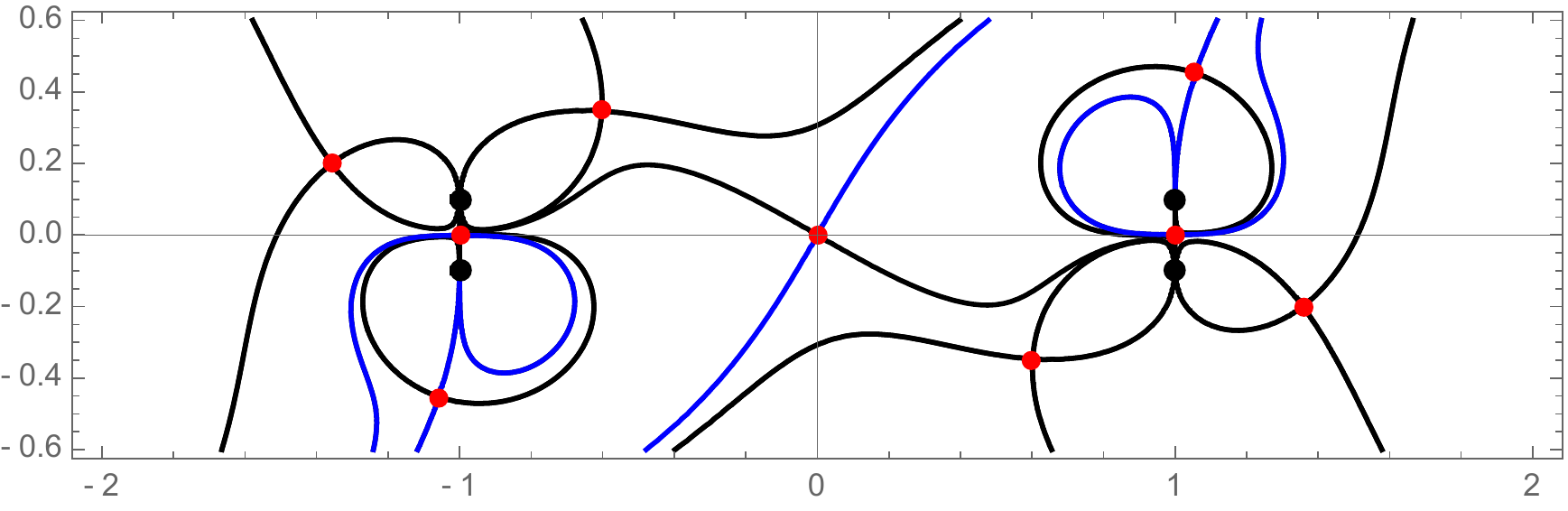}
\caption{The Picard-Lefschetz diagram for the Young experiment for $\epsilon =0.1$ from $\mu=-1.6, -1.5, -1,-0.5,$ and $\mu=0$, with the steepest ascent, descent contours (black) and thimbles (blue) corresponding to the saddle points (red) for $\hbar =1$.}\label{fig:YoungThimbles}
\end{figure}

Fig.~\ref{fig:YoungThimbles} shows the corresponding Picard-Lefschetz diagrams for various positions $\mu$ for $\hbar=1$. In the description we will for simplicity assume the left slit to be at $s_1$ and the right slit to be at $s_2$, \textit{i.e.}, $s_1< s_2$:
\begin{itemize}
\item For positions on the screen far to the left of the slits, $\mu \ll s_1$, the thimble consists of five steepest descent contours. The thimble runs from the lower left to the upper right via a complex saddle point. The thimble subsequently loops around the upper left and the upper right poles. For positions $\mu \leq \frac{s_1+s_2}{2}$, the wavefunction is dominated by the left slit. It is for this reason not surprising to see that the thimble corresponding to the right slit is representative in this regime.
\item As $\mu$ approaches $s_1$, we observe a Stokes transition after which only four saddle points are relevant. The thimble moves from the lower-left via a saddle point to the upper left pole, after which it passes through the saddle point between the two left poles. The right part of the thimble is largely unchanged.
\item For $\mu$ near $s_1$, we observe yet another Stokes transition after which only three saddle points remain relevant. The thimble runs from the lower left via the saddle point between the to left poles to the upper right.
\item When $\mu$ approaches the mid-point $\frac{\mu_1+\mu_2}{2}$, we observe that a complex saddle point becomes relevant after a Stokes phenomenon. The thimble now consists of four steepest descent contours.
\item For $\mu$ near the mid-point $\mu = \frac{\mu_1+\mu_2}{2}$, we observe that after yet another Stokes transition, we obtain a thimble consisting of five steepest descent contours. Note that the middle saddle point has moved to the origin $x=0$. When the position $\mu$ is increased further, this saddle point will move to the poles corresponding to the right slit. The corresponding Picard-Lefschetz diagrams are mirror images of the ones discussed above.
\end{itemize}
In the semi-classical limit $\hbar \to 0$, the geometry of the Lefschetz thimble is to an increasing extent determined by the Pythagorean term in equation \eqref{eq:YoungExp}. As a consequence, after a few Stokes transitions, the eight saddle points which can be associated to the poles corresponding to the two slits become tighter bound to the poles representing the geometry of the right part of the thimble in figure \ref{fig:YoungThimbles}. The remaining saddle point still moves between the poles corresponding to the two slits. However, note that the integral is increasingly dominated by the two saddle points between the four poles. These two saddle points approach the real line at $x=s_1$ and $x=s_2$ in this limit.

\begin{figure}
\centering
\begin{subfigure}[b]{0.49\textwidth}
\includegraphics[width=\textwidth]{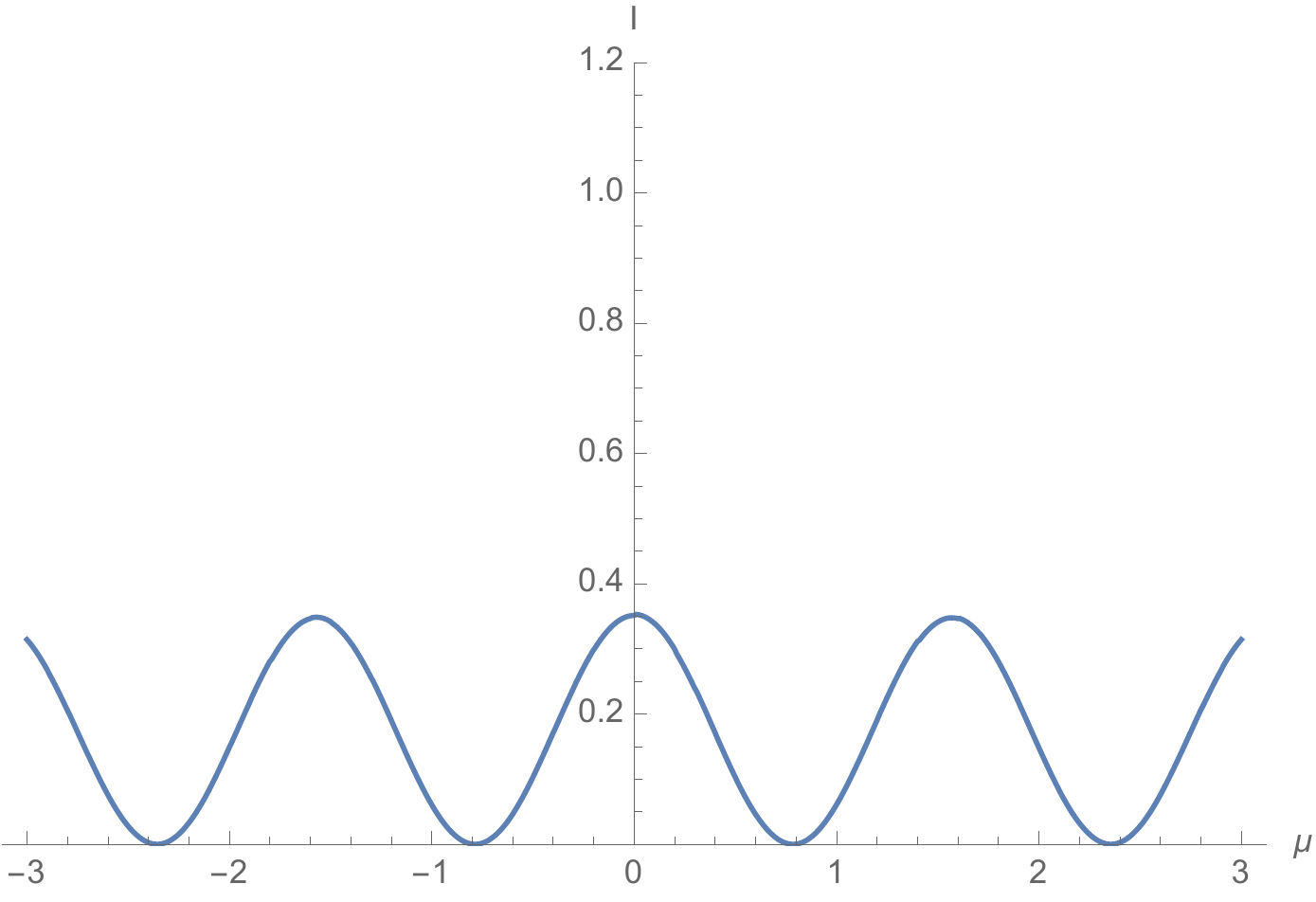}
\caption{$\hbar=1$}
\end{subfigure}
\begin{subfigure}[b]{0.49\textwidth}
\includegraphics[width=\textwidth]{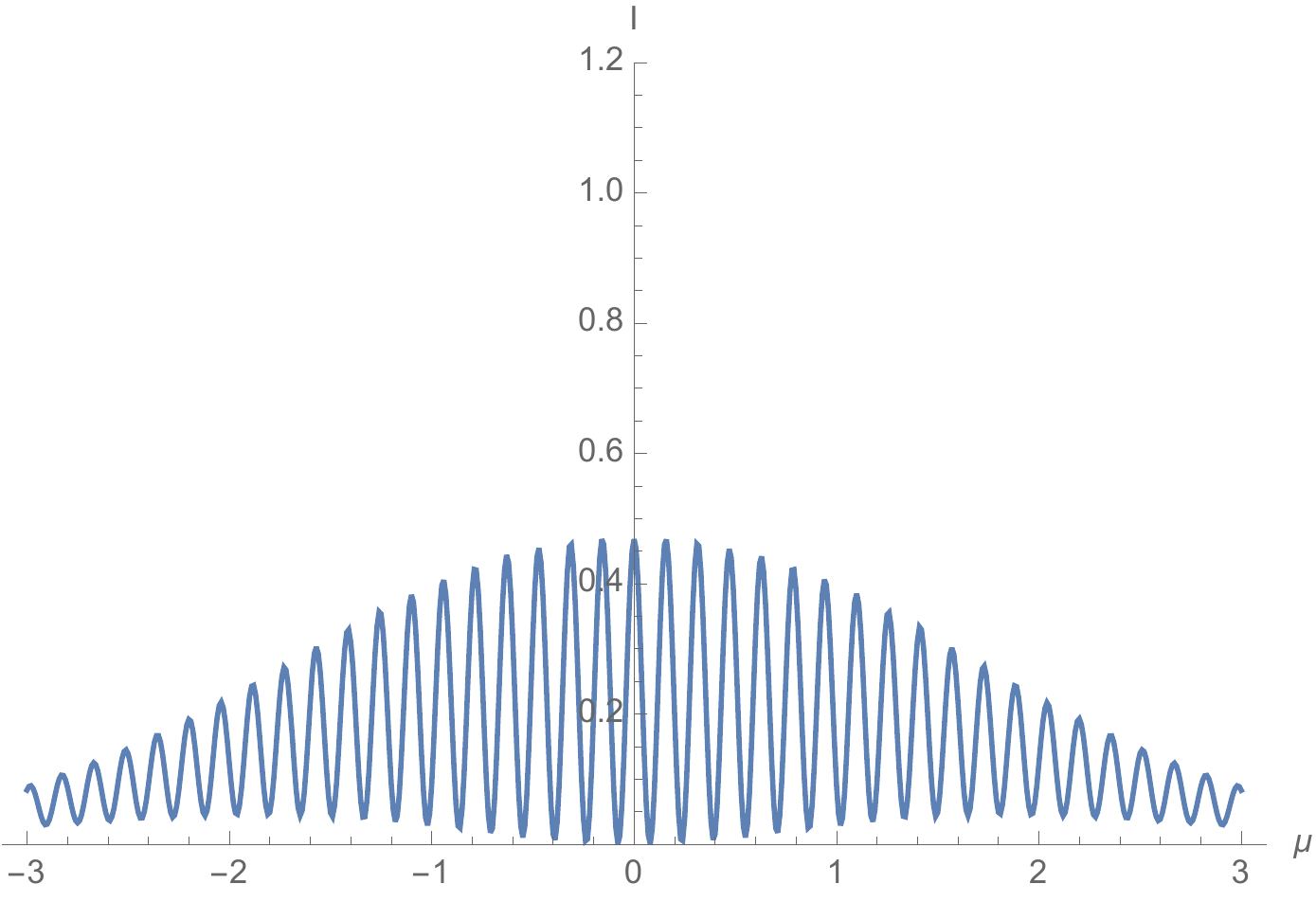}
\caption{$\hbar=1/10$}
\end{subfigure} \\
\begin{subfigure}[b]{0.49\textwidth}
\includegraphics[width=\textwidth]{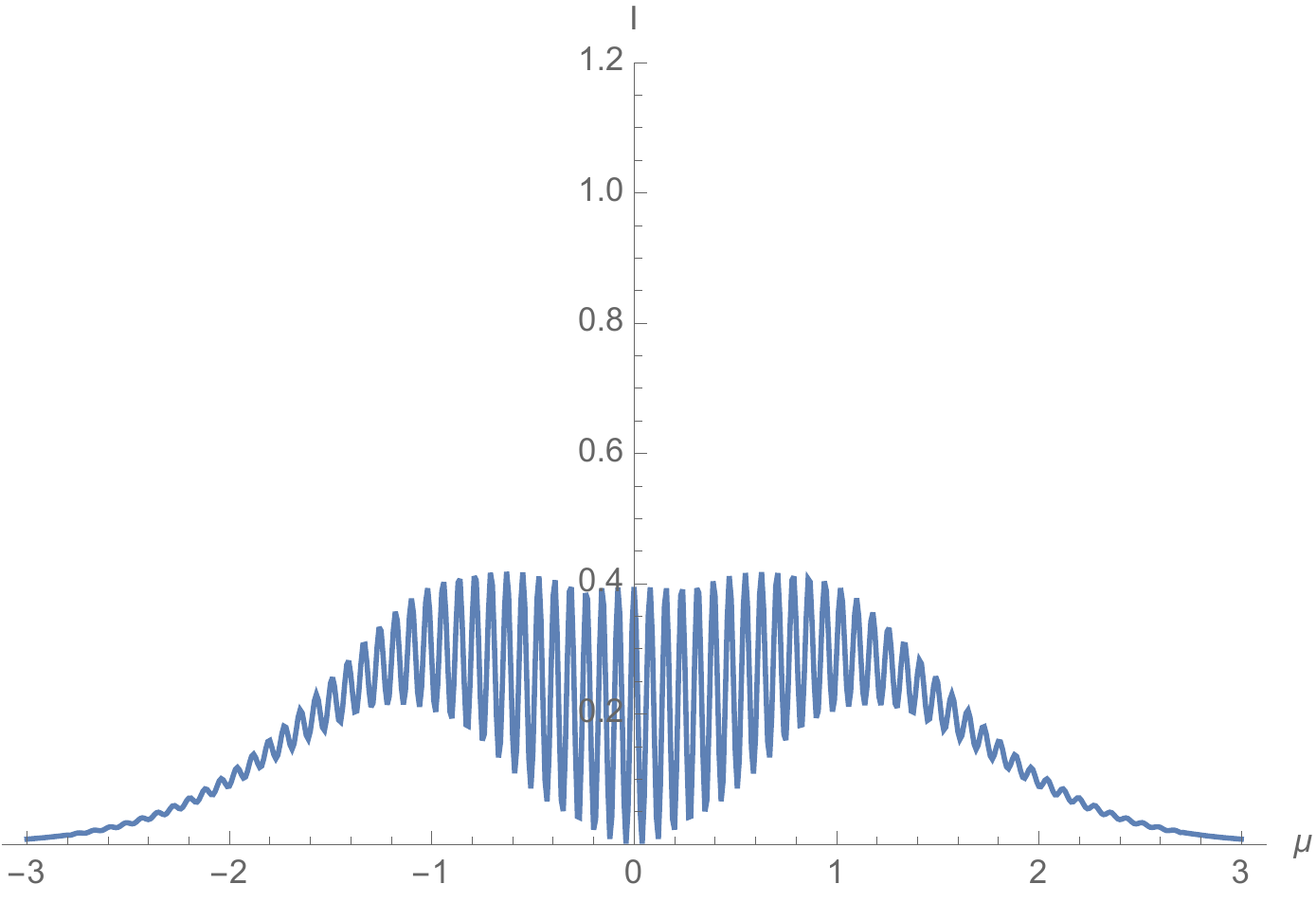}
\caption{$\hbar=1/20$}
\end{subfigure}
\begin{subfigure}[b]{0.49\textwidth}
\includegraphics[width=\textwidth]{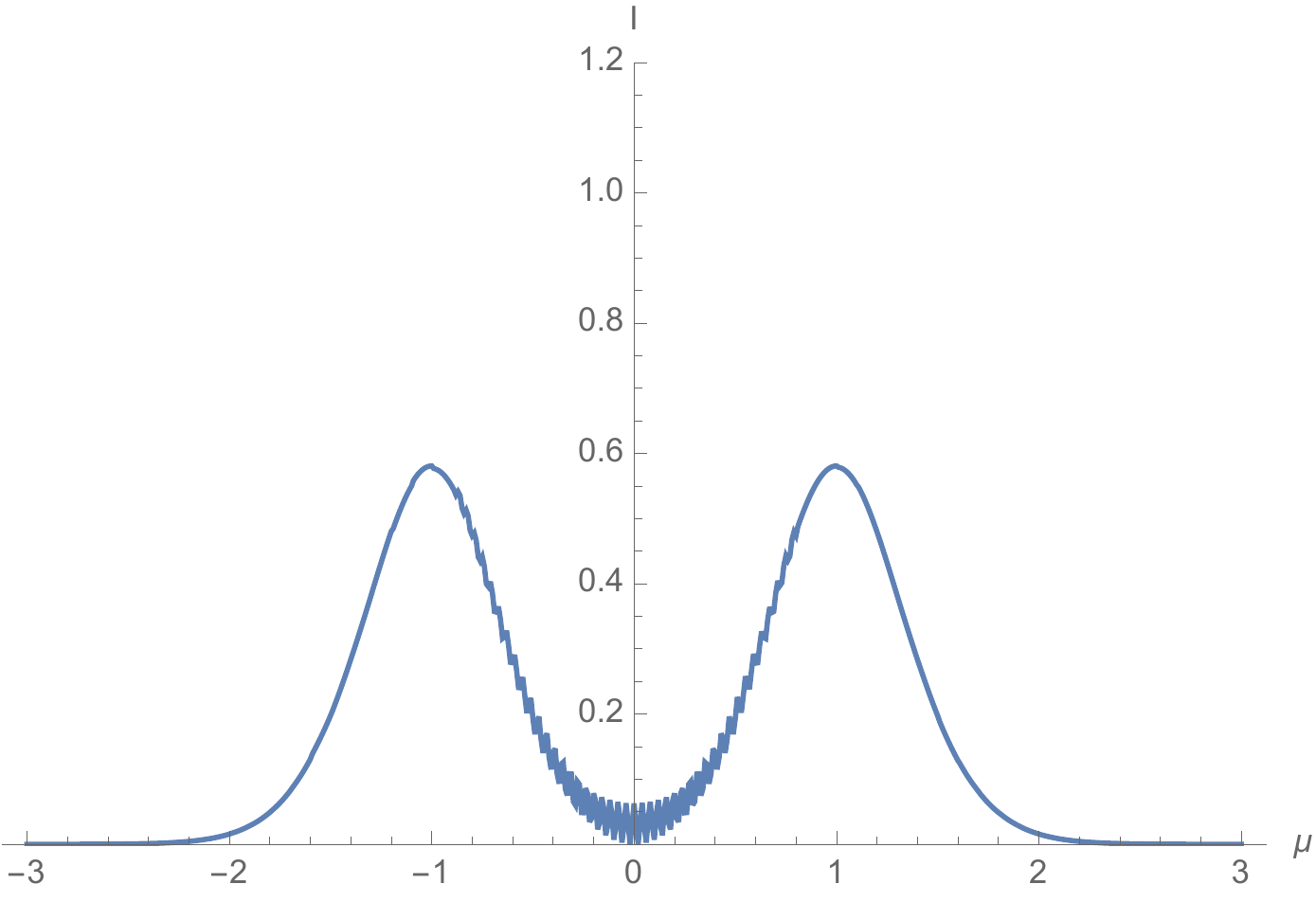}
\caption{$\hbar=1/40$}
\end{subfigure} \\
\begin{subfigure}[b]{0.49\textwidth}
\includegraphics[width=\textwidth]{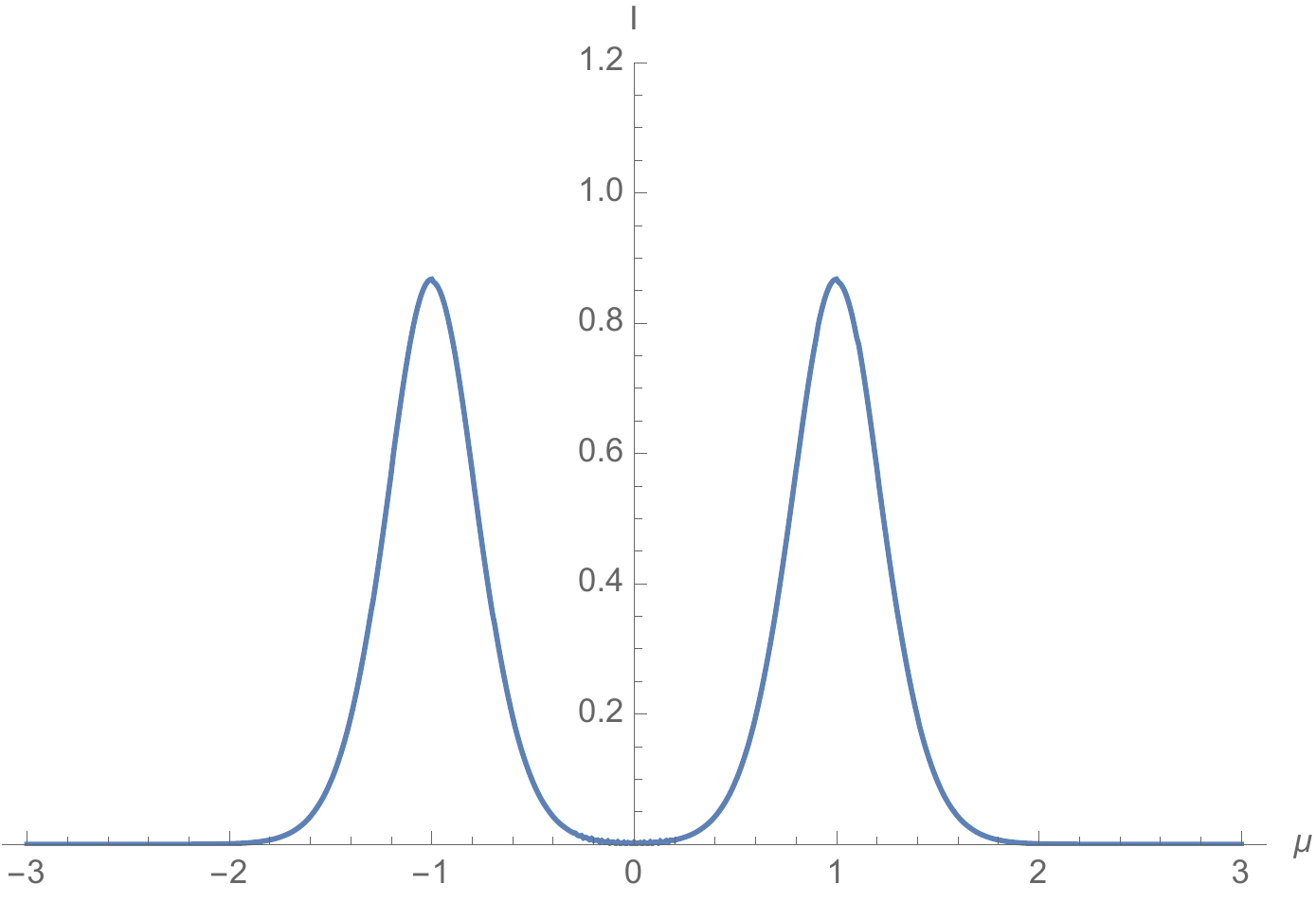}
\caption{$\hbar=1/60$}
\end{subfigure}
\begin{subfigure}[b]{0.49\textwidth}
\includegraphics[width=\textwidth]{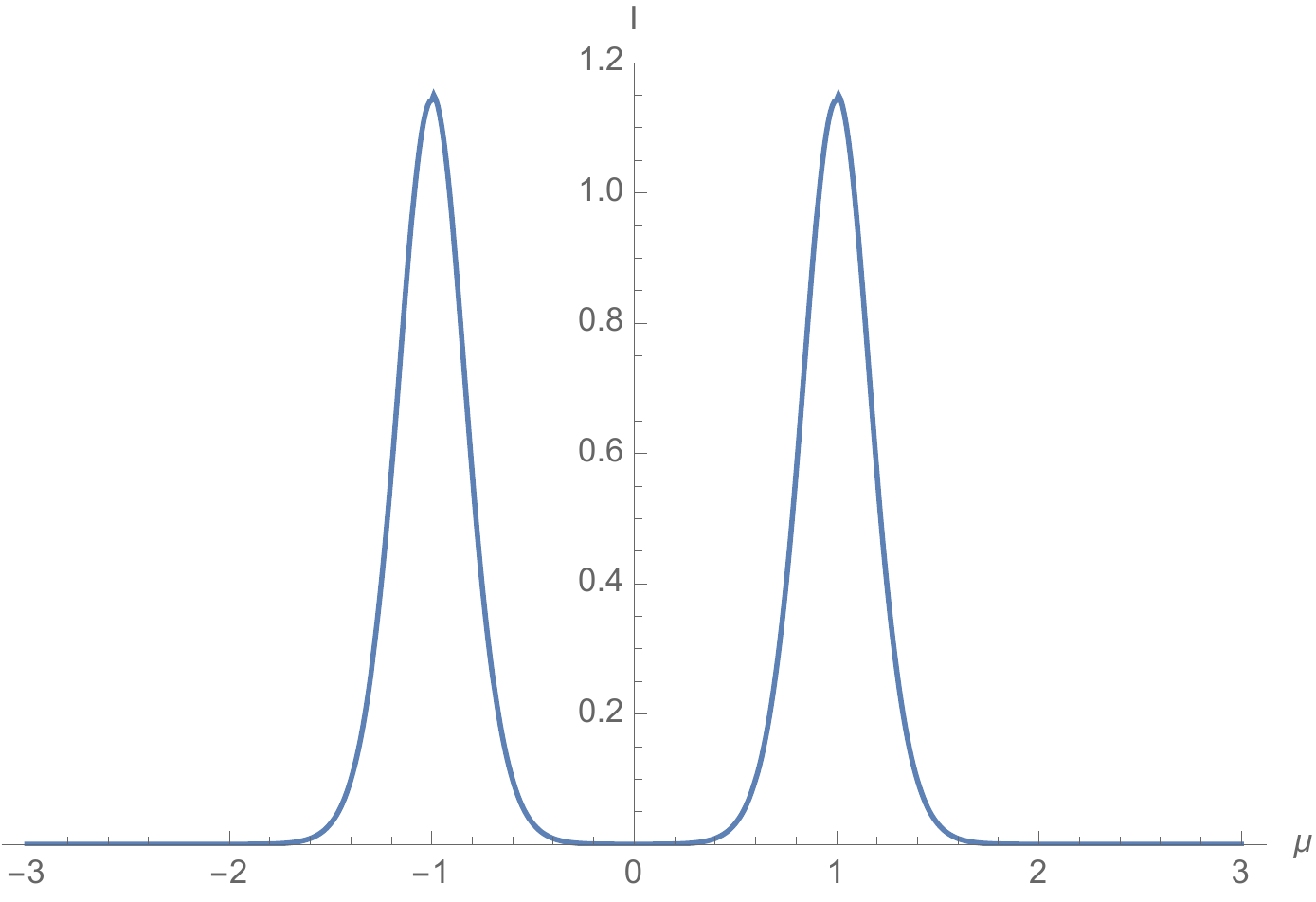}
\caption{$\hbar=1/80$}
\end{subfigure}
\caption{The intensity $I$ as a function of position $\mu$ for various $\hbar$}\label{fig:YoungInterference}
\end{figure}

Given the thimble, we can efficiently evaluate the oscillatory integral for various $\hbar$ (see figure \ref{fig:YoungInterference}). For relatively large $\hbar$, the intensity on the screen is dominated by interference effects. For both $\hbar=1$ and $\hbar=1/2$ we do not observe the classical intensity peaks corresponding to the two slits. In the semi-classical limit, $\hbar \to 0$, we the interference pattern is slowly replaced by the classical peaks. Note that this transition from the quantum to the classical regime cannot be studied in the traditional thin slit approximation.

Observe that, while the behavior of strong lenses is dominated by caustics where the saddle points become degenerate, the qualitative behavior of the double-slit experiment is completely determined by the Stokes transitions. The saddle points are everywhere non-degenerate and the $h$-function is a Morse function. In both instances, the saddle point approximation fails and the integral should be evaluated along the complete Lefschetz thimble. We expect this to be a generic feature in quantum mechanical interference phenomena.

\end{document}